%% file: PhD_Thesis_DeLorenzo.tex
\documentclass [a4paper,12pt]{book}
\usepackage[top=3cm,bottom=3cm,left=3cm,right=3cm,heightrounded,bindingoffset=5mm]{geometry}

\usepackage[utf8]{inputenc}
\usepackage[babel]{csquotes}
\usepackage[T1]{fontenc} 
\usepackage[french,english]{babel}
\usepackage[normalem]{ulem} 
\usepackage{mathptmx}

\usepackage{amsfonts,amssymb,amsmath,amsthm,amstext}
\usepackage{mathrsfs}					
\usepackage{enumitem}				
\usepackage{graphics,graphicx}						
\usepackage[table]{xcolor}					
\usepackage{subfigure}
\usepackage{float}							
\usepackage{multirow}
\usepackage{mathtools}
\usepackage {booktabs}
\usepackage {microtype}
\usepackage{tabularx}
\usepackage[innercaption]{sidecap}
\usepackage{braket}
\usepackage{cases}
\usepackage{pdfpages}
\usepackage{emptypage}
\usepackage{cite}

\usepackage[margin=10pt, font=footnotesize, labelfont=bf, labelsep=period]{caption}	[2004/07/16]
\usepackage{titlesec,titletoc}
\usepackage{fancyhdr}							
\usepackage[Lenny]{fncychap}

\usepackage[colorlinks,citecolor=blue!80!black,linkcolor=black,urlcolor=blue]{hyperref}

\titleformat{\section}[hang]{\fontsize{17}{20}\scshape}{\thesection}{15pt}{}[\vspace{-.4ex}\hrule]
\titleformat{\subsection}[hang]{\fontsize{14}{17}\scshape\bfseries}{\thesubsection}{15pt}{}
\titleformat{\subsubsection}[hang]{\fontsize{12}{15}\scshape\bfseries}{\thesubsection}{15pt}{}
\fancypagestyle{plain}{%
\fancyhf{} 						
\fancyhead[C]{} 	
 
}

\titleclass{\part}{top}
\titleformat{\part}
  {\centering\normalfont\Huge\bfseries}{}{0pt}{}
   \ChNameVar{\scshape\fontsize{16}{18}\selectfont} 
   \ChNumVar{\fontsize{76}{78}\usefont{OT1}{ptm}{m}{n}\selectfont} 
   \ChTitleVar{\fontsize{30}{32} \scshape}\fontfamily{ppl}  \selectfont
   \ChRuleWidth{1pt}

\def\be{\begin{equation}}
\def\ee{\end{equation}}

\newcommand{\M}{{\mathbb M}}
\newcommand{\R}{{\mathbb R}}
\newcommand{\N}{{\mathbb N}}
\newcommand{\va}{\scriptscriptstyle}
\newcommand{\heq}{\,\hat=\,}
\newcommand{\D}{\Delta}
\DeclareMathAlphabet{\mathfs}{U}{rsfs}{m}{n} 
\newcommand{\mfs}[1]{\mathfs {#1}}         
\newcommand{\sg}{{\kappa_{\va SG}}}
\newcommand{\scrai}{{\mathscr I}}
\def\scri{\mathscr{I}}
\newcommand{\sL}{{\mfs L}}
\newcommand{\sI}{{\mfs I}}
\newcommand{\ba}{\nopagebreak[3]\begin{eqnarray}}
\newcommand{\ea}{\end{eqnarray}}
\newcommand{\rH}{r_{\va H}}
\newcommand{\rO}{r_{\va O}}
\newcommand{\Mi}{{\va M}}

\newcommand{\xBH}{x_{\va BH}}
\newcommand{\xo}{x_{\va 0}}

\newcommand{\ddd}{d}

\newcommand{\Sch}{Schwarzschild }

\newcommand{\sH}{{\mfs H}}
\newcommand{\sR}{{\mfs R}}
\DeclarePairedDelimiter{\avrg}{\langle}{\rangle}

\DeclarePairedDelimiter{\abs}{\lvert}{\rvert}

\newcommand{\lpl}{\ell_{\va P}}
\newcommand{\mpl}{m_{\va P}}

\newcommand{\cL}{{\mathcal L}}
\newcommand{\nn}{\nonumber}
\newcommand{\w}{\wedge}
\newcommand{\f}{\frac}
\newcommand{\tl}{\tilde}
\def\p{\partial}
\newcommand{\na}{\nabla}

\def\a{\alpha}
\def\b{\beta}
\def\g{\gamma}
\def\d{\delta}

\def\eps{\epsilon}

\def\th{\theta}

\def\k{\kappa}
\def\l{\lambda}
\def\m{\mu}
\def\n{\nu}

\def\r{\rho}

\def\s{\sigma}
\def\t{\tau}

\def\om{\omega}
\def\G{\Gamma}
\def\D{\Delta}

\def\L{\Lambda}

\newcommand{\ut}[1]{ \underset{\widetilde{}}{#1}{} }
\newcommand{\oge}[1]{\overset{\scriptscriptstyle e}{#1}{}}
\newcommand{\Ref}[1]{(\ref{#1})}

\begin{document}

\frontmatter
\input{Frontispiece/Frontispiece}

\phantomsection
\chapter*{}
\null\vspace{\stretch{1}}
\begin{flushright}
{\Huge $\star$}
\end{flushright}
\vspace{\stretch{8}}\null

\input{Resume/Resume}

\renewcommand\contentsname{Table of Contents}
\vspace{-2ex}
\tableofcontents

\mainmatter
\input{Introduction/Introduction}


\part*{}
\phantomsection
\addcontentsline{toc}{part}{The Information Paradox}
\input{Part_1/Part_1}

\graphicspath{{Part_1/Volume/}}
\input{Part_1/Volume/Volume}

\graphicspath{{Part_1/Fireworks/}}
\input{Part_1/Fireworks/Fireworks}
%

\part*{}
\phantomsection
\addcontentsline{toc}{part}{Spacetime and Thermodynamics}
\input{Part_2/Part_2}

\graphicspath{{Part_2/Lightcone_Thermodynamics/}}
\input{Part_2/Lightcone_Thermodynamics/Light_cone_Thermodynamics}

\graphicspath{{Part_2/LightBH/}}
\include{Part_2/LightBH/LightBH}

\graphicspath{{Part_2/DesmoTorsion/}}
\include{Part_2/DesmoTorsion/DesmoTorsion}


\backmatter

\phantomsection
\input{Appendices/appPart1}

\phantomsection
\graphicspath{{Part_2/Lightcone_Thermodynamics/}}
\input{Appendices/appPart2}

\renewcommand{\bibname}{References}
\phantomsection\addcontentsline{toc}{chapter}{\bibname}
\bibliography{references}
\bibliographystyle{apalike}

\end{document}

%% file: Frontispiece/Frontispiece.tex
\begin{titlepage}
	\chead{}
	\pdfbookmark[0]{Frontispiece/Page de titre}{titre}
	\input{Frontispiece/titre.tex}
	\newpage
	\thispagestyle{empty}
	\input{Frontispiece/licence}
\end{titlepage}

%% file: Frontispiece/titre.tex
{\newgeometry{top=2.3cm,bottom=2cm,left=2cm,right=2cm}
\begin{center}
	\begin{minipage}[c]{0.70\linewidth}
		\raggedright\includegraphics[height=2cm]{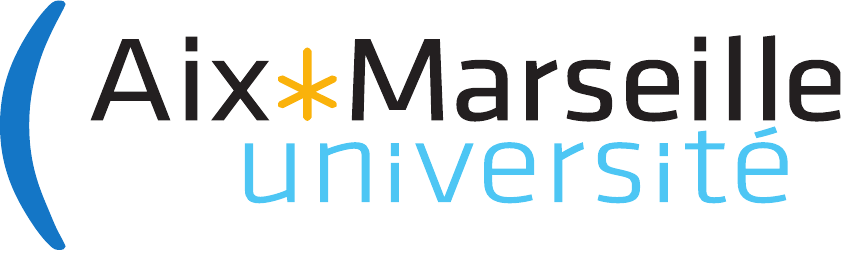}\end{minipage}\hfill
	\begin{minipage}[c]{0.30\linewidth}
		\raggedleft\includegraphics[height=2cm]{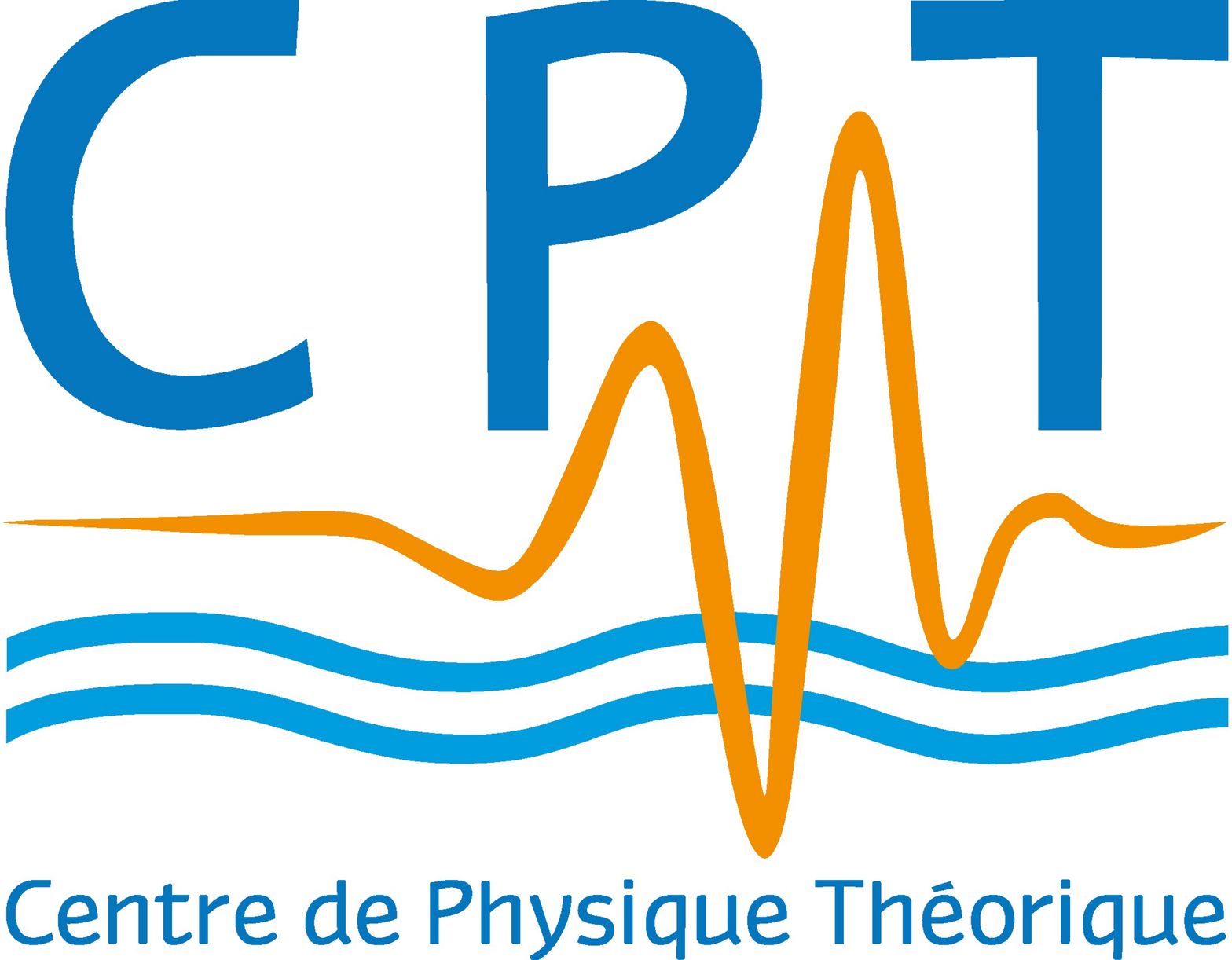}
	\end{minipage}\hfill
	\vspace{1cm}
	\LARGE {\scshape Aix-Marseille Universit\'e}\\
	\vspace{0.2cm}
	\Large{ {\scshape ED352}\\
	\vspace{0.2cm}
	\vspace{0.2cm}
	{\scshape Centre de Physique Th\'eorique	 }}\\ \normalsize{{\scshape\'Equipe de Gravit\'e Quantique}}

		\vspace{1cm}
		\large Th\`ese pr\'esent\'ee pour obtenir le grade universitaire de docteur\\ en \\ Physique et Sciences de la Mati\`ere\\
    {\em Sp\'ecialit\'e}: Physique Th\'eorique et Math\'ematique
    
        \vspace{1.1cm}
        \Large {\bf \scshape Tommaso DE LORENZO}\\
        \vspace{1.1cm}

\hrule
\vspace{2ex}

        {\LARGE \scshape Black Holes as a Gateway to the Quantum}\\
        {\Large \scshape Classical and Semi-Classical Explorations}
        \vspace{-.6ex}\\
	\rule{0.3\textwidth}{.4pt}\\
	\vspace{0.2cm}
	{\large \scshape Les Trous Noirs comme Porte d'Entr\'ee vers le Quantique\\}
	{\normalsize\scshape Explorations Classique et Semi-Classique}\\
	 \end{center}

\hrule

   	\vfill 
\begin{flushleft}

	\normalsize Soutenue le 18/09/2018 devant le jury compos\'e de:\\

\vspace{0.4cm}
\begin{tabular}{lll}
Aur\'elien BARRAU & Universit\'e Grenoble-Alpes (IUF)& Rapporteur
	\vspace{0.08cm}\\
Daniel SUDARSKY & Universidad Nacional Aut\'onoma de M\'exico &\\
&(ICN) & Rapporteur
   	 \vspace{0.08cm}  \\
Francesca VIDOTTO & Universidad del Pa\'is Vasco&\\
&(EHU) & Examinateur 
   	\vspace{0.08cm} \\
Jose-Luis JARAMILLO & Universit\'e de Bourgogne (IMB) & Examinateur  
	\vspace{0.08cm} \\
Carlo ROVELLI & Aix-Marseille Universit\'e (CPT) & Examinateur
   	\vspace{0.08cm} \\
Simone SPEZIALE & Aix-Marseille Universit\'e (CPT) & Co-Directeur de th\`ese
    	\vspace{0.08cm} \\
Alejandro PEREZ & Aix-Marseille Universit\'e (CPT) & Directeur de th\`ese
\end{tabular}\\
	\vspace{0.5cm}
Num\'ero national de th\`ese / suffixe local:  2018AIXM0264 / 023ED352
\end{flushleft}

%% file: Frontispiece/licence.tex
~\vfill
\begin{center}
	\begin{minipage}[c]{0.25\linewidth}
		\raggedright\includegraphics[height=35px]{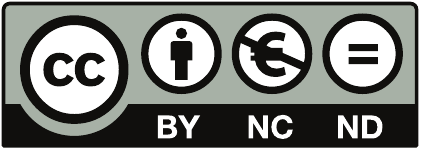}
	\end{minipage}\hfill
\end{center}

\noindent Cette oeuvre est mise \'a disposition selon les termes de la \href{https://creativecommons.org/licenses/by-nc-nd/4.0/deed.fr}{Licence Creative Commons Attribution - Pas d'Utilisation Commerciale - Pas de Modification 4.0 International}. 
}

%% file: Resume/Resume.tex
\phantomsection\addcontentsline{toc}{chapter}{R\'esum\'e}
\markboth{R\'esum\'e}{}
\chapter*{R\'esum\'e}
\vspace{-2ex}

Que se passe-t-il à l'intérieur d'un trou noir? Que devient ultimement la singularité? Est-elle résolue par les effets gravitationnels? \vspace{.5ex}

L'évaporation de Hawking est-elle le seul phénomène dictant l'évolution future des trous noirs? L'information quantique est-elle conservée au cours de ce processus? \vspace{.5ex}

Qu'est-ce que la thermodynamique des trous noirs peut-elle nous apprendre au sujet de la nature quantique de l'espace-temps? Quels sont les degrés de liberté en jeu dans l'entropie de Bekenstein-Hawking?\vspace{.5ex}

Toutes ces questions, parmi d'autres, interrogent la structure fondamentale de l'espace-temps et des champs quantiques. Elles ont émergé de l'étude des propriétés des trous noirs. Le débat à leur sujet est ouvert, intense et productif, nourri de contributions venues de nombreux physiciens théoriciens et de philosophes. Depuis leur découverte en 1916 (quelques mois après la publication de la théorie de la Relativité Générale par Einstein) les trous noirs ont continuellement interrogé notre connaissance, soulevant de fascinants problèmes à la fois techniques et conceptuels. Beaucoup d'entre eux ont déjà été résolus, mais d'autres, d'où sont issus les questions introductives, attendent encore une réponse. Et pour cause, face à nous se trouve un régime particulier de la physique, inaccessible aux autres systèmes physiques connus, là où la structure de l'espace-temps retrouve la théorie quantique. {\em La physique des trous noirs est une porte ouverte sur la nature quantique de la gravité}. Plusieurs idées nouvelles intéressantes, dont certaines ont été appliquées avec succès à d'autres domaines de la physique, ont émergé de ce débat: l'entropie d'entanglement \cite{Solodukhin:2011gn}, l'holographie \cite{Bigatti:1999dp}, ou encore la thermodynamique de l'espace-temps \cite{padmanabhan2003gravity}, etc. Cependant, les réponses complètes aux questions précédentes demeurent absentes.

Mon travail de thèse a été entièrement consacré à ce domaine central de la physique théorique. L'objectif poursuivi était de parvenir à la compréhension la plus compète et générale possible des problématiques du débat concernant les trous noirs en gravité quantique. Cet objectif a conduit à la production de résultats nouveaux qui constituent l'essentiel de ce manuscrit. J'ai principalement travaillé en utilisant la Relativité Générale classique et la Théorie Quantique des Champs (TQC) en Espace-Temps Courbe, tout en gardant constamment un œil ouvert sur les avancées des théories de gravité quantique. Je crois en effet qu'une telle interaction entre différents niveaux d'approximation peut être important pour parvenir à franchir certaines étapes significatives, et répondre ainsi aux questions initiales, conduisant à une meilleure compréhension de la physique fondamentale.

Comme le lecteur pourrait le remarquer, les questions posées en ce début de résumé sont divisées en trois catégories distinctes. Le premier groupe contient des questions traitant de la nature des singularités au centre des trous noirs. Ceci fut le sujet principale de ma thèse de master ainsi que d'autres travaux s'en rapprochant \cite{de2014investigating,DeLorenzo:2014pta,DeLorenzo:2015taa}, et ne fait donc pas parti de la présente dissertation. Ce que la dissertation contient est un ensemble de projets \cite{DeLorenzo:2015gtx,Christodoulou:2016tuu,DeLorenzo:2017tgx,LightBH,desmoTorsion}, divisés en deux parties abordant les deux autres groupes de questions.
\vspace{1ex} 

\noindent {\bf Partie I : Le paradoxe de l'information}\\
Le célèbre \emph{paradoxe de l'information} apparaît avec le résultat inattendu de Stephen Hawking selon lequel les trous noirs s'évaporent (i.e. perdent de la masse) via l'émission de radiation thermique \cite{hawking1974black}. 
Imaginez un trou noir qui se forme par l'effondrement de matière se trouvant initialement dans un état quantique pur. Si le trou noir disparaît complètement, ce qui reste est un état thermique de particules de Hawking \cite{hawking:1976breakdown}. Un état initial pur a évolué vers un état final mixte qui ne contient pas assez d'information pour reconstruire l'état initial. De l'information a donc été perdue, en contradiction avec l'évolution unitaire de la mécanique quantique. Une description plus détaillée du paradoxe, ainsi qu'une présentation de ses solutions possibles, se trouve dans \hyperref[sec:intro1]{l'introduction de la partie I}.

Parmi ses solutions, une plutôt ``naturelle'' est connue sous le nom de \emph{remnant scénario} \cite{Aharonov:1987tp}. Pour comprendre un tel scénario, il est important de remarquer que le calcul de Hawking repose sur la TQC en espace-temps courbe. Cette dernière est considérée comme valide dans un régime où la courbure est loin d'être planckienne. Cette approximation est parfaitement valide près de l'horizon du trou noir, là où le calcul est effectué. Cependant, quand le trou noir s'est évaporé jusqu'à atteindre une masse planckienne, on s'attend à ce que des effets quantiques gravitationnels importants surviennent. L'idée du remnant considère la possibilité que de tels effets préviennent l'évaporation quand l'horizon atteint le régime planckien. Pour un trou noir de masse $m$, ceci se produit en un temps de l'ordre de $m^3$. L'information tombée dans le trou noir au cours de cette période serait ``simplement'' contenue à l'intérieur du ``remnant planckien'', prêt à être soit lue, soit relâchée, soit conservée éternellement. L'objection principale à cela est que l'objet final planckien serait grosso-modo trop petit pour contenir une telle quantité d'information nécessaire pour purifier l'état extérieur. On peut formuler cela plus précisément \cite{hossenfelder2010conservative}, mais chaque fois cela dépend fortement de l'idée que l'objet final est petit.
Cependant, ``petit'' en relativité générale n'est pas un adjectif très précis. En effet, par exemple, l'intérieur d'un trou noir éternel (Kruskal) contient des hypersurfaces de genre espace de volume infini, comme les surfaces données par l'équation $r=constante$. Ainsi, si on considère un trou noir éternel d'aire planckienne, devrait-on le considérer petit parce-que son aire extérieure est petite, ou au contraire immense car son volume interne est infini? Les trous noirs physiques, formés par l'effondrement de matière n'ont pas de volume infini. Néanmoins, l'hypersurface de volume maximal peut être identifiée \cite{Christodoulou:2014yia}. Puisque l'intérieur d'un trou noir n'est pas stationnaire par rapport au temps de Killing extérieur $t$, le volume maximal disponible à l'intérieur de l'horizon (qui a une aire constante) croît avec le temps comme $V \sim m^2\, t$ en unité planckienne. 

Au chapitre~\ref{chap:volumeBH} il est montré que le résultat décrit ci-dessus se généralise aux espace-temps décrivant un trou noir qui s'évapore. Jusqu'au étapes très tardives de son évaporation, la croissance du volume suit la même loi que le cas statique $V \sim m^2\, t$, où la masse $m$ est désormais la masse initiale \cite{Christodoulou:2016tuu}. Ceci implique que si on considère par exemple un trou noir de masse solaire qui s'évapore jusqu'à, disons, 10 fois la masse de Planck, il aura alors une aire extérieure de l'ordre de 100 fois l'aire de Planck (environ $10^{-70} m^2$), masquant un volume proportionnel à $m^5$, ce qui est environ $10^5$ fois le volume de l'univers observable ! À cela s'ajoute que la surface définissant le volume est de symétrie sphérique, commençant au rayon de l'horizon planckien et s'étendant en rayon suivant $r(t) \sim 3/2 m(t)$. Tout ceci montre que les trous noirs s'évaporant sont naturellement muni d'une géométrie interne du type ``sac d'or'' \cite{wheeler1964relativity} (ou ``corne d'abondance'' \cite{Banks:1994ph}), et ceci joue un rôle clef dans les réponses qualitatives aux objections standards contre le ``remnant scénario'' \cite{hossenfelder2010conservative}. 

Comme dit précédemment, le résultat de Hawking est considéré être valide jusqu'au régime de Planck. Jusque là, aucun autre effet dérivé de la physique connue n'est attendu, ce qui aurait pu venir modifier la lente décroissance de la masse du trou noir. Néanmoins, des effets ``exotiques'' qui pourrait autoriser une libération de l'information ont été proposés dans la littérature. L'un des plus récent (bien que reposant sur d'anciens articles \cite{Hajicek:2001yd}) a été développé dans \cite{Haggard:2014rza}. L'idée est la suivante. La Relativité Générale prédit que l'effondrement d'un étoile au-delà de son propre horizon entraînerait nécessairement l'effondrement complet jusqu'à la formation d'un singularité. Cependant, la gravité quantique prévoit aussi que l'effondrement s'arrête lorsque la densité et la courbure atteignent des valeurs planckiennes. Cette vieille idée est à la base, par exemple, des solutions de trous noirs sans singularité, mentionnées un peu plus haut. Le scénario de \cite{Haggard:2014rza} suggère que l'étoile ne cesserait pas seulement de s'effondrer, mais rebondirait sur elle-même vers l'extérieur, détruisant ainsi le trou noir. Le processus surviendrait en un temps de l'ordre de $m^2$, bien plus petit que l'évaporation de Hawking de l'ordre de $m^3$. Sur une telle échelle de temps, les effets de Hawking deviennent négligeables et le paradoxe de l'information n'est même plus formulable. Un trou noir, cependant, est par définition une région de l'espace-temps d'où rien ne peut sortir. Alors comment se fait-il que l'étoile rebondisse hors de son propre horizon? Les auteurs suggèrent que de faibles effets quantiques gravitationnels à l'extérieur de l'horizon pourrait s'accumuler dans le temps, et devenir finalement suffisamment importants pour permettre la transition quantique d'une géométrie de trou noir vers celle de trou blanc. L'espace-temps total introduit dans \cite{Haggard:2014rza} ne présente pas d'horizon, ce qui montre qu'une surface de piégeage initiale suivie d'une surface d'anti-piégeage permet à l'étoile rebondissante d'émerger. Les équations classiques d'Einstein sont satisfaites partout exceptées dans une région quantique ``tampon'' qui connecte les deux phases classiques. Immédiatement après, une série d'articles ont affirmé qu'un tel scénario pourrait générer un signal observable d'ondes gamma. \cite{Barrau:2016fcg,Barrau:2014yka,Barrau:2015uca,Vidotto:2016jqx}.

Au chapitre~\ref{chap:Fireworks}, une analyse des instabilités d'un tel modèle est présentée. Il est montré que la proposition originale a besoin d'être revue, car fortement instable \cite{DeLorenzo:2015gtx} . En effet, un calcul explicite montre par exemple que l'état du vide d'un champ scalaire évoluant dans l'espace-temps proposé développe une singularité du tenseur d'énergie-impulsion. Un argument similaire s'applique si une perturbation classique (une balle de ping-pong) est lâchée depuis l'infini en direction du trou rebondissant. À première vue, ces instabilités disparaissent seulement si l'échelle du temps du processus est de l'ordre de $m$. Ces instabilités, d'autre part, peuvent être supprimées par une modification minimale du modèle, sans modifier l'échelle de temps du processus. Ce nouveau modèle est une version asymétrique par rapport au temps du modèle original, avec une échelle de temps pour la phase finale du trou blanc qui est plus courte que $m\, \log m$, tandis que le processus complet a une échelle de temps arbitraire. Néanmoins, il faut souligner que la nécessité d'une modification asymétrique du modèle semble soulever des problèmes importants qui ne peuvent pas être résolus en détail sans une théorie complète de gravité quantique. Une discussion à ce sujet se trouve à la fin du chapitre~\ref{chap:Fireworks}.
\vspace{1ex}

\noindent {\bf Partie II: Thermodynamique de l'espace-temps}\\ 
Le résultat déjà mentionné que les trous noirs émettent une radiation thermique est apparu comme une belle surprise pour la communauté des physiciens théoriciens. En effet, dans un précédent article \cite{Bardeen1973}, Hawking lui-même, avec Bardeen et Carter, avaient prouvé que la mécanique des trous noirs vérifie une intrigante analogie mathématique avec les quatre lois de la thermodynamique. L'aire $A$ et la gravité de surface $\kappa$ de l'horizon jouent les rôles, respectivement, de l'entropie et de la température.
\begin{displayquote}
[...] \emph{It should however be emphasized that $\kappa$ and $A$ are distinct from the temperature and entropy of the black hole.
In fact the effective temperature of a black hole is absolute zero.
One way of seeing this is to note that a black hole cannot be in equilibrium
with black body radiation at any non-zero temperature, \emph{because no
radiation could be emitted from the hole} whereas some radiation would
always cross the horizon into the black hole.} \cite{Bardeen1973}
\end{displayquote}
Non seulement la température de Hawking n'est pas nulle, mais elle est proportionnelle à $\kappa$, ce qui invalide complètement la citation précédente. Un trou noir \emph{est} un objet thermodynamique, se comportant comme un corps noir avec une température $T=\kappa/(2\pi)$ et une entropie $S=A/4$! 

L'entropie est un outil très utile de la thermodynamique qui mesure notre ignorance de la structure microscopique d'un système. L'apparition d'une notion naturelle d'entropie en mécanique des trous noirs a donc été immédiatement interprétée comme la manifestation de notre ignorance concernant la description quantique de l'espace-temps. En d'autres termes, la thermodynamique des trous noirs s'interprète naturellement comme une source importante d'information à propos de la théorie de la gravité quantique dont le traitement semi-classique devrait être la limite. Après 40 ans cependant, il n'y a toujours pas de consensus scientifique concernant la nature de l'entropie. Ainsi, l'étude d'exemples permettant de saisir les caractéristiques principales du problème pourrait aider à comprendre le processus.

Cette motivation nous amène au chapitre~\ref{chap:lightcone}, où il est montré que les cônes de lumières dans un espace de Minkoswki peuvent être vus comme un bel analogue aux horizons des trous noirs \cite{DeLorenzo:2017tgx}. En effet, l'intersection des cônes de lumière des espaces de Minkowski sont des horizons bifurquants de Killing conformes par rapport à des observateurs stationnaires conformes, ce qui déifinit le plus general parmi le Champ des vecteurs radials de Killing Conformes d'un espace de Minkoswki (CKCM). En utilisant la stationnarité conforme, une généralisation invariante conforme des quatre lois de la thermodynamique des trous noirs est démontrée. On définie alors une température de cône de lumière (conforme) constante, donnée par l'expression standard en terme de la gravité de surface (invariant conforme). Des échanges d'énergie (invariante conforme) à travers l'horizon conforme sont décrits, en théorie des perturbations, par une première loi où l'entropie varie selon le quart des variations de la notion invariante conforme d'aire de l'horizon. Cette analogie intéressante entre les propriétés des CKCM et la thermodynamique des trous noirs met en évidence les caractéristiques mathématiques basiques de cette dernière pour un espace-temps muni d'un champ gravitationnel trivial. Cependant, il faut nuancer le propos en précisant que les diverses notions conformes qui apparaissent dans les lois n'ont pas de signification physique claire. Néanmoins, les limitations précédentes peuvent être dépassées au moyen d'une transformation conforme envoyant $(\R^4, \eta_{ab})$ vers un espace-temps modèle $(M,\tilde g_{ab})$ avec $\tilde g_{ab}=\omega^2 \eta_{ab}$, de telle sorte que $\xi^a$ devient un véritable champ de Killing et les horizons conformes deviennent des horizons de Killing de l'espace-temps cible. Les quatre lois demeurent vraies dans l'espace-temps cible, pour de mêmes valeurs numériques, mais désormais les différentes quantités acquièrent la signification physique et géométrique habituelle qu'elles ont dans le contexte des trous noirs.

Au chapitre~\ref{chap:lightBH}, les caractéristiques génériques globales des espace-temps obtenus par la procédure précédente sont étudiées. Dans quel cas ces espace-temps représentent-ils des trous noirs? Que sont-ils sinon? Il est clair qu'il y a un nombre infini de possibilités. Néanmoins, on montre que les caractéristiques génériques globales peuvent être exhibées dans un petit nombre de cas. Le cas le plus simple correspond à $\omega=\alpha/r^2$ qui reproduit la solution de Bertotti-Robinson de la théorie d'Einstein-Maxwell \cite{Bertotti:1959pf}. Sa géométrie est connue pour encoder la géométrie poche de l'horizon des trous noirs extrêmes ou quasi-extrêmes de Reissner-Nordstrom. Un autre exemple caractéristique est la réalisation de {\em de Sitter} où les horizons de bifurcation correspondent aux horizons cosmologiques  d'intersection (il n'y a pas de trou noir dans ce cas). Les configurations de trou noir faiblement asymptotiquement {\em Anti-de Sitter} sont aussi présentés, ainsi que quelques espace-temps plus exotiques encore, avec des horizons de Killing mais sans trou noir. Ces résultats renforcent et clarifient les conclusions du chapitre précédent, et ouvrent un fenêtre sur de possibles applications de cette affinité thermodynamique entre les trous noirs et l'espace-temps de Minkowski.

Le chapitre final, chapitre~\ref{chap:desmo}, contient de récents travaux \cite{desmoTorsion} liés à l'idée de Jacobson selon laquelle la structure continue de l'espace-temps pourrait émerger comme la description de l'équilibre thermodynamique de degrés de liberté quantiques plus fondamentaux \cite{PhysRevLett.75.1260}. En partant de la platitude locale, en tout point de l'espace-temps, on peut considérer un champ local de Killing Rindler $\chi$ avec son horizon associé $\mathcal{H}$, et une petite perturbation produite par un petit flux d'énergie-impulsion $\delta T_{ab}$. Les équations d'Einstein sont déduites en imposant la relation de l'équilibre thermodynamique de Clausius $\delta Q = T \delta S$ pour l'horizon, en supposant que (i) l'entropie de l'horizon est proportionnelle à son aire (ii) une température de Unruh, et que (iii) $\delta Q=\int_\mathcal{H} T_{ab} \chi^a dS^b$. Dans cette perspective, les équations d'Einstein sont une équation d'état : c'est la thermodynamique des degrés de liberté sous-jacents  qui explique la dynamique de l'espace-temps. Partant de ces résultats de thermodynamique des trous noirs, Jacobson renverse la logique : la thermodynamique des trous noirs n'est plus la conséquence des équations d'Einstein, mais il s'agit d'une manifestation explicite d'une réalité plus fondamentale. La gravité d'Einstein apparaît comme une description émergente à l'équilibre thermodynamique de faible énergie d'une physique plus fondamentale encore inconnue. Cette idée a été appliquée à d'autres théorie de la gravité, au-delà de la Relativité Générale. Un exemple intéressant  de telles théories est celle d'Einstein-Cartan (EC), où le tenseur de torsion non-nul est considéré pour coupler la matière fermionique à la gravité  \cite{Shapiro:2001rz}. Les \'equations des champs se composent de deux \'equations tensorielles : l'une qui, en l'absence de torsion, se r\'eduit aux \'equations d'Einstein de RG, et l'autre pour le tenseur de torsion.
Une première tentative de généralisation de la dérivation de Jacobson à la théorie d'Einstein-Cartan a été obtenue récemment dans \cite{Dey:2017fld}. On prétendait qu'une restriction du tenseur de torsion et qu'un traitement hors-équilibre (comme dans \cite{PhysRevLett.96.121301,Chirco:2009dc}) était nécessaire pour atteindre ce but. Cependant une analyse plus attentive, présentée dans le chapitre final, montre qu'aucune de ces deux hypothèses n'est nécessaire, et que le raisonnement fonctionne pour la théorie EC aussi bien que dans le cas original de la GR. Il est à noter deux observations cruciales. La première est que la notion même de champ de Killing et d'horizon de Killing sont purement métriques, et donc sont insensibles à la présence de torsion. Ceci implique une grande simplification de la dérivation. En effet, une étape clef consister à calculer le changement d'aire de l'horizon $\delta A \propto \delta S$ en utilisant l'équation de Raychauduri pour la congruence des géodésiques générée par le champ de Killing sur l'horizon. Ceci étant donné, l'équation standard de Raychauduri pour les géodésiques de Levi-Civita peut donc être utilisée. La deuxième observation est que l'invariance sous difféomorphisme de la théorie EC identifie de manière unique, ``on-shell'' des \'equations pour la torsion, le Tenseur d'\'Energie-Impulsion (TEI) \emph{conservé} qui devrait être utilisé pour définir le flux d'énergie $\delta Q$. Ce tenseur n'est pas celui dérivé de la variation de l'action de la matière par rapport au tenseur métrique, mais il implique aussi des termes dépendants de la torsion. Et ces termes sont exactement ceux nécessaires pour dériver les équation dynamiques de EC, suivant la dérivation originale de Jacobson. Utiliser la composante de torsion des \'equations du champ EC pour identifier le TEI conserv\'e peut sembler inad\'equat avec l'id\'ee sous-jacente d'utiliser seulement des arguments thermodynamiques pour d\'eriver la dynamique gravitationnelle. Deux observations semblent indiquer que cette approximation est parfaitement raisonnable. Il ne semble pas y avoir d'obstacle conceptuel pour prouver la conservation du TEI \`a partir de l'invariance sous diff\'eomorphisme ``off-shell'' des \'equations du champ de torsion. L'obstacle semble \^etre seulement technique et pourrait \^etre r\'esolu. Deuxi\`emement, le traitement hors \'equilibre, n\'ecessaire si les \'equations du champ de torsion ne sont par utilis\'ees \cite{Dey:2017fld}, para\^it largement arbitraire. Je renvoie \`a la section~\ref{sec:arbitrio} pour plus de d\'etails. La philosophie qui semble donc plus raisonnable est qu'il doit exister un argument thermodynamique ind\'ependant, et encore inconnu, pour d\'eriver la composante de torsion des \'equations du champ EC. \`A partir de l\`a, l'autre est d\'erivable avec la description \`a l'\'equilibre pr\'esent\'e dans le chapitre~\ref{chap:desmo}.

\vspace{1ex}

%% file: Introduction/Introduction.tex
{\phantomsection\renewcommand\thechapter{\rm i}
\addcontentsline{toc}{chapter}{Introduction}
\markboth{Introduction}{}
\chapter*{Introduction}

What happens inside a black hole? What is the fate of the singularity? Is it resolved by quantum gravitational effects? \vspace{1ex}

Is Hawking's evaporation the only process driving black hole evolution? Is quantum information conserved in the evaporation process?\vspace{1ex}

Which insights about the quantum nature of spacetime does the thermodynamics of black holes provide? Which degrees of freedom account for the Bekenstein-Hawking's entropy?\vspace{1ex}

All these questions, among others, which interrogate the fundamental structure of spacetime and quantum fields, have arisen from the study of the properties of Black Holes (BHs). The debate around them is open, intense and productive, with contributions by many theoretical physicists and philosophers. Since their discovery back in 1916, black holes have continuously questioned our knowledge, raising both technically and conceptually fascinating problems. Many of them have been solved over the years. One representative example is the 20-year-long discussion on the (coordinate) singularity at the horizon of the \Sch solution \cite{goenner1998expanding}. Other problems, however, from which the opening questions come from, still require a solution. The main reason being that we are faced with regimes of physics inaccessible by any other system, where the not yet completely known interplay between quantum theory and spacetime is unveiled. {\em Black hole physics is a gateway to the quantum nature of gravity}. Several new interesting ideas, some of them successfully applied in other fields of physics, such as entanglement entropy \cite{Solodukhin:2011gn}, holography \cite{Bigatti:1999dp}, and spacetime thermodynamics \cite{padmanabhan2003gravity}, have arisen from this debate. However, complete answers to the above questions are still missing.

My thesis work has been completely devoted to this central domain of theoretical physics. The guiding aim has been to understand, in the most complete and widest possible manner, the problems involved in this debate surrounding those quantum gravity related questions concerning black holes. This process has produced original results that constitute the main core of this manuscript. I have mainly worked using classical General Relativity and Quantum Field Theory in Curved Spacetime, while constantly keeping an eye open to the achievements of the proposed theories of quantum gravity. I believe, indeed, that such an interplay between levels of approximation is constructive and can be important in making significant steps forward in answering the initial questions, eventually providing crucial developments in our understanding of fundamental physics. In the next Section of this Introduction, I will try to make clear what the different levels of approximation are and the assumptions one can work with, specifying which ones will be used in the main body of the thesis. The rest of the present Section is an outline of the manuscript with an overview of the results.

\section{Outline and Overview}\label{sec:overview}

As the reader may have noticed, the questions at the beginning of this Introduction are divided into three distinct but interconnected groups. The first group contains questions related to the nature of the singularities at the center of black holes. This has been the main topic of my master thesis as well as some related works \cite{de2014investigating,DeLorenzo:2014pta,DeLorenzo:2015taa}, and is therefore not part of the present dissertation. Some underlying ideas will nonetheless be introduced in next Section, since they play a role in the main discussion.

What this dissertation does contain, on the other hand, is a collection of projects carried out during the three years of my PhD \cite{DeLorenzo:2015gtx,Christodoulou:2016tuu,DeLorenzo:2017tgx,LightBH,desmoTorsion}. They are divided into two Parts addressing respectively the remaining two groups of the initial questions.
\vspace{1ex} 

\noindent {\bf Part I: The information paradox}\\
The so called \emph{information paradox} arises from Stephen Hawking's ground-breaking result that black holes evaporate (i.e. lose mass) via the emission of thermal radiation \cite{hawking1974black}. 
As a consequence, imagine a black hole that is formed by the collapse of matter initially in a pure state. If the black hole completely disappears, what remains is a thermal state of Hawking's particles \cite{hawking:1976breakdown}. An initial pure state has evolved into a final mixed state that does not contain enough information to reconstruct the initial state. Information has been lost, in contradiction with the unitary evolution of quantum mechanics. This is the usual way the problem is stated. A much more precise discussion which enlightens important details is provided in the \hyperref[sec:intro1]{introduction to Part I}.  

Among the proposed solutions to the paradox, a rather ``natural'' one is known under the name of \emph{remnant scenario} \cite{Aharonov:1987tp}. To understand such a scenario, it is important to notice that Hawking's computation relies on the framework of Quantum Field Theory in Curved Spacetime. The latter is believed to be valid in regimes where the curvature is far from being Planckian (see next Section for more details). This approximation is perfectly valid around the horizon of a macroscopic black hole, which is where the core of the computation is performed. However, when a black hole has evaporated to Planckian size, quantum gravitational effects are expected to be dominant. The remnant idea considers the possibility that such effects would stop the evaporation when the horizon has reached Planckian dimensions. If $m$ is the mass of the black hole, this happens in a time scale $\sim m^3$ (roughly $10^{55}$ times the actual
age of the Universe for a solar mass black hole!). The information that has fallen in during this period would ``simply'' be stored inside the Planckian object (a remnant), ready to be either read, released, or eternally stored. The main objection to this scenario is, roughly speaking, that the Planckian final object is too small to contain the huge amount of information needed to purify the external state. More precise statements can be formulated \cite{Banks:1994ph,hossenfelder2010conservative}, but all of them are strongly based on considering the final object as small.
However, ``small'' in general relativity is not a very precise adjective. Indeed, the interior of any eternal (Kruskal) black hole, for instance, contains spacelike hypersurfaces of infinite volume, such as any $r=const$ surface. So, if we consider a Planckian-area eternal black hole, should we consider it as small because of the small external area, or huge because of the infinite internal volume? Physical black holes formed by collapsing matter have no infinite volume. Nonetheless, the hypersurface with maximum volume can be identified \cite{Christodoulou:2014yia}. The result is that, since the interior of a black hole is not stationary with respect to the Killing exterior time $t$, the maximum volume available inside the horizon (which has constant area) grows with time as $V \sim m^2\, t$ in Planck units. 
\vspace{1ex}

\noindent
{\bf Chapter ~\ref{chap:volumeBH}.} In this first main Chapter it is shown that the result sketched above generalises to spacetimes describing an evaporating black hole. The growing of the volume follows, up to the very late stages of the evaporation, the same scaling as for the static case $V \sim m^2\, t$, where now $m$ is the initial mass \cite{Christodoulou:2016tuu}. This implies that if one considers a, say, solar mass black hole that evaporates to, say, $10$ times the Planck mass, then at that stage the hole will have an external area of the order of $100$ times the Planck area (roughly $10^{-70} m^2$) hiding a volume proportional to $m^5$, that is roughly $10^5$ times the volume of our observable Universe! Additionally and interestingly, the surface defining the volume is a spherically symmetric one starting at the radius of the Planckian horizon, and radially growing as $r(t) \sim 3/2 m(t)$--Fig.~\ref{fig:complete}. This shows that evaporating black holes are naturally endowed with an internal bag-of-gold \cite{wheeler1964relativity} (or cornucopion \cite{Banks:1994ph}) type of geometry, and this plays a key role in qualitatively answering the standard objections against the remnant scenario--see \cite{hossenfelder2010conservative} and \cite{Banks:1994ph} for details.
\vspace{1ex}

As said before, Hawking's evaporation is considered to be valid up to the Planckian regime. Until then, no other effects derived from known physics are expected to take place and disturb the slow decrease of the mass of a black hole.  Nonetheless, ``exotic'' effects that may allow the information to be released have been proposed in the literature and are sketched in the \hyperref[sec:intro1]{introduction to Part I}. One of the most recent ones (even if based on some older papers, e.g. \cite{Hajicek:2001yd}) has been put forward in \cite{Haggard:2014rza}. The idea is as follows: General Relativity predicts that a collapsing star falling whithin its own horizon will inexorably keep collapsing until it reaches a singularity. However--see next Section for details--an expected outcome of quantum gravity is that the collapse stops when densities and curvature reach Planckian values. This old idea is at the basis of, for example, the non-singular black hole solutions mentioned a few paragraphs above. The scenario of \cite{Haggard:2014rza} suggests that the star would not simply stop its collapse, but would bounce back out, thus destroying the black hole. The entire process would happen in a $m^2$ time scale, much shorter than Hawking's evaporation time scale, $m^3$. In such a time scale, Hawking's effect becomes negligible and the information paradox becomes not even formulable. However, a black hole is by definition a region from which nothing can escape. So how does it happen that the star bounces back out of its own black hole horizon? The authors suggested that tiny quantum gravitational effects outside the horizon can accumulate in time, becoming so important to allow a quantum transition from a black hole geometry to a white hole one. The global spacetime introduced in \cite{Haggard:2014rza} is horizon-less, showing an initial trapping horizon followed by an anti-trapped one, allowing the bouncing star to emerge. Classical Einstein's equations are satisfied all over except for an interpolating quantum gravity region connecting the two classical phases. Immediately after, a series of papers claimed that this model may provide an observable signal in the gamma ray-burst spectrum  \cite{Barrau:2016fcg,Barrau:2014yka,Barrau:2015uca,Vidotto:2016jqx}. It is worth mentioning that a similar scenario has been simultaneously proposed in \cite{Barcelo:2014npa,Barcelo:2015noa}. The main difference is that the time scale of the process becomes even shorter, of order $m$. The observable consequences mentioned above do not apply in this case.
\vspace{1ex}

\noindent
{\bf Chapter~\ref{chap:Fireworks}.} In this Chapter an analysis of the instabilities of the black-hole-to-white-hole model is presented. It is shown that the original proposal needs to be revised, being strongly unstable \cite{DeLorenzo:2015gtx}. Indeed, an explicit computation shows that for instance the vacuum state of a scalar field evolving in the proposed spacetime develops a singular energy-momentum tensor. A similar argument applies if a classical perturbation (a ping-pong ball) is dropped from infinity toward the bouncing hole. At first sight these instabilities disappear only if the time scale of the process is of order $m$, electing the model of  \cite{Barcelo:2014npa,Barcelo:2015noa} as the correct one--see also \cite{Barcelo:2015uff}. However, this Chapter also shows that the instabilities can be removed by a simple minimal modification of the model without the need of modifying the time scale of the process. The new model is a time-asymmetric version of the original one with a time scale for the final white hole phase that is shorter than $m\, \log m$, while the full process can have an arbitrary time scale. Nonetheless, it is worth emphasising that the need for a time-asymmetric modification of the model seems to uncover important issues that cannot be addressed in detail without a full quantum gravity treatment. A discussion about these points is provided in Section~\ref{sw}.
\vspace{1ex}

Recent developments on the black-hole-to-white-hole transition includes a series of papers where an explicit computation of the probability of such a process to happen in the framework of full covariant Loop Quantum Gravity is provided \cite{Christodoulou:2018ryl,Marios:2018hzw,Christodoulou:2016vny}. The result, even though preliminary because of the approximations used, is that the process becomes probabilistically important only when the mass of the hole is of the order of the Planck mass. It goes therefore against the initial expectation that the transition could be dominant over the Hawking's evaporation for large black holes. Nevertheless, with this result at hands a new interesting scenario has been proposed \cite{Bianchi:2018mml}. Hawking's evaporation drives a black hole to Planckian size, but stops where the transition to a white hole happens. The resulting white hole is seen as a remnant. It inherits from the evaporated black hole the large volume studied in Chapter~\ref{chap:volumeBH}. This scenario provides a realisation of the general speculative idea sketched in Section~\ref{sec:discussion} and nicely connects the topics of these first two main Chapters.\vspace{1ex}

\noindent {\bf Part II: Thermodynamics of spacetime}\\ 
The already mentioned result that black holes emit thermal radiation came as a beautiful surprise for the theoretical physicists community. In a previous paper, indeed, Hawking himself, together with Bardeen and Carter, had proven that black hole mechanics satisfies an intriguing mathematical analogue of the four laws of thermodynamics \cite{Bardeen1973}. \emph{(0th law)} The surface gravity $\kappa$ is constant on the horizon; {\em (1st law)} For perturbation of a black hole spacetime, the change in energy $\delta M$ is related to the change in area by \footnote{For simplicity I consider here a static Schwarzschild black hole, but a more general first law is valid for a rotating and charged black hole--see  the \hyperref[sec:intro2]{introduction to Part II}.}
\begin{equation*}
\delta M = \frac{\kappa}{8\pi}\,\delta A\,;
\end{equation*}
{\em (2nd law)} The area $A$ cannot decrease: $\delta A \geq 0$; {\em (3rd law)} It is impossible to form a black hole with zero surface gravity. The analogy holds when area $A$ and surface gravity $\kappa$ of the horizon play the role of, respectively, entropy and temperature. 

Such laws were seen as no more than an interesting mathematical analogue. Without Hawking's result, indeed, the effective temperature of a black hole was believed to be absolute zero. A way of seeing this is that a black hole couldn't be in thermal equilibrium with some radiation at any non-zero temperature since it can swallow it without being able to emit radiation to maintain thermal equilibrium  \cite{Bardeen1973}.
Strikingly, Hawking's temperature not only is non-zero, but it is proportional to $\kappa$, directly invalidating the above purely classical reasoning. A black hole \emph{is} a thermodynamical object, behaving like a black body with a temperature $T=\kappa/(2\pi)$ and an entropy $S=A/4$! 

Entropy in thermodynamics is a very useful tool to encode our ignorance about the microscopic structure of the system under consideration. The appearance of a natural notion of entropy in black hole mechanics, therefore, has been immediately interpreted as the manifestation of our ignorance about the fundamental quantum gravitational description of space and time. In other words, black hole thermodynamics is naturally interpreted as providing important information about the quantum theory of gravity of which the semiclassical treatment should be a suitable limit of. After 40 years, however, a common consensus of the scientific community about the nature of this entropy is still missing.  In this respect, studying examples able to capture the basic features of the problem could help the understanding process.\vspace{1ex}

\noindent
{\bf Chapter~\ref{chap:lightcone}.} This motivation leads to Chapter~\ref{chap:lightcone}, where it is shown that light cones of Minkowski spacetime can be seen as a nice analogue of black hole horizons  \cite{DeLorenzo:2017tgx}. More in details, light cones of Minkowski spacetimes are bifurcate conformal Killing horizons with respect to conformally stationary observers, the latter being the integral curves of most general radial Conformal Killing Field in Minkowski spacetime $\xi$(MCKF). Using this conformal stationarity, a conformally invariant generalisation of the four laws of black hole thermodynamics is proven. A constant light cone (conformal) temperature, given by Hawking's expression in terms of the (conformally invariant) surface gravity, is defined. Exchanges of conformally invariant energy across the conformal horizon are described, in perturbation theory, by a first law where entropy changes are given by 1/4 of the changes of a conformally invariant notion of area of the horizon. This interesting analogy between the properties of MCKFs and thermodynamics of black holes captures the basic mathematical features of the latter on a background with trivial gravitational field. However, the various conformal invariant notions entering the laws have no clear physical meaning. Nevertheless, the previous limitation can be resolved if one performs a conformal transformation sending $(\R^4, \eta_{ab})$ to a model spacetime $(M,\tilde g_{ab})$ with $\tilde g_{ab}=\omega^2 \eta_{ab}$ so that $\xi$ becomes a genuine Killing field and the conformal bifurcate horizons become bifurcate Killing horizons in the target spacetime. The four laws remain true in the target spacetime with identical numerical values, but now all the quantities involved acquire the standard physical and geometric meaning that they have in the context of black holes. 
\vspace{1ex}

\noindent
{\bf Chapter~\ref{chap:lightBH}.} In Chapter~\ref{chap:lightBH}, the generic global features of the conformally flat spacetimes obtained by the previous procedure are studied. Which are the case where these spacetimes represent black holes?  What are they in the other cases? It is clear that there is an infinite number of possibilities. Nevertheless, it is shown that the generic global features can be made apparent in a small number of representative cases. The simplest case corresponds to $\omega=\alpha/r^2$ that reproduces the Bertotti-Robinson  solution of Einstein-Maxwell's theory \cite{Bertotti:1959pf}. Its geometry has been known to encode the near horizon geometry of close-to-extremal and extremal Reissner-Nordstrom black holes--see for instance \cite[sec. 4.4.2]{fabbri2005modeling}. 
Another representative example is the de Sitter realization where the bifurcating horizons correspond to intersecting cosmological horizons (there is no black hole in this case). Weakly asymptotically Anti de Sitter black hole realizations are also presented, together with a few other more exotic spacetimes with Killing horizons but no black holes. These results clarify and strengthen the conclusions of the previous Chapter, and open possible windows on applications of this light cones-black holes thermodynamical affinity.     

\noindent
{\bf Chapter~\ref{chap:desmo}.} In the final Chapter, I present a work \cite{desmoTorsion} related to Jacobson's idea that the continuum structure of spacetime could emerge as the thermodynamical equilibrium description of more fundamental quantum degrees of freedom \cite{PhysRevLett.75.1260}. By evoking local flatness, at any point of a spacetime one can considered a local Rindler Killing field $\chi$ with its associated horizon $\mathcal{H}$, and a small perturbation given by a small flow of energy-momentum $\delta T_{ab}$. The Einstein's equation is derived by imposing the thermodynamical equilibrium Clausius' relation $\delta Q = T \delta S$ for the horizon, by assuming that (i) the horizon carries an entropy proportional to the area and (ii) a temperature $T=\kappa/(2\pi)$, and that (iii) $\delta Q=\int_\mathcal{H} T_{ab} \chi^a dS^b$. In this perspective Einstein's equation is an equation of state: it is the thermodynamics associated to the underlying degrees of freedom that gives rise to spacetime dynamics. Starting from the results of black hole thermodynamics, Jacobson reversed the logic: black hole thermodynamics is not a consequence of Einstein's equations, but it is an explicit manifestation of a more fundamental fact: the fact that Einstein's gravity is an emergent low energy thermodynamical equilibrium description of some yet unknown fundamental physics. This idea has been applied to other theories of gravity, going beyond General Relativity. An interesting example of such theories is the Einstein-Cartan's (EC) one, where a non-vanishing torsion tensor is considered with the scope of coupling fermionic matter to gravity--see \cite{Shapiro:2001rz} for a review. The field equations of the theory consist of two tensorial equations: one that in absence of torsion reduces to the GR Einstein's equation, and one for the torsion tensor. A first attempt to generalise Jacobson's derivation to the EC theory was recently made in \cite{Dey:2017fld}. It was claimed that a restriction on the torsion tensor as well as a non-equilibrium treatment--as in \cite{PhysRevLett.96.121301,Chirco:2009dc}--are needed to derive the first set of the field equations. The torsion part is left underived. However a more careful analysis shows that the two assumptions can be encompassed, and that the argument works for the EC theory just like in the GR original case. There are two crucial observations. The first one is that the very notions of Killing fields and Killing horizons are purely metric ones, and therefore insensitive to the presence of torsion. This implies a great simplification in the derivation. Indeed, a crucial step is to compute the change of the area of the horizon $\delta A \propto \delta S$ by using Raychauduri's equation for the geodesic congruence generated by the Killing field at the horizon. Given the above observation, the standard Raychauduri's equation for Levi-Civita geodesics can therefore be used. The second one is that diffeomorphism invariance of EC theory identifies uniquely, on-shell of the torsion field equations, the \emph{conserved} stress-energy tensor (SET) that should be used to define the energy flux $\delta Q$. This tensor is not the one derived from varying the matter action with respect to the metric tensor, but it involves additional torsion-dependent terms. With these two elements, it suffices to follow Jacobson's original derivation to derive the EC dynamical equations from thermal spacetime equilibrium. Having used the torsion part of the EC field equations to identify the conserved SET may be considered at odds with the underlying idea of using only thermodynamical arguments to derive the gravitational dynamics. Two considerations suggest that the approximation used are perfectly reasonable. First, there seems to be no conceptual obstacles to prove the conservation of the SET from diff-invariance off-shell of the torsion field equations. The obstacle seems to be only technical and may be soon resolved. The second point is that the non-equilibrium treatment, that is needed if the torsion field equations are not used \cite{Dey:2017fld}, appears to be largely arbitrary. I refer to Section~\ref{sec:arbitrio} for more details. The philosophy that seems more reasonable now, therefore, is that there would exist a yet unknown independent thermodynamical argument to derive the torsion part of EC field equations. From there the other set would be derivable within the equilibrium description used in the Chapter. 
\vspace{1ex}

\section[Classical, Semi-Classical, and Quantum Gravity]{Classical, Semi-Classical, and Quantum \\Gravity}\label{sec:interplay}
The main frameworks I have been working with in this thesis are classical General Relativity (GR) and Quantum Field Theory in Curved SpaceTime (QFT in CST). The former is the well-known classical theory for the gravitational interaction; the latter is a mathematically and conceptually consistent theory that generalises the particle physics QFT in flat spacetime to a general curved background. This Section is not intended to contain a review of these theories, for which many textbooks covering the topics can be found. See for instance \cite{wald2010general} for GR and \cite{birrelldavies:QFTCST} for QFT in CST. The aim of the Section is instead to clarify to which extent QFT in CST is used in the main text as the first approximation to the yet unknown theory of quantum gravity (QG).
Contextually, the sense in which, as said in the previous Section, ``expected results of Quantum Gravity'' are used in the thesis will also be clarified. And this latter point actually comes first in the next paragraph.

What this Section does not contain either is a review about the state of art of Quantum Gravity, about the different approaches and about the motivations of looking for such a theory. Also in this case the reader is referred to textbooks and reviews present in the literature, such as \cite{Rovelli:2000aw,Hedrich:2009pb,oriti2009approaches}. For the sake of the present discussion, it is enough to recall that gravity is the only known force of Nature that we are not yet able to consistently cast in the framework of quantum field theory. The straightforward application of the standard quantisation scheme to the gravitational interaction produces a quantum description of gravity mediated by a massless spin-2 boson (the graviton). This theory is  however non-renormalisable \cite{sagnotti:twoloops}, and therefore incapable of making predictions. Among many technical reasons, a conceptually simple but important one is that the techniques of QFT have been developed to describe dynamical matter fields evolving on a fixed non-dynamical flat spacetime. The dynamical field of GR to be quantised by a theory of QG, on the other hand, is spacetime itself. Finding a consistent way of quantising gravity, therefore, it is an incredibly fascinating path that may require important conceptual steps on our comprehension of both gravity itself and quantum theory. Several approaches have been developed starting from different perspectives and philosophies, but a fully consistent theory is still missing.

In my work I have not considered any particular theory of QG. At the same time, 
I've been using a prediction that is shared by all these approaches, namely the appearance of a minimal length scale given by the Planck length
\be\label{eq:lpl}
\ell_{\va P} = \sqrt{\frac{\hbar\,G}{c^3}}\sim 10^{-33}cm\,.
\ee 
 For instance, in the framework of QG I'm more familiar with, i.e. Loop Quantum Gravity (LQG) \cite{rovelli2014covariant}, quantum operators corresponding to areas and volumes of microscopic regions of spacetime are defined. Their spectrum turns out to be discrete with minimum values proportional to the Planck area and volume respectively. In String Theory \cite{polchinski1998string}, the minimal string scale is proportional to the Planck length. The fixed point of the renormalization group in the Asymptotic Safety approach \cite{percacci2017introduction} is expected to be at the Planck energy. And so on--see \cite{hossenfelder2013minimal,singh1989notes} for complete reviews on the topic. But not only: even without invoking any theory of quantum gravity, $\lpl$ as minimum length emerges from general considerations and gedankenexperiments. For instance \cite{Mead:1964zz}, imagine to include gravity in the Heisenberg's microscope thought-experiment that lead to uncertainty principle of quantum mechanics (QM). Consider therefore a photon of frequency $\omega$ which scatters with a non-relativistic particle whose position on the $x$-axis needs to be measured. The frequency of the photon implies an uncertainty on the position of the particle given by its Compton wavelength
 \be\label{eq:deltax}
 \Delta x \gtrsim \frac{c}{\omega}\,.
 \ee
Now, the photon energy will exert a gravitational acceleration on the particle given by
\be
a \sim \frac{\hbar G\,\omega}{c^2\,R^2}
\ee
for the time $\tau$ during which the photon is in the region of interaction, which is hereafter considered a ball of radius $R$,
 implying $\tau \sim R/c$. The velocity acquired by the particle is thus given by
\be
v = a\,\tau \sim \frac{\hbar G\,\omega}{c^3\,R}
\ee
implying an additional uncertainty in position given by the distance traveled by the particle
\be
L = v \, \tau \sim \frac{\hbar \,G}{c^4} \omega\,.
\ee
Combining the above equation with \eqref{eq:deltax} one finds
\be
\Delta x \gtrsim \lpl\,.
\ee
Thus, when gravity it brought into the quantum measurement game, a minimal measurement distance arises, and is given again by the Planck length. The above argument is the simplest one can derive, where only Newtonian gravitational interaction is considered, where the change in momentum of the photon by the scattering is neglected, and other approximations. Refined arguments are present in the literature, but the qualitative result does not change--see \cite{hossenfelder2013minimal} and references therein. An additional interesting argument can be made. In QM there exists a typical length associated with any particle of mass $m$: its Compton wavelength $\lambda_{\va m} = \hbar/(m\,c)$. The fundamental uncertainty principle of QM (or QFT) tells us that the energy needed to measure the particle position within its Compton wavelength is enough to trigger the quintessentially QFT phenomenon of particle creation: a new particle with the same mass is created. Also GR associates a length to a given mass $m$, namely it Schwarzschild radius $r_{\va m} = 2Gm/c^2$. Compressing the mass $m$ beyond $r_{\va m}$ triggers the quintessentially GR phenomenon of creation of a black hole. The two length scales become of the same order when $m$ is
 \be
 m \sim \mpl =  \sqrt{\frac{\hbar\,c}G}\sim 10^{-5}\,g\,:
 \ee
the Planck mass $\mpl$. When this happens, $r_{\va m}$ and $\lambda_{m}$ are equal to the Planck length. To summarise, the advent of a minimum length scale $\lpl$ where purely quantum properties of spacetime are expected to be important is strongly suggested by general arguments, and it is a genuine prediction of all the available approaches to quantum gravity \footnote{Remarkably, the Planck length (mass and time) is the \emph{unique} combination of the constants of Nature $G$, $\hbar$ and $c$ with the dimension of length (mass and time) \cite{planck1899naturlische}.}. 
\begin{displayquote}
{\em Exploring the consequences of a minimal length scale is one of the best motivated avenues to make contact with the phenomenology of quantum gravity, and to gain insights about the fundamental structure of space and time.} \cite{hossenfelder2013minimal}
\end{displayquote} 
Such consequences are particularly evident and important in the physics of BHs and in the early cosmology. For instance, if a minimum length $\lpl$ is present, the curvature $\mathbf{K}$ of any spacetime cannot exceed $\mathbf{K} \sim \lpl^{-2}$, or equivalently the typical curvature radius will not be smaller than $\lpl$. The infinite curvature singularities predicted by GR at the centre of the BH and at the Big Bang are therefore expected to be cured by quantum gravitational effects. Moreover, a mass $m$ cannot be compressed inside a volume smaller than $\lpl^3$ implying a maximum density per given mass $m$ of $\rho \sim m/\lpl^3$. This implies that the motion of a collapsing star will differ from the one predicted by GR when Planckian volume is reached, avoiding the infinite squeeze of the matter into the singularity. Clearly, the precise details of such manifestations require a complete theory of QG. Imagine that such a theory existed. As always in physics, one could decide to neglect some details of the full theory, using an effective description that captures the important points needed for the particular problem at hands. Since we still lack a consistent theory of QG, the reverse approach can be very constructive: one can \emph{ad hoc} propose effective descriptions of the expected outcomes and study their viability. This in turn can provide important hints about the full theory. An illustrative example of the same kind in particle physics is Enrico Fermi's description of $\beta$-decay \cite{Fermi:1933jpa}. Built on the basis of known physics to describe the $\beta$-decay of neutrons, it was proven experimentally to be pretty accurate in its goal. Its non-viability for the study of other processes has then guided the path to the more fundamental quantum field theoretical treatment of weak interactions. In BH physics a concrete example of such an approach is provided by the so called \emph{non-singular BH metrics}--see for instance \cite{de2014investigating} and references therein. Effective spacetimes are proposed such that no singularity is present, and such that, far away from the centre, they are almost indistinguishable from standard BHs. Their viability is then tested by studying, for instance, their Hawking's evaporation (see below). Interestingly, it results in an inconsistency with energy conservation for the most known proposed models \cite{Frolov:2017rjz,de2014investigating}. Resolving the inconsistency might provide useful hints about the full QG theory. 

Planckian regimes are remarkably extreme, particularly when compared to regimes accessible on earth: the Planck length is about $18$ order of magnitude smaller than the charge radius of a proton; the Planck mass is about $10^{14}$ times the collisions' energy at LHC; Planckian curvature radius is $10^{41}$ times smaller than the curvature radius at the earth's surface. One can therefore expect intermediate regimes in which purely quantum gravitational effects are negligible, but both GR and QFT are important. Let me borrow another example from particle physics. A wide range of phenomena in atomic physics are successfully described by treating the electrons within the QFT framework, while the electromagnetic field generated by the nucleus as external and non-dynamical. Clearly this is an approximation of the more fundamental theory of Quantum ElectroDynamics (QED), where the electromagnetic and matter fields are coupled and quantised at the same footing. This more complete theory, however, only adds corrections that for the purposes of atomic physics can be neglected. The approximation becomes inappropriate when one wants to study sub-atomic physics, for which the full quantum field theoretical treatment of the electromagnetic field is needed (as well as of the weak and the strong force). In the gravitational context, an analogous approximate theory to full QG has also been developed: \emph{Quantum Field Theory in Curved Spacetime (QFT in CST)}. It provides a quantum field theoretical description of elementary matter in a non-dynamical external gravitational field, i.e. a curved spacetime. As briefly mentioned at the beginning, for non-interacting fields such theory comes out to be well defined and conceptually consistent \cite{Kay1988}. The matter Lagrangian is coupled to the curvature produced by $g_{ab}$ usually via a minimal coupling provided by substituting the flat partial derivative $\partial_a$ with the Levi-Civita covariant derivative $\nabla_a$ associated with $g_{ab}$. This is the coupling used in the analysis of Chapter~\ref{chap:Fireworks}; other types of couplings are possible, such as conformal coupling that is used in Chapter~\ref{chap:lightcone}, and the coupling of fermions to the affine torsion-full covariant derivative, used in Chapter~\ref{chap:desmo}. 
The most interesting prediction of QFT in CST is particle creation from the vacuum state of a quantum fields due to the external background. It is the analogue of the Schwinger effect for quantum matter fields in a strong external electromagnetic field. The particle creation in curved spacetimes is of particularly importance in BH physics. It is indeed the core of Hawking's famous result that an outside observer sees the BH emitting particles at the rate one would expect if the latter was a black body at temperature \cite{hawking1974black}
\be\label{hawTe}
T_{\va H} = \frac{\hbar\,\kappa}{2\pi\,c\, k_{\va B}}\,,
\ee
where $\kappa$ is the surface gravity of the horizon. For all the technical details about QFT in CST, particles creation and Hawking's effect I refer to classical textbooks as \cite{birrelldavies:QFTCST}, and to \cite{Kay1988,Wald:1995yp} for an elegant algebraic approach. The core of Hawking's computation is also present, even if rephrased, in Section~\ref{HRCT}.

The above discussion on the minimal length already shows that one cannot trust QFT in CST at the Planck scale. But how far can one go with it? A way to estimate this is to consider its quintessential phenomenon, namely particles creation. Created particles carry energy, and energy is the source of the curvature of spacetime. Hence, the approximation of considering a fixed non-dynamical background can be considered good as far as the induced curvature is small compared to the typical curvature of the background. To estimate when this breaks down, let me consider the Hawking's effects around a spherically symmetric uncharged BH of mass $M$. Its horizon is a sphere at radius $r_s = 2G M /c^2$, and its surface gravity $\kappa$, which has dimension of acceleration, is given by
\be
\kappa = \frac{c^4}{4GM}\,.
\ee 
The Hawking's temperature \eqref{hawTe} reduces to 
\be\label{hawTeSc}
T_{\va H} =\frac{\hbar\,c^3}{4k_{\va B}GM}	\,.
\ee
The typical curvature radius \footnote{Consider a point $p$ in spacetime, and call $K$ the least upper value of the Riemann tensor $|\mathbb{R}_{abcd}|$ in any orthonormal frame. The typical curvature $R$ radius is defined to be $R = K^{-1/2}$.} is given by the \Sch radius $r_s$. Considering as first approximation the Stephan-Boltzmann law, the energy density of Hawking's radiation scales as $\rho\sim (k_{\va B} T_H)^4/(\hbar^3 c) = \hbar c/r_s^4$. The curvature induced by this energy density is $\sim G/c^4 \rho= \lpl^2/r_s^4$ with its typical curvature radius of $\sim r_s^2/\lpl$. The latter becomes of the same order of the background one for $r_s \sim \lpl$. This shows that, as efficiently expressed by Hawking himself,
\begin{displayquote}
{\em one would [...] expect that the scheme of treating the matter fields quantum mechanically on a
classical curved space-time background would be a good approximation, except in regions where the curvature was comparable to the Planck value $1/\lpl^2$.} \cite{hawking1975}
\end{displayquote}

The last step of the above reasoning is extremely non-trivial. The curvature induced by the energy density $\rho$ of the radiation, indeed, has been evaluated by considering $\rho$ as the stress energy tensor (SET) on the right-hand-side of the Einstein's equations
\be
R_{ab} -\frac{1}{2}g_{ab} = \frac{8\pi G}{c^4} T_{ab}\,.
\ee 
However, $\rho$ is a very crude approximation of the true SET of the quantum field from which the particles came. In particular, such SET should be an operator to be evaluated on quantum states, for instance considering its expectation value $ \bra{\psi} T_{ab}\ket{\psi}$ on a state $\ket{\psi}$.
This is a first glimpse of a much more complicated dilemma which is know as {\em back-reaction problem} \cite{wald1977back}. Already defining a meaningful SET operator for a quantum field in a given curved spacetime turns out to be a highly non-trivial goal. The standard procedure of normal ordering in flat space that allows a normalisation of the infinite vacuum expectation value cannot be directly applied. Even if a satisfactory definition existed, how would $\bra{\psi} T_{ab} \ket{\psi}$ couple to the curvature of the background geometry? The standard attempt is to consider the so called  \emph{semi-classical Einstein's equation}
\be\label{semiEinst}
R_{ab} -\frac{1}{2}g_{ab} = \frac{8\pi G}{c^4} \bra{\psi} T_{ab} \ket{\psi}\,,
\ee
where the gravitational field is still treated classically. For this approximation to be valid, the quantum fluctuations of $T_{ab}$ on $\ket{\psi}$ should be negligible with respect to the value of $\bra{\psi} T_{ab} \ket{\psi}$. If the fluctuations were large, indeed, the gravitational field should be also expected to have large fluctuations, requiring a full QG treatment. Giving an estimate on the validity of such approximation, however, would require a complete knowledge of the normalised SET, which, as previously mentioned, is not yet generically available. Back-reaction in QFT in CST is an open problem \cite{singh1989notes}.

The situation is greatly simplified in $1+1$ dimensions. In this setting meaningful SET operator can be defined and analytically computed for specific problems \cite{fabbri2005modeling}. Even in this case, however, explicitly solving the semi-classical Einstein's equations \eqref{semiEinst} is far from being obvious--see the discussion in Section~\ref{scs}. Explicit solutions to the back-reaction problem are available only in $1+1$ dilaton gravity, a modified theory of gravity where a scalar field (the dilaton) is coupled directly in the Einstein-Hilbert action--see \cite{Grumiller:2002nm} and references therein. Of particular interest for the present discussion is  the fact that a BH model, the {\em CGHS model} \cite{Callan:1992rs}, can be constructed in this theory and the back-reaction of Hawking's radiation analytically solved. The result is important because it confirms the more reasonable guess one could have done in the physical $4$-dimensional situation. By invoking conservation of energy, indeed, one could have guessed that the energy radiated away by Hawking's particle should have been compensated by a decrease of the mass of the BH. This is what the CGHS model has confirmed. The hole \emph{evaporates} by the emission of a thermal flux of particles! This crucial result from the analysis of the back-reaction problem is the main one used in this thesis. More details here are therefore not needed, and I refer the reader to the mentioned references for additional information. The semi-classical Einstein's equations \eqref{semiEinst} are also used in Chapter~\ref{chap:Fireworks} as a tool for the analysis of the stability, and therefore the viability, of an effective metric describing an hypothetical quantum gravitational process. I refer to Section~\ref{scs} for further details.

Some explicit numbers are useful to complete this Section. Inserting the explicit values of the constants in the Hawking's temperature for a \Sch BH, Eq.~\eqref{hawTeSc}, one finds
\be
T_{\va H} \sim 10^{-6}\frac{M_\odot}{M}\, {}^\circ K
\ee
where $M_\odot$ is the solar mass. The smallest astrophysical BH ever observed is the one in the binary system GRO 1655-40 which has a mass $M_{\va GRO} = 6.3\,M_\odot$ \footnote{The mass of the BH in the binary system XTE J1650-500 may be even smaller, with a mass estimated to be between $5$ and $10$ solar masses.}. The associated Hawking's temperature would therefore be of order $\sim 10^{-7}\, {}^\circ K$, seven orders of magnitude smaller than the cosmic microwave background temperature $T_{\va CMB} \sim 3\, {}^\circ K$. Moreover, no BHs are expected to form from a collapsing star of mass below the Chandrasekhar limit of $1.4 \,M_\odot$. Hawking's effect is therefore totally negligible for any astrophysical BH, which would be absorbing radiation faster than they emit it, therefore increasing in mass. However, the opposite is true for microscopic primordial BHs \cite{hawking1971gravitationally,Boudoul:2002cv}. If their existence was confirmed, then a BH of mass $10^{15}\,g \sim 5 \cdot 10^{-19} \,M_\odot$ created in the very early stages of the Universe would lose all its mass via Hawking's evaporation, reaching the very final stages of its life today.

\paragraph{Notation} Throughout this thesis, unless where explicitly specified, I will use metric signature $(-,+,+,+)$ and Planck units $G=\hbar=c=k_B=1$. 
}

%% file: Part_1/Part_1.tex
\thispagestyle{empty} \phantom{}

\vfill 

\begin{center}
\hrule\vspace{.2ex}\hrule\vspace{1ex}
{\huge \scshape\bfseries Part I\\}
\vspace{1ex}\hrule\vspace{.2ex}\hrule

\vspace{10ex}
{\fontsize{35}{38} \scshape\bfseries The Information Paradox\\}
\end{center}

\vfill

{
\markboth{Part I: The Information Paradox}{Introduction to Part I}
\label{sec:intro1}
Typing \emph{information paradox} on Google Scholar search bar produces $1\,920\,000$ results, \emph{black hole information problem} results in $1\,370\,000$ items, while \emph{black hole information paradox} in ``only'' $110\,000$. These numbers give an idea on how this puzzle, first introduced by Hawking in 1976 \cite{hawking:1976breakdown}, is since then central in theoretical physics. The usual way the paradox is staten is that there is an incompatibility between unitarity of quantum field theory and black holes (BHs), since a quantum field evolving in the spacetime of an evaporating BH undertakes a non-unitary transition from an initial pure state into a final mixed state.
This sentence, however, is based on a series of (sometimes subtle) assumptions and considerations, which also change depending on the background of the scientist who is tackling the problem: assumptions can be added, forgotten or over/under emphasised. A general relativist probably sees the issue in a different way than what a quantum information scientist does, the String Theory community sees the problem in a different way than the non-perturbative quantum gravity one, a philosopher sees it differently with respect to a physicist, and so on.

\newpage

In what follows, I introduce the paradox in the way I understand it, with a step by step logic that I believe  better uncovers the possible underlying assumptions one can encounter in the literature, and helps classifying the proposed solutions. Such view is mainly influenced by some recent papers on the subject \cite{Perez:2014xca,Okon:2016qlh,Unruh:2017uaw,Marolf:2017jkr}. The discussions of Sections~\ref{sec:interplay} and the \hyperref[sec:intro2]{introduction to Part II} will play an important role.

Let me therefore consider the minimum set of assumptions for a first problem to arise:
\begin{enumerate}[label=(\Alph*)]
\item A BH forms.
\label{it:collapse}
\item The conditions of the system are such that QFT in CST is valid around the horizon of the BH.\label{it:QFT}
\item The Hawking's effect results in a decrease of the mass of the BH.
\label{it:evap}
\item Physically reasonable spacetimes are globally hyperbolic.\label{it:hyp}
\end{enumerate}
When the first three conditions are satisfied, the unique resulting complete story of formation and evaporation of a black hole is depicted in Fig.~\ref{fig:paradox}: a black hole is formed by the collapse of some matter distribution--Assumption~\ref{it:collapse}--; since QFT in CST can be validly used--Assumption~\ref{it:QFT}--, Hawking's particles are created around the horizon of the BH; the Hawking's effect causes the decreasing of the mass of the hole--Assumption~\ref{it:evap}--, which eventually completely disappears--again Assumption~\ref{it:QFT}--; an initial Cauchy hypersurface $\Sigma_i$ evolves into a final hypersurface $\Sigma_f$ which is \emph{not} a Cauchy surface. In other words, the spacetime is not globally hyperbolic, in contradiction with Assumption~\ref{it:hyp}.
\begin{figure}[t]
\centering
    \includegraphics[width=0.5\textwidth, keepaspectratio]{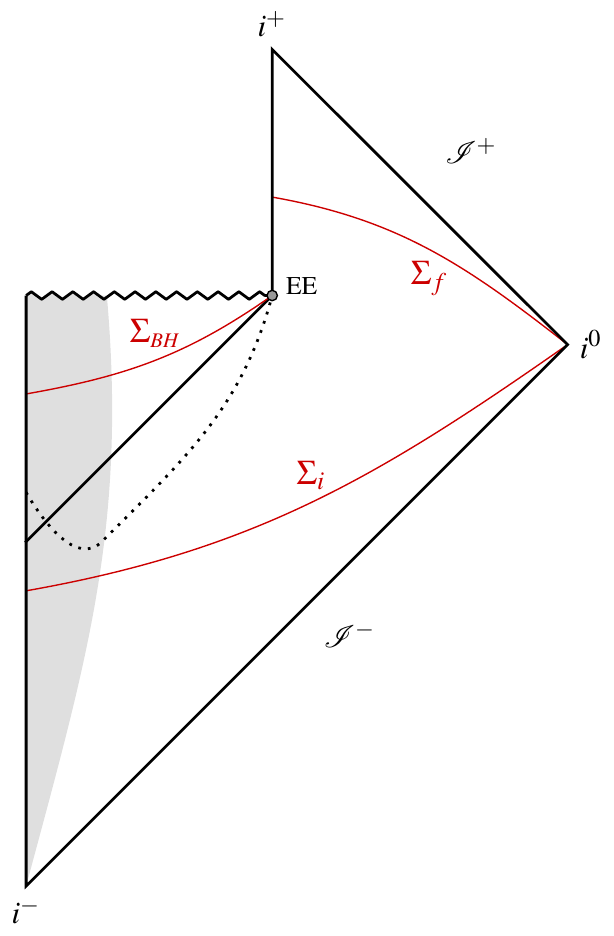}
    \caption{The Penrose diagram of the complete process of formation and evaporation of a spherically symmetric uncharged black hole, given that assumptions~\ref{it:collapse}-\ref{it:evap} are satisfied. The collapsing star is depicted by the shaded region. The dynamical horizon (dotted line) forms and grows inside the star, to then decrease in the evaporating exterior region. It meets the event horizon (thick line) at the evaporation event EE. The spacelike surface $\Sigma_i$ is a Cauchy surface for the whole spacetime, while $\Sigma_f$ and $\Sigma_{\va BH}$ are not. $\Sigma_{\va BH}$, for instance, only spans the interior of the black hole.}
\label{fig:paradox}
\end{figure}
Such inconsistency is not yet the information paradox in the way it is usually staten. 
In such a spacetime a quantum field which evolves with the laws of QFT in CST--Assumption~\ref{it:QFT}-- will indeed undertake a non-unitary evolution. However this does not imply a direct contradiction with quantum mechanics.
As clearly expressed by Unruh and Wald, indeed, 
\begin{displayquote}
{\em the pure state to mixed state evolution [...] is a prediction of quantum theory in any situation where the final ``time'' is not a Cauchy surface, not a violation of quantum theory.}  \cite{Unruh:2017uaw}
\end{displayquote}
For instance, the evolution in flat space from a $t=const$ surface (Cauchy) to a hyperboloid (non-Cauchy) is not unitary, and it is fine with quantum mechanics.
Thus, till now it is a \emph{retrodictability problem} to be faced rather than an information paradox. 
Classical data on $\Sigma_f$ are not enough to retrodict the process on $\Sigma_i$.

Let me therefore concentrate on the possible solutions to this puzzle, by discussing more in details the set of assumptions and the consequences of renouncing to each of them. 

Assumption~\ref{it:collapse} simply states that black holes can form in the Universe, typically by the collapse of some matter distribution. Dropping this assumption may seem at first sight equivalent to say ``It is hard to find a government majority in Italy? Well let say that a government in Italy has never existed'' \footnote{This part of the introductory section has been written the day after the strong disagreement between president Mattarella and Salvini-Di Maio on Paolo Savona as Minister for Economic and Financial Affairs. The clash resulted in Giuseppe Conte renouncing to become Prime Minister and Mattarella nominating Carlo Cottarelli as Prime Minister \emph{ad interim}. But the situation can still change. Stay tuned.}. However this is not the case. It has been proposed by several groups that effects such as pre-Hawking radiation arising during a gravitational collapse may avoid the formation of a horizon. The collapsing star could then bounce back to infinity \cite{Mersini-Houghton:2014zka,Mersini-Houghton:2014yq}, it could continue collapsing at a rate slower than its own  loss of mass driven by the pre-Hawking evaporation \cite{Baccetti:2017ioi,Baccetti:2016lsb,Kawai:2013mda,PhysRevD.93.044011}, or it could become some exotic object such as a \emph{fuzzball} \cite{Mathur:2005zp}. However, these proposals have been hardly, and in my opinion convincingly, questioned in \cite{Chen:2017pkl}. Therefore I will not consider further such solutions to the problem referring the reader to the mentioned references for more details, and I will continue analysing the other assumptions. 

Assumption~\ref{it:QFT} has been discussed in details in Section~\ref{sec:interplay}. In this setting it particularly means that Hawking's computation is valid for BHs of {\em any} dimensions. 

Assumption~\ref{it:evap} is related to the discussion on the back-reaction problem also analysed in Section~\ref{sec:interplay}. The bottom line is that we do not know how to fully treat the back-reaction of quantum fields on a BH background. Since we trust conservation of energy, however, we expect that if energy is radiated away from the hole, the mass of the latter has to decrease accordingly. Modifying this assumption requires a better understanding of the back-reaction problem, that we do not have yet. Therefore one can consider this assumption as dependent on the previous one: as long as QFT in CST is valid around the black hole horizon, Hawking's particles are radiated away and the mass of the hole decreases.

Assumption~\ref{it:hyp} is usually a basic axiom of GR, since it implies the Cauchy problem to be well defined, in turn making GR a predictable theory. It is related to the so called \emph{cosmic censorship conjecture} \cite{penrose1979singularities}. Here I will not discuss the possibility of directly renouncing to this assumption, but I will come back to this point in a while from a different perspective.

To avoid the problem, therefore, one can still either question Assumption~\ref{it:QFT} or directly question the conclusion. Let me start discussing the latter.
Questioning the conclusion means claiming that it is not true that such spacetime does not admit a full foliation with Cauchy surfaces. This is the point of view of a recent provocative paper by Tim Maudlin \cite{Maudlin:2017lye}. The basic claim is that one should extend the standard definition of Cauchy surfaces, by allowing them to be disconnected. In this way, one can construct a foliation of the whole formation-evaporation spacetime, by  
completing the final slices $\Sigma_f$ with slices $\Sigma_{\va BH}$ spanning the interior of the horizon. As pointed out in \cite{Manchak:2018lhx}, this extension of the definition of Cauchy surfaces has several technical problems. Revisiting them is 
beyond the scope of this Section, and I refer to the mentioned reference for details.
More interestingly for the present discussion is the following argument: imagine to consider correct the new definition of global hyperbolicity. Then the initial pure quantum state of the fields driving the evaporation is automatically pure on new final Cauchy surface $\Sigma =:\Sigma_f \cup \Sigma_f^{in}$. Correlations have therefore to be present between modes on the internal and the external part of $\Sigma$. It has been argued in \cite{Okon:2017pvc} that the presence of those correlations may force the state on $\Sigma_f$ to be non-Hadamard close to $r=0$, so developing a pathological energy-momentum tensor. In this perspective, therefore, Maudlin's proposal can also be seen as questioning the validity of QFT, falling among the solutions that refuse Assumption~\ref{it:QFT} that I will deal with in the next paragraph.

The discussion of Section~\ref{sec:interplay} shows that QFT in CST \emph{cannot} be trusted at Planckian scales. In the present context, therefore, it is a conservative option to consider Assumption~\ref{it:QFT} not to be valid close to the singularity and at the very late stages of the evaporation. There quantum gravitational effects are dominant. This results in a modification of the formation-evaporation diagram as depicted in Fig.~\ref{fig:paradigm}. 
\begin{figure}[t]
\centering
    \includegraphics[width=0.5\textwidth, keepaspectratio]{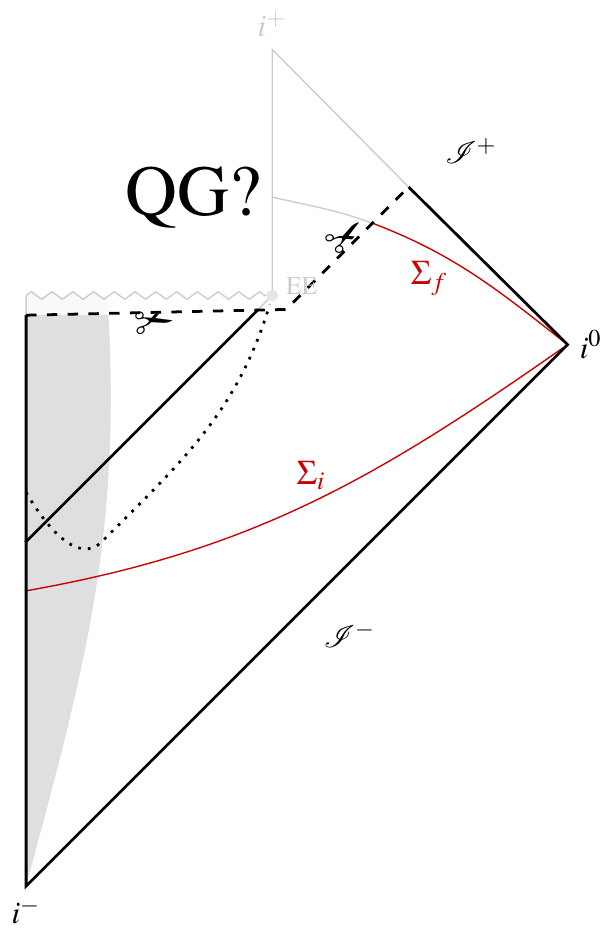}
    \caption{The description of Planckian regimes needs a full theory of quantum gravity. This picture depicts the most general paradigm arising from this idea in the case of an evaporating black hole. The spacetime of Fig.~\ref{fig:paradox} is modified by removing the regions around the singularity and at the end of the evaporation, as well as everything inside their light cone.}
\label{fig:paradigm}
\end{figure}
Pathological regions, as well as their causal future, are replaced by some quantum gravitational uncertainty. There is no evident Cauchy to non-Cauchy transition anymore, but the solution is largely unsatisfactory, since it sweeps the problem under the thick rug of our ignorance about quantum gravity. 

The standard more concrete paradigm is to consider the assumption that, after the BH has reached Planckian regime, the geometry is well described by the flat metric except for a Planckian sized region around the \emph{would-be-singularity}. The resulting picture can be depicted as in Fig.~\ref{fig:AB}. 
\begin{figure}[!h]
\centering
    \includegraphics[width=0.5\textwidth, keepaspectratio]{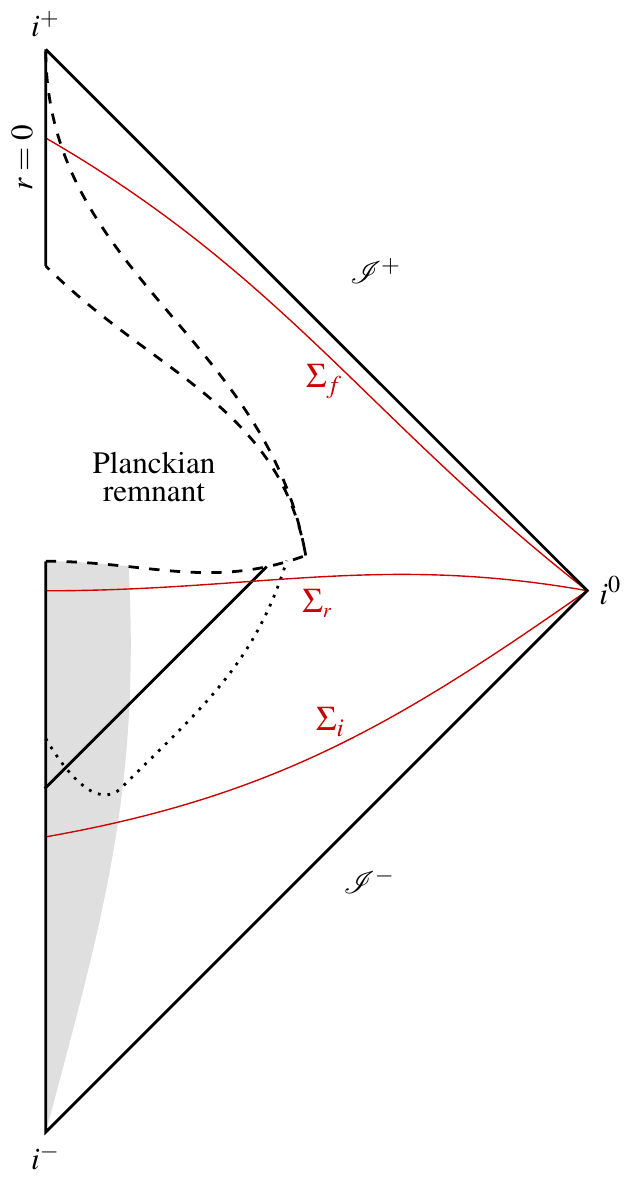}
    \caption{The Penrose diagram for the remnant scenario. Quantum gravitational effects stop the collapse and remove the singularity. The resultant Planckian object--the remnant--can decay (dashed line ending at $r=0$), eventually completely disappearing, leaving the spacetime completely flat again. The remnant can be eternal (dashed line reaching $i^+$) hiding everything has fallen inside during the evaporation period. Regions inside the dashed line need a full QG treatment to be described. In particular, their causal structure can be very different from what seems to arise from this picture. For instance, the eternal remnant can remain future causally disconnected from the exterior world. To some extent, therefore, this picture has to be interpreted as a cartoon description of the process. The hypersurface $\Sigma_r$ is a perfectly classical one, crossing the horizon when the latter is, say, $100\,\lpl$ and always remaining far from Planckian regimes. }
\label{fig:AB}
\end{figure}
An external observer following the timelike asymptotic world line sees a star collapsing, receives Hawking's radiation for a time $m^3$ until the spacetime becomes \emph{essentially} flat. Such setting is known as the \emph{remnant scenario} \cite{Aharonov:1987tp}, and it provides the basis for the discussion of Chapter~\ref{chap:volumeBH}. A precise effective description of the quantum gravitational region is usually not given. It is instead simply staten that, if such a description existed, it would result in a spacetime with a remnant interior which is either hidden behind a horizon or becomes future causally connected with the exterior world. Does this solve the retrodictability problem? As a full metric is not specified, asking whether we have a globally hyperbolic spacetime becomes a meaningless question. At this level, until a complete theory of QG becomes available, only scenarios based on hypothesis or partial results about the deep quantum regimes are possible. In this sense the retrodictability problem as presented above is transcended, being superseded by questions about the deep quantum nature of space, time and fields. Contextually, Assumption~\ref{it:hyp} looses its meaning.
The proposed scenarios are then theoretically tested with semiclassical arguments with the hope of finding new insights about QG. As will be clear soon, among these cross-checks the fate of correlations of quantum fields (i.e. information) usually plays a dominant role, giving new life to the name \emph{information paradox}. However, seen in this way, there is no evident contradiction between unitary evolution of QM and GR. The information puzzle can instead be regarded as a clear and rich arena to test ideas about quantum gravity, in the very spirit of this thesis.  

The remnant scenario is a perfect example to show what just discussed. Consider the setting in which the Hawking's radiation results from the vacuum state of a field theory evolving in the spacetime of Fig.~\ref{fig:AB}. From the time the horizon forms, Cauchy hypersurfaces in the past of the quantum region would be divided into an interior and an exterior part. The quantum state restricted to one part would be described by a density matrix, and would be in general mixed. Consider now a hypersurface, such as $\Sigma_r$ in Fig.~\ref{fig:AB}, that crosses the horizon when the latter is, say, $100\, \lpl$. The quantum state restricted to the exterior part would be a thermal state of Hawking's particles, being therefore maximally mixed. The quantum state restricted to the internal part would also be a mixed state, while the full state on the complete $\Sigma_r$ would still clearly be pure. There is therefore a large amount of correlations between the interior and the exterior which have built up in time during the Hawking's process. Such correlations are reminiscent of the one discussed above that causes Maudlin's scenario to be non-viable. In this case the state continues to be Hadamard and no divergent energies should appear. The picture on $\Sigma_r$ is now that of an essentially flat space with a Planckian ``defect'', giving life to an interior world where a large amount of information is stored. This may seem, and is currently considered by the most, paradoxical--see \cite{Chen:2014jwq} for a recent review. Chapter~\ref{chap:volumeBH} is completely devoted to the properties of surfaces such as $\Sigma_r$ in relation with the remnant scenario. Thus, I prefer to keep the discussion here as short as possible, referring to that Chapter and references therein for more details.

Hither I only discuss the interesting question of what could be the fate of the information across the quantum region. The standard hypothesis given by the remnant scenario is that, if a horizon keeps existing, than the information would simply be forever trapped in the planckian object. If the interior gets instead in causal contact with the exterior, then the information would be released to the exterior world, purifying the external state. Since the energy available is of order $\mpl$, this purification phase would be done by emitting very soft particles, during a long time that can be estimated to be of order $m^4$ \cite{Carlitz:1986ng,Preskill:1992tc}. Such a picture arises from a hidden assumption that correlations needs energy to be transferred. The results of \cite{Unruh:2012vd} show, however, that correlations can be transferred by the excitation of quantum degrees of freedom, such as spin, even without a significative energy involved. 

In a non-perturbative QG perspective, this basic idea gives rise to a rather natural interesting scenario \cite{Perez:2014xca}. The remnant would not need to last a long time, but the spacetime would be very rapidly well approximated by flat space, as in the paradigm first presented in \cite{Ashtekar:2005cj}. From a non-perturbative QG point of view flat spacetime is a mean-field approximation of a large degeneracy of fundamental pre-geometric structures. Correlations can therefore be transferred to such QG d.o.f. via the interaction with the matter quantum fields in the regions of high curvature. From a low energy description, a pure-to-mixed state transition would eventually occur, but this breakdown of unitarity is interpreted as a decoherence phenomenon, in line with the analysis of \cite{Unruh:2012vd,Unruh:2017uaw}. The QFT in CST description does not take into account purely QG degrees of freedom that can therefore play the role of a ``hidden bath'' where to store energy-free correlations. Very strong curvature regimes are needed for the QG d.o.f. to be exited. Thus, no loss of unitarity is expected to occur in standard situations. Interesting relations of the points discussed here with the {\em measurement problem} in QM are discussed in \cite{Okon:2017pvc}.

An important distinct line of though I would like to present is the very popular one of considering Assumption~\ref{it:QFT} to be violated before the Planckian regime is reached. This idea actually comes from adding another assumption to the list:
\begin{enumerate}[resume,label=(\Alph*)]
\item The Bekenstein-Hawking entropy $A/4$, where $A$ is the area of the horizon, is interpreted as an upper bound on the density of states,  in the interior of the black hole, of an underlying theory describing all possible physics.\label{it:holog}
\end{enumerate}
I refer to the \hyperref[sec:intro2]{introduction to Part II} for a discussion fully dedicated to the delicate and interesting debate around the nature of Bekenstein-Hawking entropy. For the present discussion it is enough to know that, since there is no common consensus around this assumption, the latter is tacitly not considered in all the scenarios presented above. For instance, the Planckian area of a remnant would not allow all the information needed to purify the external state to be stored in its interior. On the basis of considerations about the amount of correlations across the horizon, it was shown in \cite{PhysRevLett.71.3743} that, if the assumption is true, then information must start coming out from the BH horizon when half of the initial mass has been emitted: the \emph{Page's time}. Late Hawking's evaporation must encode correlations. At the time of complete evaporation the final exterior state would therefore be pure. It is interesting to stress how this idea is conceptually different from the ones described above. The focus here is completely on the fate of the quantum information, and a solution to the paradox is found without ever evoking any consideration about the high curvature region around the would-be-singularity.

It is usually staten that this picture has acquired a privileged status with the introduction of the \emph{$AdS$/CFT correspondence} \cite{Maldacena:1997re} in the framework of String Theory. Within this conjecture, which is briefly discussed in the \hyperref[sec:intro2]{introduction to Part II}, complicated asymptotically $A$nti-$d$e$S$itter ($AdS$) spacetimes eventually involving black holes, are mapped onto corresponding states of a--supersymmetric--conformal quantum field theory (CFT) on the boundary. Since $AdS$ is conformally flat and no breakdown of unitarity is expected in this context, there should not be any loss of unitarity in the bulk spacetime involving BH formation and evaporation neither. At this level of understanding, however, it seems that the statement can also be considered to be in agreement with the decoherence-like scenario discussed above. The departure from unitarity, indeed, was there arising from the interaction of quantum fields with QG d.o.f. in regions of high curvature. Being the $AdS$ boundary far from such regimes, the CFT evolution on it will be, to a very high accuracy, indeed unitary. This would in turn imply a unitary evolution of the bulk theory, which in $AdS$/CFT correspondence is considered to be the most fundamental theory of matter and geometry. If this were true, than QFT in CST would be interpreted as an approximate description of the full bulk theory. The boundary dual theory of such an approximate theory it is not obvious to identify. One would nonetheless expect that in some sense the d.o.f. that are not accounted for in the approximate description provided by QFT in CST would result in d.o.f. that are not accounted for in the dual approximate theory on the boundary. Such approximate theory may not be endowed with a fully unitary dynamics, therefore contradicting the conclusions arising from the scenarios provided above. See also \cite[Sec. 5]{Okon:2017pvc} for a related discussion.

Nevertheless, when these scenarios as considered, since at Page's time the BH is still macroscopic, the expected leakage of correlations to the exterior implies that one must abandon Assumption~\ref{it:QFT} in regions of arbitrarily low curvature. An intense debate has arisen with the aim of proposing a convincing mechanism in which this would happen. Novel intriguing concepts (and disputes) as complementarity \cite{susskind1993stretched}, firewalls \cite{Almheiri2013}, antipodal identification \cite{Hooft:2016itl}, etc have been and keep being proposed. At the present time, however, none of them gives a fully satisfactory answer.

Strong departures from standard physics in regions of low curvature, as the horizon of a large BH, has also been recently proposed from a completely different perspective in \cite{Haggard:2014rza}. The basic hypothesis is that tiny fundamentally quantum gravitational effects would pile up in time resulting in a departure from GR that would allow a quantum gravitational tunnelling from a black hole geometry to a white hole one. The estimated time for the transition to occur is of order $m^2$ \cite{Haggard:2014rza}, being therefore dominant over the Hawking's evaporation. This model is the starting point for the analysis of Chapter~\ref{chap:Fireworks}, to which I refer for more details.

To conclude, it is clear that a complete answer can be provided only by a full consistent theory of quantum gravity, either holographic or not. The hope is that the rapidly growing approaches and ideas, part of which has been mentioned in this Section, would eventually converge toward a shared effective solution to this fascinating puzzle. This, in turn, can provide useful guidelines for our understanding of fundamental physics.
}

%% file: Part_1/Volume/Volume.tex
\chapter{On the Volume Inside Old Black Holes}\label{chap:volumeBH}

\emph{This Chapter completely overlaps with the published paper} \cite{Christodoulou:2016tuu}.\vspace{2ex}

Since the mid-1970s, the information-loss paradox \cite{hawking:1976breakdown} has been at the center of a heated debate. The fate of the large amount of information fallen inside the hole is the main topic of several resolution proposals in the literature (for a --non-exhaustive-- review see \cite{hossenfelder2010conservative} and references therein).

In the setting in which the semi-classical approximation behind Hawking's computation remains valid up to the very late stages of the evaporation, and quantum gravitational effects play an important role only in the strong curvature regime by ``smoothing-out'' the singularity \cite{Ashtekar:2005cj}, a natural possible outcome is the formation of a \emph{remnant}: a final minuscule object that stores all the information needed to purify the external mixed state \cite{Aharonov:1987tp,Giddings:1993vj} (see \cite{Chen:2014jwq} for a recent review). 

The tiny mass and \emph{external} size of such objects are central to objections against both the existence of remnants (infinite pair production--see \cite{Giddings:1993km} and references therein--) and their impossibility of storing inside the large amount of information. The naive intuition of ``smallness'', however, can be very misleading since a remnant contains spatial hypersurfaces of very large volume, see for instance \cite{AbhayILQGS,Perez:2014xca}.

\bigskip 
Once a horizon forms, surfaces of increasingly large volume start to develop. This characteristic is naturally captured by the manifestly coordinate independent definition of volume employing maximal surfaces  recently proposed by Carlo Rovelli together with one of the authors in \cite{Christodoulou:2014yia}, where it was applied to the interior of static black holes \footnote{Other definitions for the volume have been proposed elsewhere \cite{Parikh:2006aa,1742-6596-33-1-044,DiNunno:2008id,Ballik:2010aa,Cvetic:2011aa,Gibbons:2012aa,Ballik:2013aa}.
}.

For asymptotically flat geometries, this volume can be parametrized with the advanced Eddington-Finkelstein time $v$ and is denoted as $V(v)$. In the interior of a static spherically symmetric black hole of mass $m_0$ formed by collapsing matter, the volume grows monotonically with $v$ and is given at late times $v \gg m_0$ by 
\begin{equation} \label{eq:schvolume}
V(v) \approx C\, m_0^2\, v
\end{equation} 
where $C=3 \sqrt{3}\,\pi$ for the uncharged case \footnote{The Reissner-Nordstr\"om spacetime, in which case $C$ depends on the charge $Q$, was studied in the Appendix of \cite{Christodoulou:2014yia} and similar results hold also for AdS black holes \cite{Ong:2015tua}. The Kerr case is considered in \cite{Bengtsson:2015zda}. }.

In this article, we expand upon the results in \cite{Christodoulou:2014yia} and show that the conclusions in that work extend to the case of an evaporating black hole. The volume of maximal surfaces bounded by the shrinking apparent horizon monotonically increases up to when its area has reached Planckian dimensions. Specifically, we show that, at any time, there exists a spacelike maximal surface with proper volume approximately given by \eqref{eq:schvolume} (where $m_0$ is now the initial mass), that connects the sphere of the apparent horizon at that time to the center of the collapsing object before the formation of the singularity \footnote{An argument for the persistence of the large volume in the evaporating case was discussed in \cite{Ong:2015dja}.}. The final remnant hides inside its external Planckian area a volume of order $(m_0/m_P)^5 \, l_P{}^3$.

\bigskip

We first review and clarify some aspects of the discussion given in \cite{Christodoulou:2014yia} and generalize the results presented there so that they may be used in an arbitrary spherically symmetric spacetime. In Section \ref{sec:maxSurAs2Dgeod} and the Appendix, we prove the technical result that finding the spherically symmetric maximal surfaces is equivalent to solving a two dimensional geodesic problem. In Section \ref{sec:definition} we review the definition of volume and discuss the analogy between the Minkowski and the Schwarzschild case in order to illustrate its geometric meaning. In Section \ref{sec:remnantvolume} we examine the evaporating case and calculate the volume enclosed in the horizon as a function of time at infinity. We close with a discussion on the physical relevance of our result with respect to the debate on the fate of information in evaporating black holes.

\section[Maximal Surfaces as a $1+1$ Geodesic Problem]{Maximal Surfaces as a $1+1$ Geodesic \\Problem} \label{sec:maxSurAs2Dgeod}

A general spherically symmetric spacetime can be described by a line element
\be
\ddd s^2 = g_{\alpha \beta}\ddd x^\alpha \ddd x^\beta = g_{A B}\ddd x^A \ddd x^B + r^2 \ddd \Omega^2 
\ee 
with $\ddd \Omega^2 = \sin^2\!\theta\, \ddd \phi^2 + \ddd \theta^2 $. We use the notation $x^\alpha = \{x^0,r,\theta,\phi\}$ and $x^A = \{x^0,r\}$.

Spherically symmetric hypersurfaces $\Sigma$ can be parametrically defined via a coordinate $\lambda$: 
\be
\ddd s^2_\Sigma =  (g_{A B} \dot{x}^A \dot{x}^B) \,\ddd \lambda^2 + r^2 \ddd \Omega^2 
\ee 
where $x^A = x^A(\lambda)$ and $\dot{x}^A \equiv \frac{\ddd}{\ddd \lambda}  x^A \!(\lambda)$. We have $\Sigma \sim \gamma \times S^2$, with $\gamma : \lambda \rightarrow x^A(\lambda)$ being a curve in the $x^0\textit{-}\, r$ plane. We denote as $y^a=\{\lambda,\phi,\theta\}$ and $h_{ab} = e^\alpha_a e^\beta_b g_{\alpha \beta}$ the coordinates and the induced metric on $\Sigma$ respectively, where $e^\alpha_a = \frac{\partial x^\alpha}{\partial y^a}$ provides a basis of tangent vectors on $\Sigma$.

We look for the stationary points of the volume functional: 
\begin{eqnarray}
V[\Sigma] &=& \int_\Sigma\!\ddd y^3 \sqrt{\det h_{ab}} \nonumber \\ 
&=& 4 \pi \int_{\gamma}\ddd \! \lambda \left(r^4 g_{A B} \dot{x}^A \dot{x}^B \right)^{1/2}  \nonumber \\
&=& 4 \pi \int_{\gamma} \ddd \! \lambda \left(\tilde{g}_{A B}  \dot{x}^A \dot{x}^B\right)^{1/2} \;,
\end{eqnarray}
where $\Sigma$ are spherically symmetric surfaces bounded by a given sphere $\partial \Sigma$.

Thus, the extremization of $V[\Sigma]$ is equivalent to the 2D geodesic problem for the auxiliary metric $\tilde{g}_{A B}= r^4 \, g_{A B}$. That is, $\gamma$ is a solution of 
\be \label{eq:geodEq}
\dot{x}^A \tilde{\nabla}_A \dot{x}^B = e^A_\lambda \tilde{\nabla}_A e^B_\lambda = 0    
\ee 
where $\tilde{\nabla}$ is the covariant derivative in $\tilde{g}_{A B}$ and $\lambda$ has been chosen to be an affine parameter on $\gamma$ with respect to $\tilde{g}_{A B}$. 

\bigskip

The stationary points of $V[\Sigma]$ solve the ``Plateau's problem'' or ``isoperimetric problem'' for $\partial \Sigma$. In a Euclidean context these are local minima, while in the Lorentzian context they are local maxima. It is simple to show that if the trace $K=K_{\alpha \beta} g^{\alpha \beta}$ of the extrinsic curvature of a hypersurface vanishes, the variation of the volume functional is automatically zero (see for instance \cite{Baumgart}). For this reason, in the Lorentzian context, surfaces with $K=0$ are called maximal surfaces. 

It is the authors understanding that a general proof of the opposite statement, namely that for arbitrary spacetimes extremizing $V[\Sigma]$ for a given $\partial \Sigma$ yields $K=0$ surfaces, is missing. Several precise proofs exist in the mathematical relativity literature (see for instance the seminal papers \cite{Choquet1979,Marsden1980}), that typically rely on energy conditions or other restrictions on the metric or on the surfaces. ``Physicist'' demonstrations can be found in the $3+1$ literature \cite{Gourgoulhon,Baumgart}. 

For completeness, we prove in the Appendix that, for an arbitrary metric $g_{AB}$, any surface $\Sigma \sim \gamma \times S^2$, with $\gamma$ being a solution of \eqref{eq:geodEq}, has $K=0$. From well known theorems about the geodesic equation, this also guarantees the local existence of maximal surfaces, see also \cite{CorderoCarrion:2011rf}. 

\vspace{1ex}
\centerline{\rule{2cm}{1.5pt}}
\vspace{1.5ex}

The physical relevance of maximal surfaces has long been recognised in diverse disciplines ranging from problems in mathematical physics \cite{Rassias} to architecture and the beautiful tensile structures of Frei Otto \cite{FreiOtto}. In general relativity, their usefulness for numerically solving Einstein's equations is reflected in the popular ``maximal slicing'' \footnote{The family of surfaces discussed in the next section includes the surfaces used for maximal slicing, but keep in mind that we do not restrict ourselves to surfaces satisfying the ``singularity avoidance'' or the ``nowhere-null'' condition. In fact, half of each family of $K=0$ surfaces we will study end at the singularity and become null there.} (see for instance \cite{Gourgoulhon} and references therein), which in a sense generalizes the slicing of a Newtonian spacetime by constant (absolute) time surfaces. 

Common notions of volume implicitly use maximal surfaces. These include the everyday meaning of volume, the special relativistic proper volume and the volume of the Universe, where the latter habitually refers to the proper volume of the $t=const.$ surfaces of the Friedmann-Robertson-Walker metric: spherically symmetric maximal surfaces. 

\section{Review of the Volume Definition} \label{sec:definition}
\begin{figure*}[ht]
\centering
\subfigure{
\centering
    \includegraphics[width=0.41\textwidth, keepaspectratio]{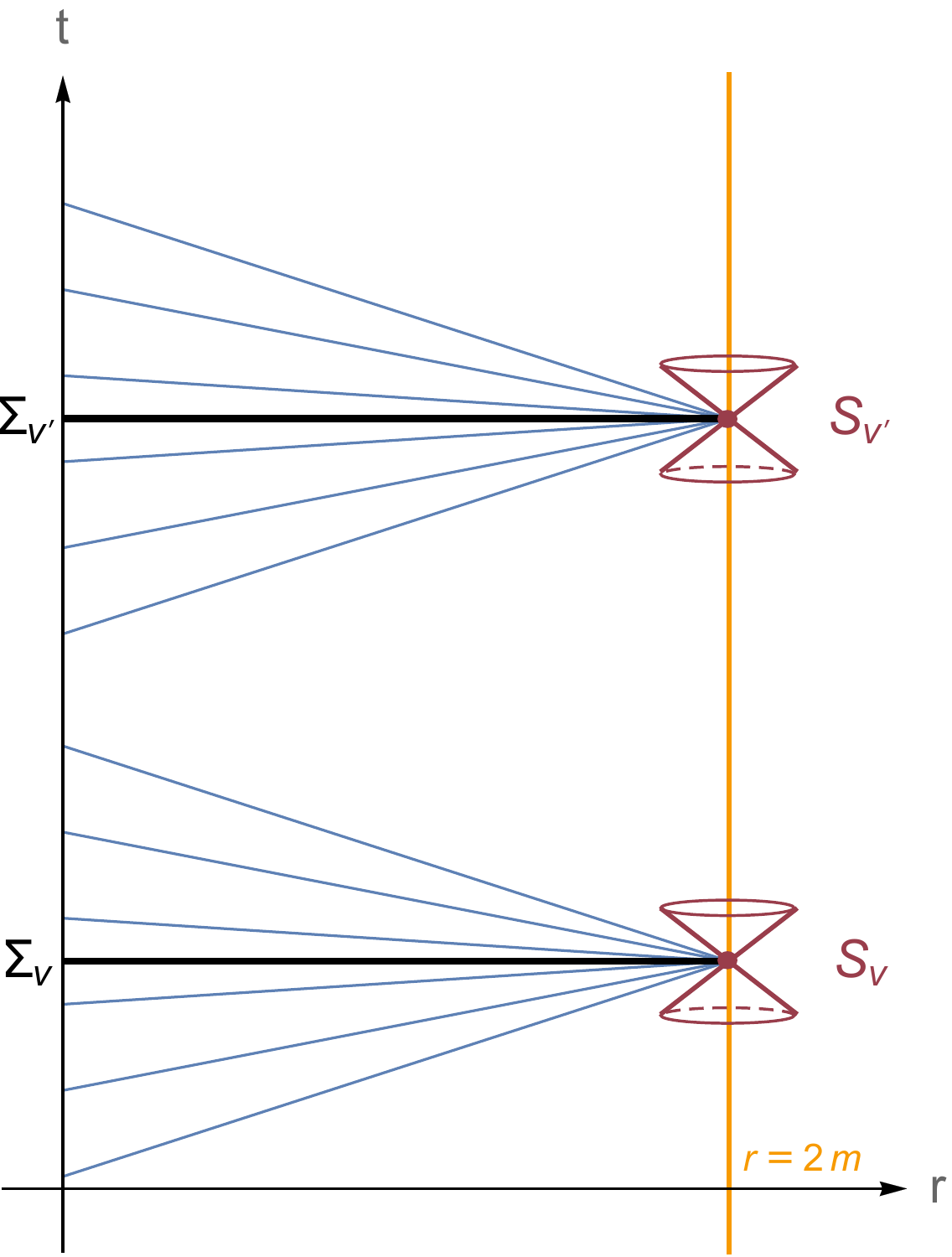}
    \label{fig:Minkowski}
}\qquad\qquad
\subfigure{
\centering
    \includegraphics[width=0.41\textwidth, keepaspectratio]{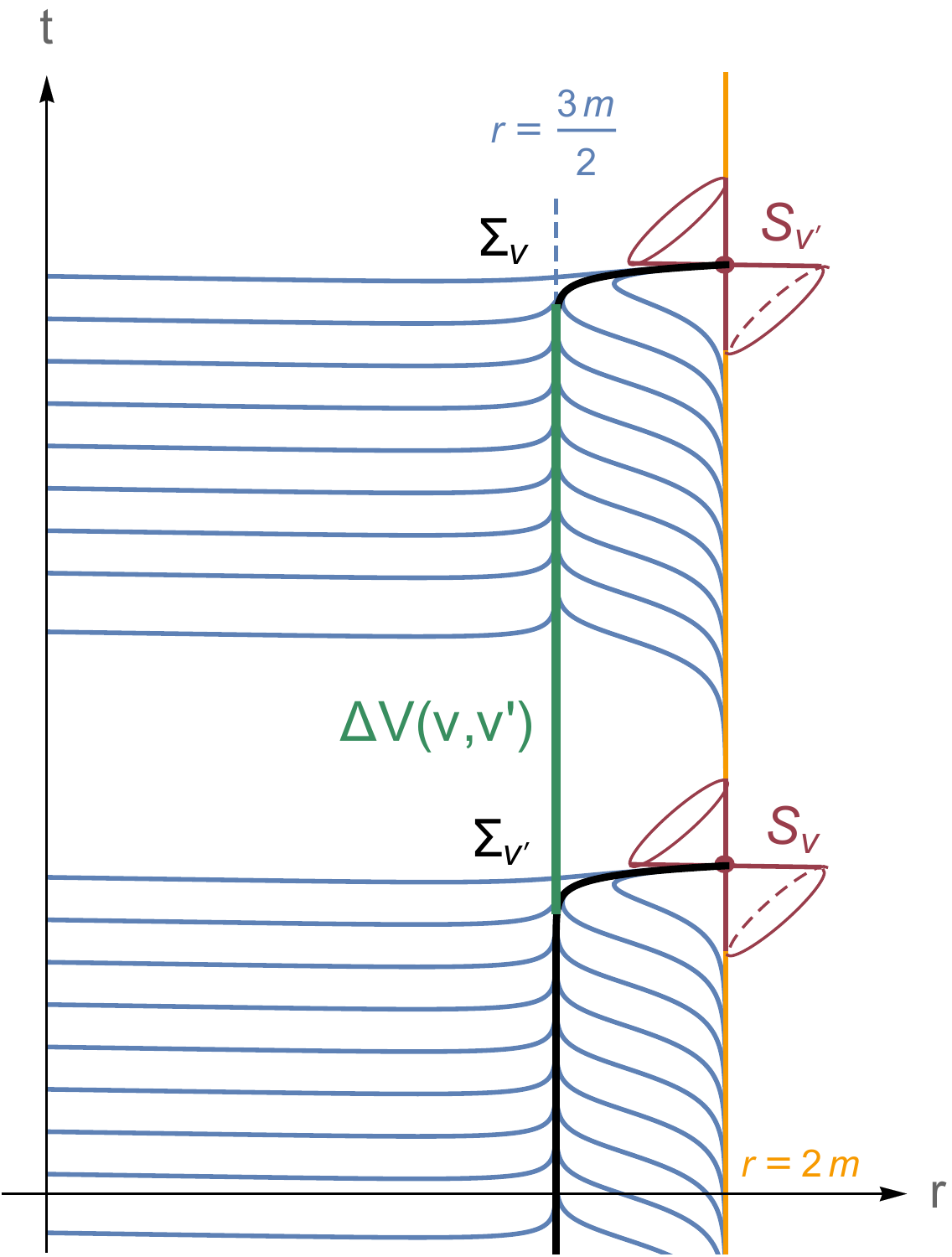}
\label{fig:Schwar}
}
\caption{{\bf Left:} Maximal surfaces (blue lines) inside a two-sphere in flat Minkowski spacetime. The largest is the $t={\rm const.}$ (bold black lines) defining its inertial frame. {\bf Right:} Maximal surfaces (blue lines) inside a two-sphere on the horizon of a static black hole. Apart from the transient part connecting it to the horizon, the largest surface (bold black lines) lies on the limiting surface $r=3/2m$. The volume \emph{difference} between the spheres $S_v$ and $S_{v'}$ is finite and given by \eqref{eq:SchVac}.}
\label{fig:mink_VS_schw}
\end{figure*}
    The volume definition given in \cite{Christodoulou:2014yia} can be stated as follows: \emph{the volume inside a sphere $S$ is defined as the proper volume of the maximal spherically symmetric surface $\Sigma$ bounded by $S$, which has the largest volume amongst all such $\Sigma$.} Note that this is a geometric statement and as such it is manifestly generally covariant.

\begin{figure}[t]
\centering
\includegraphics[width=0.4\textwidth]{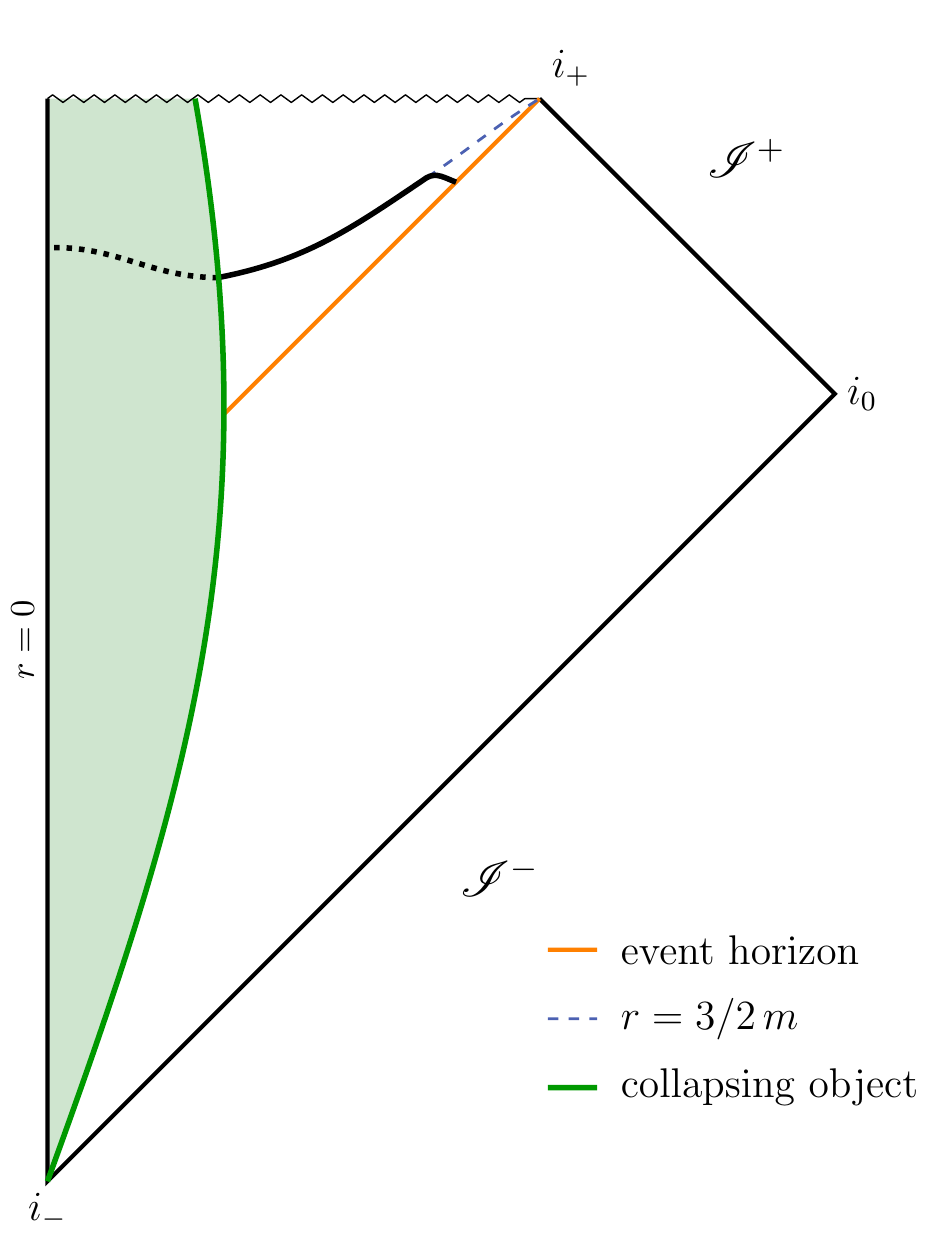}
\caption{Penrose diagram illustrating the surface defining the volume (black curve) in the case of a black hole formed by collapse. The details of the surface in the interior of the collapsing object (dotted curve) will depend on the specific metric use to describe the latter. For Oppenheimer-Snyder and null massive shell collapses, this contribution to the volume is of the order $m^3$.}
\label{fig:collapse}
\end{figure}

 In order to illustrate its geometric meaning, we examine in the rest of this section the analogy between the maximal surfaces of Minkowski spacetime and those of the Schwarzschild solution. The discussion is summarized in Figures \ref{fig:mink_VS_schw} and \ref{fig:collapse}.

 Using the advanced time $v=t+ \int \frac{\ddd r}{f(r)}$, the geometry of the two spacetimes is described by 
\be \label{eq:vacLineEl}
\ddd s^2= -f(r) \ddd v^2+  2 \ddd v \ddd r +r^2 \ddd \Omega ^2\;,
\ee 
with $f(r)=1$ and $f(r)=1- 2m/r$ respectively. Consider the sphere $S_v$ defined as the intersection of $r=2m$ and the ingoing radial null ray of constant $v$. It bounds a family of maximal surfaces, the solutions of \eqref{eq:geodEq} for different initial speeds. 

In Minkowski, these are the simultaneity surfaces of inertial observers, which are straight lines in the $t\textit{-}r$ plane. The one with the biggest volume, $\Sigma_v$, is that which defines the inertial frame of $S_v$. Its proper volume is what we call \emph{the} proper volume in special relativity; that is, $V_{\Sigma_v}=\frac43 \pi (2m)^3$. 

In Schwarzschild geometry, the maximal surfaces starting from $S_v$ approach the surface $r=3/2m$ (because of this behavior, $r=3/2m$ will be called ``limiting surface''), and become null either when they reach the singularity or when they asymptotically approach the horizon, except one that asymptotically becomes $r=3m/2$ \footnote{The existence of the limiting surface $r=3/2m$ was first pointed out in \cite{Estabrook:1973ue}. It is crucial for the singularity avoidance property of the maximal slicing, which is in fact comprised by the $\Sigma_v$ extended to infinity. Similar elongated surfaces are studied in numerical relativity \cite{Hannam:2006xw,Baumgarte:2007ht} and have been dubbed ``trumpet geometries'' \cite{Dennison:2014eta}. }. The proper volume of this surface is infinite.

This is a characteristic difference between the two geometries which underlines the common understanding that ``space and time exchange roles inside the hole''. Inside the sphere containing flat space, there are radial timelike curves of infinite length, while all radial spacelike curves have proper length at most equal to the radius of the sphere. Inside a black hole this is reversed: there are radial spacelike curves of infinite length, while radial timelike curves have proper time at most equal to $\pi m$. 

In the physical case of non-eternal black holes formed by collapse, the surface $\Sigma_v$ does not have infinite volume since it does not extend infinitely along $r=3/2m$. In fact, it connects the sphere at the horizon $S_v$ with the center of the collapsing object before the formation of the singularity, see Fig.~\ref{fig:collapse}. The surface in its interior will be given by solving \eqref{eq:geodEq} for the interior metric. For a collapse modeled by a null massive shell or \emph{\`a la} Oppenheimer-Snyder \cite{PhysRev.56.455}, the contribution to $V(v)$ will be of the order of that of the flat sphere $\sim m^3$. At late times $v >> m$, this contribution is negligible with respect to the one given by the main part lying on $r=3/2m$, and the volume is given by $\eqref{eq:schvolume}$.

This characteristic monotonic behaviour is perhaps best understood by extending the definition to the case of an eternal black hole. In this case we consider the volume \emph{difference} $\Delta V(v,v')$ between two spheres $S_v$ and $S_{v'}$ labeled by different times at infinity, in analogy to considering the proper time between any two points on a timelike curve that otherwise extends to arbitrary values of its affine parameter. 
   
In Minkowski, this difference is zero: the proper volume of the sphere of fixed radius remains constant. In Schwarzschild, by the translation invariance inside the horizon, $\Delta V(v,v')$ is given by the volume of the part of $\Sigma_{v'}$ that lies on the limiting surface $r=3m/2$ and does not overlap with $\Sigma_v$. \emph{Thus, this difference is finite, monotonically increasing and given by} 
\be \label{eq:SchVac}
\Delta V(v,v') = 3 \sqrt{3}\, \pi \, m^2 \, (v'-v)\;.
\ee
Notice that the result for a black hole formed by collapse, eq.~\eqref{eq:schvolume}, is nothing but the approximate version of the above equation with $v=0$.

The analysis presented in this section can be nicely extended to the case of an evaporating black hole to which we now turn our attention. 

\section[The Volume of an Evaporating Black Hole]{The Volume of an Evaporating Black Hole}\label{sec:remnantvolume}
The spacetime of an evaporating spherically symmetric black hole can be described by the Vaidya metric \cite{Vaidya:1951zz}, given by replacing $f(r)$ in \eqref{eq:vacLineEl} with $f(r,v)=1 - 2m(v)/r$. For our purposes it is sufficient to model the formation of the hole by the collapse of an ingoing null shell at the retarded time $v=0$, and the loss of mass due to evaporation by integrating the thermal power emission law \cite{hawking1974black}. The resulting mass function is
\be\label{eq:mv}
m(v) = \Theta(v) \big(m_0^3-3B\, v\big)^{1/3} \;,
\ee
where $\Theta(v)$ is the step function, $B \sim 10^{-3}$ a parameter that corrects for back reaction \cite{PhysRevD.52.5857} and $m_0$ the mass of the shell. The spacetime has a shrinking timelike apparent horizon given by $r_{H}(v) = 2 m(v)$.

\begin{figure}[t]
\centering
\includegraphics[width=0.4\textwidth]{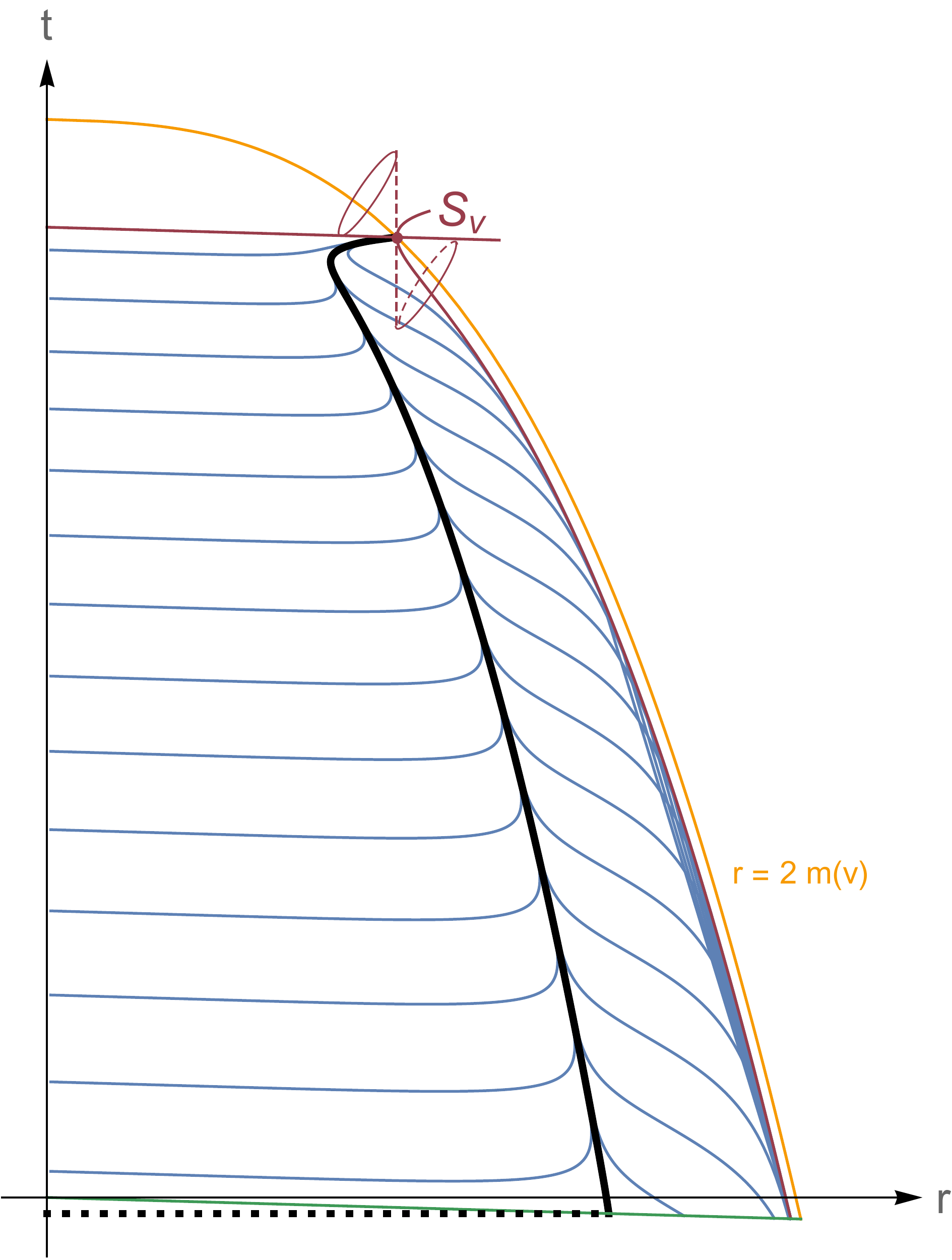}
\caption{Eddington-Finkelstein diagram of the two families of extremal volume surfaces (blue lines) inside an evaporating black hole formed by a collapsing object. The surface defining the volume is in bold black. Note the close analogy with the static case, compare with Fig.~\ref{fig:mink_VS_schw}.}
\label{fig:evaporating}
\end{figure}

By numerically solving \eqref{eq:geodEq}, we can draw the family of maximal surfaces for the spheres at the apparent horizon for different $v$. The situation, depicted in Figure \ref{fig:evaporating}, is in direct analogy with the non-evaporating case. There is again a limiting surface, persisting up to very late stages of the evaporation. Thus, as in the static case, the volume of the biggest maximal surface $\Sigma_v$ inside $S_v$ is the one connecting the latter to the center of the collapsing shell.

\begin{figure}[t]
\centering
\includegraphics[width=0.4\textwidth]{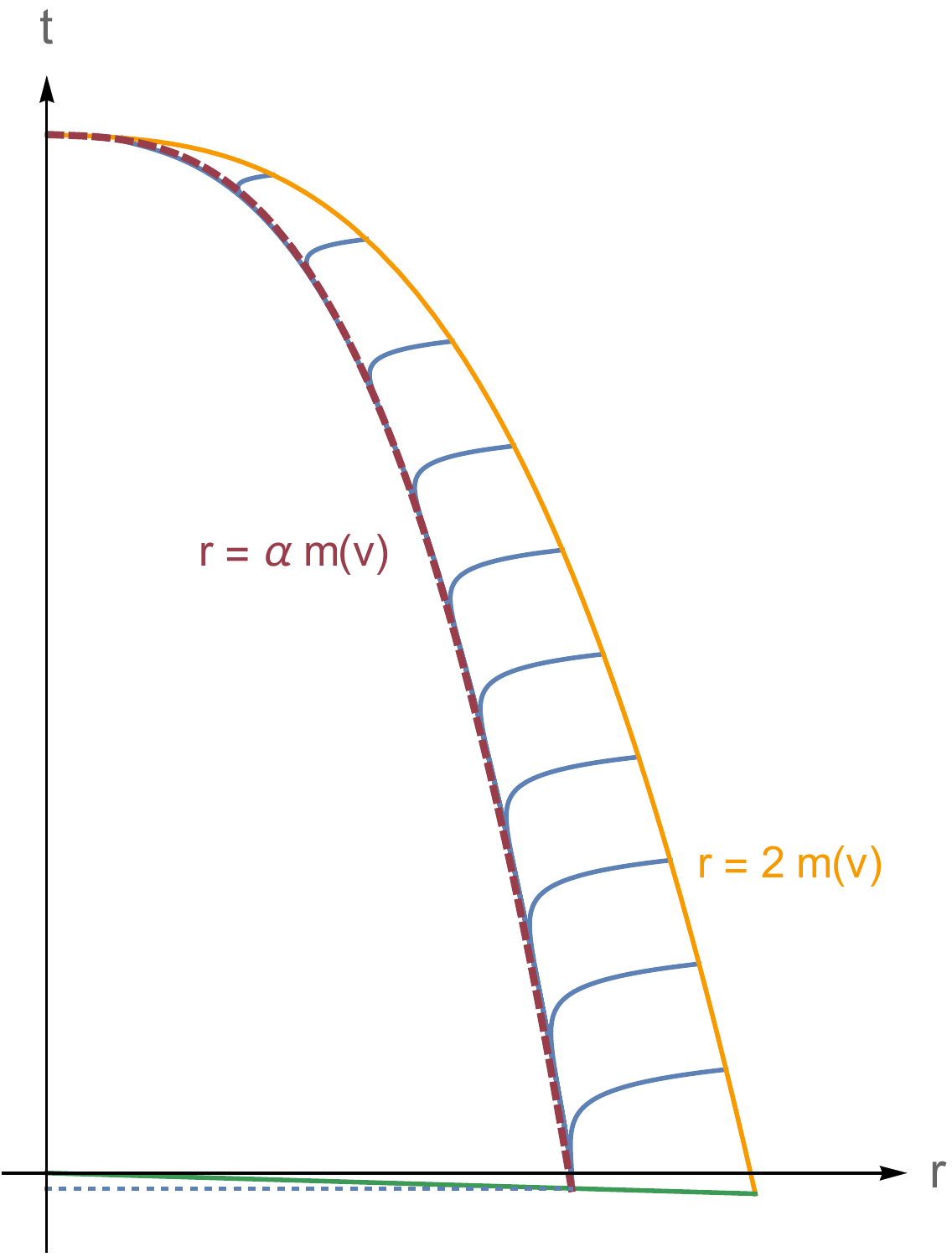}
\caption{The surfaces defining the volume enclosed in spheres at the apparent horizon of an evaporating black hole at different times (blue lines). The limiting surface lies close to $r=\alpha\,m(v)$, with $\alpha$ given by \eqref{eq:alpha} (dashed line). Here $m=10$ in Planck units.}
\label{fig:complete}
\end{figure}

We may get an estimate for the volume as a function of time and the initial mass as follows: we compute the volume of a surface $r=\alpha\, m(v)$ and find the $\alpha$ for which this is maximized:
\be\label{eq:alpha}
\alpha = \frac32 - \frac{45\, B}{8\, m_0^2} + O\left(\frac{1}{m_0^4}\right)\;.
\ee
Indeed, the limiting surface is very well approximated by $r=\alpha\, m(v)$ even for low masses, see Fig.~\ref{fig:complete}. 

Expanding the volume of $r=\alpha\, m(v)$ to leading order in $1/m_0$ we get: 
\be\label{eq:volapprox}
V(v) \approx 3\sqrt{3}\,\pi \,m_0 ^2 \,v \,\left( 1- \frac{9 \,B}{2\,m_0^2}\right) \;.
\ee

Thus, for large masses, we have again recovered \eqref{eq:schvolume}. 

A direct calculation shows that the surface $r(v) = \alpha\,m(v)$ ceases to be spacelike when the mass function takes the value
\begin{equation}
m \approx \left(3 \sqrt{B} - \frac{225\,B^{3/2}}{8\,m_0^2}\right) \,m_P < m_P / 10 \;.
\end{equation}
This provides an estimate for the regime of validity of eq.~\eqref{eq:volapprox}. Interestingly, the non-existence of large spacelike maximal surfaces appears to coincide with the regime in which the mass has become Planckian. 
These estimates agree with the numerical investigation of the actual surfaces, see Fig.~\ref{fig:complete}. We conclude that the volume increases monotonically, following the approximate behavior given in \eqref{eq:volapprox}, up to when its external area becomes Planckian. At this very late time, the internal volume is of order $m_0^5$ in Planck units.

\vspace{1ex}
\centerline{\rule{2cm}{1.5pt}}
\vspace{1.5ex}

Intuitively, the picture is the following: from the perspective of the maximal surfaces, collapse and horizon \textit{at any subsequent exterior time} are simultaneous, see Fig.~\ref{fig:complete}.
The exterior elapsed time corresponds inside the hole to the stretching of space, as given by \eqref{eq:schvolume}.
 
\subsection*{A few numbers}
Before closing this section, let us put the above in perspective: when a solar mass ($10^{30} \, kg$) black hole becomes Planckian (it needs $10^{55}$ times the actual age of the Universe), it will contain volumes equivalent to $10^5$ times our observable Universe, hidden behind a Planckian area ($10^{-70} \,m^2$). 

Perhaps more pertinent is to consider small primordial black holes with mass less than $10^{12}\, kg$. Their initial horizon radius and volume are of the order of the proton charge radius ($10^{-15}m$) and volume ($10^{-45}m^3$) respectively. They would be in the final stages of evaporation now, hiding volumes of about one liter ($10^{-3}m^3$).

\section[Remnants and the Information Paradox] {Remnants and the Information Paradox}\label{sec:discussion}

As was briefly discussed in the introduction, the results presented above can be relevant in the discussion about the loss-of-information paradox, particularly in the context of scenarios that assume the semiclassical analysis of quantum field theory on curved spacetimes to be valid in regions of low curvature and until near-complete radiation of the initial mass \footnote{Another potential application of this result is in black hole thermodynamics in view of recent results on the Von Neumann entropy associated to volumes \cite{Astuti:2016dmk}.}. Such scenarios disregard the possibility of having information being carried out of the hole by the late Hawking photons \cite{PhysRevLett.71.3743,Braunstein:2009my}, avoiding the recent firewall and complementarity debate \cite{Almheiri2013}. Another alternative that has recently aroused interest and is not considered here, is that a black hole may end its lifetime much earlier than near-complete evaporation by tunneling to a white-hole geometry. This is possible thanks to quantum gravitational effects that, due to the long times involved, can become important in low curvature regions outside the horizon \cite{Haggard:2014rza,DeLorenzo:2015gtx,Christodoulou:2016vny} \footnote{An alternative scenario in which this process happens must faster by assuming faster-than-light propagation of a shock-wave from the bounce region is considered in \cite{Barcelo:2014npa,Barcelo:2015noa}.}. 

Consider then the setting in which the semi-classical approximation behind Hawking's computation remains valid up to the very end of the evaporation. The hole will completely evaporate and the information will unavoidably be lost, as originally suggested by Hawking \cite{hawking:1976breakdown}. While it seems intuitively reasonable for what appears to be a tiny object to decay away and disappear, it is compelling to ask what became of the macroscopic region inside.

Conversely, consider the additional hypothesis that quantum gravitational effects play an important role in the strong curvature regime by ``smoothing-out'' the singularity \cite{Ashtekar:2005cj}. When the mass becomes Planckian, the semi-classical approximation underlying Hawking's computation fails and the evaporation stops (see for instance \cite{Adler:2001vs}). The hole does not completely disappear and one can consider the possibility of having a minuscule object that stores all the information needed to purify the external mixed state: a remnant \cite{Aharonov:1987tp,Giddings:1993vj,Chen:2014jwq}. 

Standard objections against the remnant scenario such as the infinite pair production \cite{Giddings:1993km} and their impossibility in storing inside a large amount of information, rely on considering the remnant as a small object. Our result shows that the remnant is instead better understood as the small throat of an immense internal region, with a volume of the order of $m_0^5$. General Relativity naturally gives a ``bag of gold'' type description of the interior of a remnant, without the need of ad-hoc spacetimes that involve some ``gluing'' of geometries \cite{wheeler1964relativity,Hsu:2007dr}. Notice that the result of the previous section is insensitive to the details of the \emph{would-be-singularity} region since the limiting surface is in a relatively low-curvature region. 

In \cite{hossenfelder2010conservative,AbhayILQGS,Perez:2014xca} the authors suggest that a large available internal space could store a sufficient amount of very long wavelength modes that carry all the information needed to purify the external mixed state, albeit the available energy being of the order of a few Planck masses. The surfaces studied here are good candidates on which this idea could be tested \footnote{In \cite{Zhang:2015gda} it is argued that these surfaces do not store enough information for purification. However, in that work the Hawking temperature is assumed constant. The computed information is therefore the one stored in a static black hole, and it is not pertinent to this discussion.}. The details of the mechanism by which information would be stored have not, to our knowledge, been made precise; demonstrating this possibility is beyond the aim of this work and, in what follows, we assume this to be possible. 
 
We can identify two characteristically distinct possibilities for the evolution of the large interior region. The bulk of these large surfaces is causally disconnected from their bounding sphere on the horizon \cite{Bengtsson:2015zda}. They can remain causally disconnected from the rest of the spacetime, which may lead to a baby universe scenario \cite{Frolov:1988vj,Frolov:1989pf}. 

\begin{figure}[!t]
\centering
\includegraphics[width=0.4\textwidth]{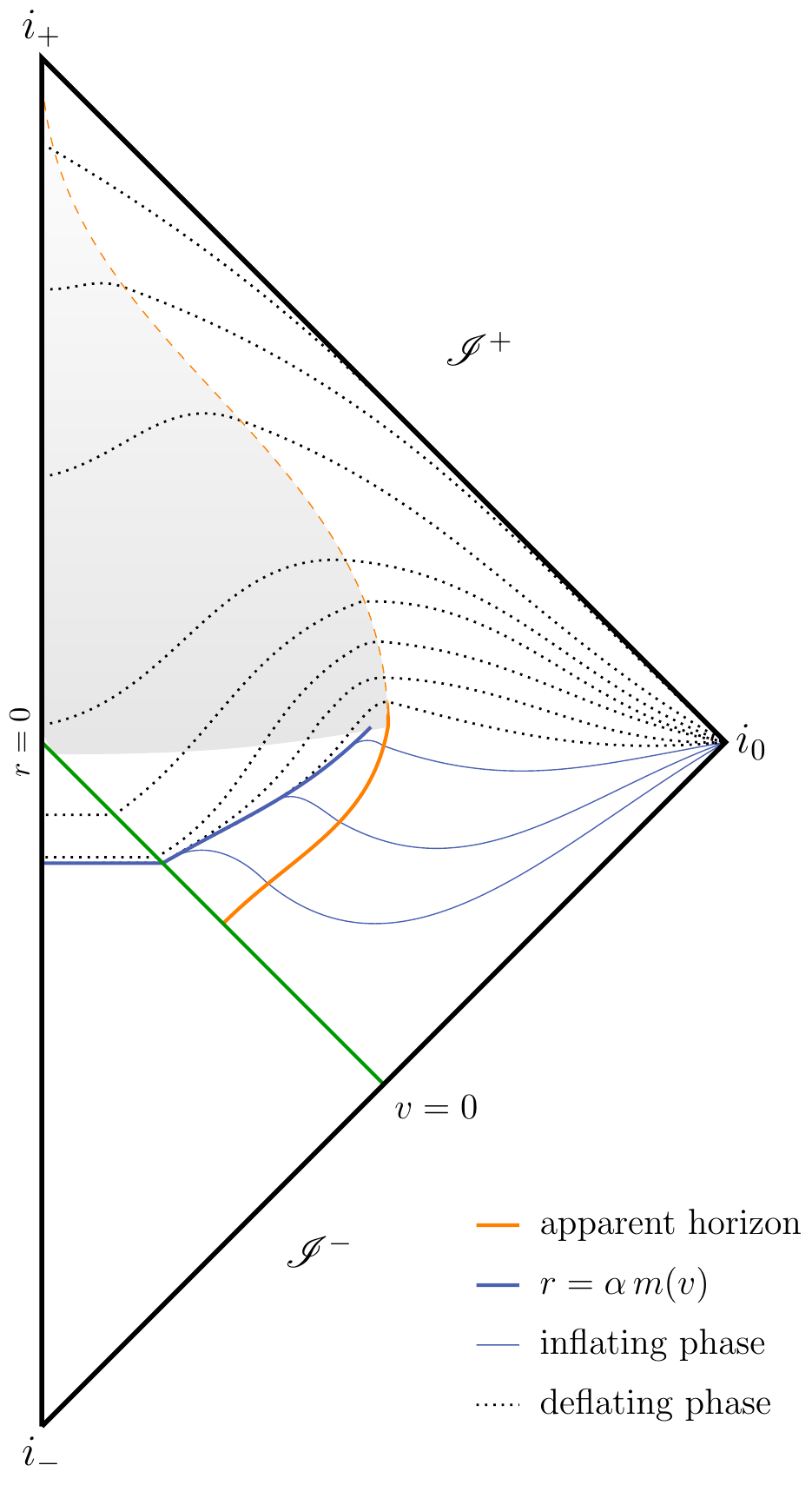}
\caption{Speculative evolution of maximal surfaces in the case of a long-lived remnant scenario. The volume acquired during the evaporation process (continuous surfaces) deflates back to flat space (dotted surfaces). This is expected to happen in a time of order $m_0^4$, during which all the information stored can be released.}
\label{fig:conjecture}
\end{figure}

On the contrary, quantum gravitational effects can modify the (effective) metric and bring these regions back to causal contact with the exterior, while deflating their volume, allowing for the emission of the purifying information to infinity (the information could also be coded in correlations with the fundamental pre-geometric structures of quantum gravity, as proposed in \cite{Perez:2014xca}). This scenario, where the inflating phase is followed by a slow deflating phase of the remnant, is sketched in Fig.~\ref{fig:conjecture}.

We expect this deflating process to be slow, in accordance with bounds on the purification time \cite{Carlitz:1986ng,Bianchi:2014bma} and the lifetime of long-lived emitting remnants, estimated to be of order $m_0^4$. The latter scenario can be made precise by constructing an effective metric describing this process through the evolution of maximal surfaces in the sense of Fig.~\ref{fig:conjecture}. It then suffices to numerically solve equation \eqref{eq:geodEq} in order to study the evolution.


%% file: Part_1/Fireworks/Fireworks.tex
\chapter[Improved Black Hole Fireworks]{Improved Black Hole Fireworks: Asymmetric Black-Hole-to-White-Hole Tunneling Scenario}\label{chap:Fireworks}

\emph{This Chapter completely overlaps with the published paper} \cite{DeLorenzo:2015gtx}.\vspace{2ex}

Regular collapse models where the black hole singularity is replaced by some smooth geometry have a long history \cite{bardeen1968non, Frolov:1981mz, Roman:1983zza,Casadio:1998yr, AyonBeato:2000zs,Mazur:2001fv, dymnikova2002cosmological,Ashtekar:2005cj,hayward2006formation,ModestoLQGBH, Visser:2009pw,Modesto:2010rv,LitimASBH,BambiModestoKerr,Bambi:2013caa,Rovelli:2014cta,Frolov:BHclosed, Mersini-Houghton:2014yq,ModestoNonLocalBH,DeLorenzo:2014pta,Saueressig:2015xua,Frolov:2015bta,DeLorenzo:2015taa}. The {\em leitmotiv} of these models is the attempt to understand issues related to the Hawking information loss paradox
on an effective background spacetime capturing  the idea that black hole singularities must be resolved by quantum gravity effects.
Ideally one would want to justify the relevant physical features of these models in terms of a fundamental quantum theoretical description.
Lacking a precise dynamical description of quantum gravity, their key features are often justified in terms of generic behaviour  that leads to singularity avoidance 
in simplified symmetry reduced models of quantum gravity \cite{Ashtekar:2003hd, Ashtekar:2011ni, Gambini:2013ooa, Gambini:2014qga}.  As one would also expect QFT on curved spacetimes  
to be a valid approximation to quantum dynamics in regions where the gravitational degrees of freedom are well described by a classical background metric of low curvature in Planck units, valuable insights should be accessible through semiclassical methods.
Along these lines a necessary viability criterion for these models is their {\em semiclassical stability}: contributions of quantum fluctuations of a test field in suitable quantum states\footnote{Those satisfying the correct 
boundary conditions that define gravitational collapse.} to the expectation value of the energy momentum tensor must remain small (in Planck units) in semiclassical regions.  
In this article we study the {\em semiclassical stability} of the recently introduced bouncing black hole model proposed in \cite{Haggard:2014rza}.\footnote{A similar scenario in which the same bouncing process happens in much shorter timescales by assuming faster-than-light propagation of a shock-wave from the bounce region is considered in \cite{Barcelo:2014npa,Barcelo:2015noa}.} We find the model to be strongly unstable under small perturbations and consequently we propose a simple but nontrivial modification that avoids these instabilities without modifying the key features of the original idea. 

The paper is organized as follows. In Section \ref{sec:halcarlomodel} we review the definition of the fireworks model.
In Section \ref{scs} we study the semiclassical stability of the fireworks spacetime by computing the expectation value of the energy momentum tensor in a suitable state of a quantum test field on that background.  In order to produce analytic expressions, and thus make clearer our presentation, we will assume that our quantum test field is a massless scalar field and those calculations on the Schwarzschild background will be first done in the approximation where back-reaction is neglected; see Section \ref{next}. We will argue at the end of this section that the result remains valid in the $3+1$ framework where backscattering is taken into account. In Section \ref{corr} we propose a way in which the background of \cite{Haggard:2014rza} could be modified in order to avoid these instabilities as well as other ones described in Section \ref{babacul}. The new model is a time-asymmetric version of the original one, where the black hole phase is followed by an extremely fast explosion with time scale shorter than $m\log m$ in Planck units. Finally, we discuss the implications of such modifications in Section \ref{sw}.

\section{The Fireworks Model}\label{sec:halcarlomodel}
\begin{figure}[t]
\center
\includegraphics[width=0.6\textwidth]{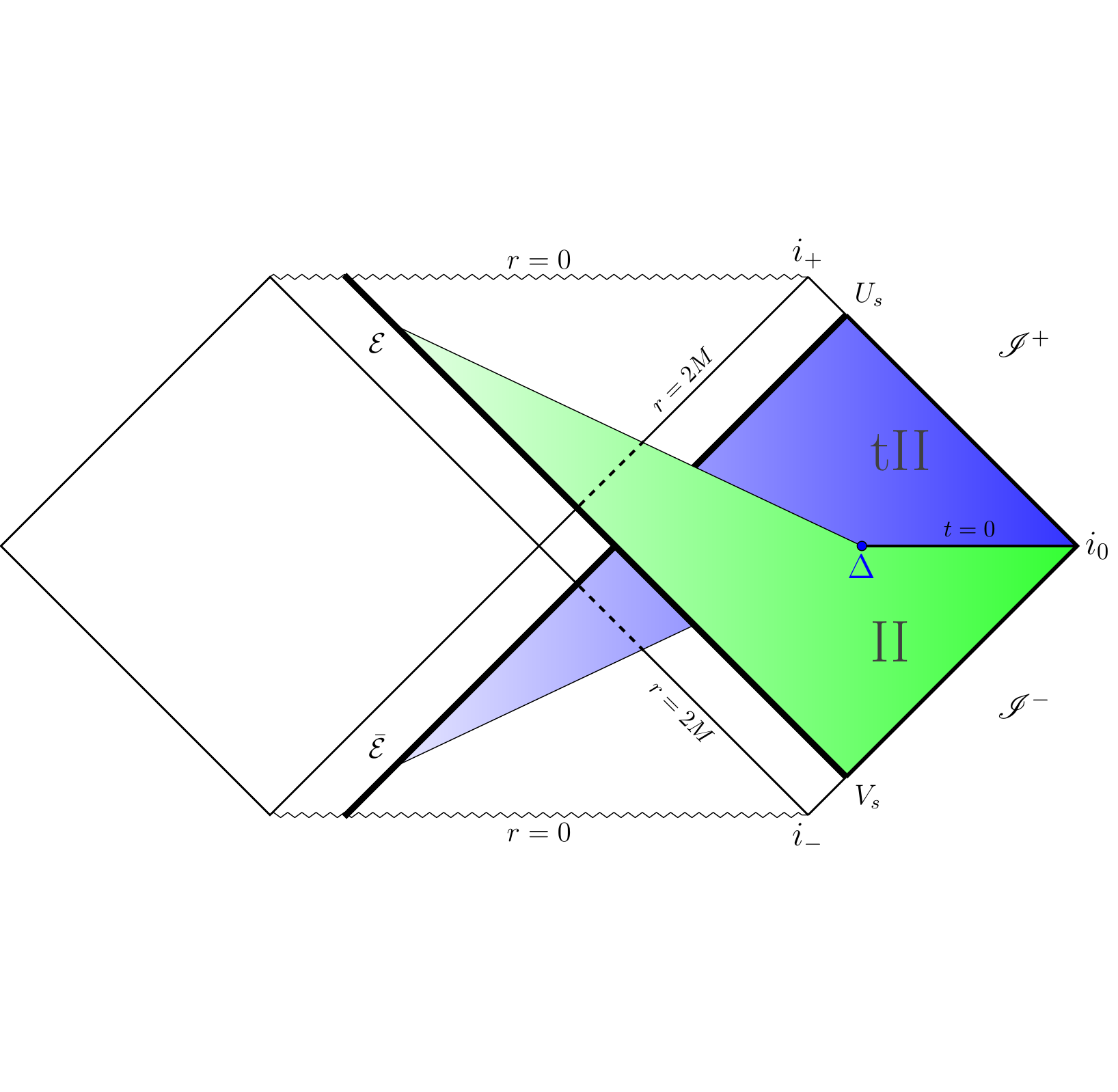} \hfill
\includegraphics[width=0.3\textwidth]{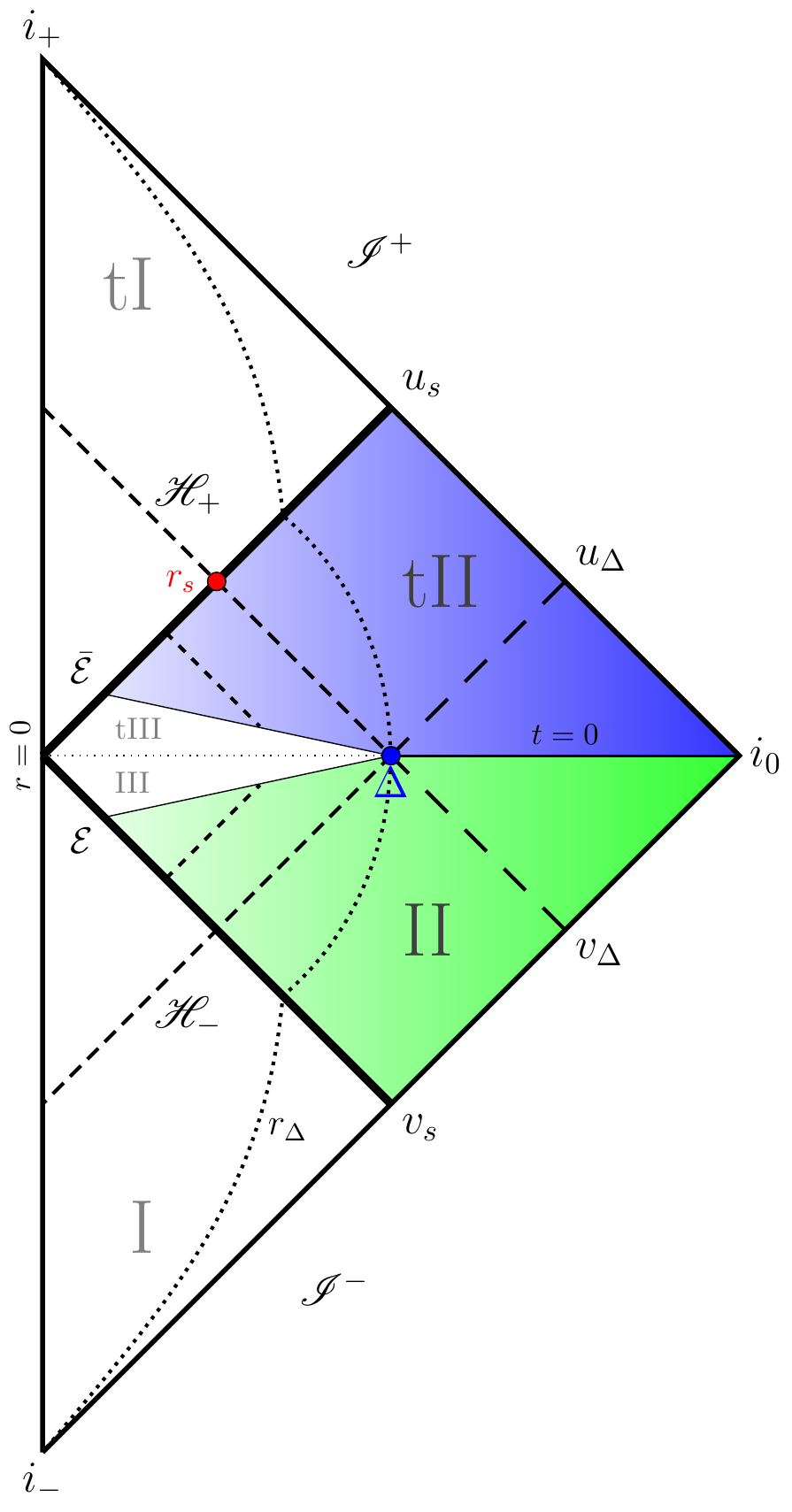}
\caption{Geometry of the black hole fireworks scenario. {\bf Left:} Kruskal-Szekeres diagram, with the two interesting overlapping regions shaded with different colors. {\bf Right:} The resulting completely time-symmetric Carter-Penrose diagram.}
\label{fig:penroseA}
\end{figure}

The Penrose diagram of the Haggard-Rovelli \cite{Haggard:2014rza} proposal for a bouncing black hole is shown in Fig.~\ref{fig:penroseA}. 
This spacetime corresponds to the collapse of a spherical shell of mass $m$, and it is constructed in terms of patches that are isometric to the Schwalzschild, Minkowski, and an unspecified quantum effective geometry glued together through transition hypersurfaces. In the last region Einstein's equations are not satisfied with any form of classical matter; its presence is interpreted as a modification of the classical dynamics induced by the effect of quantum gravity fluctuations. 

The model can be obtained from the cutting and pasting of regions easily identified in the Penrose diagram of the maximally extended Schwarzschild solution of mass $m$ as follows:  
One first identifies a point $\Delta$ with Kruskal-Szekeres coordinates $(U_\Delta=-V_\Delta, V_\Delta)$ with $V_\Delta>0$ so that $\Delta$ lies in the exterior of the white as well as the black hole regions. One then chooses a null surface $V=V_s$ such that $V_\Delta > V_s$ and a  point $\mathcal{E}$ with coordinates $(U_\mathcal{E},V_\mathcal{E}=V_s)$ and $U_\mathcal{E}>0$, i.e., $\mathcal{E}$ lies on the null surface $V=V_s$ and in the interior of the black hole region. Finally one picks a space-like hypersurface $\Sigma_{ \mathcal{E}\to \Delta}$ connecting $\Delta$ to $\mathcal{E}$ and extends this space-like hypersurface to space-like infinity $i_0$ along the hypersurface $\Sigma_{\Delta\to i_0}$ defined by the condition $t=0$ in Eddington-Finkelstein coordinates. One names Region II the spacetime region bounded by the null surface $V=V_s$ in the past and $\Sigma_{ \mathcal{E}\to \Delta}\cup\Sigma_{\Delta\to i_0}$ in the future. There is a partner Region tII defined in analogy to Region II  by the time reflection $(U,V)\to (-U,-V)$. 
See Fig.~\ref{fig:penroseA}-Left. 
The Carter-Penrose diagram of the fireworks model (Fig.~\ref{fig:penroseA}-Right) is obtained by inserting the interpolating Regions III+tIII that complete the spacetime to the future of $\Sigma_{\mathcal{E}\to \Delta}$ in Region II up to $\Sigma_{\bar{\mathcal{E}}\to \Delta}$  in Region tII.
The regions $v\le v_s$ and  $u\ge u_s$ are described by Minkowski Region I and Region tI respectively.
The gluing across the null surfaces is done by demanding continuity of the metric; this leads to a distributional energy momentum tensor and the standard interpretation of the null gluing surface as a spherical shell of mass $m$ collapsing to $r=0$ in the past and then bouncing out in the future. The geometry in Region III+tIII is not explicitly defined in the model; however, the absence of singularities require the putative energy-momentum tensor to violate  energy conditions in Region III+tIII. This is interpreted as a spacetime region where quantum gravity effects are large.

The resulting spacetime represents the dynamics of a null in-falling shell of total mass $m$ that bounces at $r=0$ and comes out as a null outgoing shell of the same mass. 
The point $\mathcal{E}$ is the point where the ingoing shell enters (or touches) the quantum Region III, while $\Delta$ is considered as the outmost boundary of the quantum Region III+tIII.  As we will recall below, the time scale of the bounce is argued to be of the order of $m^2$ (in Planck units). This `fast' process makes the dissipation effects of Hawking radiation negligible. This is argued to justify the time-symmetric character of the bouncing scenario.

The spacetime is event-horizon-free, but displays a trapping and an anti-trapping surface. Notice that the past directed outgoing null rays from $\Delta$---defining a null surface that approaches exponentially the trapping surface in the past---represents what we will call the {\em past classicality horizon}, denoted $\sH_{-}$: any observer crossing $\sH_{-}$ will end up falling into the quantum Region III+tIII. More precisely, the domain of dependence $D({\rm III+tIII})$ has a boundary defined by two null surfaces. We call $\sH_{-}$ (resp. $\sH_{+}$) the past (resp. future) null component of that boundary.

To completely specify the model, one has to fix $V_{\Delta}>0$ in order to fix the position of the point $\Delta =(-V_\Delta, V_\Delta)$, and duration of the process which is parametrized by $\Delta V=V_{\Delta}-V_s$.  The condition $V_{\Delta}>0$ implies that the quantum region III+tIII extends outside the Schwarzschild trapping horizon.  This is a central conceptual point in the proposed model: one is allowing large quantum effects to leak out of the Schwarzschild horizon where curvature is low and far from Planckian (here $m\gg1$ in Planck units). In the original paper \cite{Haggard:2014rza}, this is stated by saying that ``there is no reason to trust the classical theory outside the horizon for arbitrarily long times and sufficiently close to $r = 2m$''.  
The authors of \cite{Haggard:2014rza} propose that quantum gravitational effects can be accumulated with ``time'' and become nonnegligible outside the horizon. Accordingly, they  introduced a \emph{nonclassicality parameter} defined along the world-line of a stationary observer sitting at $r=r_\Delta$ for a proper time $\tau$ as
\be\label{qqq}
q=\ell_p^{2-b} \sR\tau^b
\ee
where $\sR$ is a measure of spacetime curvature defined for concreteness in terms of the  Kretschmann invariant $\sR^2=R_{abcd} R^{abcd}=48 m^2/r^6$ and $b$ is a phenomenological parameter of order unity.  For concreteness we take $b=1$ following \cite{Haggard:2014rza}. The parameter $\tau$ is the proper time of the stationary observer from the crossing of the collapsing shell to the point $\Delta$ (see Fig.~\ref{fig:penroseA}), that is
\be\label{eq:tau}
\tau = \sqrt{1- \frac{2m}{r_\Delta}} \, \Delta v\;,
\ee
where $v$ is the standard advanced inertial time at $\sI^-$.
The quantity $q$ is maximized for 
\be\label{threes}
\begin{split}
r_\Delta = \frac{7}{3}m.
\end{split}
\ee
 This means that the quantum Region III+tIII extends macroscopically outside the BH horizon.  The bouncing time is defined to be the value of $\Delta v$ for which the 
nonclassicality parameter (linear in $\Delta v$) becomes of order unity. This happens for
\be\label{eq:deltav}
\Delta v\equiv v_\Delta-v_s \sim \tau \sim m^2\;.
\ee
Due to the time symmetry of the construction, the observer at $r_\Delta$ sees the entire bouncing process happening in a proper time $\tau_{\rm tot} = 2 \tau \sim m^2$.
This time scale is very important in what follows and is argued to produce possible experimental observations \cite{Barrau:2014hda,Barrau:2014yka,Barrau:2015uca}.

\section{Semiclassical Stability}\label{scs}

The question any classical ansatz spacetime has to be confronted with is whether it admits a physically reasonable quantum state for the test fields living on it.
This requirement represents the first step toward addressing the problem of \emph{back-reaction}. More precisely, in those regions where we can trust the validity of QFT in curved spacetime one expects the quantum dynamics to be well approximated by the semiclassical Einstein's equation  
\be\label{eq:semee}
G_{ab}(g_{ab})=8 \pi\, \avrg{\,T_{ab}(g_{ab})} \;,
\ee
where $\avrg{\,T_{ab}(g_{ab})\,}$ represents the expectation value of the stress-energy tensor of the quantum matter fields propagating on the metric $g_{ab}$. 

The most famous example is the effect of Hawking evaporation on a black hole background \cite{hawking1974black,hawking1975}. The original computation has been made in the fixed background approximation, completely neglecting the back-reaction. However, this leads to an infinite amount of radiated energy from the hole, clearly in contradiction with energy conservation. Intuitively, one expects the energy radiated to be balanced by a reduction of the Bondi mass of the black hole, leading to the evaporation of the hole and consequently the well-known loss of information paradox \cite{hawking:1976breakdown}.
There are both analytical and numerical works indicating some general features of the evaporation problem \cite{Hajicek19809,Bardeen:1981zz,Parentani:1994ij,Massar:1994iy}; nevertheless, a complete description remains unsolved even in the semiclassical regime of equation \eqref{eq:semee}.

Indeed, the complete backreaction problem could be framed in a formal approximation procedure where one starts by evaluating $\avrg{T_{ab}(g_{ab}^0)}$  on a seed background $g_{ab}^{0}$, and then inserts the result into semiclassical Einstein equations~\eqref{eq:semee} in other to find a new metric $g_{ab}^{1}$: the first-order quantum corrected background metric. Iterating the process one can try to find higher-order corrected line elements eventually converging to a consistent solution $g_{ab}$ of equation \eqref{eq:semee}. Every single step is in general a really difficult task to achieve and the final convergence is not even guaranteed. 

Fortunately, for the present analysis it will be sufficient to solve a much simpler problem. Indeed, the classical initial background $g_{ab}^{0}$---solution of the classical Einstein equations---is a good zeroth approximation of the quantum dynamics only if the quantum corrections coming from $\avrg{T_{ab}(g_{ab}^0)}$  are small in semiclassical regions. This stability of the seed background under the effects of the propagation of quantum test fields living on it will be called {\em quantum-stability} property. In the following of this Section, we will compute $\avrg{T_{ab}(g_{ab}^0)}$ for the model of reference \cite{Haggard:2014rza} and show that it diverges in Region tII. The quantum-stability property, therefore, is not satisfied by the fireworks model.

The computation of $\avrg{\,T_{ab}(g_{ab}^0)}$ on a given unperturbed geometry can be already a very difficult task. In fact, there is in general uncertainties related to the choice of the appropriate physical state for the quantum fields and, at the same time, one needs to appeal to renormalization techniques to eliminate usual $UV$  divergences of QFT in a way that is consistent with general covariance \cite{Wald:1995yp}. Both issues are more subtle and difficult when the background spacetime is not flat. However, the great symmetry of our example and its direct relationship with the well-studied Schwarzschild geometry will allow us to make very precise statements.
 
\section{Analytic Calculation in the $1+1$ Setting}\label{next}
In this section we use spherical symmetry and we neglect back-scattering as well as the influence of modes other than $s$-modes. This allows for an effective description in terms of a $1+1$ theory. These simplifications make possible the analytic computation of effects that qualitatively remain valid in the $3+1$ framework. More precisely, we show that the computation of $\avrg{\,T_{ab}(g_{ab}^0)}$ in the framework of the fireworks background presents a divergent behaviour. Quantum fields are represented by a single massless scalar $\phi$ satisfying the Klein-Gordon equation 
\be\label{uno}
g_{0}^{ab} \nabla_a\nabla_b \phi=0
\ee 
with $g_{0}^{ab}$ the background geometry of the fireworks model in the $r-t$ space. In more detail, the metric in Region II+tII is given by 
\be\label{schw}
ds_{0}^2=-\left(1-\frac{2m}{r}\right) dvdu,
\ee
where $v=t+r_*$ and $u=t-r_*$, with $t$ the Killing parameter and 
\be r_*=r+2m\log\left(\frac{r}{2m}-1\right).\ee
In Region I the metric is 
 \be\label{Mink}
ds_{0}^2=-dvdu_{in},
\ee
where $u_{in}=t_M-r$ and $v=t_M+r$ and $t_M$ is the inertial Minkowski time defined by an observer at the center of the shell. 
The explicit relation between $u_{in}$ and Schwarzschild coordinates can be computed from the matching conditions that follow from
demanding continuity of the metric across the shell, namely
\be\label{doce}
u=u_{\rm in}-4m \log \left( \frac{v_s-u_{\rm in}-4m}{4m}\right).
\ee

\vspace{2ex}\noindent{\scshape\bfseries The state representing gravitational collapse.}\hspace{3ex} The fireworks model describes the physics of a collapsing shell that would classically lead to the formation of a spherical black hole spacetime. This physical situation imposes  clear-cut constraints on the initial conditions of the quantum state of the field $\phi$. On the one hand,
the state for the in-modes of the quantum fields on $\sI^-$ must not be substantially excited. In other words, aside from the zeroth order matter 
distribution defining the collapsing shell that will lead to the formation of the trapped regions in the future, no substantial amount of energy momentum of $\phi$ is poured in from $\sI^{-}$.\footnote{In Appendix~\ref{app:HH} we study the contrasting situation where an infinite amount of radiation is sent from infinity: the Hartle-Hawking state.}
This is translated into the demand that the in-modes of the quantum field on  $\sI^{-}$ must be in the vacuum state. A similar boundary condition must hold also for the out-modes in the flat interior of the collapsing shell (Region I). Small perturbations of these conditions could be admitted yet, and this would not change the conclusions that will follow.  

These two conditions are satisfied by the so-called vacuum in-state $\ket{in}$ \cite{Unruh:1976db}, defined as the unique vacuum state of the Fock space where positive frequencies are defined with respect to the mode expansion of solutions of \eqref{uno} of the form
\be
\phi_{in}=e^{i \omega v}, \qquad \phi_{out}=e^{i \omega u_{in}} \;.
\ee
This state corresponds to the required physical condition that there is no incoming radiation from $\sI^-$ as well as no outgoing radiation from inside the shell.
This state represents the idealized physical situation one wants to describe in the context of gravitational collapse.  

\vspace{2ex}\noindent{\scshape\bfseries The region of applicability.}\hspace{3ex} There is uncertainty on the features of the quantum fields in the future domain of dependence of Region III+tIII as the effective $1+1$ geometry is expected not to capture all the physics of the dynamics of the field through that part of the spacetime.  Therefore, all of the components of $\avrg{\,T_{ab}(g_{ab}^0)}$ that we want to compute can be used to describe the energy momentum expectation value only in Region I and in the portion of Region II in the past of $\Sigma_{ \mathcal{E}\to \Delta}$ union the null outgoing ray $u=u_\Delta$ starting at $\Delta$ and reaching $\scrai^+$.

Nevertheless, whatever might be the dynamics in the strong quantum region, we expect to be able to predict without uncertainties at least some of the components of $\avrg{\,T_{ab}(g_{ab}^0)}$ for those points to the future of the horizon $\sH_{+}$. A closer look shows that, due to the decoupling of in and out modes for a conformal theory in the present $1+1$ context, the component $\avrg{\,T_{vv}(g_{ab}^0)}$ is independent of the features of the quantum Region III+tIII. Both $\avrg{\,T_{uu}(g_{ab}^0)}$ and $\avrg{\,T_{uv}(g_{ab}^0)}$, on the other hand, will be modified by quantum gravity effects.
In those regions of applicability, the computation comes out to be a standard computation \cite{PhysRevD.13.2720,birrelldavies:QFTCST}, well illustrated for instance in \cite{fabbri2005modeling}. 

With these preliminary considerations stated, we are now ready to compute the expectation value of the energy momentum tensor in the vacuum in-state defined on the background geometry of the fireworks spacetime. In the region of interest, and for $\avrg{\,T_{vv}(g_{ab}^0)}$ we can simply import the results from the standard calculation on a background given by the gravitational collapse of a shell of mass $m$.   
Following for instance \cite{fabbri2005modeling}, see Appendix~\ref{app:HH}, the components of the covariant quantum stress-energy tensor are given by
\be\label{jiji}
\begin{split}
\bra{in}T_{uu}\ket{in} &= \frac{\hbar}{24 \pi}\left[ -\frac{m}{r^3} + \frac{3}{2}\frac{m^2}{r^4} -\frac{8m}{(u_{in}-v_s)^3}-\frac{24 m^2}{(u_{in}-v_s)^4}\right]\\
\bra{in}T_{vv}\ket{in} &= \frac{\hbar}{24 \pi}\left[ -\frac{m}{r^3} + \frac{3}{2}\frac{m^2}{r^4} \right]\\
\bra{in}T_{uv}\ket{in} &= -\frac{\hbar}{24 \pi}\left( 1- \frac{2m}{r} \right) \frac{m}{r^3} \;.
\end{split}
\ee
While the above equations seem to show that $\avrg{\,T_{ab}(g_{ab}^0)}$ is finite everywhere, they do not. The problem is that the Eddington-Finkelstein coordinates used to compute them are not well defined at the trapping horizons: the modes are infinitely oscillating there. A clear analysis of the divergence behavior of the tensor  $\avrg{\,T_{ab}(g_{ab}^0)}$ can be achieved by using good coordinates close to the trapping horizons. The expectation value of the energy momentum tensor in our state can be shown to be regular in whole Region II, see for instance \cite{fabbri2005modeling}. What about Region tII?

Only $\avrg{\,T_{vv}(g_{ab}^0)}$ is relevant for the rest of our analysis: as mentioned above, indeed, it is the only component of the energy momentum tensor for which \eqref{jiji} can be trusted in the future of $\sH_{+}$  independently of the unknown geometry of Region III+tIII. A suitable choice of good coordinates are the Minkowski null coordinates $(u,v_{\rm out})$ in terms of which the metric in Region tI takes the form
\be
ds_0^2=-dudv_{out}.
\ee
 Continuity of the metric across the outgoing shell implies
\be
v=v_{\rm out}-4m \log \left( \frac{u_s-v_{\rm out}-4m}{4m}\right).
\ee
Since, by definition, $\avrg{\,T_{ab}(g_{ab})}$ is covariant, one finds
\be\label{eq:Tkrus}
\bra{in}T_{v_{\rm out}v_{\rm out}}\ket{in} = \left( \frac{dv}{dv_{\rm out}} \right)^2 \bra{in}T_{vv}\ket{in}=\left(\frac{u_s-v_{\rm out}}{u_s-v_{\rm out}-4m} \right)^2 \bra{in}T_{uu}\ket{in}\;.
\ee

In these coordinates and on the outgoing shell, $u_s-v_{\rm out}=2r$. The above quantity  diverges at the white hole trapping horizon $r=2m$ (which in the patchwork construction of \cite{Haggard:2014rza} is close to $\sH_+$) as $(r-2m)^{-2}$. This divergence of $\avrg{T_{ab}}$ is, as we have just shown, explicit in the simplified $1+1$ context.\footnote{In the same way one can show that all the components of the renormalized energy momentum tensor remain finite at the future horizon (close to $\sH_-$).} However, it is a general feature that remains valid in the physical $3+1$ context. Some references where explicit calculations are given are \cite{PhysRevD.15.2088, PhysRevD.21.2185,Balbinot1999301}. All this is implied by the very general result implying that the Hartle-Hawking state is the only globally nonsingular state---satisfying the Hadamard condition that implies the regularity of $\avrg{T_{ab}}$---on the maximally extended Schwarzschild spacetime which is invariant under Killing time translations \cite{Wald:1995yp}. 

We conclude that in the vacuum in-state the expectation value of energy-momentum tensor diverges at the trapping horizon $r=2m$ close to $\sH_+$. However, this horizon is outside the region of validity of our calculation as defined above: it is completely inside the future domain of dependence of the quantum Region III+tIII.\footnote{One can try to interpolate the black hole patch with the white hole one by an effective metric, see for example \cite{Barcelo:2015noa}. This is however not relevant for our discussion.} Nonetheless, the would-be-divergent component is still problematic. The reason is that the trapping horizon and $\sH_{+}$ get exponentially close to each other along the generators of $\sH_{+}$.  

More precisely, let us call   $r_s$ the value of the radius at the intersection of $\sH_+$ and the outgoing shell; see Fig.~\ref{fig:penroseA}. 
From the integration of the null geodesic equation, one finds
\be\label{eq:rs}
\begin{split}
r_s &= 2m \left( 1 + W \left[\frac{r_\Delta - 2m}{2m} \exp \left\{\frac{r_\Delta -2m}{2m} - \frac{\Delta u}{4m}\right\} \right] \right) 
\end{split}
\ee 
where $W[x]$ is the Lambert function and $\Delta u = u_s - u_\Delta$.
Clearly, $r_s$ represents the closest point to the past horizon for which we can trust the expression of  $\bra{in}T_{v_{\rm out}v_{\rm out}}\ket{in}$  given in equation eq.~\eqref{eq:Tkrus}. Consequently, it  also gives the largest possible value of that component of the energy momentum tensor. At that point we have
\be
\begin{split}
\bra{in}T_{v_{\rm out}v_{\rm out}}\ket{in}\rvert_{r_s} &= \frac{\hbar}{24\pi} \left( \frac{r_s}{r_s-2m}\right)^2  \left[ -\frac{m}{r_s^3} + \frac{3}{2}\frac{m^2}{r_s^4} \right]\\
& \sim - \frac{\hbar}{192\pi } \left(\frac{\exp \left \{ \Delta v/(4m) - (r_{\Delta}-2m)/(2m) \right\}}{r_{\Delta}-2m}\right)^2 \;,
\end{split}
\ee
where we used the fact that $r_s \to 2m$ and that, by construction, \be \Delta u = \Delta v .\label{eq:deltas} \ee
Demanding the quantum energy-momentum tensor to be sub-planckian everywhere, we can find a relation between the two parameters of the models, namely $r_{\Delta}$ and $\Delta v$. In fact, ($\hbar = 1$)
\be
\abs{\bra{in}T_{v_{\rm out}v_{\rm out}}\ket{in} } < 1
\ee
implies
\be\label{eq:condition}
\left(\frac{r_{\Delta}}{2m}-1\right) e^{\frac{r_{\Delta}}{2m}-1} > \frac{1}{\sqrt{768 \pi}\,}m e^{\frac{\Delta v}{4m}} \quad \Rightarrow \quad
\frac{r_{\Delta}}{2m}-1 > W\left[ \frac{1}{\sqrt{768 \pi}\, m} e^{\frac{\Delta v}{4m}}\right]\,.
\ee
The longer the lifetime $\Delta v$ of the hole, the more the quantum region must extend out of the classical horizon (as parametrized by $r_{\Delta}$) in order for the stress-energy tensor to be subplanckian along $\sH_+$. In particular, if, as estimated in \cite{Haggard:2014rza}, $r_{\Delta}=\frac{7}{3} m$ (see eq. \eqref{threes}), condition~\eqref{eq:condition} implies
\be\label{eq:bound}
\Delta v \lesssim m \log m\;.
\ee
That is, {if we do not want trans-Planckian behaviors of the renormalized quantum stress-energy tensor, the lifetime of the hole has to be so short that the model would already be ruled out by present observations.  For instance, the characteristic time $\tau=m\log(m)$ would be of about $10$ minutes for the central supermassive black hole in our Milky Way. For the same black hole one could try to tune the parameter $r_{\Delta}$ to allow a lifetime of order $m^2$; however, a simple look at equation \eqref{eq:condition} shows that this would imply extending the quantum region outside of the horizon to include almost the whole of the observable universe. 

\section{Asymmetric Fireworks}\label{corr}

The issues presented in the previous section constrain the white hole lifetime to be much shorter than the one defined in the original paper. 
Similar constraints can be found from simple classical considerations.\footnote{Personal communications with Eugenio Bianchi and Matteo Smerlak.} 
In all cases the problems are related to the instability due to the presence of a white hole horizon: infinite blueshift of perturbations that are well behaved at $\sI^-$. 
Our argument is related to those classical instabilities if we replace the concept of perturbations by quantum fluctuations in the in-vacuum.
However, an important point is that, in all cases, the constraints concern the lifetime of the white hole horizon only. The lifetime of the black hole horizon (which is the one constrained by observations) can be freely set without running into the present type of instabilities. 

This can be easily seen from eq.~\eqref{eq:rs}. The relevant parameter for our discussion is the $\Delta u$ that we identified with $\Delta v$, due to the choice made originally in \cite{Haggard:2014rza} to place the point $\Delta$ on the surface $t=0$. Discarding the identification \eqref{eq:deltas} and following exactly the same procedure, the crucial bound in eq.~\eqref{eq:bound} now becomes
\be\label{eq:conditionNew}
\Delta u \lesssim m \log m\;.
\ee
A possible way out, therefore, is to abandon the time-symmetric nature of the bounce in the original form of the fireworks model.
More precisely, to avoid the time-symmetric condition $\Delta u = \Delta v$ one can modify the construction of the spacetime (Section ~\ref{sec:halcarlomodel}) by choosing the outgoing bouncing shell to come out at a retarded time $U_s$ different from $-V_s$. The resulting spacetime, depicted in Fig.~\ref{fig:a-penrose}, differs from the original one as if the point $\Delta$ has been moved away from the $t=0$ surface along a curve $r=r_{\Delta}$.

\begin{figure}[h]
\center
\includegraphics[width=0.4\textwidth]{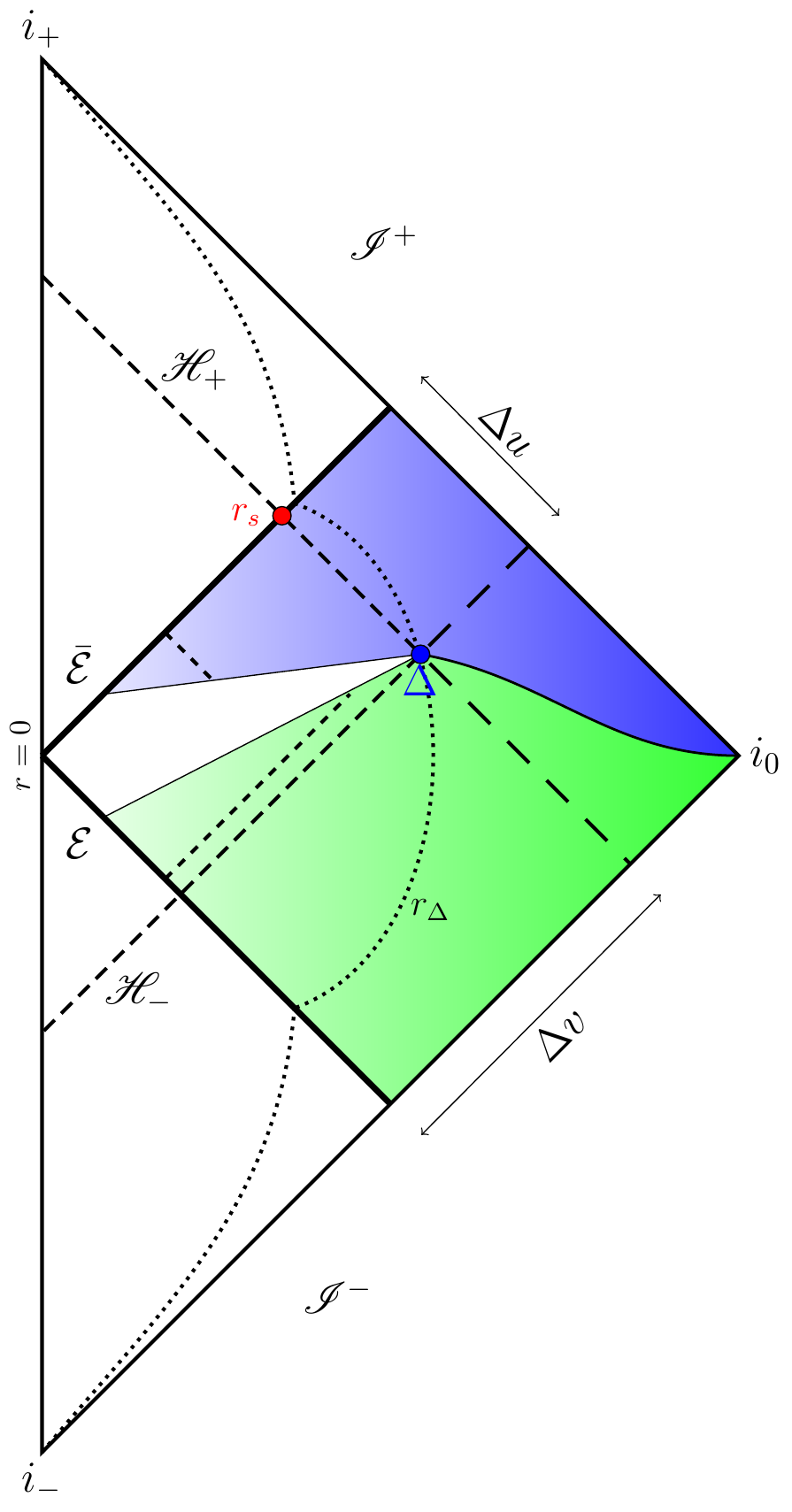}
\caption{Geometry of a asymmetric bouncing scenario. The parameter $\Delta v \sim m^2$ captures both the accumulation time for quantum gravitational effect outside the horizon and the lifetime of the hole. The parameter $\Delta u$, on the other hand, represents the lifetime of the white hole and it is forced by the arguments presented in the text to be of order $m \log m$.}
\label{fig:a-penrose}
\end{figure}

In particular, one can choose the value of $U_s$ such that the quantum stability requirement, expressed by eq.~\eqref{eq:conditionNew}, is satisfied. Moreover, the analysis of the nonclassicality parameter presented at the end of Section~\ref{sec:halcarlomodel} is still precisely valid, and so are eq.~\eqref{threes} and \eqref{eq:deltav}. The accumulation of quantum gravitational effects outside the horizon that allows the black-hole-to-white-hole transition has not been modified, and the above instabilities are removed simply by shortening the lifetime of the white hole horizon.

In Section~\ref{sw} we will largely discuss the nature and the consequences of time asymmetry introduced in our modification of the model. Here we just want to emphasize that the lifetime of the whole process (from collapse to annihilation) remains of the order of $m^2$ as in the original model, much shorter than the $m^3$ time scale predicted by Hawking evaporation.\footnote{In doing this simple comparison between time scales we are making a little abuse of notation. For a more precise statement, see the precise analysis reported in Appendix~\ref{app:lifetime}.} This implies that the nature of the time asymmetry is not a dissipative effect due to the Hawking evaporation as one could intuitively expect: the energy radiated after a time of the order of $m^2$ is just of the order of the Planck mass $m_P$. The Hawking effect is negligible and the processes discussed here are basically nondissipative.

\section[Black-Hole-to-White-Hole Instability]{Black-Hole-to-White-Hole  Instability} \label{babacul}

The modification proposed also removes another related type of instability  studied in \cite{PhysRevLett.33.442,lake1978white,Blau:1989zs,PhysRevD.47.2383,PhysRevD.50.6150}.
The idea is the following. Since a white hole is attractive, any small perturbation of ambient matter will be accelerated toward it. At the same time, since no matter can cross the white hole horizon, after a sufficiently long time, a macroscopic mass will be accreted onto an arbitrarily thin shell close to the horizon and will produce, when interacting with any object coming out from the white hole, a new collapse into a future singularity.
\begin{figure}[t]
\center
\includegraphics[width=0.25\textwidth]{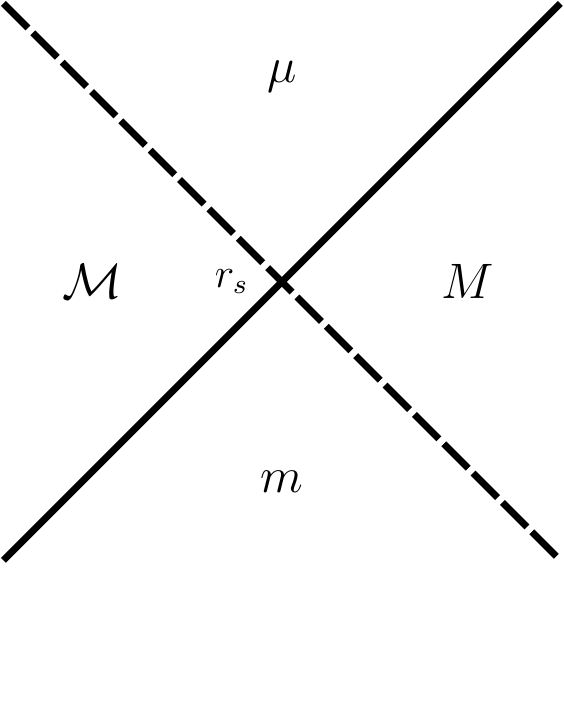} \qquad\qquad\qquad\quad
\includegraphics[width=0.4\textwidth]{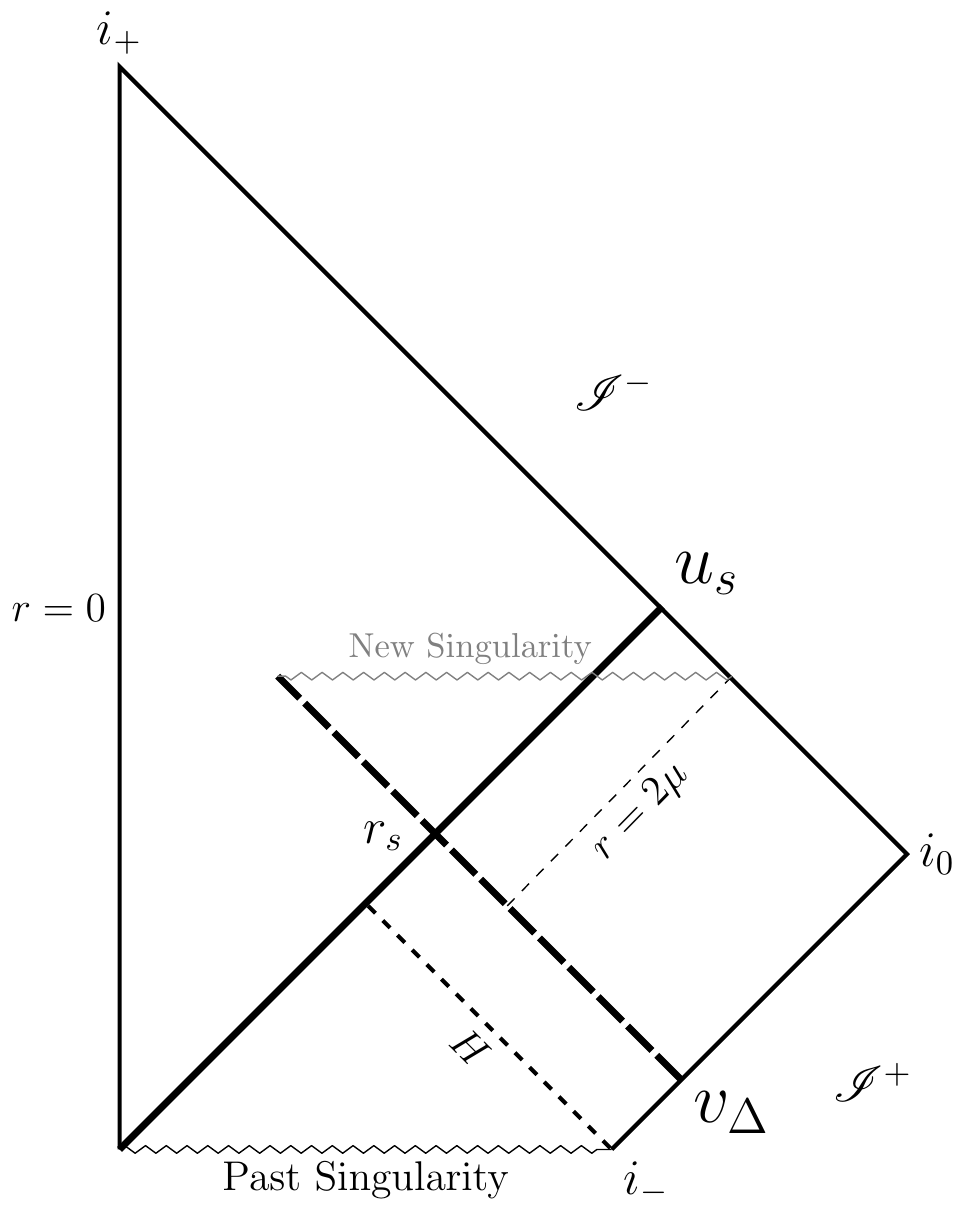}
\caption{
{\bf Left:} The Dray-'t Hooft geometry. Four Schwarzschild patches with different masses are glued together along null shells which intersect at a radius $r_s$. The condition for the geometries to be glued smoothly generates a relation between the four masses and $r_s$. {\bf Right:} The death of a white hole. A white hole emits all its mass $m$ along a massive null shell at the retarded time $u_s$. A small massive perturbation $\nu$ is sent at the advanced time $v_\Delta$ into the white hole geometry and interact with the outgoing shell at $r=r_s$. At the interaction point we have a Dray-'t Hooft geometry with $m=m$, $\mathcal{M}=0$, $M = m+\nu$. The last mass $\mu$ is uniquely determined by the constraint which leads to eq.~\eqref{eq:mu}: $\mu = r_s / (r_s-2m)\nu$. If $r_s$ is lower then $2 \mu$, the emerging shell is captured in the future black hole horizon of the new geometry generated by the interaction and cannot escape to infinity. The white hole is dead recollapsing into a black hole.}
\label{fig:deathWH}
\end{figure}

The interaction between any small matter perturbation of mass $\nu$ sent for instance along the null geodesic $v=v_\Delta$ and the outgoing mass $m$ shell at $r=r_s$ can be described by a Dray-'t Hooft geometry \cite{Dray:1985yt} (Fig.~\ref{fig:deathWH}-Left).  The spacetime for $v>v_\Delta$ and $u>u_s$ is a Schwarzschild geometry with mass $\mu$, given as a function of the initial parameter $m$, $\nu$ and the radius $r_s$ by
\be\label{eq:mu}
\mu = \frac{r_s}{r_s-2m}\,\nu\;.
\ee
It is clear now that if $r_s < 2 \mu$ the outgoing shell is captured inside the new black hole horizon and cannot escape to infinity: any small perturbation $\nu$ interacting with the bouncing shell will cause the system to recollapse into a black hole of mass $\mu$; see Fig.~\ref{fig:deathWH}-Right. Thus, no fireworks can be seen from infinity. The model is, however, still valid if $r_s > 2 \mu$ or equivalently  if
\be
r_s > 2(m+\nu).
\ee
From equation~\eqref{eq:rs} we get
\be
\frac{\Delta u}{4m}  < \left(\frac{r_\Delta-2(m+\nu)}{2m}\right) \log \left(\frac{ {r_\Delta}-2m}{2\nu}\right)\;,
\ee
and assuming $\nu \ll m$ we find again
\be
\Delta u \lesssim m \log \left(\frac{m}{\nu}\right)\;.
\ee
The tunneling process is strongly unstable under perturbations if $\Delta u$, the lifetime of the white hole, is bigger then of order $m \log m$. This argument could surely be discussed together with the other issues that have forced us to consider an asymmetric bouncing scenario, but presented in this way the different time scales involved become clear. This  has been extensively discussed in a recent paper by Barcel\'o et al. \cite{Barcelo:2015uff} in the context of the original symmetric model. The asymmetric modification that we have introduced here also cures this instability. 

\section{Smashing Watches}\label{sw}

In this section we want to discuss the physical consequences of the introduction of a time asymmetry in the model. The bouncing process can be described by a quantum field in the $\ket{in}$ vacuum state on $\sI^{-}$ evolving into a final state $\ket{out}$ on $\sI^{+}$. Both states represent an idealized flat initial geometry with an infinitely diluted, but sharply defined, spherical shell carrying mass $m$. More precisely, from the point of view of an observer at infinity, the {\em in} and the {\em out} classical data are just equivalent.

On the other hand, the semiclassical analysis of the dynamics of the state $\ket{in}$ across the spacetime tells us that the state in the future must be very different from what it was in the past. We have actually shown that the components of $\avrg{\,T_{ab}(g_{ab}^0)}$ are perfectly smooth for $u<u_{\Delta}$ while they are dangerously diverging in some regions to the future $u>u_{\Delta}$.  These divergences can be cured by modifying the background in consistency with this time asymmetry. We have achieved this by shortening the lifetime of the white hole in Section \ref{corr}; see Figure \ref{fig:a-penrose}. 

Nevertheless, in doing so we have preserved the equivalence of the past and future classical data. The point we want to emphasize here is that the time-asymmetric nature of the inner spacetime needed to avoid instabilities should imply strong modifications also in the classical final \emph{out} description of the model, that can be very different from the simple mean field approximation proposed by the fireworks model.

One can illustrate the point in terms of the nonclassicality parameter $q$, eq.~\eqref{qqq}. Recall that the idea is that quantum effects accumulate from $v=v_s$ along the world line of a stationary observer at $r_{\Delta}$ until the quantity $q$ becomes of order one at $v=v_{\Delta}$. This happens after a time of the order of $m^2$. Let us now run the process backward in time. This inverse process is still a bounce now described by an initial state given by $\ket{out}$ evolving into $\ket{in}$. Its dynamics is given by the time reversal of the original one. However, as $q$ only knows about the local geometry, one finds that for the reverse process $q$ is far from unity at $\Delta$. This means that something must be very different for the later observer; something else (not explicitly stated in the model of Figure~\ref{fig:a-penrose}) must contribute to the nonclassicality so that it builds up very much quicker in the inverse process.

If correct, the cause of the shortening of time scales in the future of the bounce must be found in the details of the quantum state of the system beyond the mean-field approximation implicitly used when proposing a background geometry. Notice that the future observer is exposed to quantum gravitational effects coming from the {\em would-be-singularity}---whatever replaces the singularity predicted by the classical theory, i.e. Region III+tIII. These effects must be important enough to drastically reduce the lifetime of the white hole from $m^2$ to $m\log m$. 

But then if these quantum gravitational effects are so strong, why should we trust a semiclassical description at all in the vicinity of the white hole?  
Why should the spacetime become classical again so quickly with the mass $m$ entirely carried by a spherical bouncing shell?
It is hard to address these questions without a full quantum dynamical treatment. 

Nevertheless, the standard collapse process strongly suggests irreversibility already at the classical level. Gravitational collapse is like breaking a watch. This can be intuitively seen, from the classical point of view, by considering the standard spacetime depicting the gravitational collapse of a spherical shell (put the diagram on the right of Figure \ref{fig:deathWH} upright). Initial states given by the shell plus smooth matter and geometry perturbation at $\sI^-$ are special,  they are `low-entropy' states representing our `watches'. They come in different types depending on the details of the initial state. This states are bound to evolve into very complicated final states: smashed watches. This is clear from the fact that only a very precise fine tuning of the features of the state at $\sI^+\cup i_+\cup H$ would evolve backwards to our nice watch at $\sI^-$ (those final states are measure zero in the phase space of possible final states). 

The previous irreversibility mechanism becomes even more apparent if quantum gravity is brought into the discussion. 
Everything that crosses the horizon $\sH_{-}$ will end up at the would-be-singularity exciting degrees of freedom that were not available at low energies.
\begin{figure}[t]
\center
\includegraphics[width=0.4\textwidth]{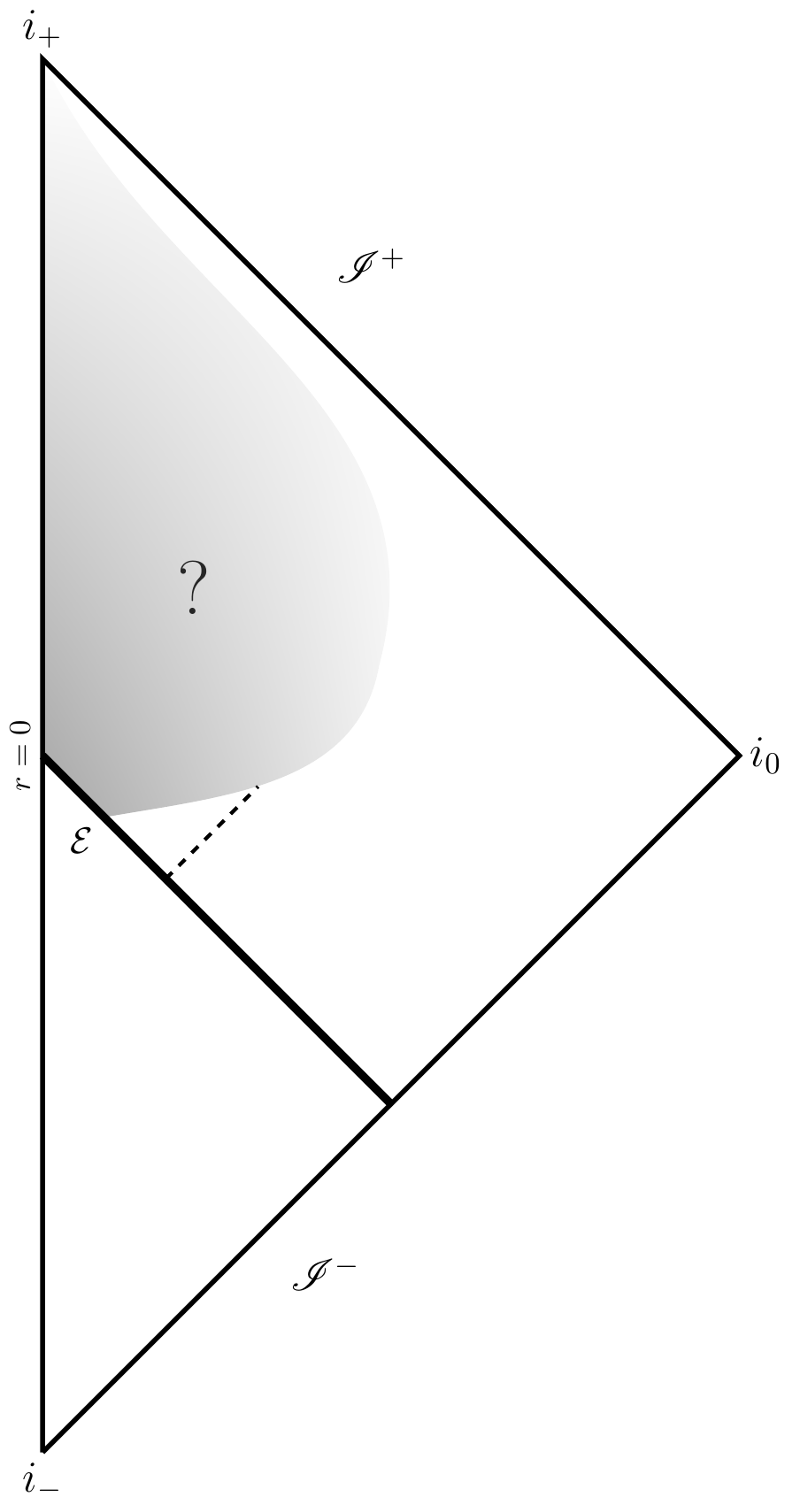}
\caption{Geometry of a general asymmetric scenario. The region where the semiclassical analysis breaks down is shaded. The metric outside is isomorphic to Schwarzschild with mass $m$.}
\label{fig:QuantumFoam}
\end{figure}
The phase space regions available for these falling degrees of freedom can become dramatically larger with the potential effect of further increasing the irreversibility of the overall process. Concretely, as the shell approaches $r=0$ more and more degrees of freedom get excited: from known standard model degrees of freedom (quark-gluon plasma phase, Hagedorn exponential growth of available degrees of freedom, etc.) to beyond standard model degrees of freedom and all the way down to Planck scale. At that ultimate fundamental level, in an approach like LQG, quantum geometries are degenerate: the phase space of available `geometries' at the Planck scale includes a huge number of configurations (microstates) which are simply overlooked in the low energy coarse graining associated with the semiclassical background geometry proposed to describe the process.\footnote{These microstates are responsible for black hole entropy in LQG \cite{G.:2015sda} and have been argued to provide a simple, natural resolution of Hawking's information loss paradox in \cite{Perez:2014xca} in the more conservative framework where Hawking evaporation is the main quantum effect for BHs with $m\gg 1$ \cite{Ashtekar:2005cj}.} All this implies the type of irreversibility proper to systems that satisfy the second law of thermodynamics.

In view of all this we find no reason to discard scenarios where the spacetime does not become semiclassical so quickly to the future of the bounce and where the initial mass $m$ shell dissolves into a quantum substance after the bounce. The details can only be described in the context of full quantum gravity. This very uncertain state of affairs is represented in Figure \ref{fig:QuantumFoam}. 

\section{Conclusions}\label{diss}

We have explored certain instabilities of the fireworks scenario proposed in \cite{Haggard:2014rza} and have proposed a simple way to resolve them.
These instabilities are all associated with the presence of a white hole trapping horizon that is sufficiently long lived.
General considerations demand the gravitational collapse (even in the fast scenario of fireworks where Hawking radiation does not play an important dynamical role) to be time-asymmetric and it  is precisely by allowing such asymmetry that the instabilities are resolved. In this way the black hole phase lasts a time of order $m^2$ followed by an extremely fast explosion where the mass $m$ is radiated back to infinity in a time shorter than $m\log(m)$ in Planck units ($10^{-4} s$ for a solar mass BH, $10^{-9}s $ for a lunar mass BH). The same considerations of the irreversible nature of the gravitational collapse lead to uncertainties in the description of the details of this late bounce. A more precise (not yet available) quantum gravity description of the dynamics across the would-be-singularity could shed light on these details.
It is possible that, despite these uncertainties, the scenarios discussed here could
lead to some generic observable phenomenology (for instance the $m\log(m)$ explosion scale). We leave this question to the experts. 

%% file: Part_2/Part_2.tex
\thispagestyle{empty} \phantom{}
\vfill

\begin{center}
\hrule\vspace{.2ex}\hrule\vspace{1ex}
{\huge \scshape\bfseries Part II\\}
\vspace{1ex}\hrule\vspace{.2ex}\hrule

\vspace{10ex}
{\fontsize{35}{38} \scshape\bfseries Spacetime and Thermodynamics\\}
\end{center}
\vfill

{\markboth{Part II: Spacetime and Thermodynamics}{Introduction to Part II}
\phantomsection\label{sec:intro2}

In the \hyperref[sec:intro1]{introduction to the previous Part}, I have briefly mentioned that the debate around the information paradox is crucially related to what is the deep meaning of the Bekenstein-Hawking entropy. In this introductory section to the second part of the manuscript I would like to discuss what is the BH entropy, how it arises and its importance for the physics of gravitation, as well as giving an overview on the different approaches to its explanation. To do so, it is useful to abandon Planck units for a while and keep all the constants of Nature explicit.

\newpage

The idea of assigning an entropy to BHs was first introduced by Jacob Bekenstein \cite{Bekenstein:1972tm,PhysRevD.7.2333}. His proposal was grounded on previous results that suggested an analogy between a BH and a thermodynamical system. A first result is the \emph{no-hair theorem} \cite{Carter:1971zc,Hawking:1971vc}, stating that any stationary axisymmetric BH can by described, no matter how it has been formed, by means of only three parameters: the mass $M$, the angular momentum $J$ and the charge $Q$. This is reminiscent of a description of a system with only macroscopic thermodynamical variable such as energy, temperature and pressure. A second result is Hawking's \emph{area law theorem} \cite{Hawking:1971vc}, showing that the area of the horizon can never decrease when classical processes, such as matter satisfying the weak energy falling into the hole, are considered. Again a reminiscence of the second law of thermodynamics for which the entropy of a system can only increase. 
If an external observer releases a box of particles into the horizon of a BH, the initial entropy of the box becomes inaccessible to him. Given the no-hair theorem, moreover, the BH can be described by only three parameters, and there is no way for the external observer to determine its interior entropy. He might therefore conclude that the entropy of the Universe has decreased, in contradiction with the second law of thermodynamics. The problem is resolved if one assigns an entropy to the BH so that it compensates the loss of the entropy of the box \cite{Bekenstein:1972tm,PhysRevD.7.2333}. Inspired by the area law theorem, Bekenstein suggested to assign a dimension-less entropy to the BH proportional to its horizon area
\be\label{eq:bek}
S_{\va BH} = \eta \frac{A}{\lpl^2}\,,
\ee
where $\eta$ is a dimension-less parameter. 
The Planck area at the denominator was introduced only on dimensional grounds. At the same time he proposed a \emph{generalised second law (GSL)} which states that the sum of the BH entropy and the total exterior world entropy does not decrease. The BH-thermodynamics analogy was then completed in \cite{Bardeen1973}, where the four laws of BH mechanics were precisely staten.
\begin{itemize}
\setlength{\itemindent}{2em}
\item[\emph{0th law}:] The surface gravity $\kappa$ of the horizon is constant on the horizon itself.
\item[\emph{1st law}:] For two stationary black holes differing only by small variations in the parameters $M$, $J$, and $Q$,
\be
c^2\,\delta M = \frac{c^2\,\kappa}{8\pi \,G}\delta A +\Omega_{\va H} \,\delta J +\Phi_{\va H} \delta Q
\ee
where $\Omega_{\va H}$ and $\Phi_{\va H}$ are respectively the angular velocity and the electric potential at the horizon.\\
\item[\emph{2nd law}:] The area of the horizon can never decrease
\be
\delta A \geq 0\,.
\ee
\item[\emph{3rd law}:] It is impossible by any procedure to reduce the surface gravity $\kappa$ to zero in a finite number of steps.
\end{itemize}
For the derivation of the four laws I refer to the original paper \cite{Bardeen1973}, to \cite{Wald:1995yp} for an alternative physical derivation of the first law, as well as to Chapter~\ref{chap:lightcone}, where the latter is generalised to light cones in Minkowski spacetime. The zeroth and the first law identify (a multiple of) the surface gravity $\kappa$ as the temperature, reinforce the analogy area-entropy, and designate the change in charge and angular momentum as work terms on the system.

As the treatment of \cite{Bardeen1973} is made within purely classical GR, no $\hbar$ enters the game. Moreover, let me quote a very explicative passage of the original paper
\begin{displayquote}
\emph{It should however be emphasised that $\kappa$ and $A$ are distinct from the temperature and entropy of the black hole.
In fact the effective temperature of a black hole is absolute zero.
One way of seeing this is to note that a black hole cannot be in equilibrium
with black body radiation at any non-zero temperature, \emph{because no
radiation could be emitted from the hole} whereas some radiation would
always cross the horizon into the black hole.} \cite{Bardeen1973}
\end{displayquote}
This view drastically changed with Hawking's result that black holes radiate particles as a black body at temperature
\be\label{eq:HT}
T_{\va H}=\frac{\hbar \,\kappa}{2\pi c\,k_{\va B}}
\ee
would do.
When the above expression is inserted into the first law one finds
\be
\frac{c^2}{k_{\va B}}  \delta M= T_{\va H}\frac{c^3 {}\delta A}{4\,\hbar\,G}+\frac{\Omega_{\va H}}{k_{\va B}} \,\delta J +\frac{\Phi_{\va H}}{k_{\va B}} \delta Q
\ee
where both sides have now dimensions of a temperature. This equation allows to identify a dimension-less entropy for a black hole as 
\be
S_{\va BH} =\frac{c^3 A}{4 G\, \hbar}=\frac{A}{4 \lpl^2}\,.
\ee
Bekenstein's idea became a fact of semi-classical gravity: BHs have an entropy which is proportional to the area of the horizon. The proportionality constant $\eta$ in \eqref{eq:bek} is fixed to be $1/4$. The above expression in Planck units becomes the familiar $S_{\va BH} = A/4$. 

Let me emphasise three interesting features. First of all, $S_{\va BH}$ is huge! For a solar mass black hole it turns out to be $10^{18}$ times the thermal entropy of our burning sun. Second, imagine dropping a mass of $1\,Kg$ into the supermassive BH at the centre of our galaxy. Its entropy change in this process is of the same order of (again) the thermal entropy of the sun, and, more importantly, it does not depend on the initial entropy of the mass that fell in. These observations give a first idea about the mysterious nature of such entropy. Third, $S_{\va BH}$ scales with the 2-dimensional area, contrary to the 3-dimensional volume scaling of the entropy of  standard thermodynamical systems.

As for the information paradox described in the \hyperref[sec:intro1]{introduction to Part I}, a complete understanding of $S_{\va BH}$ would probably need a complete theory of quantum gravity.
A first way to see this is the following: imagine to divide the area of the horizon in triangular cells of area $s$. One has therefore $N=A/s$ number of cells. By for instance colouring such cells black or white, one can encode information on it. The maximum number of combinations is $2^N$, therefore allowing the encoding of $ \sim N=A/s$ bits of information. If this information were to be lost, this would result in a entropy $S \sim A/s$. This shows that to match the BH entropy, the dimension of the cells needs to be $s=\lpl^2$, therefore appealing to quantum gravity for its explanation. Another way is to push the analogy with thermodynamics further and ask the central question: can this entropy be thought as a counting of the total number of possible microscopic states corresponding to the same macroscopic parameters $M$, $J$ and $Q$? The definition of micro-states for a geometric objet such a BH seems to point again toward a quantum description of spacetime, and therefore a theory of QG. 
Once more, ``black holes as a gateway to the quantum''. And once more, since a full QG theory is not available, it is crucial to try to find as more insights as possible from semi-classical reasonings, in the very spirit of this thesis \footnote{More recently, a new mathematically elegant and clean derivation of the first law identified $S_{\va BH}$ with the Noether charge associated with the diffeomorphism generated by the horizon Killing field \cite{Wald:1993nt}. Such an additional classical identification may provide useful insights for a more fundamental statistical explanation of the entropy.}.

The debate around the nature of the BH entropy presents two main conceptually different voices. On one side is the proposal that $S_{\va BH}$ counts in some way the number of micro-states of the BH that can communicate with the world outside the horizon \cite{Jacobson:2005kr}.  This idea, which I would call \emph{horizon interpretation}, is realised in two different ways. In the first approach, $S_{\va BH}$ is considered to be given by the entanglement entropy (EnEn) of quantum fields across the horizon \cite{PhysRevD.34.373}. Being related to the Ultra-Violet (UV) structure of fields in the proximity of the entangling surface, the EnEn is proportional to the area of the latter. The entropy of the BH is therefore interpreted as a semi-classical feature which does not need to be necessarily related to quantum gravitational degrees of freedom. Nonetheless, the EnEn is a UV divergent quantity; to match the BH entropy, therefore, one has to evoke a UV cutoff at Planck scale, bringing QG back in play. Moreover, the EnEn is also proportional to the number of different field species which exist
in Nature, at odd with the fact that $S_{\va BH}$ seems to have a universal character \cite{Solodukhin:2011gn}. The other manifestation of the horizon interpretation is that $S_{\va BH}$ counts the quantum geometrical degrees of freedom sitting on the horizon surface. This view is supported by computations in Loop Quantum Gravity, where an explicit counting of the vast degeneracy of quantum surface states compatible with the horizon of a BH can be performed \cite{Perez:2017cmj}. The incompleteness of LQG, as well as approximations and some ambiguities, make these computations inconclusive.

On the other side of the debate one finds the idea of interpreting the fact that $S_{\va BH}$ scales as in a lower dimensional system as an indication of some new fundamental aspects of Nature. This led to the formulation of the so called \emph{holographic principle} \cite{Susskind:1994vu}. It is a postulated new principle of Nature, at the level of the equivalence, the gauge or the uncertainty principles. It roughly states that there exists a fundamental description of the classical physical world in terms of a hologram on a lower dimensional screen. 
More precisely, it is a tentative way of rising up to a fundamental principle the so called \emph{entropy bounds}. The first of such bounds was proposed by Bekenstein in \cite{PhysRevD.23.287} with the aim of proving the generalised second law. Indeed, If it were possible to have matter systems with arbitrarily large entropy at a given mass and size, the GSL would be violated simply by throwing such matter into a BH. The entropy bound roughly states that there exists a maximum value of the entropy that a matter system can have, which is proportional to its energy $E$ times the radius $R$ of the smallest sphere that fits around the matter system: $S \lesssim E\,R$. In this way the GSL is safe. The bound is corroborated by some examples \cite{PhysRevD.23.287}, but it has been disproven by several counterexamples. The simplest one is to observe that a collapsing spherical star of given initial entropy and mass will always reach a small enough radius for the bound to be violated. Several tentative refinements eventually brought to a formulation of the bound that it has not been disproven yet, namely Bousso's covariant formulation.
\begin{displayquote}
{\em Let $A(B)$ be the area of an arbitrary co-dimension two
spatial surface $B$ (which need not be closed). A
co-dimension one hypersurface $L$ is called a \emph{light-sheet}
of $B$ if $L$ is generated by light rays which begin at $B$,
extend orthogonally away from $B$, and have non-positive
expansion, $\theta \leq 0$ everywhere on $L$. Let $S$ be the entropy on any light-sheet of $B$. Then $S \leq A(B)/4$. [...] The light-sheet construction is well-defined in the limit where geometry can be described classically. It is conjectured to be valid for all physically realistic matter
systems.} \cite{Bousso:2002ju}
\end{displayquote}
Explaining in full details the Bousso's bound goes clearly beyond the aim of this Section, and I refer to \cite{Bousso:2002ju} and to a series of enlightening on-line lectures \cite{BoussoYouTube} for more details. At the same time, the Bousso's bound is the most structured basis of the holographic principle, which in turn takes on a dominant role in the debate around the questions about BH physics presented in this thesis.
I would like therefore to clarify some points, which I believe are important to better understand the different views on the issues. 

First, reading the statement for the first time one could ask why is the entropy on (the elegant but not obvious geometric notion of) light-sheets considered, instead of, for instance, the more natural entropy on any spacelike surface enclosed in $B$. The answer is that tentatives of defining entropy bounds using other surfaces have been disproved by simple counter examples, as for instance the one reported above for the Bekenstein's bound. The geometric definition of light-sheets allows to avoid those problems. For a hint on how light-sheets look like, see the highly symmetric examples in Fig.~\ref{fig:bousso}. 
\begin{figure}[!h]
\centering
    \includegraphics[width=0.4\textwidth, keepaspectratio]{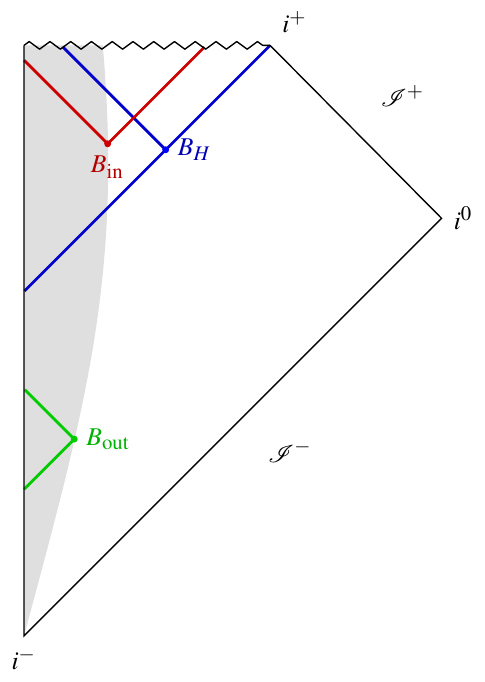}
\caption{The light-sheets are easily identified in highly symmetric situations. (a) A surface $B$ of a sphere in flat spacetime has two light-sheets given by the past and the future ingoing light cones shining from $B$. In other terms, the boundaries of the future and past Cauchy development of $B$. 
The entropy on any light sheet in this case is therefore the same as the entropy contained in the 3-dimensional sphere. This is not the case in general curved geometries, as for instance in (b) a spacetime describing a spherically symmetric collapsing star forming a BH. The light sheet for three different spherical 2D-surfaces are depicted. For a sphere $B_{\rm out}$ outside the horizon, the two light-sheets are the analogue for the case in flat spacetime and again the entropy on them is the same as for the 3-dimensional sphere enclosed. For a sphere $B_{\rm in}$ in the interior part of the black hole, on the other hand, the light-sheets are the two future directed light cones. In this picture $B_{\rm in}$ is the outer boundary of the collapsing star at a given time. The entropy of matter on the out-going one is zero, thus different from the entropy of the enclosed 3-d sphere. The in-going light-sheet, on the other hand, intersect all the collapsing matter. If imagined at later times, at some point such light-sheet would not intersect all the matter anymore, terminating at the singularity. This indicates how the Bousso's bound avoids the simple counterexample to the Bekenstein's bound reported in the main text. See \cite{Bousso:2002ju} for more details. Finally the most interesting surface depicted is the one on the horizon, $B_{\va H}$. The light-sheets associated to it are three: the two pieces of horizon emanating from it, past and future, and the in-going light cone terminating either at the regular centre of the star, or at the singularity depending on the cross-section.}
\label{fig:bousso}
\end{figure}
On the other hand, identifying them in general curved spacetimes and for general surfaces $B$ is usually hard if not impossible. Due to this difficulty, the Bousso's bound has been shown to hold only in highly symmetric situations, and no general proof is available. Second, as written in the statement, the definition of light-sheets is possible only for classical spacetimes and the entropy which is bounded is the entropy given by classical matter degrees of freedom on it. It is valid therefore for a non-evaporating black hole formed by a collapsing mass. In spherical symmetric case, there are three light-sheets associated with a cross-section $B$ of the event horizon at a given time: the pieces of event horizon to the past and to the future of $B$, and the ingoing spherical light front reaching the singularity, see Fig.~\ref{fig:bousso}. The piece of horizon to the past of $B$ is crossed by the matter that formed the BH. The Bousso's bound tells us that the entropy of such matter cannot be bigger than one-fourth the area of the horizon. The future piece does not encounter any matter, and therefore is trivial. The ingoing light front might, depending on the cross section, encounter all the matter, part of the matter to then meet the singularity, or just go straight to the singularity with no matter in it \footnote{Notice that the validity of the construction of such light-sheet is not ensured close to the singularity, since it is not a well defined classical region.}. The entropy of matter on these last two light-sheets is therefore clearly less or equal to the one in the first one. No more information is gained by looking at them. To this point, the bound only gives us information about the entropy of the matter felt into the hole, but it does not give any hint about the nature of the BH entropy in the spirit of the discussion above. It does so when the Bousso's bound is erected to the holographic principle: 
\begin{displayquote}
{\em The area of any surface $B$ measures the information content of an underlying theory describing all possible physics on the light-sheets of $B$.} \cite{Bousso:2002ju}
\end{displayquote}
Given this postulate, the BH entropy acquires now the meaning of measure of all possible degrees of freedom an underlying quantum theory of everything may provide. This includes eventual quantum gravitational effects at the singularity, correlations between fields outside and inside the horizon, and so on. Thus providing an \emph{interior interpretation} of $S_{\va BH}$, in contrast with the horizon interpretation discussed above. Coming now from a fundamental principle, moreover, this interpretation can be applied to cross-sections of the horizon of an evaporating BH, with important consequences regarding the information paradox. The proposed scenarios arising from this line of though are discussed at the end of the  \hyperref[sec:intro1]{introduction to Part I}. To conclude, let me mention that there are no explicit realisations of the holographic principle yet, except for the so called  \emph{$AdS$/CFT correspondence} \cite{Maldacena:1997re}. Roughly speaking, in few particular realisations of String Theory on asymptotically high-dimensional $A$nti-$d$e$S$itter ($AdS$) spacetimes--such as $AdS_5 \times S^5$--, one can find a slicing of the spacetime such that the state of strings on each slide is fully described by data not exceeding $A$ bits, where $A$ is the area of the boundary of the slice. On this boundary one can define an auxiliary lower dimensional supersymmetric Conformal Field Theory (CFT)--such as $N=4$ super Yang-Mills--which generates the unitary evolution of boundary data from slice to slice, and therefore of the bulk spacetime. The idea that such correspondence might be true also in more physical situations has earned a huge consensus in the String Theory community, bringing to conjecture the so called \emph{gauge/gravity duality}, and to the original paper \cite{Maldacena:1997re} being the most cited one in the field of high energy physics. However, even the less ambitious $AdS$/CFT correspondence has not yet been formulated in sufficient detail and with sufficient precision to make a clear argument.

I hope this discussion helps clarifying what are the basis on which the two main different perspectives (with their sub-classes) about the nature of $S_{\va BH}$ rely. None of them gives yet a definitive answer. Additional classical and semiclassical investigations in the spirit of this thesis may help choosing between them, or maybe finding a completely new paradigm.

\bigskip

The ideas discussed above are not the only ones that have arisen from the suggestive relation between thermodynamics and black holes. In Chapter~\ref{chap:desmo}, this thesis also deals with the so called \emph{thermodynamics of spacetime} approach originated in \cite{PhysRevLett.75.1260}. The basic idea is to turn the logic around, in the following sense. The thermodynamical properties of BHs, with the consequent proportionality between area and entropy, are derived from the Einstein's equations. In \cite{PhysRevLett.75.1260}, instead, the relation between area and entropy is taken as an assumption (among others), and the Einstein's equations are derived from the equilibrium Clausius' relation $\delta S = \delta Q/T$  between heat $Q$, entropy $S$ and temperature $T$. One of the assumptions can be relaxed by mean of a non-equilibrium description in which the generalised Clausius' relation is used $\delta S = \delta Q/T + \delta S_i$  \cite{PhysRevLett.96.121301,Chirco:2009dc}. In this more general setting $\delta S_i$ represents the irreversible part of the entropy due to non-equilibrium heat fluxes associated with internal degrees of freedom. In Chapter~\ref{chap:desmo} the framework is discussed to derive the Einstein-Cartan's equations describing a theory of gravity with non-vanishing torsion. In doing so, subtle aspects of both the equilibrium and non-equilibrium settings are discussed, and the ambiguity of some  assumption underlined. Therefore I refer the reader to that final Chapter for more details. Here I would only like to conclude this introductory section with a brief observation. Even though the derivation is based on many assumptions and it is in my opinion not completely clean, the underlying idea is quite interesting. From a non-perturbative quantum gravitational perspective, indeed, one expects that a full QG theory will provide quantum states describing discrete quantum geometries, and a dynamics given by a sort of quantum version of the Einstein's equation. In LQG, for instance, the dynamics is provided by the quantum version of the Hamiltonian constraint, which classically is equivalent to the Einstein's equation. An interesting point commended in \cite{PhysRevLett.75.1260} is that the result can be interpreted as suggesting that there may actually not be quantum version of the Einstein's equation, since they will arise only as an equation of state for the thermodynamical equilibrium description of fundamental degrees of freedom.

}

%% file: Part_2/Lightcone_Thermodynamics/Light_cone_Thermodynamics.tex
\chapter{Light Cone Thermodynamics}\label{chap:lightcone}

\emph{This Chapter completely overlaps with the published paper} \cite{DeLorenzo:2017tgx}.\vspace{2ex}

\section{The Results in a Nutshell}\label{sec:nutshell}

Classical Black Holes behave in analogy with thermodynamical systems \cite{Bardeen1973}. According to general relativity they satisfy the four laws of black hole mechanics. 
The surface gravity $\kappa_{\va SG}$ of a stationary black hole is constant on the horizon: {\em the zeroth law}. Under small  perturbations,  stationary black holes---which are characterized by a mass $M$, an angular momentum $J$,  and a charge $Q$---satisfy the {\em first law}
\be\label{1st}
\delta M=\frac{\kappa_{\va SG}}{8\pi} \delta A+\Omega \delta J+\Phi \delta Q, 
\ee 
where $\Omega$, $\Phi$, and $A$ are the angular velocity, electrostatic potential, and area of the stationary (Killing) horizon.  The Hawking area theorem \cite{hawking1973large}
\be\label{arealaw}
\Delta A\ge0
\ee
is regarded as the {\em second law}. The {\em third law}---expected to be valid from the cosmic censorship conjecture  \cite{Penrose:1969pc}---corresponds to the statement that extremal black holes, for which $\kappa_{\va SG}=0$, cannot be obtained from a non extremal one by a finite sequence of physical processes. 
When quantum effects are considered these analogies become facts of semiclassical gravity. Stationary black holes radiate particles in a thermal spectrum with temperature $T=\kappa_{\va SG}/(2\pi)$ \cite{hawking1974black,hawking1975} so that the first term in \eqref{1st} can be interpreted as a heat term expressed in terms of changes in the black hole entropy $S=A/4$ in Planck units. The area law \eqref{arealaw} is promoted to the generalized second law \cite{PhysRevD.7.2333}.  
In the quantum realm another statement that could be associated to the  third law is that extremal black holes should have vanishing entropy, an argument for which can be found in \cite{Hawking:1994ii}.

All this is naturally interpreted as providing valuable information about the quantum theory of gravity of which the semiclassical treatment should be a suitable limit of. In this respect it is useful to have examples of a similar behaviour in simplified situations. The Fulling-Davies-Unruh thermal properties associated to quantum field theory in flat spacetimes \cite{Fulling:1972md,Davies:1974th,Unruh:1976db} is often used as an example illustrating, in a simplified arena, some of the aspects behind the physics of black holes. In this respect, the Rindler (Killing) horizon associated with a family of constantly accelerated observers in Minkowski spacetime is taken as an analogue of the black hole horizon. This analogy is supported further by the 
statement that the near horizon geometry of a non-extremal black hole can be described, in suitable coordinates, by the metric
\be\label{nhl}
ds^2=-\kappa_{\va SG} R^2 dt^2+dR^2+(r_{\va H}^2 + a^2) dS^2+O(R^3)
\ee
where $dS^2 = d\vartheta^2 + \sin^2\vartheta d\varphi^2$ is the metric of the unit two-sphere, $a\equiv J/M$. At least in the $(R,t)$ ``plane'', the above equation matches the 2d Rindler metric where $R=0$ is the location of the black hole horizon. The thermodynamical properties of Rindler horizons have been extensively discussed in the literature \cite{PhysRevD.87.124031,PhysRevLett.75.1260,PhysRevD.82.124019,PhysRevD.82.024010,Padmanabhan:2009vy}. 
However, strictly speaking, the near horizon geometry is not Rindler due to the presence of the term $r_{H}^2 dS^2$ in the previous equation that makes the topology of the horizon $S^2\times \R$ instead of $\R^3$, which implies the area of the black hole horizon to be finite $A=4\pi (r^2_{H}+a^2)$ instead of infinite. Another obvious difference is that, in contrast with Rindler, the Riemann curvature is non zero at the black hole horizon. Only in the infinite area limit the local geometry becomes exactly that of a Rindler horizon. 
Furthermore, in contrast with black hole horizons, the Rindler horizon has a domain of dependence that includes the whole of what one would regard as the {\em outside} region. In fact the Rindler horizon defines  a good initial value characteristic surface.
More precisely, any regular initial data for an hyperbolic equation  such as a Klein-Gordon or Maxwell field with support on the corresponding wedge---what would be the {\em outside}---can be encoded in data on the Rindler horizon \cite{wald2010general}. An implication of this is that no energy flow can actually escape to infinity without crossing the Rindler horizon. No notion analogous to the asymptotic observers outside of the black hole exists when considering the Rindler wedge and its Killing horizon boundary. A geometric way to stating this is that the Rindler horizon is given by the union of the past light cone of a point at $\sI^+$ with the future light cone of a point at $\sI^-$. In this sense the Rindler wedge is better described as a limiting case of the interior of finite diamonds---see next paragraph---rather than representing faithfully the outside region of a black hole spacetime. 
In this work we show that there exists a more complete analogue of black holes in Minskowski spacetime. 

There is a natural interest in double cone regions in Minkowski spacetime, also called diamonds, in algebraic quantum field theory \cite{Hislop:1981uh} or in the link between entanglement entropy and Einstein equations \cite{PhysRevLett.116.201101}. The conformal relationship with the Rindler wedge has been used in order to define the corresponding modular Hamiltonian and study thermodynamical properties in \cite{Martinetti:2002sz, Martinetti:2008ja}.
Here we concentrate on the causal complement of the diamond, and show that it shares several analogies with the exterior region of a stationary spherically symmetric black hole.
  
Such flat spacetime regions as the diamond and its complement are directly related to the geometry of radial Minkowski  Conformal Killing vector Fields (MCKFs). What we shall show is that radial MCKFs can be classified in a natural correspondence with black holes spacetimes of the Reissner-Nordstrom (RN) family (i.e. $J=0$). They can be timelike everywhere in correspondence with the naked singularity case $M^2 < Q^2$ where the stationarity Killing field is timelike everywhere. But more interestingly, radial MCKFs can become null, and being surface forming, generate conformal Killing horizons. As we will show in Section \ref{MCKF} these are conformal bifurcate Killing horizons analogue, in a suitable sense, to the black hole horizons in the RN family. The results of this paper can be summarised as follows:

\begin{enumerate}
\item   \label{item} {\em Radial MCKFs define conformal Killing horizons:} Radial MCKFs become null on the light cones of two events on Minkowski spacetime that are separated by a timelike interval. By means of a Lorentz transformation these two events can be located on the time axis of an inertial frame. A further time translation can place the two events in a time reflection symmetric configuration so that the symmetry $t\to -t$ of Reissner-Nordstrom spacetimes is reproduced.

\item {\em They have the same topology as black hole Killing horizons:} The topology of the conformal Killing horizon in Minkowski spacetime is $S^2\times \R$ as for the Killing horizons of the RN spacetime. 

\item {\em These horizons are of the bifurcate type:} Radial MCKFs vanish on a 2-dimensional sphere of radius $r_H$ and finite area $A=4\pi r_H^2$ that is the analogue of the minimal surface where the Killing horizon of the RN black hole vanishes. The bifurcate surface is the intersection of the two light cones described above.

\item  {\em They separate events in spacetime as in the BH case:} The global structure of the radial MCKF is closely analogous to the one of the Killing horizon of the RN spacetime. 
More precisely, there are basically the same worth of regions where the radial MCKF and the RN time translational Killing vector field is timelike and spacelike respectively.  In the non-extremal case there are outer and inner horizons in correspondence to the non-extremal RN solution. One of the two asymptotically flat regions of the maximally extended RN spacetime corresponds to the points in the  domain of dependence of the portion of the $t=0$ hypersurface in Minkowski spacetime inside the bifurcate sphere, namely {\em the diamond}; the other asymptotically flat region corresponds to the domain of dependence of its causal complement, namely {\em the black hole} exterior in our analogy. There are regions where the radial MCKF becomes spacelike. These  too are in correspondence with regions in the non-extremal RN black hole, namely the regions between the inner and the outer horizons.  In the extremal limit the regions where the radial MCKF is spacelike, as well as one of the asymptotic region, disappear and the correspondence with the extremal RN solution is maintained.
All this will be shown in detail in the following section; the correspondence is illustrated in Figure~\ref{fig:penrose}.

\item {\em They satisfy the zeroth law:} The suitably generalized notion of surface gravity $\kappa_{\va SG}$ is constant on the conformal Killing horizon: {\em the zeroth law}.  Extremal Killing horizons have $\sg=0$. 

\item     {\em They satisfy the first law:} Considering the effects of matter perturbations described by a conformally invariant matter model, one can show that radial conformal Killing horizons satisfy the balance law $\delta M=\kappa_{\va SG}\, \delta A/ (8 \pi)+\delta M_\infty$: \emph{the first law}. Here $\delta M$ is the conformally invariant mass of the perturbation, $\delta M_\infty$ is the amount of conformal energy flowing out to future null infinity $\sI^+$, and  $\delta A$ is what we call  {\em conformal area change}. The name stems from the fact that it corresponds to the change of a geometric notion with the meaning of horizon area in the appropriate conformal frame.

\item  {\em They satisfy the second law:} In the type of processes considered above and assuming the usual energy conditions, $\delta A\ge 0$: {\em the second law}. 

\item {\em They have constant (conformal) temperature:} When quantum fields are considered, a constant Hawking-like temperature $T=\kappa_{\va SG}/(2\pi)$ can be assigned to radial MCKFs. In view of this, $\delta S=\delta A/4$ in Planck units acquires the meaning of entropy variation of the conformal horizon.

\item {\em They satisfy a version of the third law:} Extremal radial MCKFs have vanishing temperature as well as vanishing entropy: {\em the third law}.

\item {\em Minkowski vacuum is the associated Hartle-Hawking state:} The Minkowski vacuum of any conformally invariant quantum field can be seen as the state of thermal equilibrium---usually called Hartle-Hawking state---in the Fock space defined with respect to the MCKF.

\item The near MCKF horizon limit matches the expression \eqref{nhl} with $a=0$. The Rindler horizon limit could be obtained by sending the events mentioned in Item \ref{item} suitably to future $i^+$ and past $i^-$ timelike infinity respectively. In that limit $r_{\va H}\to \infty$, the near horizon metric becomes the Rindler metric, and the corresponding radial MCKF becomes the familiar boost Killing field.
   
\end{enumerate}

The properties listed above will be discussed in more detail in the sections that follow. In Section \ref{MCKF} we construct general radial MCKFs
from the generators of the conformal group $SO(5,1)$, and explain their geometry. The causal domains they define and the analogy with black holes shown in Figure~\ref{fig:penrose} will be clarified there. The analogue of classical laws of black hole thermodynamics are shown to hold in a suitable sense for radial MCKFs in Section \ref{sec:LCthermo}. Finally we show, in Section \ref{HRCT}, that a semiclassical temperature can be assigned to radial MCKFs and we discuss the physical meaning of that temperature. 

\section[Conformal Killing Fields in Minkowski Spacetime]{Conformal Killing Fields in Minkowski\\Spacetime}\label{MCKF}
\begin{figure}[!h]
\center
\begin{center}
\begin{minipage}[c]{.33\textwidth}
\centering
	\includegraphics[height=8cm]{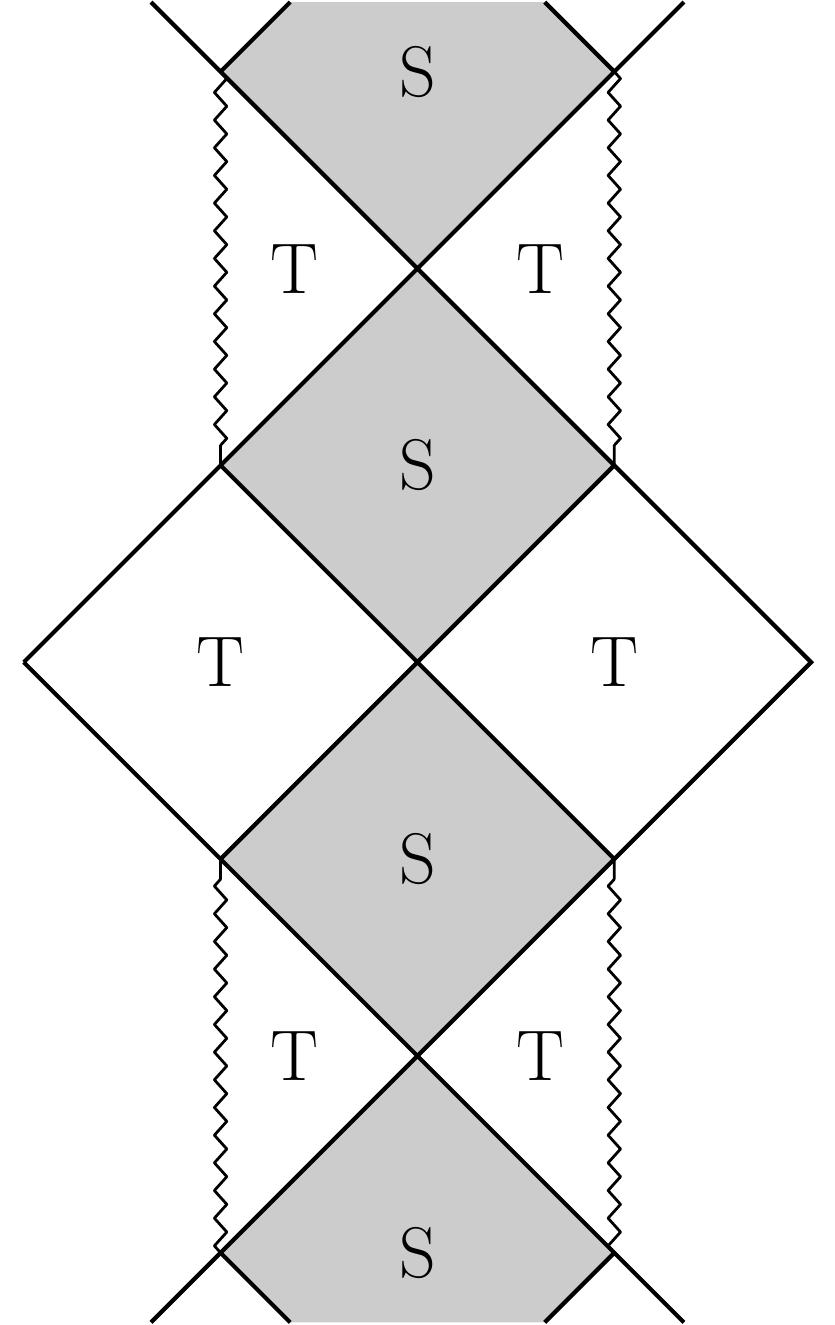}
\end{minipage}%
\begin{minipage}[c]{.6\textwidth}
\centering
          \includegraphics[height=8cm]{light-cones}
\end{minipage}\\
\vspace{2cm}
\begin{minipage}[c]{.33\textwidth}
	\hspace{1.4cm}
	\includegraphics[height=8cm]{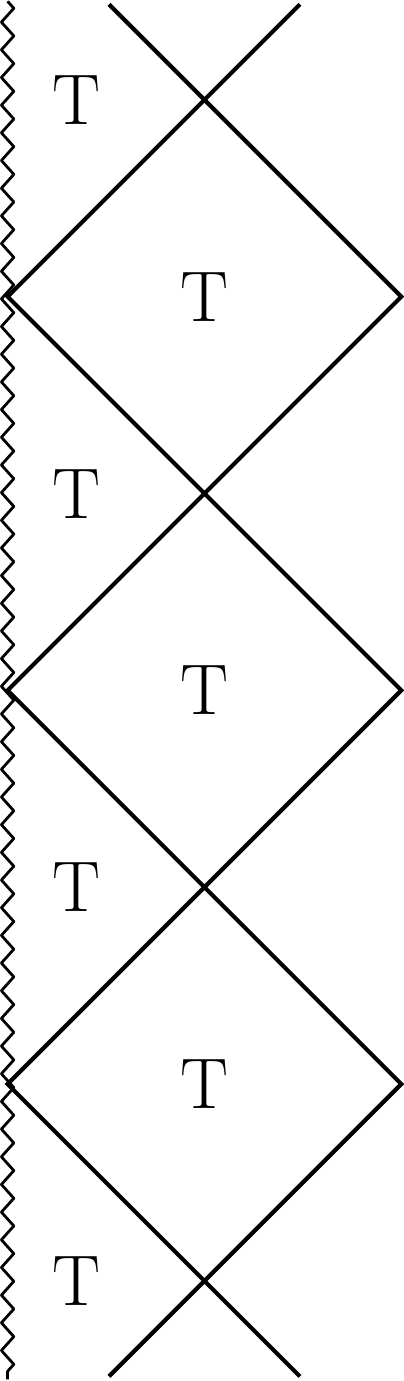}
\end{minipage}
\begin{minipage}[c]{.6\textwidth}
\centering
          \includegraphics[height=8cm]{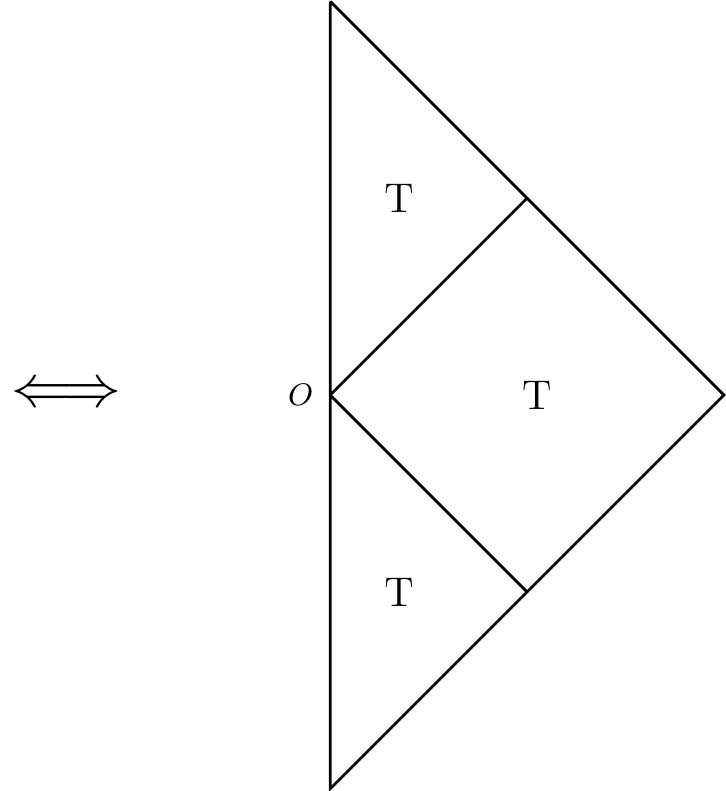}
\end{minipage}
\end{center}

\caption{The Penrose diagram of the Reissner-Nordstrom black hole on the left compared with the causal structure of the radial CKF in Minkowski spacetime on the right, in both the non-extremal $\Delta>0$ and extremal $\Delta =0$ case. The letters $S$ and $T$ designate the regions where the Killing or conformal Killing fields are spacelike or timelike respectively. The light cone emanating from the points $O^\pm$ (and $O$ in the extremal case) are the hypersurface where the MCKF is null.}
\label{fig:penrose}
\end{figure}

The conformal group in four dimensional Minkowski spacetime $\M^4$  is isomorphic to the group $SO(5,1)$ with
its 15 generators given explicitly by \cite{francesco2012conformal}
\be
\begin{split}
& P_\mu=\partial_{\mu} \ \ \ \ \ \ \ \ \ \ \ \ \ \ \ \ \ \ \ \ \ \ \ \ \ \ \ \ \ \ {\rm Translations} \\
& L_{\mu\nu}=\left(x_\nu\partial_{\mu} -x_\mu\partial_{\nu}\right)  \ \ \ \ \ \ \ \ \ \ \ \ \ {\rm Lorentz\ transformations} \\
& D= x^\mu\partial_{\mu}  \ \ \ \ \ \ \ \ \ \ \ \ \ \ \ \ \ \  \ \ \ \ \ \ \ \ \ \  {\rm Dilations}\\
& K_{\mu}=\left(2 x_\mu x^\nu \partial_{\nu} -x\cdot x\, \partial_{\mu}\right)\ \ \ \ \ \ \, {\rm Special\ conformal\ transformations,} 
\end{split}
\ee
where $f\cdot g \equiv f^\mu g_\mu$.
Any generator defines a Conformal Killing Field in Minkowski spacetime (MCKF), namely a vector field $\xi$ along which the metric $\eta_{ab}$ changes only by a conformal factor:
\be\label{lietext}
\sL_\xi \,\eta_{ab}=\nabla_a\xi_b+\nabla_b\xi_a=\frac{\psi}{2} \eta_{ab}
\ee
with
\be
\psi = \nabla_a \xi^a\,.
\ee
Consider now the Minkowski metric in spherical coordinates
\be\label{eq:min}
ds^2 = -dt^2 + dr^2 + r^2 dS^2\,.
\ee
Then dilations can be written as
\be
D=r\partial_r+t\partial_t
\ee
and $K_0$ as
\be
K_0=-2t D-(r^2-t^2) P_0\,.
\ee
Together with $P_0=\partial_t$, those are the only generators that do not contain angular components. Hence the most general radial MCKF has the form 
\be \label{explicit}
\begin{split}
\xi =&-a K_0+b D+c P_0\\
 =&(2at+b) D+ [a(r^2-t^2)+c]  P_0, 
\end{split}
\ee
with $a,b,c$ arbitrary constants.
Explicitly
\be\label{eq:CKF}
\begin{split}
\xi^\mu \partial_\mu &= \big[a(t^2+r^2)+bt+c\big]\partial_t + r(2at + b)\partial_r\\
&=(av^2+bv+c)\partial_v+(au^2+bu+c)\partial_u\,,
\end{split}
\ee
where $v=t+r$, $u=t-r$ are the standard null coordinates. The norm of $\xi$ is easily computed to be
\be
\label{eq:normabc}
\xi \cdot \xi= - (av^2+bv+c)(au^2+bu+c)\,.
\ee
Its causal behaviour, therefore, can be studied introducing the quantity
\be\label{Delta}
\Delta \equiv b^2 - 4ac\,.
\ee
The complete classification of such MCKFs is given in \cite{doi:10.1063/1.532903}. Here we are interested in the case $a \neq 0$, where we have three different types of behaviour depending on the sign of the parameter $\Delta$. 
When $\Delta < 0$, the MCKF is timelike everywhere, like the stationarity Killing field in the Reissner-Nordstrom solutions with a naked singularity $M^2<Q^2$. When $\Delta > 0$, on the other hand, the MCKF is null along two constant $u$ and two constant $v$ null hypersurfaces respectively given by 
\be
u_\pm = \frac{-b \pm \sqrt{\Delta}}{2a} \ \ \ \ {\rm or} \ \ \ \  v_\pm = \frac{-b \pm \sqrt{\Delta}}{2a}\,.
\ee
In other words, the MCKF is null on the past and future light cones of two points $O^{\pm}$, with coordinates given respectively by $O^{\pm} : (t = u_{\pm}, r = 0)$. These two light cones divide Minkowski spacetime into six regions. In those regions the norm of the MCKF changes going from timelike to spacelike as depicted on the top right panel of Figure~\ref{fig:penrose}. The boundary of these regions are null surfaces generated by the MCKF; they define conformal Killing horizons. The vector field $\xi$ vanishes at the bifurcate 2-dimensional surface defined by the intersection of the previous null surfaces, namely at the sphere
\be\label{rh}
t=t_H=-\frac{b}{2a}\,, \qquad r=r_H=\frac{\sqrt{\Delta}}{2a}\,.
\ee
The {\em ``extremal''} case $\Delta=0$ is a limiting case between the other two: the MCKF is null on the light cones $u_0=v_0=-b/(2a)$ emanating from a single point $O$, and timelike everywhere else. This is depicted in the bottom right panel of Figure~\ref{fig:penrose}.

It is interesting to notice that the four regions around the bifurcate sphere in the non-extremal case are in one-to-one correspondence with the corresponding four regions around the bifurcate sphere in the case of stationary black holes of the Reissner-Nordstrom family. The correspondence is maintained in the extremal limit where the bifurcate sphere degenerates to a point and the four regions collapse to a single one. In the black hole case the bifurcate sphere is pushed to infinity and one of the asymptotically flat regions disappears. In our case the bifurcate sphere is shrunk to a point at the origin and the region in the interior of the light cones, the diamond, disappears. The analogy is emphasised in Figure~\ref{fig:penrose}.

\begin{figure}[t]
\center
\subfigure[]{
   \includegraphics[height=9cm]{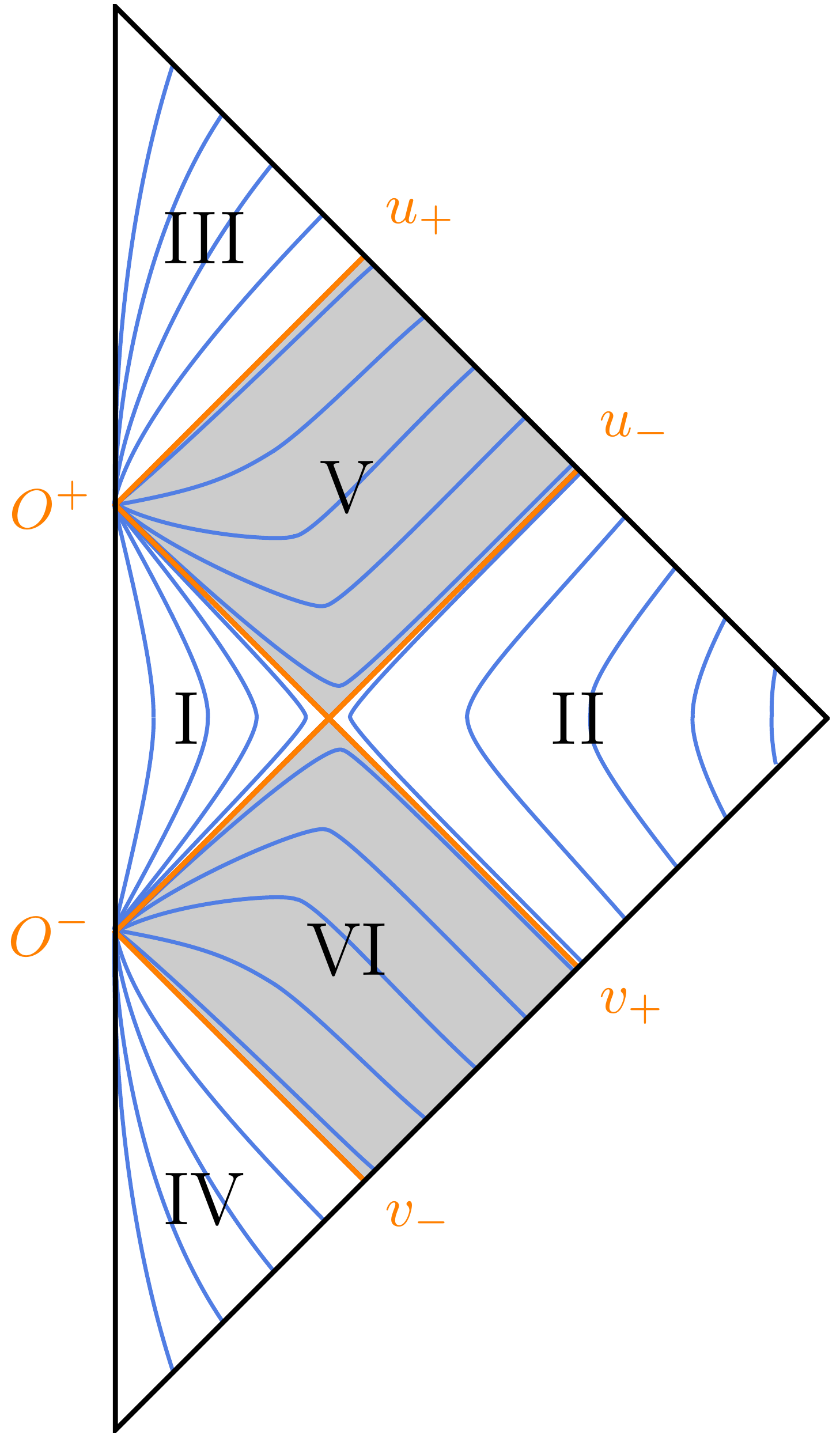}}
\hspace{15ex}
\subfigure[]{
   \includegraphics[height=9cm]{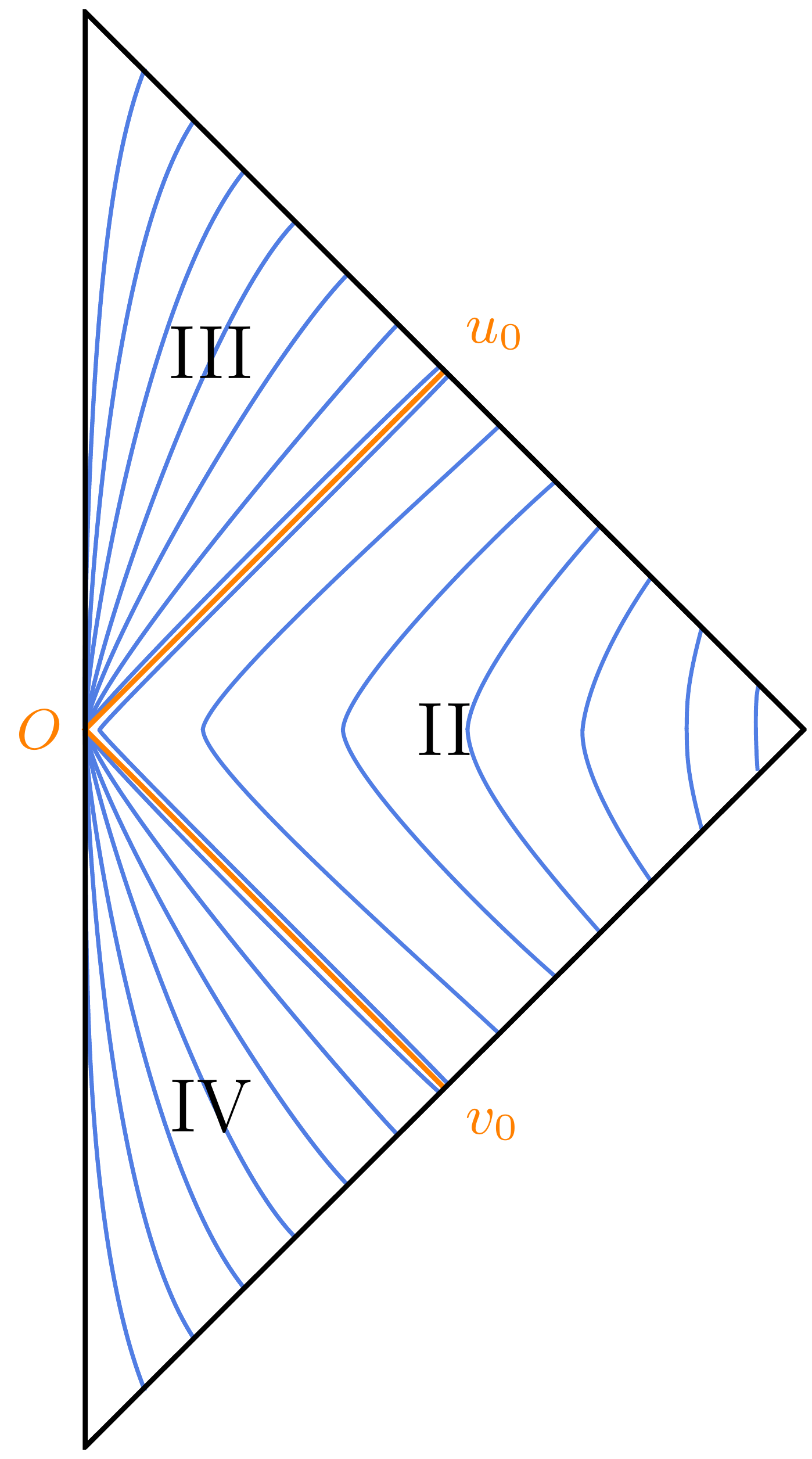}}
\caption{The flow of the radial MCKF, depending on the value of the parameter $\Delta$, Eq.~\eqref{Delta}. When $\Delta < 0$: the MCKF is timelike everywhere. a) $\Delta > 0$: the Minkowski spacetime is divided into 6 different regions, where the norm of the MCKF changes from being timelike to being spacelike, through being null along the four light rays $u_\pm$, $v_\pm$. b) $\Delta = 0$: the MCKF is everywhere timelike except for the two null rays $u_0=v_0$ where it is null.}
\label{fig:families}
\end{figure}

The flow of $\xi$ describes uniformly accelerated observers, with integral curves being a one parameter family of rectangular hyperbolas given by \cite{doi:10.1063/1.532903}
\be\label{eq:hyperbolas}
t^2 - \left(r+\frac{\zeta}{2a}\right)^2 = \frac{\Delta - \zeta^2}{4a^2}\,,
\ee
where $\zeta$ is the parameter labeling members of the family. The complete situation is depicted in Figure~\ref{fig:families}. From the picture it is clear that, seen from the point of view of the observers that follow the MCKF in Region II, the boundary of the region is a {\em bifurcate conformal Killing horizon} with topology $S^2 \times \R$. This is the same topology as the one of {\em bifurcate Killing horizons} of stationary black holes in the asymptotically flat spacetime context.
To summarise: radial conformal Killing fields in Minkowski spacetime generate bifurcate conformal Killing horizons that reproduce the main topological features of stationary spherically symmetric Killing horizons. This is the first obvious indication that makes MCKFs interesting for drawing analogies with black holes. The aim of what follows is to show that this analogy is more profound and extends very nicely to the thermodynamical properties of black hole Killing horizons. In Section~\ref{sec:LCthermo}, indeed, we will be able to define, in a suitable sense that will become clear, the four laws of thermodynamics for  bifurcate MCKF horizons.

\subsection{Introducing two geometric scales defining the \\radial MCKF} 

Here we associate the parameters $a, b$, and $c$ in \eqref{explicit} with geometric notions. First we set $b=0$ by means of a time translation $t\to t-b/(2a)$. In this way the points $O^{\pm}$ are placed on the time axis in the future and the past of the origin at equal timelike distance, and the distributions of regions become $t$-reflection symmetric. We make this choice from now on. Notice in addition that the parameter $a$ has dimension $[length]^{-2}$, while $c$ is dimensionless. We can therefore rewrite those constants in terms of two physical length scales. The first one is the radius of the bifurcate sphere $r=r_{\va H}$, Eq.~\eqref{rh}:
\be\label{rH}
r_{\va H}^2=\frac{\Delta}{4a^2} = -\frac{c}{a}\,.
\ee
There is also another natural geometric scale associated to the radius of the sphere $r_{\va O}$ at $t=0$ where we demand $\xi$ to be normalized. Such sphere represents the ensemble of events where the MCKF can be associated with the orbits of observers. We call this sphere the {\em observers sphere}. The normalization condition at $r_{\va O}$ is the analogue of the normalization condition for the stationarity Killing vector field at infinity in asymptotically flat  stationary spacetimes, e.g. stationary black holes, or the selection of a special observer trajectory when normalizing the boost Killing field in the Rindler wedge. Therefore, we demand the condition  $\xi\cdot\xi|_{t=0,r=r_{\va O}}=-1$ which, together with Eq.~\eqref{rH}, allows to determine both $a$ and $c$ as a function of $r_{\va H}$ and $r_{\va O}$. Explicitly one finds
\be
a = \frac{1}{r_{\va O}^2-r_{\va H}^2}\,, \ \ \ \ \ 
c = -\frac{r_{\va H}^2}{r_{\va O}^2-r_{\va H}^2}\,.
\ee
The radial conformal Killing field takes then the form
\be\label{eq:CKFb0}
\begin{split}
\xi^\mu \partial_\mu &= \frac{1}{r_{\va O}^2-r_{\va H}^2}\Big[(t^2+r^2-r_{\va H}^2)\partial_t + 2tr\,\partial_r\Big]\\
&=\frac{v^2-r_{\va H}^2}{r_{\va O}^2-r_{\va H}^2}\,\partial_v+\frac{u^2-r_{\va H}^2}{r_{\va O}^2-r_{\va H}^2}\,\partial_u\,;
\end{split}
\ee
its norm becomes
\be\label{eq:norm}
\xi \cdot \xi = - \frac{(v^2-r_{\va H}^2)(u^2-r_{\va H}^2)}{(r_{\va O}^2-r_{\va H}^2)^2}\,;
\ee
the parameter $\Delta$
\be
\Delta =\frac{4r_{\va H}^2}{(r_{\va O}^2-r_{\va H}^2)^2}\,,
\ee
which implies
\be
u_\pm = v_\pm = \pm r_{\va H}\,.
\ee
From equation \eqref{eq:CKFb0} we clearly see that $\xi$ vanishes at the bifurcate sphere $r=r_{\va H}$ and that $\xi=\partial_t$ at the observers sphere $r=r_{\va O}$; both spheres are defined to be on the $t=0$ surface. The vector field vanishes also at $O^{\pm}$. These two length scales completely determine the radial MCKF forming conformal Killing horizons.

\section{Light Cone Thermodynamics}\label{sec:LCthermo}
\begin{figure}[t]
\center
\includegraphics[height=9cm]{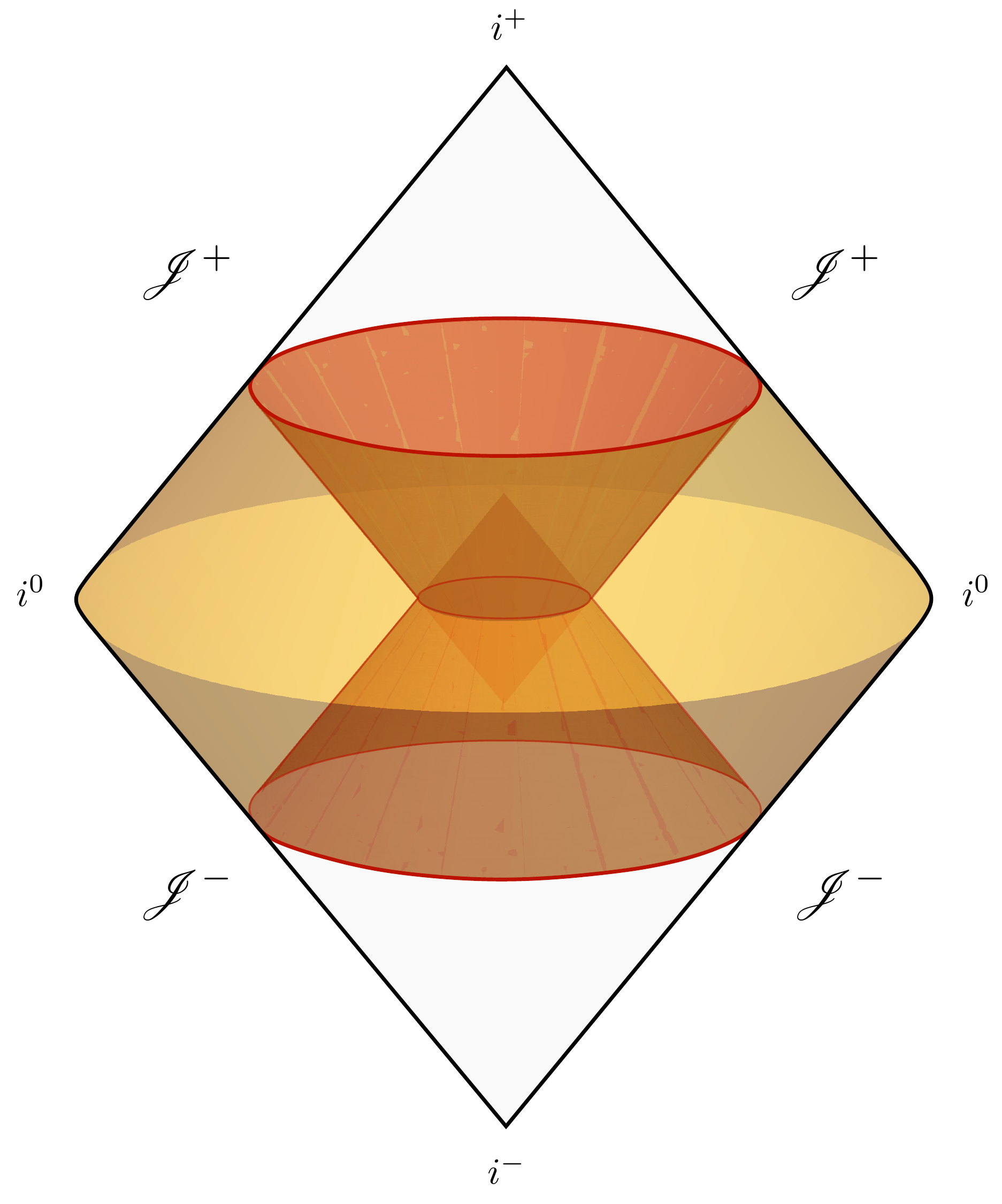} 
\caption{A $(2+1)$ dimensional diagram depicting the regions of interest in Minkowski spacetime. The two cones truncated at the sphere of radius $r=r_{\va H}$ represent the bifurcate conformal Killing horizon. They meet future and past null infinity on spherical cross-sections, represented by the two bigger rings. The central ring is the bifurcate sphere $r = r_H$, where $\xi=0$. The horizon is therefore a sphere with radius growing at the speed of light. In the center, one can also see the Cauchy development of the bifurcate sphere, \emph{the diamond}, Region I. Shaded in yellow is Region II, the region representing the outside of the horizon. Geometrically, it is the Cauchy development of the complement of Region I. The remaining part of Minkowski spacetime is occupied by Regions III to VI, which, to simplify the picture, are not clearly depicted here.}
\label{fig:Penrose_CD}
\end{figure}

In this central section of the paper, we will formulate the laws of thermodynamics for the bifurcate conformal Killing horizon generated by the radial MCKF. The horizon is defined by the two pieces of light cones meeting at the bifurcate sphere of radius $r_H$. It is the boundary of the causal complement of the diamond, Region II: the analogue of the exterior region of a stationary black hole spacetime. In Figure~\ref{fig:Penrose_CD}, the $S^2 \times \R$ topology of the horizon, together with the structure of Region II, is emphasised.

As already mentioned, see Eq.~\eqref{lietext}, a MCKF $\xi$ satisfies
\be
\sL_\xi \,\eta_{ab}=\nabla_a\xi_b+\nabla_b\xi_a=\frac{\psi}{2} \eta_{ab}
\ee
with
\be
\psi = \nabla_a \xi^a\,.
\ee
A conformal Killing horizon $H$ is defined as the surface where the MCKF is null, $\xi \cdot \xi=0$. Therefore, the gradient of $\xi\cdot\xi$ must be proportional to the normal to the horizon $\xi$. The proportionality factor defines the surface gravity $\sg$ \footnote{\label{fofo}It is easy to check that $\sg$ is invariant under conformal transformations $\eta_{ab}\to g_{ab}= \Omega^2 \eta_{ab}$ \cite{Jacobson:1993pf}. Under such transformations we have \be\nabla_a(g_{bc} \xi^b\xi^c)\hat=\nabla_a(\Omega^2 \eta_{bc} \xi^b\xi^c)\hat=\Omega^2 \nabla_a(\eta_{bc} \xi^b\xi^c)\hat=\sg\Omega^2 \eta_{ab}\xi^b\hat=\sg g_{ab}\xi^b,\ee 
where we used that $\xi\cdot \xi=0$ on the horizon.} 
via the equation
\be\label{susu}
\nabla_a (\xi\cdot \xi)\,\hat=-2\sg \eta_{ab} \xi^b\,.
\ee
The symbol $\hat=$ stands for relations valid only at the horizon; we will use this notation whenever stressing such property is necessary. The CKF is also geodesic at the horizon \cite{sultana2004conformal,dyer1979conformal}, so that one can define the function $\kappa$ as
\be\label{eq:inaffinity}
\xi^a \nabla_a \xi^b {\,\hat=\,}\kappa\,\xi^b\,.
\ee
Thus $\kappa$ is the function measuring the failure of $\xi$ to be an affine geodesic on the horizon.
While for Killing horizons $\kappa = \sg$, for conformal Killing horizons the following relation is valid:
\be
\sg  = \kappa - \frac{\psi}{2}\,.
\ee
We will now use these relations in the special case of the MCKF defined in the previous sections.

\subsection{The zeroth law}
It is immediate to show that, for a general CKF, the quantity $\sg$ is Lie dragged along the field itself \cite{sultana2004conformal,dyer1979conformal}; namely
\be
\sL_{\xi}\sg = 0\,.
\ee
In our case, by spherical symmetry, this implies that $\sg$ is actually constant on the horizon $H$ \footnote{It can however be shown for a general CKF under some assumptions \cite{sultana2004conformal}.}. This proves \emph{the zeroth law} of light cone thermodynamics.

\subsection{The first law}\label{sec:firstlaw}
Let us now define the energy-momentum current
\be\label{cucur}
J^a = T^{ab}\xi_b \,.
\ee
Conformally invariant field theories satisfy $T^a_{\;a}=0$ on shell \cite{francesco2012conformal}. For these theories, the current $J^a$ is conserved. Indeed
\be
\nabla_a J^a = \frac{\psi}{4} \;T^a_{\;a}=0 \,.
\ee
In such cases the current defines a conserved charge
\be\label{M}
M = \int_\Sigma T_{ab} \, \xi^a d\Sigma^b
\ee
with $d\Sigma^a$ being the volume element of a general Cauchy hypersuface $\Sigma$.

We want now to study the analogue of what is called the ``physical process version'' \cite{Hawking1972,Gao:2001ut} of the first law of black hole thermodynamics, which is also valid for more general bifurcate Killing horizons \cite{Amsel:2007mh} \footnote{There are works in the literature where the mechanical laws for conformal Killing horizons are investigated from the purely Hamiltonian perspective \cite{Chatterjee:2015fsa, Chatterjee:2014jda}. The strategy used in our specific and simple flat spacetime example seems more transparent for a deeper geometric insight.}. This will define the \emph{first law of light cone thermodynamics}. 
Let us therefore consider the process in which a small amount $\delta M$ of such ``energy'' \footnote{See Subsection~\ref{sec:energy} for a discussion on the meaning of such a conserved quantity.} passes through what plays the role of the future horizon $H^+$, namely the future light cone $u=u_-$. The passage of the matter will perturb the horizon. We will show that, at first order in linearised gravity, there is a balance law relating $\delta M$ with the conformal area change of the horizon $H^+$ (see below). 
The crucial difference in proving this relation with respect to the black hole case is that here the cross-sectional area of the horizon is changing along the affine geodesic generators of the light cone, even when no perturbation is considered. More technically, if we take the advanced time $v=t+r$ as an affine parameter along the null generators $\ell=\partial_v$ of $H^+$ in the flat background geometry, then one can show that the expansion $\theta$ is
\be\label{eq:MinExp}
\theta = \frac{1}{r} = \frac{2}{v-u}\,.
\ee
By definition, the expansion is the rate of change of the cross-sectional area with respect to the parameter \cite{poisson_2004}, namely,
\be\label{eq:expdS}
\theta =\frac{1}{dS} \frac{\partial dS}{\partial v}\,.
\ee
In our case $dS = 1/4\,(v-u)^2 \sin\vartheta d\vartheta d\varphi = $ and the two above equations are indeed in agreement.
The non vanishing of those quantities, therefore, implies that the area of the horizon is constantly increasing even if no flux of energy is considered. Consequently, we need to distinguish the two different changes in area: the background one that we call $dS_0$, and the one induced by the passage of the perturbation that we call $dS_1$. When a generic flux of energy is considered, we can write the expansion as
\be
\theta =\frac{1}{dS_0+dS_1} \frac{\partial}{\partial v} ( dS_0 + dS_1)\,.
\ee
In the approximation in which $dS_1 / dS_0  \ll 1$, we can expand it up to first order and find
\be\label{eq:ExpExp}
\begin{split}
\theta &= \frac{1}{dS_0}\frac{\partial dS_0}{\partial v} + \frac{1}{dS_0} \left(\frac{\partial dS_1}{\partial v} - \frac{1}{dS_0} \frac{\partial dS_0}{\partial v} dS_1\right)+ O\left( \frac{dS_1}{dS_0}\right)^2\\
&\equiv\theta_0 +\theta_1+ O\left( \frac{dS_1}{dS_0}\right)^2,
\end{split}
\ee
where $\theta_0$ is the unperturbed expansion given by Eqs.~\eqref{eq:MinExp}-\eqref{eq:expdS} and where we defined the perturbation 
\be\label{eq:th1}
\theta_1 \equiv \frac{1}{dS_0} \left(\frac{\partial dS_1}{\partial v} - \theta_0 dS_1\right)= \frac{\partial}{\partial v} \left(\frac{dS_1}{dS_0}\right)\,.
\ee
Moreover, by the Raychauduri equation one has that the variation of the expansion is connected with the flux of energy as \cite{wald2010general} 
\be\label{raycha}
\ell(\theta) = -\frac{1}{2} \theta^2 - \sigma_{ab}\sigma^{ab} + \omega_{ab}\omega^{ab} -8\pi T_{ab} \ell^a\ell^b\,;
\ee
where $\sigma_{ab}$ and $\omega_{ab}$ are the shear and twist tensors respectively, and $\ell(f) = \partial_v f$ for any function $f$.
Using Eq.~\eqref{eq:ExpExp} and the fact that, in this case $\omega^{0}_{ab}=\sigma^0_{ab}=0$, we get
\be
\ell(\theta_0) + \ell(\theta_1) = -\frac{1}{2} \theta_{0}{}^{2} -\theta_0\theta_1 -8\pi \delta T_{ab} \ell^a\ell^b+ O(\theta_1)^2\,,
\ee
where $\delta T_{ab}$ represents a small energy perturbation justifying the use of the perturbed equation \eqref{eq:ExpExp} \footnote{One may wonder whether $\ell$ is still an affinely parametrised generator of the horizon in the perturbed spacetime. This would be the case if the $\ell=du$ continues to be a null one-form for the metric $g_{ab}=\eta_{ab}+\delta g_{ab}$. Using the gauge symmetry of linearized gravity $\delta g_{ab}^\prime=\delta g_{ab}+2 \nabla^{\va (0)}_{(a}v_{b)}$ (for a vector field $v^a$), the condition $g^{ab} du_adu_b=0$ is equivalent to
 $\delta g^{uu}+\ell(\ell\cdot v)=0$, which can be solved for the gauge parameter $v$. }. 
Since $\theta_0$ is, by definition, the solution of the unperturbed Raychauduri equation---Eq.~\eqref{raycha} with $T_{ab} =0$---, the above equation reduces at first order to
\be\label{eq:rayTh1}
\ell(\theta_1) +\theta_0\theta_1 = - 8\pi \delta T_{ab} \ell^a\ell^b\,.
\ee
The last elements we need before starting the proof of the first law are the following: first notice from Eq.~\eqref{eq:CKFb0} that, on the horizon, our MCKF $\xi$ is parallel to $\ell$, namely
\be\label{alphaell}
\xi \heq \alpha \ell\,,
\ee
with $ \alpha = (v^2-r_{\va H}^2)/(r_{\va O}^2-r_{\va H}^2)$.
Substituting the above equation into definition \eqref{eq:inaffinity}, one can show that
\be\label{kappaell}
\kappa \,\hat=\, \ell(\alpha)=\partial_v \alpha \,.
\ee

Now we are ready to prove the first law of light cone thermodynamics. We consider a generic perturbation $\delta M$ of conformally invariant energy defined by \eqref{M}. Since this quantity is conserved, we can choose the Cauchy surface to integrate over in the more convenient way. For our purposes, we choose the union of the future horizon $H^+$ with the piece of $\sI^+$ contained in Region II, the latter being $\Sigma_\infty \equiv \rm {II} \cap \sI^+$. Eq.~\eqref{M} therefore becomes
\be\label{42}
\begin{split}
\delta M &=\int\limits_{H^+} \delta T_{ab} \, \xi^a d\Sigma^b_{0}+\int\limits_{\Sigma_\infty} \delta T_{ab} \xi^a d\Sigma_\infty^a=\\
 &= \int\limits_{H^+} \alpha \delta T_{ab} \,\ell^a\ell^b \,dS_0 dv +\delta M_{\infty}\\
 &= -\frac{1}{8\pi} \int\limits_{H^+} \alpha \Big( \partial_v(\theta_1) +\theta_0\theta_1 \Big) dS_0 dv+\delta M_{\infty} \\
 &= -\frac{1}{8\pi} \int\limits_{H^+} \alpha \Big( \partial_v(\theta_1) dS_0 +\partial_v (dS_0)\theta_1 \Big) dv+\delta M_{\infty} \\
 &= -\frac{1}{8\pi} \left[ \oint_{H^+(v)} \alpha\,\theta_1 dS_0 \Bigg{|}_{v_+}^{\infty} - \int\limits_{H^+} \kappa\, \theta_1 dS_0 dv \right]+\delta M_{\infty} \\
 &= \frac{\sg}{8\pi} \int\limits_{H^+} \frac{\kappa}{\sg} \,\theta_1 dS_0 dv+\delta M_{\infty}
\end{split}
\ee
To go from the first to the second line, we have used the fact that the unperturbed surface element of the horizon is given by $d\Sigma_0^a=\ell^a dS_0dv$, the fact that we defined the energy flux at infinity as
\be
\delta M_{\infty}\equiv\int\limits_{\Sigma_\infty} \delta T_{ab} \xi^a d\Sigma_\infty^b\,,
\ee
and the proportionality between $\xi$ and $\ell$, Eq.~\eqref{alphaell}. The third line follows from the perturbed Raychauduri equation \eqref{eq:rayTh1}. Using the definition of $\theta_0$ \eqref{eq:expdS} we obtain the fourth line. 
Integrating by parts and using (\ref{kappaell}) leads to line five. The boundary term vanishes from the fact that $\alpha(r_H)=0$, i.e. $\xi^a=0$ at the bifurcate surface, and that we choose initial condition at infinity in the usual teleological manner \cite{Gao:2001ut}, namely $\theta_1(\infty)=0$. 

We define the conformal area change $\delta A$ of the horizon as
\be \label{eq:area}
\delta A \equiv \int\limits_{H^+} \frac{\kappa}{\sg} \,\theta_1 dS_0 dv= \frac{8\pi}{\sg}{\int\limits_{H^+} \delta T_{ab} \,\xi^ad\Sigma^b},
\ee
in terms of which the first law follows
\be\label{firstLaw}
\delta M = \frac{\kappa_{\va SG}}{8\pi} \, {\delta A}+\delta M_{\infty}\,.
\ee

Some remarks are in order concerning the interpretation of equation \eqref{eq:area} and (\ref{firstLaw}):

First notice that the definition of $\delta A$ reduces to the standard expression when $\xi$ is a Killing field. Indeed, in that case $\kappa=\sg$ and the unperturbed area of the horizon is constant, i.e. $\theta_0=0$.  From the definition \eqref{eq:th1}, \eqref{eq:area} is therefore the change in area of the Killing horizon. 

Now we argue that, in a suitable sense, \eqref{eq:area} retains the usual interpretation in the case of the expanding ($\theta_0\not=0$) conformal Killing horizon associated to a MKCF when perturbed with conformal invariant matter.
Under such circumstances, it can be verified that all the quantities appearing in the first law (\ref{firstLaw}) are conformal invariant; this follows directly from the conformal invariance of  $\sg$ (see Footnote \ref{fofo}), and that of the flux density $ \delta T_{ab} \,\xi^a d\Sigma^b$ when conformal matter is considered \footnote{\label{FOOT}This a consequence of the fact that under a transformation $ g'_{ab}=\Omega^2 g_{ab}$ the energy momentum tensor of conformally invariant matter $\delta T_{ab}$ transforms as $\delta T'_{ab}=\Omega^{-2}\delta T_{ab}$ \cite{wald2010general}, and the volume element transforms as $d\Sigma'^a = \Omega^2 d\Sigma^a$, see Eq.~\eqref{sigmas}.}. Therefore, $\delta A$ is conformally invariant and this is the key for its geometric interpretation. To see this one can conformally map Minkowski spacetime to a new spacetime $g_{ab}=\Omega^2 \eta_{ab}$ where $\xi$ becomes a {\em bonafide} Killing vector field. Under such conformal transformation $\kappa \to \kappa_{\va SG}$ and $\theta_0 \to 0$, and thus the conformal invariant quantity $\delta A$ acquires the standard meaning of horizon area change, thus justifying its name. 
We show an explicit realization of such conformal map in Appendix~\ref{AP2}.

Finally, as $\xi$ diverges at $\sI^+$ one might be worried that $\delta M_{\infty}$ might be divergent.
However, for massless fields the peeling properties of $T_{ab}$ are just the right ones for $\delta M_{\infty}$ to be convergent. Indeed this follows from the fact that $$T_{uu}=\frac{T_{uu}^0}{v^2}+O(v^{-3})\ \ \ \ \ \ \ T_{uv}=\frac{T_{uv}^0}{v^4}+O(v^{-5})\,,$$ and the form of $\xi$ given in \eqref{eq:CKFb0}. For a massless scalar field this is shown in \cite{wald2010general}; 
for Maxwell fields this can be seen in \cite{Adamo2012}. All this is expected from the fact that the current \eqref{cucur}
is conserved.

\subsection{The second law}

The quantity $\delta A$ is strictly positive in the context of first order perturbations of Minkowski spacetimes. 
This follows directly from the first law and the assumption that the conformal matter satisfies the energy condition 
$T_{ab} \ell^a\ell^b\ge 0$. It is the standard manifestation of the attractive nature of gravity in its linearized form. Needless is to say that this version of the second law is somewhat trivial in comparison with the very general area theorem for black hole \cite{hawking1973large}, as well as generic Killing \cite{chrusiel2001regularity}, horizons.

\subsection{The third law}

In our context the third law is valid in a very concrete and strict fashion. In the limit of extremality, $r_H\to 0$, the surface gravity $\kappa_{\va SG}\to 0$ and, the analogue of the entropy, the area $A$, goes to zero as well. This version of the third law is the analogue of the statement that at zero temperature the entropy vanishes, which is only true for systems with non degenerate ground states. No dynamical-process version of the third law  appears to make sense in our context. This might resonate at first sight with the statement \cite{Hawking:1994ii} that extremal BHs must have vanishing entropy, but the similarity is only in appearance as the area of the bifurcate sphere remains non-vanishing in the BH case.

\section[Quantum Effects: ``Hawking radiation'' and Conformal Temperature]{Quantum Effects: ``Hawking radiation'' and\\Conformal Temperature} \label{HRCT}
When the laws of black hole mechanics where discovered, they were thought as a mere analogy with the one of thermodynamics. It is only after Hawking's discovery of semiclassical radiation \cite{hawking1974black,hawking1975} that they assumed a proper status of laws of black hole thermodynamics. In the previous Section, we established  the equivalent of the early analogy for the case of light cones in Minkowski spacetime and their gravitational perturbation. In what follows, we show that, also in this case, a semiclassical computation can be performed to give a thermodynamical meaning to those laws. In a suitable sense, radial MCKFs can be assigned a temperature
\be\label{temperature}
T = \frac{\sg}{2\pi}\,.
\ee 
Thus the first law \eqref{firstLaw} becomes
\be
\delta M = T\delta S+\delta M_{\infty}\,.
\ee
with
\be
T = \frac{\sg}{2\pi} \ \ \ \ \ \ {\rm and} \ \ \ \ \ \ \delta S= \frac{\delta A}{4}\,,
\ee
exactly as for stationary black holes.

To do so, let us start by noticing that for each region Minkowski spacetime is divided into by our radial MCKF $\xi$, there exists a coordinate transformation $(t,r,\vartheta,\varphi) \to (\tau,\rho,\vartheta,\varphi)$ adapted to the MCKF in the sense that $\xi(\tau)=-1$. The explicit maps are written in Appendix~\ref{app:coordinate}. Here we report the one for the region of interest, namely Region II. It reads \cite{HalILQGS}
\be
\label{eq:cootran}
\begin{split}
t &=\frac{\sqrt{\Delta}}{2a} \frac{\sinh(\tau \sqrt{\Delta})}{\cosh(\rho \sqrt{\Delta}) - \cosh(\tau \sqrt{\Delta})}\\
r &=\frac{\sqrt{\Delta}}{2a} \frac{\sinh(\rho \sqrt{\Delta})}{\cosh(\rho \sqrt{\Delta}) - \cosh(\tau \sqrt{\Delta})}\,,
\end{split}
\ee 
with $0\leq \rho < +\infty$ and $|\tau|<\rho$. Defining the null coordinates $\bar{v} = \tau + \rho$ and $\bar{u}=\tau - \rho$, the following relation with Minkowskian $u$ and $v$ is valid \cite{PhysRevD.26.1881}:
\be\label{uv}
\begin{split}
v &= t+r = -\frac{\sqrt{\Delta}}{2a} \coth\frac{\bar u\sqrt{\Delta}}{2}\\ 
u &= t-r = -\frac{\sqrt{\Delta}}{2a} \coth\frac{\bar v\sqrt{\Delta}}{2}\, ,
\end{split}
\ee
where, given the above mentioned restrictions on the coordinate, we have $\bar{u} \in (-\infty,0)$ and $\bar{v} \in (0,+\infty)$. The Minkowski metric \eqref{eq:min} becomes
\be \label{eq:confmetric}
ds^2 = \Omega^2 \left(-d\tau^2 + d\rho^2 +\Delta^{-1} \sinh^2(\rho \sqrt{\Delta})dS^2\right)
\ee
where the conformal factor takes the value
\be\label{eq:confFac}
\Omega = \frac{\Delta/2a}{\cosh(\rho \sqrt{\Delta})-\cosh(\tau \sqrt{\Delta})}\,.
\ee
As anticipated, the metric depends on the coordinate $\tau$ only through the conformal factor. The vector $\partial_\tau$, therefore, is a conformal Killing field for the Minkowski spacetime which can be shown to coincide with Eq.~\eqref{eq:CKFb0}. Explicitly
\be\label{eq:CKFtau}
\begin{split}
\xi^a \partial_a = \partial_\tau &= \left(a v^2 -\frac{\Delta}{4a}\right)\partial_v +  \left(a u^2 -\frac{\Delta}{4a}\right)\partial_u \\
&= (av^2 + c)\, \partial_v + (au^2 + c)\, \partial_u\\
&= \big(a(t^2+r^2)+c\big)\partial_t + 2art \,\partial_r\,.
\end{split}
\ee

\subsection{Bogoliubov transformations}

Consider now a scalar field $\phi$ evolving in Minkowski space. We will  define a vacuum state $\ket{0}$ and its corresponding Fock space $\mathcal{F}$ using the notion of positive frequency compatible with the notion of energy entering the first law Eq.~\eqref{firstLaw}. 
Making more precise what we anticipated in the first lines of this section, the remarkable result is that the standard Minkowski vacuum state is seen, in the Fock space $\mathcal{F}$, as a thermal state at the constant conformal temperature
\be
T = \frac{\sg}{2\pi}\,.
\ee
The term conformal temperature is used because the state of the radiation looks thermal in terms of time translation notion associated to the conformal Killing time. See Section~\ref{sec:energy} for a detailed discussion, where the relationship between this notion of temperature and the physical temperature measured by a thermometer is also addressed.

Let us start by defining a meaningful notion of Fock space related to our conformally static observers. As for the discussion in the previous sections, Eq.~\eqref{firstLaw} holds only for conformally invariant matter models. For concreteness, here we consider a conformally invariant scalar field $\phi$ satisfying the conformally coupled Klein-Gordon (KG) equation  
\be\label{eq:KG}
\left(\Box^2-\frac16 R\right) \,\phi = 0\,,
\ee
where $\Box = g_{ab}\nabla^a\nabla^b$, with $\nabla^a$ the covariant derivative with respect to a general metric $g_{ab}$, and $R$ the  Ricci curvature scalar. The previous equation is conformally invariant in the sense that under a conformal transformation $g_{ab} \to g'_{ab} = C^2 g_{ab}$ solutions of \eqref{eq:KG} defined in terms of $g_{ab}$ are mapped into solutions of the same equation in terms of $g'_{ab}$ by the rule $\phi \to \phi'=C^{-1}\phi$ \cite{wald2010general,birrelldavies:QFTCST}.
 
As Eq.~\eqref{eq:confmetric} shows, the complement of the diamond in Minkowski space is conformally related to a region of a static Friedmann-Robertson-Walker (FRW) spacetime with negative spatial curvature $k=-|\Delta|$; see Appendix~\ref{app:FRW} for further details. The FRW Killing field $\partial_\tau$ corresponds to the Minkowski conformal Killing field in Eq.~\eqref{eq:CKFtau}. The strategy is therefore to find a complete set of solutions $U_i(x)$ of the KG equation in the static FRW spacetime, to deduce the one in our region using conformal invariance; here $i$ is a generic index labeling the modes. The Klein-Gordon equation \eqref{eq:KG} in the FRW spacetime under consideration reads
\be\label{eq:KG-exp}
\begin{split}
0 &= \left( \Box^2 -\frac16 R \right) U_i(x)\\
&= \left[ \frac{1}{\sqrt{-g}}\partial_\mu \left(\sqrt{-g}\partial^\mu \right) -\frac16 R \right] U_i(x)\\
&= \left[ -\partial_\tau^2 + \frac{1}{\sinh^2 (\rho\sqrt{\Delta})} \left( \partial_\rho \sinh^2(\rho\sqrt{\D})\, \partial_\rho + \frac{\Delta}{\sin\vartheta} \partial_\vartheta \sin\vartheta\, \partial_\vartheta + \frac{\D}{\sin^2\vartheta}\partial^2_\varphi \right) + \D\right] U_i(x) \,,
\end{split}
\ee
where $g = \det g_{ab}$ and we have used the fact that $R = -6 \D$. One can solve the previous equation by the ansatz 
\be
U^{\ell m}_\omega(x)=  \exp(- i \omega \tau) \frac{R_{\leftrightarrow \omega}^{\ell}(\rho)}{\sinh (\rho\sqrt{\Delta})} Y^{\ell m}(\vartheta, \varphi)\,,
\ee
which after  substitution  in  \eqref{eq:KG-exp}  gives
\be\label{eq:KGrho}
\left( \partial^2_\rho + \omega^2 -\frac{\ell(\ell+1) \, \D}{\sinh^2 (\rho\sqrt{\Delta})}\right) R^\ell_{\leftrightarrow \omega}(\rho) =0 \,,
\ee
where $\leftrightarrow$ denotes the two possible solutions: out-going modes will be denoted by a right arrow ($\rightarrow$) while in-going modes by a left arrow ($\leftarrow$).
Notice that we have substituted the generic index $i$ with the more specific $\omega$, $\ell$ and $m$. A complete set of solutions to this equation is given in \cite{birrelldavies:QFTCST}. These modes are  positive frequency modes with respect to the notion of time translation defined by the Killing time $\tau$, and they are orthonormal with respect to the Klein-Gordon scalar product, namely
\be
\begin{split}
( U_{\leftrightarrow \omega'}^{\ell' m'},U_{\leftrightarrow\omega}^{\ell m})  &= -i \int_\Sigma \left( U_{\leftrightarrow \omega}^{\ell m} \partial_a \bar{U}_{\leftrightarrow\omega'}^{\ell' m'} -\bar{U}_{\leftrightarrow\omega'}^{\ell' m'}\partial_a U_{\leftrightarrow\omega}^{\ell m}\right) d\Sigma^a \\
&= \delta_{\leftrightarrow}\delta^{\ell\ell'} \, \delta^{mm'} \, \delta(\omega,\omega')\,,
\end{split}
\ee
where $\delta_{\leftrightarrow}$ means that outgoing modes are orthogonal to ingoing ones.
As said at the beginning of the subsection, due to conformal invariance the set of modes \footnote{Such modes are the ``sphere modes'' considered in \cite{HalILQGS}.} defined by
\be\label{eq:modes}
u_{\leftrightarrow\omega}^{\ell m}(x) = \Omega^{-1}(x) U_{\leftrightarrow\omega}^{\ell m}(x) = \Omega^{-1}(x) \,e^{-i\omega \tau}  \frac{R_{\leftrightarrow\omega}^{\ell}(\rho)}{\sinh (\rho\sqrt{\Delta})} Y^{\ell m}(\vartheta, \varphi)
\ee
with $\Omega(x)$ given by Eq.~\eqref{eq:confFac}, are a complete set of solutions of the Klein-Gordon equation in our region of interest, the complement of the diamond in Minkowski space. Moreover, they satisfy
\be\label{eq:orto}
\begin{split}
(u_{\leftrightarrow\omega'}^{\ell' m'}, u_{\leftrightarrow\omega}^{\ell m}) &= (\Omega^{-1} \, U_{\leftrightarrow\omega}^{\ell m},\Omega^{-1} \,U_{\leftrightarrow\omega'}^{\ell' m'}) \\
&= -i \int_\Sigma \left[ \Omega^{-1}\, U_{\leftrightarrow\omega}^{\ell m} \partial_a \Big(\Omega^{-1} \bar{U}_{\leftrightarrow\omega'}^{\ell' m'}\Big) -\Omega^{-1} \bar{U}_{\leftrightarrow\omega'}^{\ell' m'}\partial_a \Big(\Omega^{-1}\,U_{\leftrightarrow\omega}^{\ell m} \Big)\right] d\Sigma_{\rm\va II}^a \\
&= -i \int_\Sigma \Big[ \Omega^{-2} \Big(U_{\leftrightarrow\omega}^{\ell m} \partial_a \bar{U}_{\leftrightarrow\omega'}^{\ell' m'} -\bar{U}_{\leftrightarrow\omega'}^{\ell' m'}\partial_a \,U_{\leftrightarrow\omega}^{\ell m}\Big) \\
&\phantom{= -i \int_\Sigma \Big[ \Omega^{-2} \Big(U_{\leftrightarrow\omega}^{\ell m} \partial_a \bar{U}_{\leftrightarrow\omega'}^{\ell' m'} -\bar{U}_{\leftrightarrow\omega'}^{\ell' m'}}+ \Omega^{-1} \bar{U}_{\leftrightarrow\omega'}^{\ell' m'} U_{\leftrightarrow\omega}^{\ell m} \Big(\partial_a \Omega^{-1} - \partial_a \Omega^{-1}\Big)\Big] d\Sigma_{\rm\va II}^a\\
&= -i \int_\Sigma \left( U_{\leftrightarrow\omega}^{\ell m} \partial_a \bar{U}_{\leftrightarrow\omega'}^{\ell' m'} -\bar{U}_{\leftrightarrow\omega'}^{\ell' m'}\partial_a \,U_{\leftrightarrow\omega}^{\ell m}\right) d\Sigma^a\\
&= (U_{\leftrightarrow\omega'}^{\ell' m'},U_{\leftrightarrow\omega}^{\ell m}) \,.
\end{split}
\ee
Here $\Sigma$ is a Cauchy surface shared by the two conformally related spacetimes; $d\Sigma^a$ and $d\Sigma_{\rm\va II}^a$ are the volume elements of $\Sigma$ in the static FRW spacetime and in the complement of the diamond respectively. The above result is given by the fact that the two volume elements are related by \footnote{This is generically true for any hypersurface $\Sigma$ shared between two conformally related spacetimes $g'_{ab}=C^{2} g_{ab}$. Indeed, if $n^a$ is the unit normal to $\Sigma$ with respect to $g_{ab}$, then $n'^a=C^{-1} n^a$ is the unit normal to $\Sigma$ with respect to $g'_{ab}$. The 3-dimensional volume elements, at the same time, are related by $\sqrt{h'}=C^{3} \sqrt{h}$. It follows that $d\Sigma'^a=C^2 d\Sigma^a$.}
\be\label{sigmas}
d\Sigma^a_{\rm\va II} = n_{\rm\va II}^a \sqrt{-h_{\rm\va II}} \, d^3 y = \Omega^2 n^a \sqrt{h} \,d^3y = \Omega^2 d\Sigma^a\,,
\ee
where $n^a$ is the normal to $\Sigma$, $h$ is the determinant of the intrinsic metric defining $\Sigma$ it self, and $y^i$ are the coordinate describing the latter. The subscript $\rm II$ indicates objects defined in Region II of Minkowski spacetime; the same objects without any subscript are in FRW. Eq.~\eqref{eq:orto} shows that the modes $u^{\ell m}_\omega$ provide a complete set of solutions inducing a positive definite scalar product, namely everything one needs to perform the standard quantisation procedure. Hence,
one can write the field operator in Region II of Minkowski spacetime as
\be
\begin{split}
\phi(x) &= \int_0^{+\infty} d\omega \sum_{\ell m}  \Big(a_{\leftarrow \omega}^{\ell m} u_{\leftarrow \omega}^{\ell m}(x) + a_{\leftarrow \omega}^{\ell m\,\dagger} {\bar u}_{\leftarrow \omega}^{\ell m}(x)\Big)+\Big(a_{\rightarrow \omega}^{\ell m} u_{\rightarrow \omega}^{\ell m}(x) + a_{\rightarrow \omega}^{\ell m\,\dagger} {\bar u}_{\rightarrow \omega}^{\ell m}(x)\Big) \\
 &= \int_0^{+\infty} d\omega\  \Omega^{-1}(x) \Big[\sum_{\ell m} \Big(a_{\leftarrow \omega}^{\ell m} U_{\leftarrow \omega}^{\ell m}(x) + a_{\leftarrow \omega}^{\ell m\,\dagger} U_{\leftarrow \omega}^{\ell m}(x)^*\Big) \\
 &\phantom{= \int_0^{+\infty} d\omega\  \Omega^{-1}(x) \Big[\sum_{\ell m} \Big(a_{\leftarrow \omega}^{\ell m} U_{\leftarrow \omega}^{\ell m}(x) + a_{\leftarrow \omega}^{\ell m\,\dagger}} + \Big(a_{\rightarrow \omega}^{\ell m} U_{\rightarrow \omega}^{\ell m}(x)+ a_{\rightarrow \omega}^{\ell m\,\dagger}{\bar U}_{\rightarrow \omega}^{\ell m}(x)\Big) \Big] \,,
 \end{split}
\ee
where $a_{\leftrightarrow \omega}^{\ell m \dagger}$ and $a_{\leftrightarrow \omega}^{\ell m}$ denote the creation and annihilation operators in the corresponding modes. The vacuum state $\ket{0}$ defined by $a_{\leftrightarrow \omega}^{\ell m} \ket{0}=0$ is usually called the  \emph{conformal vacuum} \cite{birrelldavies:QFTCST}. This state is highly pathological from the perspective of inertial observers. Indeed, it has vanishing entanglement with the interior of the diamond and would lead to a divergent energy momentum tensor at $H^+$. More precisely, this is not a Hadamard state. The same thing happens when one considers the Rindler vacuum defined by the boost Killing field.  

Let us notice now that Eq.~\eqref{eq:KGrho} is simplified in the limit $\rho \to +\infty$. This limit corresponds, in Region II and for $\tau > 0$, to the limit $\bar{v} \to +\infty$ or more clearly $u \to u_-$ with $v$ free to span the whole range $[v_+,+\infty)$. That is to say a ``near horizon limit'' \footnote{In the bottom part of our region, $\tau < 0$, this limit corresponds to $v \to v_+$, while $u$ free to vary. That is to say a near past horizon limit.}. 
In this limit the last term of Eq.~\eqref{eq:KGrho}, the one dependent on $\ell$, can be neglected. Solutions $R(\rho)$, therefore, do not depend on $\ell$ in such near horizon approximation and are simply given by $\exp(\pm i\omega \rho)$. The modes \eqref{eq:modes}, consequently, behave as
\be\label{modes}
{ u^{\ell m }_{\leftarrow \omega }}(x) +{ u^{\ell m }_{\rightarrow \omega }}(x) \approx  \frac{1}{ \sqrt{\omega}}\, \frac{e^{-i\omega \bar{u}} + e^{-i\omega \bar{v}}}{\Omega \, \sinh(\rho\sqrt\D)}Y^{\ell m}(\vartheta, \varphi)
 = \frac{\sqrt\D}{\sqrt{\omega}}\, \frac{e^{-i\omega \bar{u}} + e^{-i\omega \bar{v}}}{r}Y^{\ell m}(\vartheta, \varphi)
\,,
\ee
where $r$ is the Minkowskian radial coordinate and we have used definition \eqref{eq:cootran}.

Clearly, the solution of the Klein-Gordon equation and the consequent quantisation of the field can be carried out also in the whole Minkowski spacetime by considering inertial $r= const$ observers. This defines positive frequency modes $u^M_\omega$ with respect to the Killing field $\partial_t$, as well as a decomposition of the field as
\be
\phi(x) = \int_0^{+\infty} d\omega \Big({b^{\ell m}_{\leftarrow \omega }} {u^{\ell m M}_{\leftarrow \omega }}(x) + {b_{\leftarrow \omega }^{\ell m \dagger}}\, {\bar u^{\ell m M}_{\leftarrow \omega }}(x)\Big)+\Big({b^{\ell m}_{\rightarrow \omega }} {u^{\ell m M}_{\rightarrow \omega }}(x) + {b_{\rightarrow \omega }^{\ell m \dagger}}\, {\bar u^{\ell m M}_{\rightarrow \omega }}(x)\Big)\,.
\ee
In the limit $r\to +\infty$ the Minkowskian solutions can be approximated by 
\be\label{Mmodes}
{ u^{\ell m M}_{\leftarrow \omega }}(x) +{ u^{\ell m M}_{\rightarrow \omega }}(x) \approx   \frac{1}{\sqrt{\omega}}\, \frac{e^{-i\omega u} + e^{i\omega v}}{r}Y^{\ell m}(\vartheta, \varphi) \,.
\ee 
The standard Minkowski vacuum state $\ket{0}_M$ of the Fock space is defined by $b^{\ell m}_{\leftrightarrow \omega}\ket{0}_M=0$. The Minkowski modes are also orthonormal with respect to the Klein-Gordon scalar product, namely
\be\label{eq:ortoM}
(u^{\ell m M}_{\leftrightarrow \omega},u^{\ell' m' M}_{\leftrightarrow \omega'}) = \delta_{{\leftrightarrow}}\delta_{\ell\ell'}\delta_{mm'}\delta(\omega,\omega')\,,
\ee
which is immediately verified for outgoing and infalling modes by integrating on $\sI^+$ and $\sI^-$ solutions in the form \eqref{Mmodes}. 
The two different vacua are in general non-equivalent and one vacuum state can be a highly exited state in the Fock space defined by the other, and viceversa. This idea is formalised by introducing the so-called Bogoliubov transformations between the two complete sets of modes $u^{\ell m}_{\leftrightarrow \omega}$ and $u^{\ell m M}_{\leftrightarrow \omega}$. Briefly---for more details see for example \cite{wald2010general}---, since the two sets are complete, one can expand one set in terms of the other. From now on we concentrate on the outgoing modes ($\rightarrow$). We get
\be
u^{\ell m}_{\rightarrow \omega}=\int d\omega' \,\Big( \alpha^{\ell m\omega}_{\ell' m'\omega'}  \, u^{\ell' m' M}_{\rightarrow \omega'} +\beta^{\ell m\omega}_{\ell' m'\omega'} \bar u^{\ell' m' M}_{\rightarrow \omega'}\Big)\,,
\ee
where the $\alpha^{\ell m\omega}_{\ell' m'\omega'}$ and $\beta^{\ell m\omega}_{\ell' m'\omega'}$ are called Bogoliubov coefficients. Taking into account the orthonormality conditions \eqref{eq:orto}-\eqref{eq:ortoM} we get
\be
\alpha^{\ell m\omega}_{\ell' m'\omega'} = (u^{\ell' m' M}_{\rightarrow \omega'},u^{\ell m}_{\rightarrow \omega})\quad ,\quad
\beta^{\ell m\omega}_{\ell' m'\omega'}= - ( \bar u^{\ell' m' M}_{\rightarrow \omega'},u^{\ell m}_{\rightarrow \omega})\,,
\ee
and
\be\label{normBC}
\sum_{\ell'\in \N}\sum_{m'=-\ell'}^{\ell'} \int d\omega'\, \Big(\alpha^{\ell m\omega}_{\ell' m'\omega'} \alpha^{\ell' m'\omega'}_{\ell'' m''\omega''}  - \beta^{\ell m\omega}_{\ell' m'\omega'}  \bar \beta^{\ell' m'\omega'}_{\ell'' m''\omega''} \Big) = \delta (\omega ,\omega'')\,.
\ee
Moreover, defining the particle number operator for the mode $(\ell, m, \omega)$ in the $u^{\ell m}_{\rightarrow \omega}$-expansion in the usual form $N^{\ell m}_{\rightarrow \omega}=a^{\ell m \dagger}_{\rightarrow \omega} a^{\ell m}_{\rightarrow \omega} $, its expectation value on the Minkowski vacuum can generically be written as
\be
\bra{0} N^{\ell m}_{\rightarrow \omega} \ket{0} = \sum_{\ell'\in \N}\sum_{m'=-\ell'}^{\ell'}  \int d\omega' \, |\beta^{\ell m \omega}_{\ell' m' \omega'}|^2
\ee
This object is what we are mainly interested in. It tells us the expectation value of the number of excitations defined with respect to the conformal vacuum $\ket{0}$ that are present in the Minkowski quantum vacuum $\ket{0}_M$. The remarkable fact is that the computation of such object mimics exactly the one for the Hawking's particle production by a collapsing black hole.

Let us choose $\scri^+$ as the hypersurface over which we perform the integral for the computation of scalar products at least for the outgoing modes. $\scri^-$ would be the choice for the ingoing ones. In order to be able to use the near horizon approximate solutions \eqref{modes}, we introduce a complete set of outgoing wave packets on $\scri^+$ localized in retarded time $\bar u$ and near the horizon $\bar u\to+ \infty$ \cite{hawking1975}; see also \cite{fabbri2005modeling}. Concretely,  
\be\label{packets}
u^{\ell m; jn} = \frac{1}{\sqrt{\epsilon}} \int_{j\epsilon}^{(j+1)\epsilon} d\omega \, e^{2\pi i \omega n/\epsilon}\, u^{\ell m}_{\rightarrow \omega}
\ee
with integers $j \geq 0$, $n$, and where \be\label{outmodes}
u^{\ell m}_{\rightarrow \omega}=\frac{\sqrt\D}{ \sqrt{\omega}}\, \frac{e^{-i\omega \bar{u}}}{r} Y^{\ell m}(\vartheta, \varphi)\,.
\ee
 The wave packets $u^{\ell m; jn}$ are peaked around $\bar{u} \simeq 2\pi n/\epsilon$ with width $2\pi/\epsilon$. When $\epsilon$ is small, the wave packet is narrowly peaked about $\omega \simeq \omega_j = j\epsilon$ and localised near the horizon. 
 The facts that, due to spherical symmetry, the modes \eqref{modes} and \eqref{Mmodes} have exactly the same angular dependence, together with the fact that, in the region where the wave packets are picked, the behaviour in $\bar u$ and $u$ is independent of $\ell$, tells us that particle creation will be the same in all angular modes.
 
The surface element of $\scri^+$ is given by $d\Sigma^a = r^2 du\, dS^2 \delta^a_u $. The Bogoliubov coefficients of interest can therefore we written as
\be
\begin{split}
\beta^{\ell m ; jn}_{\ell' m' \omega'} &= - ( \bar u^{\ell' m' M}_{\rightarrow \omega'},u^{\ell m; jn})\\
&= i \int_{\scri^+} du\, dS^2\, r^2 \Big(u^{\ell m; jn}\partial_u u^{\ell' m' M}_{\rightarrow \omega'} -u^{\ell' m' M}_{\rightarrow \omega'} \partial_u u^{\ell m; jn}\Big)\,.
\end{split}
\ee
Since the wave packets vanish for $u \to -\infty$ and for $u>u_-$, we can integrate by part finding
\be
\beta^{\ell, m; jn}_{\ell' m' \omega'} = 2i \int_{-\infty}^{u_-} du\, dS^2\, r^2 \,u^{\ell m; jn}\partial_u u^{\ell' m' M}_{\rightarrow \omega'} \,.
\ee
We now need to insert in the equation the explicit form of the modes \eqref{packets}-\eqref{outmodes} and the outgoing part of the Minkowskian ones \eqref{Mmodes}. However, before doing that, let us recall that we are working and are mainly interested in the near horizon limit $\bar u \to +\infty$. In this limit, the inverse of the relation \eqref{uv} between the conformal retarded time $\bar{u}$ and the Minkowski $u$ simplifies into
\be
\bar{u} = \frac{2}{\sqrt\D} \operatorname{arcoth}\left(-\frac{u}{u_-}\right) \simeq \frac{1}{\sqrt\D} \log\left(\frac{u-u_-}{2 u_-} \right)\,.
\ee
So we can write
\be
\beta^{\ell, m; jn}_{\ell' m' \omega'} = \frac{\sqrt\D\delta_{\ell,\ell'}\delta_{m,m'}}{2\pi  \sqrt\epsilon} \int_{-\infty}^{u_-} du\, \int_{j\epsilon}^{\epsilon} d\omega \, e^{2\pi i \omega n/\epsilon} \sqrt{\frac{\omega'}{\omega}}\,e^{-i\omega \frac{1}{\sqrt\D} \log\left(\frac{u-u_-}{2u_-}\right)-i\omega' u }\,.
\ee
Defining now $x = u_- - u$ we get
\be
\beta^{\ell, m; jn}_{\ell' m' \omega'} = \frac{\sqrt\D\delta_{\ell,\ell'}\delta_{m,m'}}{2\pi\sqrt\epsilon}e^{-i \omega' u_-} \int_{0}^{+\infty} dx\, \int_{j\epsilon}^{\epsilon} d\omega \, e^{2\pi i \omega n/\epsilon} \sqrt{\frac{\omega'}{\omega}}\,e^{-i\omega \frac{1}{\sqrt\D} \log\left(\frac{x}{2u_-}\right) + i\omega' x }\,.
\ee
The integral over the frequency can be performed considering that $\omega$ varies in a small interval around $\omega_j$
\be
\begin{split}
\beta^{\ell, m; jn}_{\ell' m' \omega'} &= \frac{\sqrt\D\delta_{\ell,\ell'}\delta_{m,m'}}{\pi  \sqrt\epsilon} e^{-i \omega' u_-} \sqrt{\frac{\omega'}{\omega_j}}\int_{0}^{+\infty} dx\, e^{+i\omega' x} \,\frac{\sin (\epsilon L/2)}{L}\,e^{i L \omega_j}\\
&=\frac{\sqrt\D\delta_{\ell,\ell'}\delta_{m,m'}}{\pi  \sqrt\epsilon} e^{-i \omega' u_-} \sqrt{\frac{\omega'}{\omega_j}} \;I(\omega')\,,
\end{split}
\ee
where we have defined
\be
L(x) = \frac{2\pi n}{\epsilon}-\frac{1}{\sqrt\D}\log\left(\frac{x}{2u_-}\right)
\ee
and $I(\omega')$ as the integral over $x$. The computation of $\alpha_{jn,\omega'}$ gives a similar result
\be
\begin{split}
\alpha^{\ell, m; jn}_{\ell' m' \omega'}  &= \frac{\sqrt\D\delta_{\ell,\ell'}\delta_{m,m'}}{\pi  \sqrt\epsilon}e^{i \omega' u_-} \sqrt{\frac{\omega'}{\omega_j}}\int_{0}^{+\infty} dx\, e^{-i\omega' x} \,\frac{\sin (\epsilon L/2)}{L}\,e^{i L \omega_j}\\
&=\frac{\sqrt\D\delta_{\ell,\ell'}\delta_{m,m'}}{\pi  \sqrt\epsilon} e^{i \omega' u_-} \sqrt{\frac{\omega'}{\omega_j}} \;I(-\omega')\,.
\end{split}
\ee
Apart from different constants, these objects coincide with the ones defined in \cite{fabbri2005modeling}, and therefore can be solved using exactly the same techniques and procedure. We refer to the book for details and we give here only the final result. 

The important result is that the relation between $\alpha^{\ell, m; jn}_{\ell' m' \omega'} $ and $\beta^{\ell, m; jn}_{\ell' m' \omega'}$ comes out to be
\be
|\beta^{\ell, m; jn}_{\ell' m' \omega'}| = e^{-\frac{\pi \omega_j}{\sqrt\D}}\;|\alpha^{\ell, m; jn}_{\ell' m' \omega'} |\,.
\ee
Inserting this into Eq.~\eqref{normBC}, one can write
\be
-\left[1-\exp\left(\frac{2\pi \omega_j}{\sqrt\D} \right)\right] \sum_{\ell'\in \N}\sum_{m'=-\ell'}^{\ell'}   \int_0^{+\infty} d\omega'|\beta^{\ell, m; jn}_{\ell' m' \omega'}|^2 = 1
\ee
and therefore
\be\label{planck}
\bra{0} N^{\ell m}_{\omega_j} \ket{0}  = \frac{1}{\exp\left(\frac{2\pi \omega_j}{\sqrt\D} \right)-1}\,.
\ee
The above expression coincides with the Planck distribution of thermal radiation at the temperature
\be\label{temp}
T = \frac{\sqrt\D}{2\pi}\,.
\ee
To relate this result to the first law, Eq.~\eqref{firstLaw}, it is enough to notice that the explicit value of the conserved quantity $\sg$ in our case is
\be
\sg = \sqrt{\Delta}\,.
\ee
We have shown what we anticipated at the very beginning of this section: the light cone $u=u_-$ is seen as a (conformal) horizon with an associated temperature $T$ given by expression \eqref{temperature}.

The first law can therefore be rewritten as
\be
\delta M = T\delta S+\delta M_{\infty}\,.
\ee
with
\be
T = \frac{\sg}{2\pi} \ \ \ \ \ \ {\rm and} \ \ \ \ \ \ \delta S= \frac{\delta A}{4}\,.
\ee
The laws of light cone thermodynamics are now not simply a mere analogy, but they acquire a precise semiclassical thermodynamical sense, which is better discussed in the following subsection. This is, to our knowledge, the first precise implementation of the idea \cite{sultana2004conformal,dyer1979conformal} that the quantity $\sg$ should play the role of temperature for conformal Killing horizons.

 \subsection{On the meaning of conformal energy and \\temperature}\label{sec:energy}

In asymptotically flat stationary spacetimes, the time translational Killing field can be normalized at infinity in order to give the analogue of \eqref{M} the physical interpretation of energy as seen from infinity. On the other hand in our case the vector field $\xi$ is normalized only on the observer sphere $r=r_{\va O}$ and $t=0$. Thus $M$ has not the usual physical meaning for any observer in Minkowski spacetime. Nevertheless, for conformally invariant matter the mass $M$ as defined in (\ref{M}) is conformally invariant (see footnote~\ref{FOOT}). Using  \eqref{eq:confmetric}, and the fact that $\xi$ is actually a normalized Killing field of the static FRW metric, one can interpret $M$ as energy in the usual physical manner in that spacetime. 
This interpretation is compatible with the notion of frequency we used to compute the Planckian distribution \eqref{planck}. Indeed, the frequency $\omega$ is the one that would be measured by an observer moving along the Killing field $\partial_\tau$ in the static FRW space. 
For such notion of frequency $\omega$, the energy quanta $\varepsilon=\hbar \omega$ correspond to the same physical notion of energy that defines $M$.

Such interpretation carries over to its thermodynamical conjugate: the temperature. That is the reason why we call {\em conformal temperature} the temperature appearing in the first law. It carries the physical notion of temperature, namely the one measured by thermometers, only for observers in the FRW spacetime where $\xi$ is an actual time translational Killing field. In this way both energy and temperature have their usual interpretation in a spacetime that is conformally related to Minkowski.

\subsection{The Hartle-Hawking-like state}

Let us now define a new radial coordinate 
\be
R=\frac{\sqrt{\Delta}}{a} \exp(-\rho\sqrt{\Delta})\,.
\ee
The near horizon limit $\rho \to +\infty$ corresponds now to $R \to 0$. In these new coordinates, the metric \eqref{eq:confmetric} can be expanded around $R=0$ finding
\be\label{eq:metricR0}
ds_E^2=-R^2 d(\sqrt{\Delta}\tau)^2+dR^2+r_{\va H}^2 dS^2+O({R}{r^{-1}_H}\, dR^2, R r_H dS^2)\,,
\ee
where $O({R}{r^{-1}_H}\, dR^2, R r_H dS^2)$ denotes subleading terms of each component of the metric that do not change the nature of the apparent singularity present at $R=0$. Notice that the leading order of the local metric and the topological structure at the point $r=r_{\va H}$ are exactly the same as the one in the Reissner-Nordstrom metric, Eq.~\eqref{nhl}.

Moreover, the metric \eqref{eq:confmetric} can be continued analytically to imaginary conformal Killing time by sending $\tau\to- i\tau_E$. As for the case of static black holes \cite{wald2010general}, the result is a real Euclidean metric, explicitly given by
\be\label{metricEuclidean}
ds_E^2 = \Omega_{ E}^2 \left(d\tau_E^2 + d\rho^2 +\Delta^{-1} \sinh^2(\rho \sqrt{\Delta})dS^2\right)
\ee
with
\be
\Omega_{E} = \frac{\Delta/2a}{\cosh(\rho \sqrt{\Delta})-\cos(\tau_E \sqrt{\Delta})}.
\ee
Defining again the new coordinate $R$ and carrying out the limit to $R=0$, which corresponds to the Euclidean analogue of the horizon, we find the Euclidean version of \eqref{eq:metricR0}
\be
ds_E^2=R^2 d(\sqrt{\Delta}\tau_E)^2+dR^2+r_{\va H}^2 dS^2+O({R}{r^{-1}_H}\, dR^2, R r_H dS^2)\,.
\ee
The coordinate singularity at $R=0$ can be resolved by defining new coordinates $X=R \cos(\sqrt{\Delta }\tau_E)$ and $Y=R\sin(\sqrt{\Delta} \tau_E)$. In order to avoid conical singularities one must identify $\tau_E$ with a periodic coordinate such that
\be\label{periodictime}
0\le \tau_E \sqrt{\Delta}\le 2\pi\,.
\ee
This removes the apparent singularity by replacing the first two terms in the previous metric by the regular $dX^2+dY^2$ transversal metric. This periodicity in time is what is used in the black hole case to suggest the existence of a state---known as the Hartle-Hawking state---of thermal equilibrium of any quantum field at a temperature given by
\be\label{HHtemp}
T=\hbar \frac{\sqrt{\Delta}}{2\pi}\,,
\ee
which coincides with the one found in the previous section, Eq.~\eqref{temp}.
This tells us that the Minkowski vacuum can be regarded as the Hartle-Hawking type of vacuum of Region II for conformally invariant theories. 

Indeed, instead of using the near horizon approximation (and wave packets peaked there) in the previous section, one could in principle compute the Bogoliubov coefficients exactly between the Minkowski and conformal Fock spaces. 
This should lead to the conclusion that Minkowski vacuum is everywhere a thermal state with temperature \eqref{HHtemp}, as suggested by the previous analysis. As such computation might be rather involved and in view of keeping the presentation as simple as possible, one can find additional evidence for this  by computing the expectation value of the normal ordered stress energy ``tensor'' $:T_{ab}:$. The quotation marks on the word tensor are because the object $:T_{ab}:$ does not transform as a tensor under a coordinate transformation and cannot be interpreted physically as real. In fact the physical and covariant energy momentum tensor has vanishing expectation value in the Minkowski vacuum \cite{Wald:1995yp}.  

However,  $:T_{ab}:$ can be interpreted as encoding the particle content of the Minkowski vacuum as seen from the perspective of the MCKF that are of interest in our analysis. For conformal fields it can be analytically computed  in the $s$-wave approximation $\ell=0$ which reduces the calculation to an effective 2-dimensional system. The 2-dimensional $:T^{\va (2)}_{ab}:$ can be  explicitly evaluated via the Virasoro anomaly \cite{birrelldavies:QFTCST,fabbri2005modeling}. Given two sets of double null coordinates, like the two we have $(u,v)$ and $(\bar{u},\bar{v})$, $:T^{\va (2)}_{ab}:$ transforms as
\be
\begin{split}
:T^{\va (2)}_{\bar{u}\bar{u}}: &= \left(\frac{du}{d\bar{u}}\right)^2 :T^{\va (2)}_{uu}: -\frac{\hbar}{24\pi} \{u,\bar{u}\}\\
:T^{\va (2)}_{\bar{v}\bar{v}}: &= \left(\frac{dv}{d\bar{v}}\right)^2 :T^{\va (2)}_{uu}: -\frac{\hbar}{24\pi} \{v,\bar{v}\}
\end{split}
\ee
where 
\be
\{x,y\} =\frac{\dddot x}{\dot x} -\frac{3}{2} \left( \frac{\ddot x}{\dot x}\right)^2
\ee
is the Schwarzian derivative with dot representing $d/dy$. It is therefore simple to evaluate the expectation value of this object on the Minkowski vacuum $\ket{0}_M$ in our case. Since $_M\bra{0}:T_{ab}:\ket{0}_M=0$, we simply have
\be
\begin{split}
_M\bra{0}:T^{(\va 2)}_{\bar{u}\bar{u}}:\ket{0}_M = -\frac{\hbar}{24\pi } \{u,\bar{u}\} &= \frac{\hbar \Delta}{48 \pi } \\
_M\bra{0}:T^{\va (2)}_{\bar{v}\bar{v}}:\ket{0}_M =  -\frac{\hbar}{24\pi } \{v,\bar{v}\} &= \frac{\hbar \Delta}{48 \pi } \,.
\end{split}
\ee
The result indicates that the Minkowski state produce a constant ingoing and outgoing thermal bath at the temperature \eqref{temp} everywhere in Region II, which is what we expected from a thermal equilibrium state. The near horizon approximation in the computation of the previous section simplifies the relation between the two sets of double null coordinates making the computation analytically simpler, but as discussed above, it should give the same result everywhere in Region II. As mentioned above, the expectation value of the covariant stress energy tensor does not coincide with the normal ordered one. The former is simply vanishing in this case $_M\bra{0}T_{ab}\ket{0}_M=0$. 

\begin{figure}[t]
\center
\includegraphics[height=10cm]{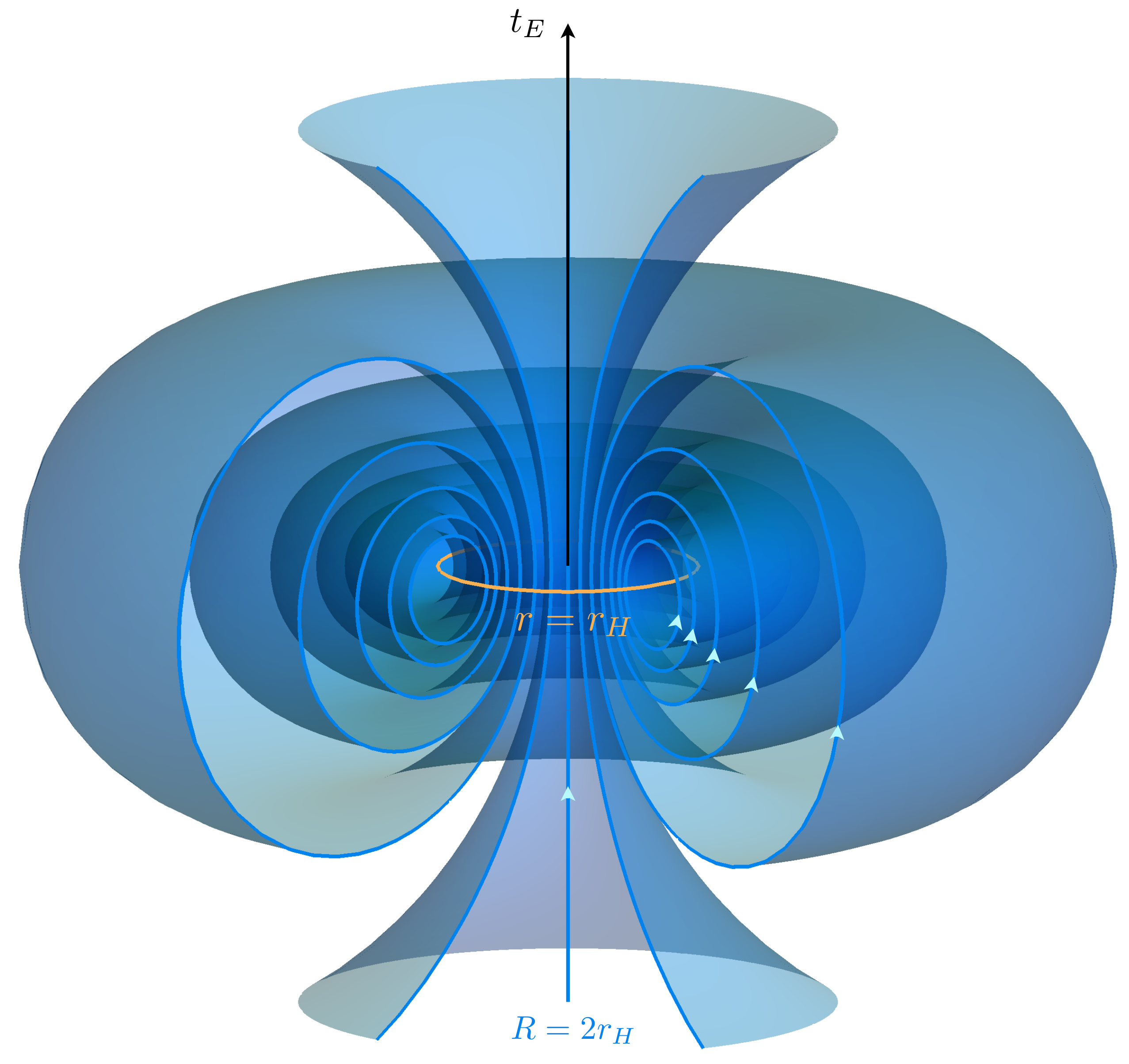}
\caption{Three dimensional representation of the flow of the conformal Killing field in the Euclidean spacetime $\R^4$. The orbits in this one-dimension-less representation are non-concentric tori around the bifurcate sphere $r=r_H$---here represented as a circle. They degenerate into the $t_E$ axis for $R=2r_H$.}
\label{fig:Euclidean}
\end{figure}

As a final remark, let us get more insight into the geometry of the Euclidean continuation \eqref{metricEuclidean} by writing the coordinate transformation to the flat Euclidean coordinates $(t_E, r,\vartheta, \varphi)$ covering $\R^4$. Defining the angular coordinate $\alpha_E\equiv \tau_E \sqrt{\Delta}$ one finds
\be
\label{eq:cootranEu}
\begin{split}
t_E &=\frac{R \sin(\alpha_E)}{1-\frac{R}{2r_{H}} \cos(\alpha_E) +\frac{R^2}{4r^2_{H}}}\\
r &= r_H \frac{1-\frac{R^2}{4r^2_{H}}}{1-\frac{R}{2r_{H}} \cos(\alpha_E) +\frac{R^2}{4r^2_{H}}}\,.
\end{split}
\ee
The bifurcate sphere in the Euclidean continuation corresponds to the sphere $r=r_H$ at $t_E=0$. The orbits of the Wick rotated radial conformal Killing field are orbits of the radial conformal Killing field of $\R^4$ with fixed points given by the Euclidean shining sphere. These orbits correspond, on the $(t_E, r)$ plane, to close loops around the bifurcate sphere, which degenerate into the $r=0$ line (the Euclidean $t_E$-axis) for $R=2r_H$; see Figure~\ref{fig:Euclidean}. The  coordinates $(\tau_E, R, \vartheta, \varphi)$ become singular there. The qualitative features of the Euclidean geometry of the MCKF is just analogous to that of the stationarity Killing field in the Euclidean RN solutions.

\section{Discussion}

We have studied in detail the properties of radial Conformal Killing Fields in Minkowski spacetime and showed that they present in many respects a natural analogue of black holes in curved spacetimes. The global properties of radial MCKFs mimic exactly the causal separation of events in the spacetime of static black holes, i.e. those in the Reissner-Nordstrom family; see Figure \ref{fig:penrose}. Event Killing horizons in the latter are replaced by conformal Killing horizons in the former. The extremal limit maintains the correspondence. 

Linear perturbations of flat Minkowski spacetime in terms of conformally invariant matter models, allow us to consider and prove suitable analogues of the laws of black hole mechanics. When quantum effects are considered, thermal properties make the classical mechanical laws amenable to a suitable thermodynamical interpretation, where entropy variations are equal to $1/4$ of the conformal area changes of the horizon in Planck units. The near horizon and near bifurcate surface features of the geometry of the radial MCKF have the same structure of the stationarity Killing field for static black holes. The Minkowski vacuum state is the analogue of the Hartle-Hawking thermal state from the particle interpretation that is natural to the MCKF. 

This work represents another simple setting where the relationship between thermality, gravity and geometry is manifest in the semiclassical framework. It gives a simple and complete example in which thermal properties analogous to those of black holes are manifest in flat spacetime. It improves the standard analogy given by the study of gravity perturbations and quantum field theory of the Rindler wedge.
On a more speculative perspective, we think that even when the interpretation of temperature, energy, and area entering the thermodynamical relations is subtle, this simple example could shed some light into a more fundamental description of the link between black hole entropy and (quantum) geometry. But this is something we will investigate in the future.

%% file: Part_2/LightBH/LightBH.tex
\chapter{Light Cone Black Holes}\label{chap:lightBH}
\emph{This Chapter overlaps with the first draft of a paper in preparation} \cite{LightBH}{\em. It contains all the main results of the latter. Additional details, connections with other works in the literature and minor results will appear in the final published version.}\vspace{2ex}


The boundary of the causal complement of a spherical ball of radius $\rH$ at time $t$ in Minkowski spacetime is a bifurcate conformal Killing horizon \cite{doi:10.1063/1.532903}. The associated conformal Killing vector field becomes null on the light cones of the two events that intersect the sphere defined by the boundary of the ball \footnote{These events are the past and future `centers' of the ball itself.}. Such null surfaces separate the whole of Minkowski in regions where the conformal Killing field $\xi^a$ is either timelike or spacelike. These regions are in direct correspondence with the different regions defined by the outer and inner horizons of non-extremal Reissner-Nordstrom black holes. When $\rH\to 0$ some regions collapse and the features of the conformal Killing field now correspond to those of extremal Reissner-Nordstrom black holes.

Minkowski Conformal Killing Fields (MCKFs) admit a conformally invariant  \cite{Jacobson:1993pf} definition of surface gravity $\kappa_{\va SG}$
\be
\nabla_a (\xi\cdot \xi)\,\hat=-2\sg  \xi_a\,, 
\ee
where $\hat=$ means that the equality holds at the Killing horizon. All four laws of black hole thermodynamics have a suitable version for the conformal killing horizons defined by MCKFs:  The surface gravity $\kappa_{\va SG}$ is constant on the horizon and it is associated to a mathematical notion of temperature (conformal temperature) \be \label{zero}{T=\frac{\kappa_{\va SG}}{2\pi}={\rm constant}}.\ee Under linear perturbations induced by conformally invariant matter fields the current $J_a=\delta T_{ab}\xi^b$ is conserved, namely
\be
\nabla^a J_a=\nabla^a (\delta T_{ba} \xi^b)=0. 
\ee
The previous equation can be used to establish a suitable version of the first law for MCKF 
\be\label{firstLaw2}
{\delta M = \frac{\kappa_{\va SG}}{8\pi} \, {\delta A}+\delta M_{\infty}\,,}
\ee
where
\be
\delta M = \int_\Sigma J_a d\Sigma^b
\ee
is the conformally invariant mass of the perturbation evaluated at an initial Cauchy surface $\Sigma$, $\delta M_{\infty}$ is the conformal mass flow at $\sI^+$, and $\delta A$ is a conformally invariant notion defined as
\be
 \delta A \equiv \int\limits_{H^+} \frac{\kappa}{\sg} \,\delta \theta dS^2 dv\,.
\ee
Here $\delta \theta$ is the first order perturbation of the expansion of the generator of the horizon, $v$ is the advanced Minkowski time (a natural affine parameter for the generators), $dS^2$ the flat background area measure of the spherical cross section $v=$constant of the unperturbed light cone, and $\kappa$ is defined by
\be
 \xi^a \nabla_a \xi^b {\,\hat=\,}\kappa\,\xi^b\,.
\ee
Unlike $\kappa_{\va SG}$, the function $\kappa$ is not constant and is not conformally  invariant. 
Finally, provided that $\delta T_{ab}$ satisfies the weak energy condition, which is equivalent to the strong one for conformally invariant matter, the second law holds, namely
\be\label{area}
{\delta A\ge 0.}
\ee
 In the `extremal' limit $\rH\to 0$ the temperature (\ref{zero}) goes to zero as well as the area of the bifurcate sphere $A=4\pi \rH^2$. We can interpret this as a form of third law of thermodynamics. 
This, plus equations (\ref{zero}), (\ref{firstLaw2}), and (\ref{area}), are the analog of the laws of black hole mechanics for the light cones in flat spacetimes that define the conformal Killing horizons associated to MCKFs \cite{DeLorenzo:2017tgx}--see previous Chapter. 

The previous formal analogy between the properties of MCKFs and thermodynamics of black holes captures the basic mathematical features of the latter on a  background with trivial gravitational field. However, the reverse side of it is that the various conformal invariant notions entering the laws have no clear physical meaning: $\delta M$ is not an energy measured by any real physical observer, the conformal temperature is not the one detected by any physical thermometer, and $\delta A$ is not really the change of the geometric area of the bifurcating sphere. Perhaps the most disturbing aspect of the previous analogy is precisely the definition of $\delta A$ entering (\ref{firstLaw2}) with its non-obvious geometric meaning due to the fact that $\kappa_{\va SG}/\kappa\not=1$ 
for conformal Killing fields in general and  the MCKFs in particular.

Nevertheless, the previous limitation can be resolved if one performs a conformal transformation sending $(\R^4, \eta_{ab})$ to a model spacetime $(M,\tilde g_{ab})$ with $\tilde g_{ab}=\omega^2 \eta_{ab}$ so that $\xi^a$ becomes a genuine Killing field. The associate conformal bifurcate horizons will be mapped into bifurcate Killing horizons in $g_{ab}$. If $\delta T_{ab}$ comes from a conformally invariant matter model, then $\delta \tilde T_{ab}=\omega^{-2} \delta T_{ab}$ the conservation of the associated current $\tilde J_a$ holds  in the new spacetime. Equations (\ref{zero}), (\ref{firstLaw2}), and (\ref{area}) remain true in the target spacetime with identical numerical values for a given perturbation. However, all the quantities involved acquire the standard physical and geometric meaning that they have in the context of black holes. 

The question we want to explore here is what are the generic global features of the spacetimes obtained by the previous procedure. Which are the cases where these spacetimes represent black holes? What are they in the other cases? Even when demanding that the procedure does not break spherical symmetry there is clearly an infinite number of possibilities. Indeed, if  $\omega_1$ is a solution, $\omega_2$ is a new solution as long as $\xi(\omega_2/\omega_1)=0$.  We will see that the generic global features can be made apparent in a small number of representative cases. The simplest case corresponds to $\omega=\alpha/r^2$ that reproduces the Bertotti-Robinson solution \cite{Bertotti:1959pf, Robinson:1959ev} of Einsteins-Maxwell theory--Section~\ref{BBRR}. Such solution has been know to encode the near horizon geometry of close-to-extremal and extremal Reissner-Nordstrom black holes. 
Another representative example is the {\em de Sitter} realization where the bifurcating horizons correspond to intersecting cosmological horizons (there is no black hole in this case)--see Section~\ref{sub:desitter}. Weakly asymptotically {\em (Anti)-de Sitter} black hole realizations are also presented--Section~\ref{sub:asyAdS}--, together with a more exotic asymptotically flat spacetime with Killing horizons but no black holes--Section~\ref{originalone}. 
 
Radial MCKF with conformal Killing horizons associated to light cones bifurcating at a sphere generalize to arbitrary dimensions. As long as the matter perturbing the geometry is conformally invariant,  the generalization of equations (\ref{zero}), (\ref{firstLaw2}), and (\ref{area}) is also valid. The Bertotti-Robinson representation, which in arbitrary dimensions is given by $AdS_2\times S_{d-2}$, remains the simplest one.
For $d=2$ the light cone black hole corresponds to the Jackiw-Teitelboim solution \cite{jackiw1984quantum}.

\section[Radial Conformal Filling Fields in Minkowski Spacetime]{Radial Conformal Killing Fields in\\Minkowski Spacetime}
 
\begin{figure}[t]
\center
\includegraphics[height=9cm]{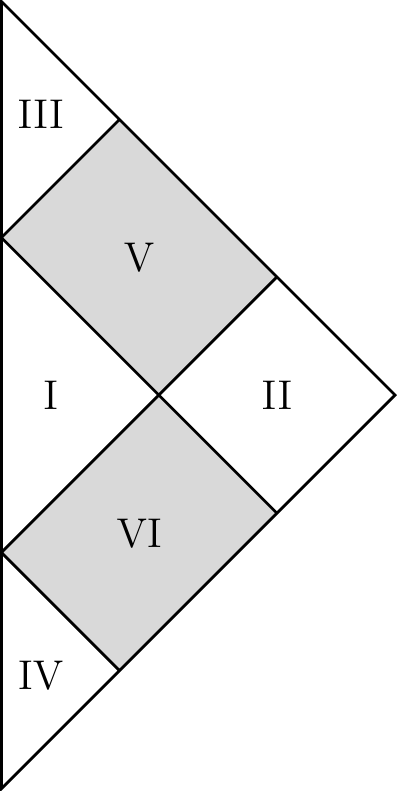} 
\caption{The most general radial conformal Killing field $\xi$ in Minkowski spacetime divided the latter in six regions. The field $\xi$ is spacelike in the shaded regions, timelike elsewhere. It becomes null on the lightcones separating the regions. It vanishes at the tips of the ligthcones and at their intersection.}
\label{fig:Penrose_CD2}
\end{figure}

Consider Minkwoski spacetime in spherical coordinates
\be
\begin{split}
ds^2_\Mi = \eta_{\mu \nu}dx^\mu dx^\nu &= -dt^2 + dr^2 + r^2 dS^2\\
&= -dvdu + \frac{(v-u)^2}{4} dS^2
\end{split}
\ee
where $dS^2$ is the unit-sphere metric, while $v=t+r$ and $u=t-r$ the standard Minkowskian null coordinates. The conformal group in four dimensional Minkowski spacetime $\M^4$  is isomorphic to the group $SO(5,1)$. Any generator defines a Conformal Killing Field in Minkowski spacetime (MCKF), namely a vector field $\xi$ along which the metric $\eta_{ab}$ changes only by a conformal factor:
\be
\sL_\xi \,\eta_{ab}=\nabla_a\xi_b+\nabla_b\xi_a=\frac{\psi}{2} \eta_{ab}
\ee
with
\be
\psi = \nabla_a \xi^a\,.
\ee
The most general radial MCKF can be written (up to Poincar\`e transformations) as 
\be
\begin{split}
\xi^\mu \frac{\partial}{\partial x^\mu} &=
\frac{1}{r_{\va O}^2-r_{\va H}^2}\left((t^2+r^2-r_{\va H}^2)\frac{\partial}{\partial {t}} + 2\, t\, r\,\frac{\partial}{\partial r}\right)
\\
&=\frac{1}{r_{\va O}^2-r_{\va H}^2}\left((v^2-r_{\va H}^2)\,\frac{\partial}{\partial v} + (u^2-r_{\va H}^2)\,\frac{\partial}{\partial u}\right)
\end{split}
\ee
where $\rH$ and $\rO$ are two constants defining the family of Killing field. They are defined to be respectively the value of the radius of the sphere where the Killing field is zero $\xi|_{r=r_{\va H}}=0$ and the radius of the sphere on the surface $t=0$ where the Killing field is normalised, $\xi\cdot \xi|_{r=r_{\va O}}=-1$. Indeed the norm is given by
\be
\xi \cdot \xi = -\frac{(v^2-r_{\va H}^2)(u^2-r_{\va H}^2)}{(r_{\va O}^2-r_{\va H}^2)^2}
\ee
and at $t=0$, we have $v=-u=r$.
This also shows that the causal behaviour of $\xi$ changes from being timelike to spacelike, passing through being null along two constant $u=u_{\pm}$ and two constant $v=v_{\pm}$ rays. This behaviour divides Minkowski spacetime in six regions, Fig.~\ref{fig:Penrose_CD2}.

Consider now any spacetime conformally related to Minkowski
\be
g_{ab} = \omega^2 \,\eta_{ab}\,.
\ee
In such a spacetime, the conformal Killing field $\xi$ remains so. Indeed
\be
\sL_{\xi}g_{ab}=\sL_{\xi}(\omega^2 \eta_{ab})=\left(\frac{\psi}{2} + \xi^a\partial_a \, (\log \,\omega^2)\right)g_{ab}.
\ee
In particular, there exist conformal transformations such that $\xi$ becomes a proper Killing field. From the above equation, it follows that those are given by conformal factors satisfying
\be\label{eq:conftokill}
\frac{\psi}{2} + \xi^a\partial_a \, (\log\omega^2)=0
\ee

Let us define the following coordinates transformation
\be\label{eq:CooTra}
\begin{split}
\tau&= \frac{(\rO^2-\rH^2)}{4\,\rH}\log\frac{(u-\rH)(v-\rH)}{(u+\rH)(v+\rH)}\\
x &=  \frac{2(\rH^2-u\, v)}{v-u}.
\end{split}
\ee
The Minkowski metric becomes
\be\label{eq:RN}
ds_\Mi^2=\frac{r^2}{\xo^2} \left(-\frac{x^2-x_{\va BH}^2}{\xo^2}\,d\tau^2+\frac{\xo^2}{x^2-x_{\va BH}^2} \, dx^2+x_{\va 0}^2 \,dS^2\right)\,.
\ee
where
\be
\xBH\equiv {2 \,\rH},\ \ \ \xo^2\equiv\rO^2-\rH^2,
\ee
and where $r$ is now a function of $\tau$ and $x$. The outer and inner horizons are located respectively at $x_\pm=\pm \xBH$. Minkowskian $\scri^\pm$ are located at finite $x$, while $i^0$ at $x \to +\infty$. Finally, the origin $u=v$ corresponds to $x \to {\rm sign}(\rH^2-t^2)\,\infty$. 
In these coordinates, our radial MCKF reduces to 
\be\label{eq:killtau}
\xi^\mu \frac{\partial}{\partial x^\mu} =  \frac{\partial}{\partial \tau} \,,
\ee
simplifying Eq.~\eqref{eq:conftokill} as
\be\label{rescaling}
\frac{\partial}{\partial \tau} (\log \omega^2) = -\frac\psi2 \,.
\ee
The general solution to the above equation is
\be
\log(\omega^2) =-\int \frac\psi{2} \,d\tau + F(x,\theta,\varphi)
\ee
where $F(x,\theta,\phi)$ is a general dimension-less function of $(x,\theta,\phi)$. The function $\psi$ can be explicitly computed and the integral performed, giving
\be
\log(\omega^2) = \log\left(\frac{\xo^2}{r^2}\right) + F(x,\theta,\varphi)\,,
\ee
where again the Minkowskian radial coordinate $r$ is a function of $\tau$ and $x$.
Redefining for convenience $F \equiv -\log\,P^2$, we find 
\be\label{eq:omega}
\omega^2 = \frac{\xo^2}{r^2}\,\frac{1}{P^2(x,\theta,\varphi)}\,.
\ee
The above equation defines an infinite family of conformal transformations of Minkowski spacetime, such that the target spacetime admits a genuine Killing field corresponding to MCKF. Each member of the family is defined by a choice of the function $P(x,\theta,\phi)$. The metric of such spacetimes is given by
\be\label{22}
ds^2=\omega^2 ds_\Mi^2=\frac{1}{P^2(x,\theta,\varphi)} \left(-\frac{x^2-x_{\va BH}^2}{\xo^2}\,d\tau^2+\frac{\xo^2}{x^2-x_{\va BH}^2} \, dx^2+x_{\va 0}^2 \,dS^2\right)\,.
\ee
As expected, it does not depend on the Killing time $\tau$. Cleary, any additional coordinates transformation that does not depend on $\tau$ sends the metric in an equivalent $\tau$-independent form. In Appendix~\ref{CooTra} some interesting example is presented. In particular, it is shown the precise relation between the above coordinates and the one presented in Eq.~\eqref{eq:cootran} \cite{DeLorenzo:2017tgx}.

\subsection{The Hartle-Hawking temperature}

By the standard Wick rotation $\tau \to -i \bar\tau$ the metric becomes Euclidean
\be
ds^2=\frac{1}{P^2}\left( \frac{x^2-x_{\va BH}^2}{\xo^2}\,d\bar\tau^2+\frac{\xo^2}{x^2-x_{\va BH}^2} \, dx^2+x_{\va 0}^2 \,dS^2\right)\,.
\ee
The apparent singularity at $x=+\xBH$ can be removed in the usual way by introducing a new set of coordinates
\be
\begin{split}
\rho^2&=\frac{\xBH^2}{\xo^2}(x^2-\xBH^2)\\
\varphi&=\frac{\xBH}{\xo^2}\,\tau\,,
\end{split}
\ee
in which the metric becomes
\be\label{brk}
ds^2= \frac{1}{P^2}\left(\rho^2 d\varphi^2+\frac{\xo^2}{\xo^2+\rho^2}\ d\rho^2+\xo^2 dS^2\right)\,.
\ee
Assuming that $P$ is non vanishing at $\rho=0$, the previous metric  would have a conical singularity at $x=+\xBH$ unless $0\le\varphi\le 2\pi$.
The quantum state of fields compatible with this topology of the Euclidean continuation is a thermal state with temperature \be
T_{\va HH}=\frac{1}{2\pi}\frac{\xBH}{\xo^2}=\frac{\rH}{\pi (\rO^2-\rH^2)}
\ee
which is the exactly Hartle-Hawking temperature found in \cite{DeLorenzo:2017tgx}, Eq.~\eqref{HHtemp}. Even when the local physical temperature will depend on the function $P$,  the above temperature--which we termed conformal temperature--is conformally invariant and it is related to the conformally invariant notion of surface gravity defined in the introduction via the standard relation $T_{\va HH}=\kappa_{\va SG}/(2\pi)$.  In the  following sections  we will study the spacetime realizations corresponding to different choices of the function $P(x,\theta,\varphi)$.

\section{Light Cone Black Holes}\label{sec:LCBH}

The causal structure of a spacetime $g_{ab}$ is easily readable once the Carter-Penrose diagram for $g_{ab}$ is found. In our case the procedure to find it is straightforward. Indeed, we already
know \cite{wald2010general} that the coordinate transformation 
\be
\label{eq:einsteinuniverse}
\begin{split}
T+R &=2 \,\arctan \left(\frac{v}{\rH}\right) \\
T-R &= 2 \,\arctan \left(\frac{u}{\rH}\right) \,.
\end{split}
\ee
is such that the Minkowski metric $\eta_{ab}$ becomes conformally related to the Einstein Universe metric $g_{ab}^{\va EU}$, i.e.
\be
\begin{split}
g_{ab}^{\va EU} &= \Omega^2_{\Mi}\, \eta_{ab}\\
-dT^2+dR^2 + \sin^2 R \,dS^2 &= \Omega^2_{\Mi} (-dt^2+dr^2+r^2 dS^2)
\end{split}
\ee
with
\be
\Omega^2_{\Mi} = \frac{4\rH^2}{(v^2+\rH^2)(u^2+\rH^2)}\,.
\ee
From the transformation \eqref{eq:einsteinuniverse} one can see that the Minkwoski spacetime covers only a portion of the Einstein's Universe spacetime. Such portion gives the Carter-Penrose diagram of $(\R^4, \eta_{ab})$. In particular, one can notice that infinite physical distances in $(\R^4, \eta_{ab})$ are mapped into finite distances in the Einstein's Universe, by the fact that $\Omega^2_{\Mi}$ vanishes for $u,v\to \pm \infty$. 

Any conformally flat spacetime $g_{ab}= \omega^2\,\eta_{ab}$ will be conformally mapped to the Einstein Universe by the same coordinate transformation:
\be
g_{ab}^{\va EU} = \Omega^2_{\Mi}\, \eta_{ab} =  \frac{\Omega^2_{\Mi}}{\omega^2}\,g_{ab}\equiv\Omega^2 \,g_{ab}\,.
\ee
Using \eqref{eq:omega}, the conformal factor $\Omega$ mapping the generic metric \eqref{22} to the Einstein's universe is found to be
\be\label{eq:omegaEUs}
\Omega^2 = \frac{4\,r^2}{(v^2+\rH^2)(u^2+\rH^2)} \, P^2(x,\theta,\varphi)=\frac{(v-u)^2}{(v^2+\rH^2)(u^2+\rH^2)} \, P^2(x,\theta,\varphi)\,.
\ee
As for Minkowski, vanishing of $\Omega$ implies infinite distances in the physical spacetime $g_{ab}$, defining therefore the boundary of it. 

Let us now analyze different interesting choices of the function $P(x,\theta,\varphi)$.

\section{The Bertotti-Robinson Realization}\label{BBRR}

The most interesting realization is also the simplest one: $P(x,\theta,\varphi) = 1$. From eq.~\eqref{eq:omega}, one can see that this spacetime is simply found dividing the Minkowski metric by $r^2$. Such an apparently simple operation has striking consequences. 
The metric is
\be\label{BR}
ds^2= -\frac{x^2-x_{\va BH}^2}{\xo^2}\,d\tau^2+\frac{\xo^2}{x^2-x_{\va BH}^2} \, dx^2+x_{\va 0}^2 \,dS^2\,.
\ee

The Ricci and the Kretschmann scalars come out to be
\be\begin{split}
\mathbf{R}&=0 \\
\mathbf{R}_{abcd}\mathbf{R}^{abcd}&=\frac{8}{\xo^4}\,.
\end{split}\ee
Since the metric is diagonal, we can easily define a tetrad $\mathbf{e}_a{}^I$ as
\be\label{eq:tetrad}
\begin{split}
\mathbf e_\mu^0 \,dx^\mu &= \sqrt{-g_{\tau\tau}} \,dt\\
\mathbf e_\mu^1 \,dx^\mu&= \sqrt{g_{xx}} \,dr\\
\mathbf e_\mu^2\,dx^\mu &= \sqrt{g_{\theta\theta}} \,d\theta\\
\mathbf e_\mu^3 \,dx^\mu&= \sqrt{g_{\varphi\varphi}} \,d\varphi\,.
\end{split}
\ee
In this tetrad, the Einstein tensor is diagonal and given by
\be\label{eq:ene}
\mathbf{G}_{IJ}=\mathbf{G}_{ab}\,\mathbf e^a_I \,\mathbf e^b_J=\frac1{\xo^2} {\rm diag}(1,-1,1,1)\,.
\ee

\begin{figure}[t]
\center
\includegraphics[width=.4\textwidth]{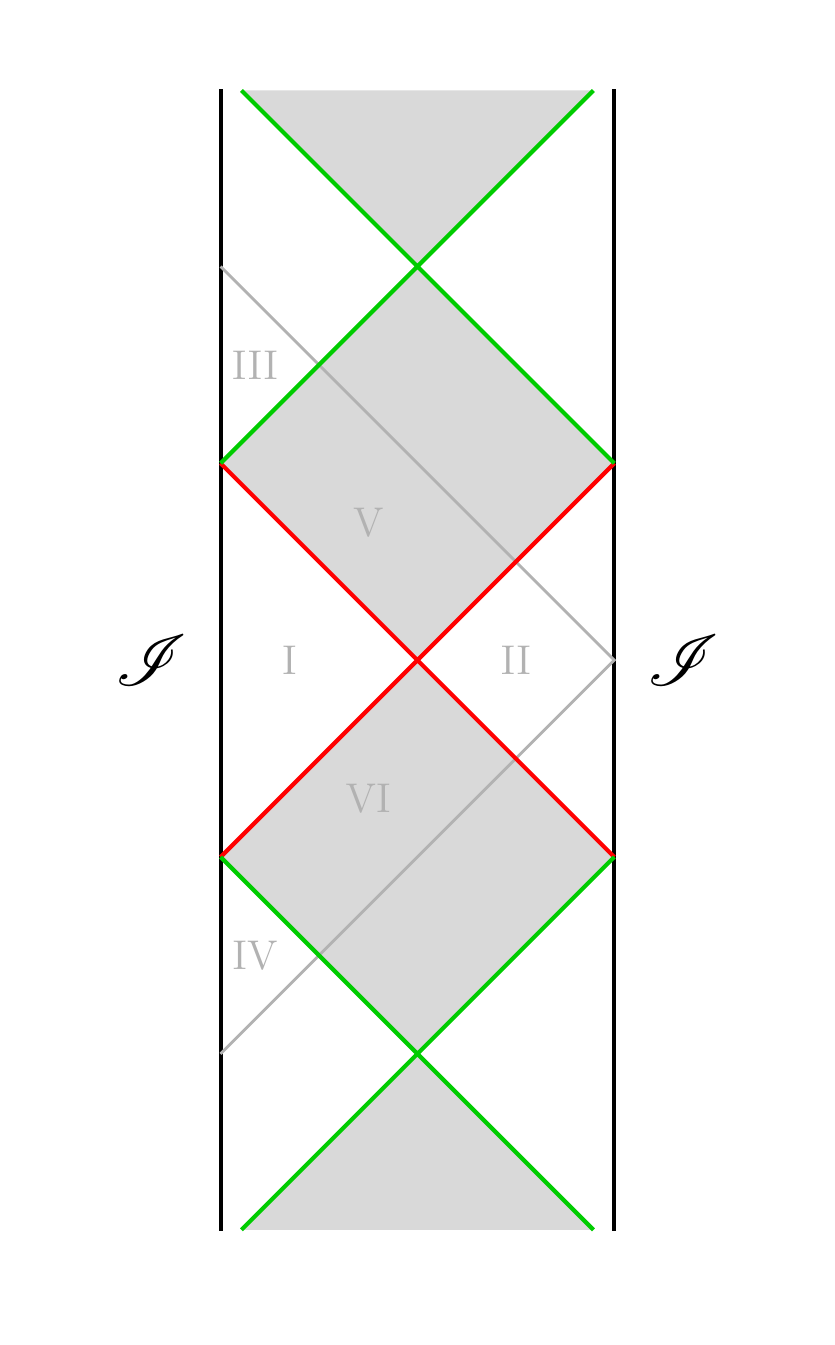} 
\caption{The causal structure of the Bertotti-Robinson spacetime. The two spherical dimensions are suppressed, so that each point represent a sphere. The Killing field is spacelike in the shaded regions and timelike elsewhere. The boundary is timelike and no singularities are present.  Grey lines and numbers show how the 6 regions of Minkowski spacetime, see Fig.~\ref{fig:Penrose_CD2}, are conformally mapped into the bulk of the target spacetime.}
\label{fig:Penrose_Bertotti}
\end{figure}

The metric \eqref{BR} can be also found as a solution to the Einstein-Maxwell equations for a the vector potential given by
\be\label{eBR}
\mathbf A=\frac{x}{x_{\va 0}} dt\,,
\ee
from which the electromagnetic tensor
\be
\mathbf F=d \mathbf A=\frac{1}{x_{\va 0}} dx\wedge dt\,.
\ee
The solution is static and spherically symmetric with a constant radial electric field whose flux defines the charge of the spacetime
\be
Q=\frac{1}{8\pi} \int_S \mathbf{\epsilon}_{abcd} \mathbf F^{cd}=x_{\va 0}\,.
\ee
The stress-energy tensor satisfies the weak, strong and dominant  energy conditions. 
The spacetime is topologically $AdS_2\times S^2$. Such solution is known in the literature as the Bertotti-Robinson spacetime \cite{Bertotti:1959pf, Robinson:1959ev}. Its Carter-Penrose diagram is depicted in Figure \ref{fig:Penrose_Bertotti}.
The geometry is everywhere regular. 
There are no singularities, despite the presence of trapped surfaces and the fact that the usual energy conditions are satisfied. Singularity theorems--see for instance \cite{wald2010general}--are avoided due to the fact that 
the spacetime is not globally hyperbolic, which rules out some versions of the theorems, and the generic null geodesic congruence condition is not satisfied \footnote{Null geodesic violating the null generic geodesic condition are those generating $\sI^{\pm}$ in Minkowski spacetime, which now pass through the bulk of the RB solution.}, which rules out those that do not require global hyperbolicity. 
%


\subsection[Close to extremal Reissner-Nordstrom near-horizon geometry]{Close to extremal Reissner-Nordstrom\\near-horizon geometry}

The Bertotti-Robinson solution is known to correspond to the near horizon geometry of a Reissner-Nordstrom (RN) black hole close to extremality--see for instance \cite{fabbri2005modeling}.
This fact in turn provides a simple interpretation to the laws of light cone mechanics  \cite{DeLorenzo:2017tgx}--Chapter~\ref{chap:lightBH}--in terms of the standard laws of black hole thermodynamics. The RN metric is given by
\ba
ds^2&=& -\left(1-\frac{2M}{r}+\frac{Q^2}{r^2}\right) dt^2+\left(1-\frac{2M}{r}+\frac{Q^2}{r^2}\right)^{-1} dr^2+r^2 dS^2 \\
&=&-\frac{(r-r_+)(r-r_-)}{r^2} dt^2+ \frac{r^2}{(r-r_+)(r-r_-)}dr^2+r^2 dS^2, 
\ea
and the associated electromagnetic field by
\be
\mathbf A=-\frac{Q}{r} dt,
\ee
where $r_{\pm}=M\pm\sqrt{M^2-Q^2}$. The near extremal case corresponds to  $M=Q+\delta M$ with $\delta M^2 \ll Q^2$, for which
the near horizon metric and electromagnetic field is obtained by expanding in the new coordinate $x$ defined by $r=x_{\va 0}+x$.
The leading order gives the metric \eqref{BR} and electromagnetic field \eqref{eBR} with $x_{\va 0}=Q$ and
$\xBH=\sqrt{2\,Q\,\delta m}$. We can relate the physical parameter or the RN solution to the parameters of the MCKF in flat spacetime via
\be
Q^2={\rO^2-\rH^2} \qquad {\rm and} \qquad \delta m = \frac{2 \rH^4}{(\rO^2-\rH^2)^{\frac32}}\,.
\ee
This shows that the limit $\rH\to 0$ corresponds exactly the extremal limit of the RN solution. On the RN side the temperature goes to zero and the bifurcating sphere goes away to infinity. On the Minkowski side, the radius of the bifurcating sphere $\rH$ shrinks to zero and the conformal Killing horizon becomes the light cone of 
a single event. 
\section[{\em De Sitter} and Asymptotically $dS$ and $AdS$ Realizations]{{\em De Sitter} and Asymptotically $dS$ and $AdS$\\Realizations}

Another interesting case arises by sending a constant $x=x_*$ Killing observer to infinity. From the discussion under Eq.~\eqref{eq:omegaEUs}, this is achieved by choosing a function $P(x,\theta,\phi)$ which vanishes at $x=x_*$. The simplest choice admitting a regular differential structure at infinity is
\be
P(x) = \frac{x_*-x}{\xo}\,.
\ee
The corresponding metric is therefore
\be
ds^2=\frac{\xo^2}{(x_*-x)^2} \left(-\frac{x^2-x_{\va BH}^2}{\xo^2}\,d\tau^2+\frac{\xo^2}{x^2-x_{\va BH}^2} \, dx^2+x_{\va 0}^2 \,dS^2\right)\,.
\ee
In the new coordinate
\be\label{X}
X^2 = \frac{\xo^4}{(x_*-x)^2}
\ee
the metric takes the simple form
\be\label{ds}
ds^2=-F(X)\,d\tau^2+\frac{1}{F(X)} \, dX^2+X^2 \,dS^2
\ee
with
\be
F(X) = 1-\frac{2\,x_* }{\xo^2}\,X+\frac{(x_*^2-\xBH^2)}{\xo^4} \,X^2\,.
\ee
The observer sent to infinity corresponds now to $X\to +\infty$.
The Ricci scalar is
\be
\mathbf{R}=-12 \left(\frac{x_*^2-\xBH^2}{\xo^4}+\frac{x_*}{\xo^2\,X}\right)
\ee
which tends to a constant as $X \to +\infty$
\be\label{eq:lambda}
4\,\Lambda = \lim_{X\to +\infty} \mathbf{R}=-12\,\frac{x_*^2-\xBH^2}{\xo^4}\,.
\ee
Such constant is positive if $x_*$ is chosen in between the horizons and negative elsewhere.
Moreover, for $x_*\neq 0$, it diverges as $X$ approaches zero
\be
\mathbf{R}|_{X\to 0} = -12\, \frac{x_*}{\xo^2\,X}+ O(1)\,.
\ee
As in the previous case, we can define a diagonal tetrad as in \eqref{eq:tetrad}, which gives the diagonal Einstein's tensor
\be\label{eq:eneAdS}
\begin{split}
\mathbf G_{00}&=-3\,\frac{x_*^2-\xBH^2}{\xo^4}+4\,\frac{x_*}{\xo^2\,X}\\
\mathbf G_{11}&=3\,\frac{x_*^2-\xBH^2}{\xo^4}-4\,\frac{x_*}{\xo^2\,X}\\
\mathbf G_{22}&=3\,\frac{x_*^2-\xBH^2}{\xo^4}-2\,\frac{x_*}{\xo^2\,X}\\
\mathbf G_{33}&=3\,\frac{x_*^2-\xBH^2}{\xo^4}-2\,\frac{x_*}{\xo^2\,X}\,.
\end{split}
\ee
The metric can therefore be interpreted as a solution to the Einstein's equation with a cosmological constant given by \eqref{eq:lambda} and a stress-energy tensor given by
\be
\mathbf T_{IJ} = \frac{2\,x_*}{\xo^2\,X} \,\text{diag}(2,-2,-1,-1)\,.
\ee
The global as well as the local nature of these spacetimes depends on the explicit value of $x_*$.
\subsection{{\em De Sitter} realization, $x_*=0$}\label{sub:desitter}

\begin{figure}[t]
\center
\includegraphics[width=.6\textwidth]{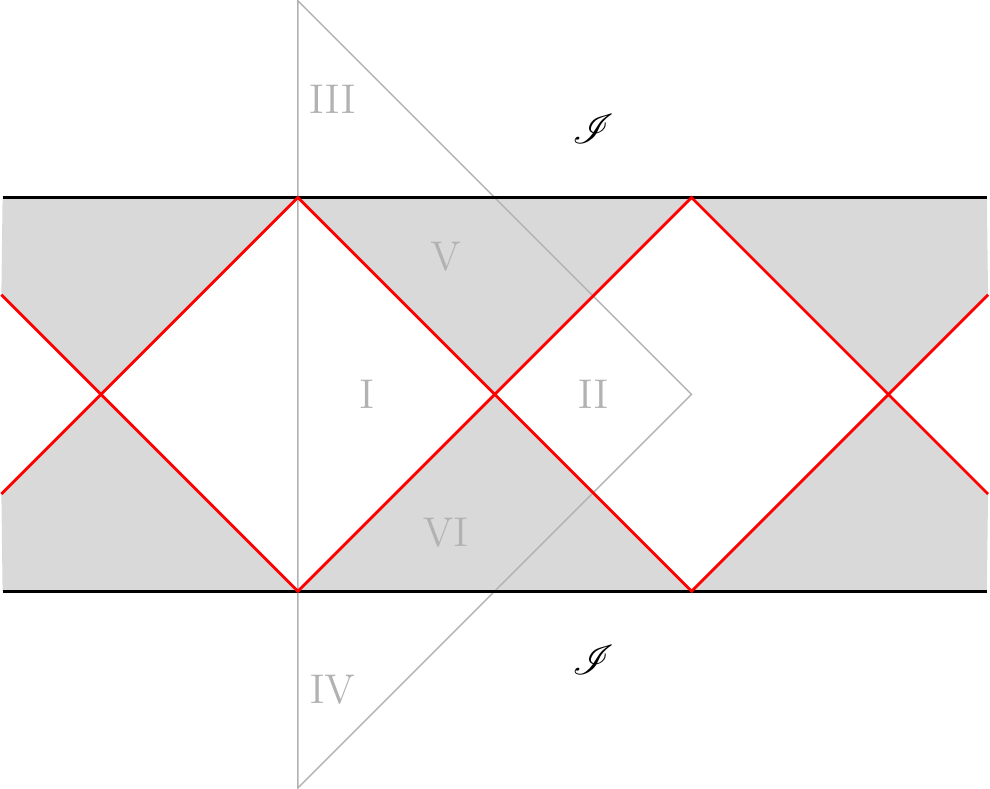}
\caption{The causal structure of the {\em de Sitter} spacetime. The two spherical dimensions are suppressed, so that each point represent a sphere. The Killing field is spacelike in the shaded regions and timelike elsewhere. The boundary is spacelike and no singularities are present. Grey lines and numbers show how the 6 regions of Minkowski spacetime depicted in Fig.~\ref{fig:Penrose_CD2} are conformally mapped into the bulk of the target spacetime.}
\label{fig:Penrose_dS}
\end{figure}

For $x_*=0$, $\mathbf T_{ab}=0$, the Ricci scalar is non-diverging, and the Einstein tensor \eqref{eq:eneAdS} corresponds to that of a positive cosmological constant term $\Lambda \,g_{ab}$ with
\be
\Lambda=3\frac{\xBH^2}{\xo^4}\,.
\ee
For this choice of $x_*$, the metric \eqref{ds} is manifestly that of {\em de Sitter} spacetime in terms of static coordinates. The bifurcating Killing horizon corresponds to the union of a past and future cosmological horizons intersecting at the bifurcating sphere, as shown in the Carter-Penrose diagram in Figure \ref{fig:Penrose_dS}. 

\subsection{Asymptotically {\em(Anti)-de Sitter} realizations}\label{sub:asyAdS}

If $x_*< 0$ one has a positive cosmological constant for $|x_*|>|\xBH$ and a negative one for $|x_*|<|\xBH$  with a $\mathbf T_{ab}$ violating all the standard energy conditions. In these cases the metric is asymptotically $dS$ and $AdS$ respectively. In the $AdS$ case the decay rate to the asymptotic geometry is slower with respect to the one imposed by the standard reflecting boundary conditions \cite{Ashtekar:1999jx} \footnote{We thank A. Ashtekar for clarifications about this point.}. This implies that a well defined notions of conserved charges at infinity is not possible. The spacetime is therefore weakly asymptotically $AdS$. 

\begin{figure}[t]
\centering
\subfigure[Weakly asymptotically AdS.]{
   \includegraphics[width=.4\textwidth]{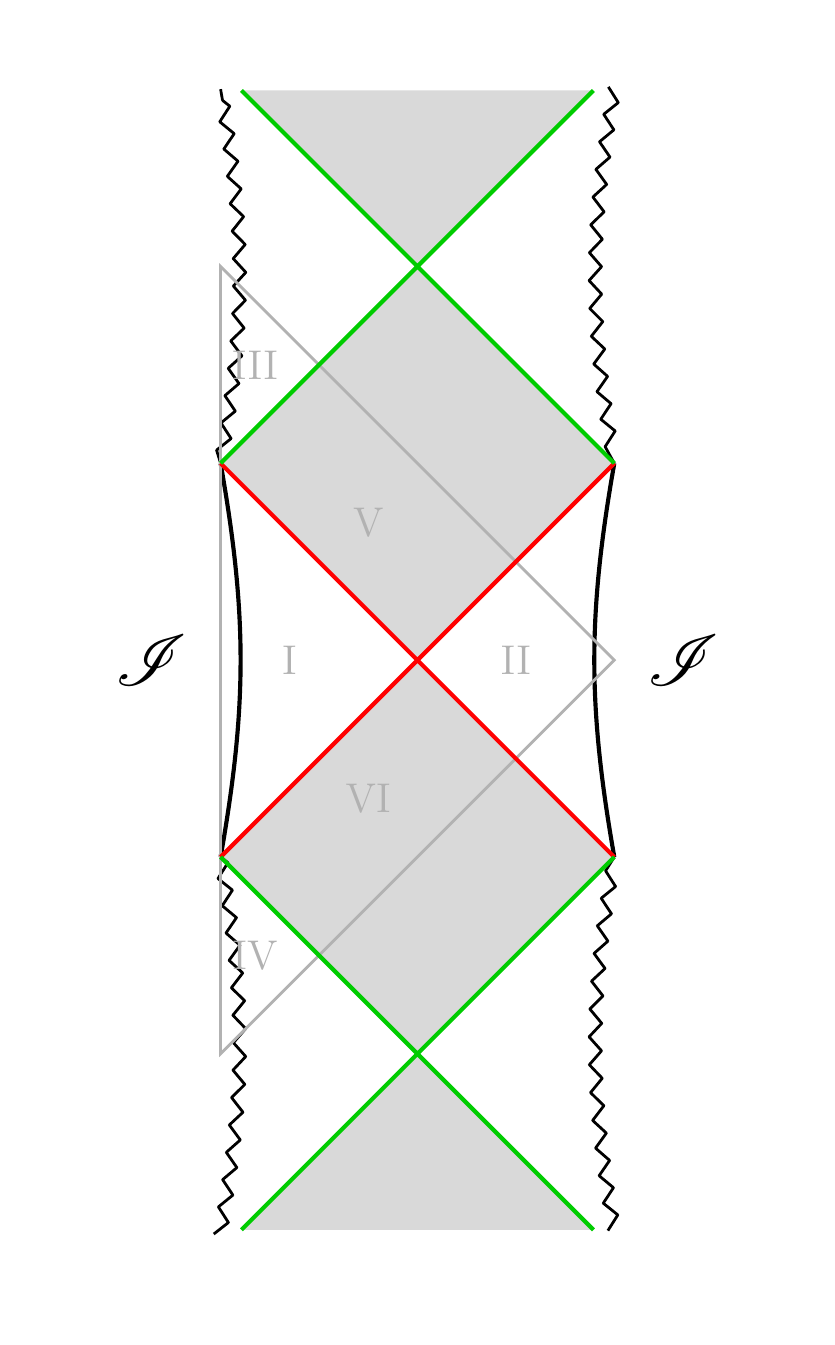}}\hspace{2cm}
\subfigure[Asymptotically dS.]{\raisebox{1cm}{
   \includegraphics[width=.26\textwidth]{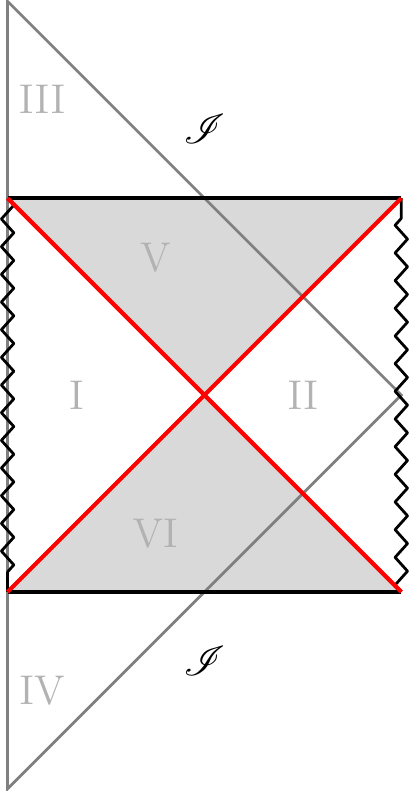}}}
\caption{The causal structure of the different spacetime realisations when an observer is sent to infinity. The two spherical dimensions are suppressed, so that each point represent a sphere. The Killing field is spacelike in the shaded regions and timelike elsewhere. Grey lines and numbers show how the 6 regions of Minkowski spacetime depicted in Fig.~\ref{fig:Penrose_CD2} are conformally mapped into the bulk of the target spacetime.}
\label{fig:Penrose_(A)dS}
\end{figure}

If $x_*>0$, instead, one has a positive asymptotic cosmological constant for $x_*<\xBH$ and a negative one for $x_*>\xBH$. Now $\mathbf T_{ab}$ satisfies the {\em weak} ($\rho\ge 0, \rho+p_i\ge 0$) and the {\em dominant} ($\rho\ge |p_i|$) energy condition but not the {\em strong} one ($\rho+p_i\ge 0, \rho+\sum_i p_i\ge 0$). For negative asymptotic cosmological constant, there are black hole regions, plus inner and outer Killing horizons. The boundary is again weakly $AdS$. The new feature with respect to the Betrotti-Robinson realization is the appearance of a time like curvature singularity at $X=0$. There is no black hole region in the $dS$ realization and the time like curvature singularities at $X=0$ remain.
The Carter-Penrose diagrams corresponding to these cases are shown in Figure \ref{fig:Penrose_(A)dS}. 

\section{An Asymptotically Flat Realisation}\label{originalone}
The inner horizons of the Bertotti-Robinson realization of Section \ref{BBRR} can become the boundary of the spacetime via a particular choice of $P(x,\theta,\varphi)$ in \eqref{22}.
Such realization was already exhibited in  \cite{DeLorenzo:2017tgx}--see Appendix~\ref{AP2}--to illustrate some aspects of the light cone thermodynamical laws.
In this case the metric is
\be
g_{ab} = \frac{16\,\rH^4}{(u-\rH)^2(v+\rH)^2}\,\eta_{ab}\,.
\ee
In Appendix~\ref{CooTra}, a coordinate transformation is found such that the metric looks
\be\label{aaaa}
ds^2=\frac{1}{\Delta}\left(-(1-2\,z)\,d\bar{\tau}^2+\frac{1}{1-2\,z} \,dz^2+z^2\,dS^2\right)\,,
\ee
where the new coordinates are dimensionless and $\Delta = \frac{4\rH^2}{\xo^4}$.
In these coordinates the horizon is located at $z=1/2$. Moreover, $z$ is positive and greater then $1/2$ outside, and decreases to zero at the Minkowskian $i^0$ and origin. Inside the horizon, on the other hand, increases from $z=1/2$ to $z\to \infty$, the latter corresponding to the inner horizon.

Using the technique of Section~\ref{sec:LCBH}, we can draw its Carter-Penrose diagram and study the properties of its boundary.
The conformal factor $\Omega^2$ mapping \eqref{aaaa} to the Einstein's Universe $g_{ab} = \Omega^2\,g_{ab}^{\va EU}$ is given by
\be\begin{split}
\Omega^2 &= \frac{(u-\rH)^2(v+\rH)^2}{4\rH^2\,(u^2+\rH^2)(v^2+\rH^2)}\\
&=\frac{(1-\sin U)(1+\sin V)}{4\rH^2}\\
&=\left(\frac{\cos T+\sin R}{2\,\rH}\right)^2
\end{split}
\ee 
where $U$ and $V$ are the Einstein's Universe null coordinates. The boundary $\scri$ is given by the condition $\Omega=0$, and therefore
\be
\scri:\,\qquad T-R=U=\frac{\pi}{2} \qquad{\rm and} \qquad T+R=V=-\frac{\pi}2\,.
\ee
which is equivalent to $u=\rH$ and $v=-\rH$, or simply $z\to \infty$. The boundary is made of two constant $U$ or $V$ surfaces, being therefore null.

\begin{figure}[t]
\center
\includegraphics[width=.2\textwidth]{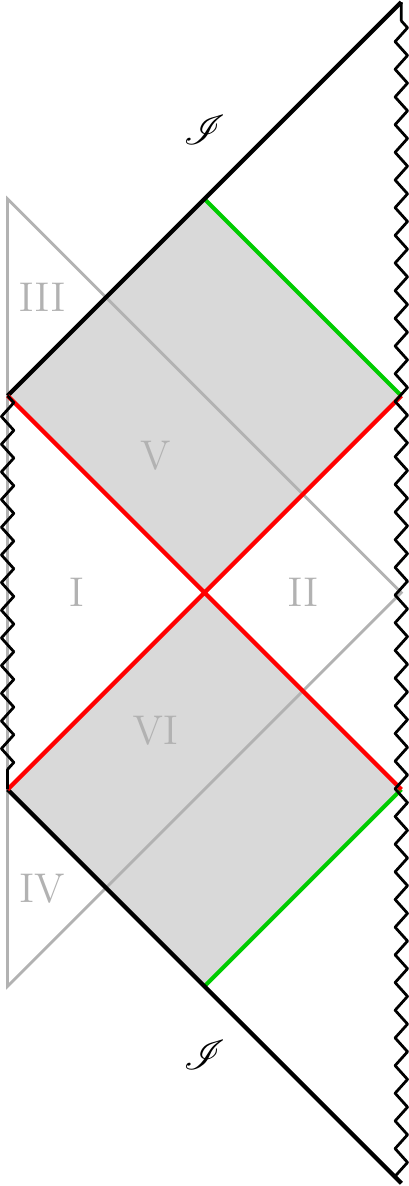} 
\caption{The causal structure of the conformally flat realization. The two spherical dimensions are suppressed, so that each point represent a sphere.  The grey lines and numbers show how the 6 regions of Minkowski spacetime depicted in Fig.~\ref{fig:Penrose_CD2} are conformally mapped into the bulk of the target spacetime.}
\label{fig:Penrose_butterfly}
\end{figure}

The gradient of the conformal factor is found to be
\be
\big(\tilde\nabla_\mu \omega\big)\,dx^\mu =-\frac{\sin T}{2 \rH}\, dT+\frac{\cos R}{2 \rH}\,dR
\ee
which is non-zero at $\scri$. The Ricci tensor in the diagonal tetrad is
\be
\mathbf{R}_{IJ} = \mathbf{R}_{ab}\,\mathbf{e}^a_I \mathbf{e}^b_J = \frac{2\Delta}{z}{\rm diag}(-1,1,2,2)
\ee
which vanishes in a neighbourhood of $\scri$, i.e. for $z\to \infty$. Our spacetime fulfils all the conditions of the definition of conformally flatness \cite{frauendiener2004conformal}.
Finally, the Ricci scalar is
\be
\mathbf{R} = \frac{12 \, \Delta}{z}
\ee
showing a curvature singularity at $z=0$. The resulting causal structure is shown in Fig.~\ref{fig:Penrose_butterfly}.

For this last case we made explicit the analysis for the construction of the Carter-Penrose diagram. The very same strategy is used for the previous cases as well, but, for brevity, we preferred not to present explicitly.

\section{Conclusions}
In Chapter~\ref{chap:lightBH}, a mathematical analogy of the laws of black hole thermodynamics has been proven for light cones in Minkowski spacetime  \cite{DeLorenzo:2017tgx}. This has been possible by observing that intersecting light cones are bifurcating conformal Killing horizons for the most general radial conformal Killing field $\xi$. The causal behavior of $\xi$ closely resemble the causal behavior of the Killing field of a Reissner-Nordstrom black hole--see Fig.~\ref{fig:penrose}. Such conformal stationarity allowed us to prove the laws of light cone thermodynamics, which however describe the properties of notions which have no direct physical meaning in flat space. At the same time, they all are conformally invariant, acquiring therefore the standard geometric meaning in conformally flat spacetime where the conformal Killing field becomes a genuine Killing field. In this Chapter we have studied the properties of the spacetime satisfying this condition. The most interesting case is the simplest one. It turns out that the conformal Killing horizon structure of light cones in Minkowski spacetime is the conformal partner of the Killing horizon structure of the Bertotti-Robinson spacetime. The latter is known to encode  the near horizon geometry of close-to-extremal and extremal charged black holes. This result completely clarifies the nature of light cone thermodynamics, that can now be seen as arising from a conformal transformation of the standard laws of BH thermodynamics. This in turn strengthen our initial claim that light cones in Minkowski spacetime encode, in a suitable sense, the main properties of BH horizons, thus providing an intriguing analogue of BHs in a spacetime with trivial curvature. The analogy is more strict and direct than the one usually considered between near horizon BH geometry and Rindler spacetime. As explained in Section~\ref{sec:nutshell}, indeed, the second analogy is strictly valid only for the $(t,r)$ plane of a BH spacetime, or in the infinite area limit. The BH and Rindler horizons, indeed, have different topologies, being respectively $ S^2 \times \mathbb{R}$ and $\mathbb{R}^2\times \mathbb{R}$. Additionally, the Rindler wedge cannot be seen as the region outside the horizon, since it lies itself in the domain of dependence of the latter. This in turn implies that no finite energy flux can escape the Rindler horizon, and no notion of asymptotic observer can be defined. These difficulties are not present in the light cone case presented in these last two Chapters. The light cone topology is $ \mathbb{R}\times S^2$ as for the BH case, and energy can be sent to infinity without crossing the horizon from the complement of the diamond, Region II in Fig.~\ref{fig:Penrose_CD2}, which therefore plays the role of the outside region. The analogy is indeed so strict that for conformally invariant matter the light cone structure is indistinguishable from the near horizon geometry of a close-to-extremal Reissner-Nordstrom black hole.
Other interesting conformally flat spacetime where the conformal Killing horizon structure becomes a proper Killing one have been presented. 

Possible applications of these intriguing analogy are currently under investigation.

%% file: Part_2/DesmoTorsion/DesmoTorsion.tex
\chapter{Spacetime Thermodynamics with Contorsion}\label{chap:desmo}
\emph{This Chapter overlaps with a paper in preparation} \cite{desmoTorsion}.\vspace{2ex}

In a famous paper \cite{PhysRevLett.75.1260}, Ted Jacobson proposed that Einstein equations could have a thermodynamical origin, compatible with the thermodynamical interpretation of the laws of black hole mechanics \cite{Bardeen1973}. His argument, based on a geometric interpretation of Clausius relation,
has been later extended to include non-equilibrium terms and higher derivative gravity theories \cite{PhysRevLett.96.121301,Chirco:2009dc,Guedens:2011dy}, and more recently to spacetimes with non-propagating torsion, namely Einstein-Cartan first-order gravity \cite{Dey:2017fld}.
This last paper motivates the study presented here.
The main difficulty of extending Jacobson's idea to Einstein-Cartan gravity is that there are two sets of independent field equations to be derived: the torsion equations as well as the Einstein equations.
The authors of \cite{Dey:2017fld} show that it is possible to derive the latter ser for a special type of torsion, and by identifying the torsional terms as a non-equilibrium contribution to Clausius relation. The torsion equations are not derived, and whether they can have also a thermodynamical origin is left as an open question. In our paper we also do not provide a derivation of the torsional equations, but we show that if they hold, the tetrad Einstein equations can be derived without the need of non-equilibrium terms nor restrictions on torsion. The technical result that allows us to achieve this is the identification of the conserved energy-momentum tensor.

The last point is crucial: in Einstein-Cartan theory, there is no conserved energy-momentum tensor that appears as source of the field equations. Nonetheless, if one restricts to invertible tetrads (and this appears necessary to connect with the metric theory and the familiar notions used in Jacobson's argument), the connection can always be written as a Levi-Civita one plus a contorsion tensor. Using this well-known decomposition, the tetrad Einstein equations can be written as the Levi-Civita Einstein tensor on the left hand side, and a torsion dependent effective energy-momentum tensor $T^{\va eff}$ on the right hand side. By taking the Levi-Civita covariant derivate of both sides, the left one vanishes due to Bianchi's identities. This in turn implies the vanishing of the right hand side, allowing to identify the conserved energy-momentum tensor also in the presence of torsion. 
For the thermodynamical argument, on the other hand, one needs to identify a conserved energy-momentum tensor without using the field equations, since these are to be derived. The first result of our paper is to show that the conservation of $T^{\va eff}$ in the Einstein-Cartan theory can be derived \emph{without} using the tetrad field equations. The proof is simple although rather lengthy, and best done using differential forms. It follows from the Noether identities of the theory, and requires the matter and torsion field equations to be satisfied.

Our second result is to use this conserved energy-momentum tensor and the contorsion description to show that the tetrad Einstein equations can be derived from the Clausius relation with the same assumptions and hypothesis of the metric case \cite{PhysRevLett.75.1260}, without the need of the non-equilibrium terms and the  restrictions on torsion used in \cite{Dey:2017fld}. 
This is possible because the starting point of Jacobson's argument, a Killing horizon associated with a locally boosted observer, is a notion which is insensitive to the presence of torsion. 
In particular, the generators of the Killing horizons follow the Levi-Civita geodesic equation. This turns out to suffice to recover the tetrad Einstein equations from the equilibrium Clausius relation, since the torsion terms are identified by the effective energy-momentum tensor.
A further advantage of our derivation is that it includes also the Immirzi term in the Einstein-Cartan theory.
Our results build on the discussion of \cite{Dey:2017fld}.
Our title is motivated by this paper, and meant to stress the role that the contorsion decomposition plays in the derivation.

To complete our discussion, we also look at the laws of black hole mechanics in the presence of torsion. The zeroth law is unaffected, and it can be proven exactly as in the metric case, provided that the energy conditions are imposed on $^{\va eff}$. The first law on the other hand depends on torsion. We consider here the `physical process' version of the first law \cite{Wald:1995yp}, which is closely related to Jacobson's argument run backwards. Using the same contorsion decomposition as before, the formal expression of the first law is unchanged, but the quantities appearing depend on torsion through the effective energy-momentum tensor.
The second law has a more marginal dependence, in the sense that torsion simply enters the inequalities on the energy conditions required.

Finally we give in one Appendix a brief comparison of two slightly different versions of Jacobson's argument \cite{PhysRevLett.75.1260,Guedens:2011dy}, and present an alternative derivation technically closer to the first law.

We use metric signature with mostly plus, and natural units $G=c=\hbar=1$.

\section{Einstein-Cartan Field Equations and Matter Sources}
Let us begin by briefly reviewing the field equations of Einstein-Cartan theory and the contorsion decomposition. We refer the reader to \cite{Hehl:1994ue} for more details, and to the Appendix~\ref{AppA} for definitions and notation.
We consider the following first-order action,
\be\label{SEC}
S_{\va EC}(e,\om) = \f1{16\pi\,} \int P_{IJKL} \Big(e^I\w e^J\w F^{KL}(\om) - \frac\L6 e^I\w e^J\w e^K\w e^L\Big),
\ee
where
\be
P_{IJKL} := \f1{2\g}(\eta_{IK}\eta_{JL}-\eta_{IL}\eta_{JK}) +\f12\eps_{IJKL},
\ee
and $\g$ is the Immirzi parameter. 
We restrict attention to invertible, right-handed tetrads. The action is then equivalent to first-order general relativity \footnote{Sometimes called Einstein-Palatini general relativity because proving its equivalence to general relativity uses the Palatini identity.} 
\be
S_{\va EP}(g,\G) = \f1{16\pi\,} \int [\sqrt{-g}(g^{\m\r}g^{\n\s}R_{\m\n\r\s}(\G) - 2\L) +\f1\g\tl\eps^{\m\n\r\s}R_{\m\n\r\s}]d^4x,
\ee
with initially independent metric and connections, which are related to the fields of \Ref{SEC} by  the familiar formulas
\be\label{gG}
g_{\m\n}=e_\m^I e_\n^J\eta_{IJ}, \qquad \G^\r_{\m\n} = e_I^\r D_\m e^I_\n:=e_I^\r (\p_\m e^I_\n+\om^{IJ}_\m e_{J\n}).
\ee

We collectively denote the matter fields as $\psi$, and consider a general matter 
Lagrangian $L_m(e,\om,\psi):={\cal L}_m(e,\om,\psi) d^4x$.
Varying the matter action we have
\begin{align}\label{dSm}
&\d S_m = \int \d L_m= \int \left(2 \tau^\m{}_I \d e^I_\m + \s^\m{}_{IJ} \d\om^{IJ}_\m + E_m\d\psi\right) e \, d^4x,
\end{align}
where $E_m$ denotes the matter field equations, and we defined the source terms
\begin{align}
&\tau^\m{}_I := \f1{2e}\f{\d \cL_m}{\d e^I_\m}= - \f1{2e}\f{\d \cL_m}{\d e^{\n}_J} e^\n_I e^\m_J =: \t^J{}_\n e^\n_I e^\m_J, 
\qquad \s^\m{}_{IJ} = \f1{e}\f{\d \cL_m} {\d \om^{IJ}_{\m}}. \label{deftausigma}
\end{align}
The sign choice in the definition of $\t$ is not universal in the literature. We picked it this way in analogy with the metric energy-momentum tensor $T_{\m\n}^\G$,  
\be\label{Tt}
T^{\G}_{\m\n} := -\f2{\sqrt{-g}} \f{\d \cL_m(g,\G)}{\d g^{\m\n}}  = - \f1{e} \f{\d \cL_m(e,\om)}{\d e^{I(\m}}e^I_{\n)} = 2\t^I{}_{(\m} e_{\n)I},
\ee
which coincides with the one of general relativity in the absence of torsion.

The field equations obtaining varying \Ref{SEC} and the matter action are 
\begin{subequations}\label{FE}\begin{align}\label{FEE}
& G^\m{}_I(e,\om) +\L e^\m_I +\f1{2\g } \eps^{\m\n\r\s}e^\a_I R_{\a\n\r\s}(e,\om) =  16\pi\, \tau^\m{}_I, \\\label{FET}
& P_{IJKL} \eps^{\m\n\r\s} e^K_\n T^L_{\r\s} = -16\pi\, \s^\m_{IJ}.
\end{align}\end{subequations}
Here
\be\label{Gmi}
G^\m{}_I(e,\om)  := \f1{4}\eps_{IJKL}\eps^{\m\n\r\s} e_\n^J F_{\r\s}^{KL}(\om) = G^{\m\n}(e,\om)e_{\n I}
\ee
is the first-order Einstein tensor, the Riemann tensor and curvature are related by $R_{\m\n\r\s}(e,\om)=e_{\m I} e_{\n J}F^{IJ}_{\r\s}(\om)$, and $T^I:=d_\om e^I$ is the torsion.
The first set \Ref{FEE}  contains the ten Einstein equations, plus six redundant equations. Although $G_{\m\n}(e,\om)$ is not symmetric a priori, 
it is easy to show that the Noether identity associated with invariance of the action under internal Lorentz transformations (see \Ref{N2} below)  implies that the equations for $G^\m{}_{[I}e_{J]\m}$ are automatically satisfied. The relevant content of \Ref{FEE} is therefore just its symmetric part, which in turn gives the tetrad Einstein equations
\be
 G_{\m\n}(e,\om) +\L g_{\m\n} +\f1{2\g } \eps_{(\m}{}^{\l\r\s} R_{\n)\l\r\s}(e,\om) =  8\pi\, T^{\G}_{\m\n}, \label{ECG}
\ee
or equivalently as functions of $(g,\G)$ via \Ref{gG}.

It is often convenient to write the field equations using the language of differential forms, as we did in the action \Ref{SEC}. To that end, we use the
Hodge dual $\star$ mapping $p$-forms to $(4-p)$-forms (see Appendix~\ref{AppA} for conventions). 
This allows us to define the Einstein 3-from
\be
\star\!G_I(\om):= - \f12\eps_{IJKL}e^J\w F^{KL}(\om), 
\ee
where the opposite sign with respect to \Ref{Gmi} is a consequence of Lorentzian signature, and equivalently the dual source forms $\star\t_I$ and $\star\s_{IJ}$. The field equations \Ref{FE} then read
\begin{subequations}\label{FFE}\begin{align}
& \star \!G_I(\om) + \L \star \!e_I - \f1\g e^J\w F_{IJ}(\om) = 16\pi\,\star\! \tau_I, \\\label{FFET}
& P_{IJKL} \, e^K\w T^L = 8\pi\,\star\! \s_{IJ}.
\end{align}\end{subequations}

\subsection{The contorsion tensor}
Although connections form an affine space with no preferred origin, the presence of an invertible tetrad suggests a natural origin: the Levi-Civita connection $\om_\m^{IJ}(e)$ associated with the tetrad. We can then always decompose an arbitrary connection into Levi-Civita plus a contorsion tensor $C_\m^{IJ}$ as 
\be\label{defC}
\om_\m^{IJ} = \om_\m^{IJ}(e) +C_\m^{IJ}.
\ee
Torsion and curvature are related to the contorsion as follows:
\begin{align}
& T^I=C^{IJ}\w e_J, \\ \label{FdC}
& F^{JK}(\om) = F^{JK}(e) + d_{\om(e)}C^{JK} + C^{JM}\w C_M{}^K = F^{JK}(e) + d_{\om}C^{JK} - C^{JM}\w C_M{}^K,
\end{align}
where $d_{{\om(e)}} $ is the  exterior derivative with respect to the Levi-Civita connection. 
Plugging this decomposition into the field equations we find
\begin{align}\label{FEC1}
& \star \!G_I(e) + \L \star \!e_I = 16\pi\,\star\! \tau_I + P_{IJKL}( d_{\om(e)}C^{JK} + C^{JM}\w C_M{}^K), \\
& P_{IJKL} e^K\w C^{LM}\w e_M = 8\pi\, \star\! \s_{IJ}.\label{FEC2}
\end{align}

The fact that the field equations for the Einstein-Cartan theory can be recasted as in \Ref{FEC1} is the source of an old debate in the literature about the role of torsion \cite{hehl2007note}: if we forget about the notion of affine parallel transport defined by $\om^{IJ}$, and use simply the one defined by $\om^{IJ}(e)$ in the sector of invertible tetrads, then the theory is indistinguishable from ordinary metric theory with some non-minimal matter coupling. The non-minimality is captured by the effective energy-momentum tensor sourcing \Ref{FEC1}, i.e.
\be\label{teff}
\star\! \tau^{\va eff}_I:=\star \tau_I + \f1{16\pi\,}P_{IJKL}( d_{\om(e)}C^{JK} + C^{JM}\w C_M{}^K).
\ee 
While we take no stand in the debate, we will heavily use this fact in the thermodynamic discussion below.
Before getting there, we need to review in the next Section the relation between the conservation of the energy-momentum tensor and the Bianchi identities.

For convenience of the reader, we report the relation between torsion and contorsion in tensor language, 
\begin{align}
& T^\r{}_{\m\n}:=e^\r_I \, T^I{}_{\m\n} = -2 C_{[\m,\n]}{}^\r=2\G^\r_{[\m\n]}, \\ & C_{\m,\n\r} = \f12 T_{\m,\n\r} - T_{[\n,\r]\m},\\& C_{(\m,\n)\r} = T_{(\m,\n)\r}.
\end{align}
Here and in the following we use a comma between indices to bundle up those with special symmetry properties. Derivative operators will always be explicitly written.
The Einstein equations \Ref{FEE} read
\begin{align}\label{FEEC}
& G_{\m\n}(e)+\L g_{\m\n} = 8\pi\, T^{\va eff}_{\m\n}, \\
& T^{\va eff}_{\m\n} = 2\t^I{}_{(\m} e_{\n)I} + \f1{16\pi\,}\Big(6g_{\a(\m}\d^{\a\r\s}_{\n)\g\d}-\f2\g g_{\g(\m}\eps_{\n)\d}{}^{\r\s}\Big) 
\Big(\oge{\na}_\r C_{\s,}{}^{\g\d} + C_{\rho,}{}^{\gamma \lambda} C_{\s,\l}{}^\d\Big).\label{Teff}
\end{align}
We refrained from expanding the completely antisymmetric $\d^{\a\r\s}_{\n\g\d}$ since no useful simplification occurs.

\section{Noether Identities and Conservation Laws}

The gravity action \Ref{SEC} is invariant under internal Lorentz transformations
\be\label{gauge}
\d_\l e^I =\l^I{}_J e^J, \qquad \d_\l\om^{IJ} = -d_\om \l^{IJ},
\ee
as well as diffeomorphisms,\footnote{Note that the Lie derivatives \Ref{diffeos} are not gauge-covariant objects. It is often convenient to consider the linear combination of transformations $L_\xi = \pounds_\xi + \d_{\om\lrcorner\xi}$ which is covariant.
}
\begin{subequations}\label{diffeos}\begin{align}
& \d_\xi e^I=\pounds_\xi e^I = de^I\lrcorner\xi+d(e^I\lrcorner\xi) = d_\om e^I\lrcorner\xi+d_\om(e^I\lrcorner\xi) - (\om^I{}_J\lrcorner\xi) e^J, \\
& \d_\xi \om^{IJ}= \pounds_\xi \om^{IJ} = d\om^{IJ}\lrcorner\xi+d(\om^{IJ}\lrcorner\xi)=F^{IJ}\lrcorner\xi+d_\om(\om^{IJ}\lrcorner\xi).
\end{align}\end{subequations}
Specializing the variation of the action \Ref{SEC} to \Ref{gauge} and \Ref{diffeos} respectively, and integrating by parts, one obtains the following Noether identities,\footnote{To obtain \Ref{NG1}, we used the identity \Ref{cyc1} below. For the reader's convenience, we report the identities also in the more common $\g$-less case,
\begin{align}
& \star\! G_{[I}\w e_{J]} = - \f12 \eps_{IJKL} e^K\w d_\om  T^L,\qquad
& d_\om \star\! G_I = -\f12\eps_{IJKL} T^J\w F^{KL}.
\end{align}
}
\begin{subequations}\label{NG}\begin{align}\label{NG1}
& P_{IJKL} e^K\w F^{LM} \w e_{M} = P_{IJKL} e^K\w d_\om T^L,\\
& d_\om (P_{IJKL} e^J\w F^{KL}) = P_{IJKL} T^J\w F^{KL}.
\end{align}\end{subequations}
These are nothing but contracted forms of the Bianchi identities $d_\om F^{IJ}=0$, $d_\om T^I=F^{IJ}\w e_J$.
Using the field equations \Ref{FFE} in \Ref{NG} one finds additional relations for the matter sources,
\begin{subequations}\label{NM}\begin{align}
\label{NM2} & d_\om\star\!\s_{IJ} = 2\star\!\t_{[I}\w e_{J]}, \\
\label{NM1} & d_\om\star\!\t_I = \f12 F^{JK}\lrcorner e_I\w \star\s_{JK} + T^{J}\lrcorner e_I \w\star\t_{J}.
\end{align}\end{subequations}
These matter Noether identities can also be derived without reference to the field equations  \Ref{FFE}: they follow from invariance of the matter action  \Ref{dSm} under \Ref{gauge} and \Ref{diffeos}, on-shell of the matter field equations.
See \cite{Hehl:1985vi,Hehl:1994ue,Barnich:2016rwk} for more details.

Recall now that, in the metric formalism, invariance of the matter Lagrangian under diffeomorphisms guarantees the conservation of the energy-momentum tensor,
\be
\d_\xi L_m = d(L_m\lrcorner \xi) \quad \Rightarrow \quad \na_\m T^{\m\n} = 0,
\ee
on-shell of the matter field equations. 
In the first-order formalism with tetrads, the energy-momentum tensor does not appear immediately in the field equations: the closest object we have is the source $\t$ of the tetrad Einstein equations \Ref{FEE}. This quantity is however not conserved, as we can see from \Ref{NM1}, whose right-hand side does not vanish on-shell.
Nevertheless, although $\t$ is not conserved, it is easy to identity an effective energy-momentum tensor which is conserved, thanks to the contorsion decomposition \Ref{FEC1}. If we take the Levi-Civita  exterior derivative $d_{{\om(e)}} $ 
on both sides of \Ref{FEC1}, the left-hand side vanishes identically. This in turns implies the vanishing of the right-hand side, which gives a local conservation law 
\be\label{dt0}
d_{{\om}(e)}\t^{\va eff}_I = 0
\ee
valid \emph{also in the presence of torsion}. Equivalently in terms of tensors, the object with vanishing (Levi-Civita) divergence is $T^{\va eff}_{\m\n}$ as defined in \Ref{Teff},
and can be bona-fide considered as the genuine energy-momentum tensor of the theory.\footnote{Another way to identify this conserved object is to 
solve the torsion equation -- which in the case of Einstein-Cartan is simply algebraic since torsion does not propagate, and plug the solution back into the action. Varying the resulting matter action with respect to the tetrad then immediately gives the effective energy-momentum tensor \Ref{Teff}. }
This equation provides the basis of energy conservation in Einstein-Cartan theory.

This fact is well-known in the literature \cite{Hehl:1976vr}, and the discussion above on the effective energy-momentum tensor should also be closely related to what Hehl calls the Freund superpotential in \cite{Boehmer:2017tvw}. 

For later purposes, we are interested in whether it is possible to derive the conservation law \Ref{dt0}
\emph{without} using the tetrad Einstein equations. This is a bit of a strange question if one starts from an action principle, but it is crucial to Jacobson's thermodynamical argument, where this is not the case. We could not find the answer to this question in the literature, which turns out to be affirmative.
The result is the following:
\smallskip

{\bf Proposition 1:} \emph{
The matter Noether identities {\rm\Ref{NM}} on-shell of the matter and torsion field equations imply the conservation law for the effective energy-momentum tensor {\rm \Ref{dt0}.}
}
\smallskip

The proof is a somewhat lengthy exercise in algebraic identities, and we leave it to Appendix 2.
We also looked for a stronger result, namely whether \Ref{dt0} also holds without imposing the torsion equation, but we did not succeed. The proof in the Appendix shows explicitly the step in which we use the torsion field equation. It should be emphasized, however, that the obstacle we see in proving such a stronger result is technical but not conceptual.

\smallskip

In tensorial language, the Noether identities for a generic gauge and diff-invariant Lagrangian density $\cal L$ read (see e.g. \cite{Barnich:2016rwk})
\begin{subequations}\label{N}\begin{align}
\label{N2} & D_\m \f{\d \cL}{\d \om^{IJ}_\m} + \f{\d \cL}{\d e^{[I}_\m} e_{J]}^\m = 0,\\\label{N1} &
\f{\d \cL}{\d \om^{IJ}_\m} F^{IJ}_{\n\m}(\om)+ \f{\d \cL}{\d e^{I}_\m} T_{\n\m}^I - e_\n^I D_\m \f{\d \cL}{\d e^{I}_\m}  = 0,
\end{align}\end{subequations}
on-shell of the matter field equations. For the Lagrangian density in \Ref{SEC}, these give respectively contractions of the algebraic and differential Bianchi identities,
\begin{align}
& \label{Bianchi2} 2R_{[\m\n]} = -\na_{\r} T^\r{}_{\m\n} -2\na_{[\m} T^\r{}_{\n]\r} + T^\r{}_{\r\s}T^\s{}_{\m\n}, 
\\ & \label{Bianchi1}
\na_\n G^\n{}_\m = T^\r{}_{\m\s} R^\s{}_\r - \f12 T^\n{}_{\r\s} R^{\r\s}{}_{\m\n},
\end{align}
from the $\g$-less terms, and 
\begin{align}
& \eps^{\a\n\r\s}R_{\m\n\r\s} = \eps^{\a\n\r\s}(\na_\n T_{\m,\r\s} + T_{\m,\l\n} T^\l{}_{\r\s} ), \\
& \eps^{\a\b\r\s}\na_\b R_{\m\n\r\s} = \eps^{\a\b\r\s} T^\l{}_{\b\r} R_{\m\n\s\l}
\end{align}
for the part in $1/\g$. As for the matter action,
\begin{align}
& D_\m(e\s^\m_{IJ}) = -2e\t^\m{}_{[I}e_{J]\m},\\
& D_\m(e\t^\m{}_I) = ee^\m_I \left( \f12 F_{\m\n}^{JK}\s^\n_{JK} + T^J_{\m\n}\t^\n{}_J  \right).\label{Nm1}
\end{align}

\section{Einstein Equations from Thermodynamics}
We now come to the main motivation for our paper: show that Proposition 1 allows us to run Jacobson's argument with the usual equilibrium assumptions.
To better appreciate our point, let us briefly recall the key steps of the metric case, referring the reader to \cite{PhysRevLett.75.1260} for more details.

\subsection{The metric case}
Consider an arbitrary metric $g_{\m\n}$ on a manifold, a  point $P$ and a neighbourhood sufficiently small for spacetime to be approximately flat. 
Denote by $\xi^\m$ the future-pointing (approximate) Killing vector generating a Rindler horizon $\cal H$ within the approximately flat region, with bifurcating surface $\cal B$ through the point $P$. This is by construction hypersurface orthogonal, null at the horizon but not outside, and vanishing at $\cal B$:
\be
\xi^2\stackrel{\cal H}{=}0, \qquad \p_\m \xi^2 =:-2\, \k\, \xi_\m, \qquad \xi^\m\stackrel{\cal B}{=}0.
\ee
Since it is Killing, it is also geodesic, 
\be\label{geoLC}
\xi^\n\oge{\na}_\n\xi^\m = - \f12 \p_\m\xi^2 = \k\, \xi^\m. 
\ee
The inaffinity $\k$ can be proven to be constant on the horizon, and referred to as the horizon surface gravity. For a Rindler horizon, constancy of $\k$ follows immediately from the vanishing of the Riemann tensor.\footnote{For a stationary black hole horizon, this is the content of the zeroth law of black hole mechanics, which requires to impose the dominant energy condition to be shown. See \cite{Bardeen1973} for the proof.}
It is useful to introduce an affine parameter $\l$ along the null geodesics,  with origin at the point $P$. It can be easily shown that 
\be\label{xil}
\xi^\m=-\l \, \k \, l^\m, \qquad l^\m\p_\m = \p_\l.
\ee

\begin{figure}[h!]
\begin{center} \includegraphics[width=.25\textwidth]{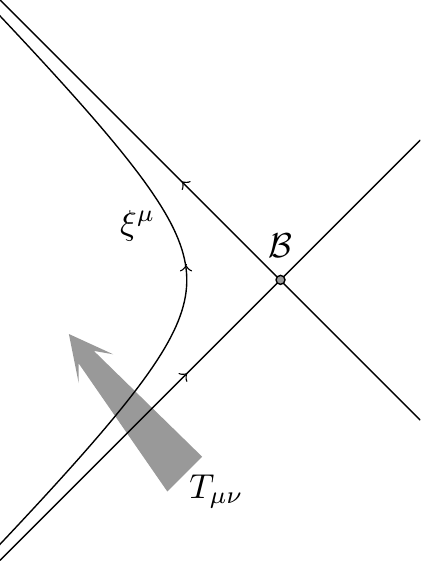} \end{center}
\caption{\small{\emph{The set-up thermodynamical derivation of Einstein's equation as proposed in \cite{PhysRevLett.75.1260}. Local flatness allows to consider approximate Rindler observers $\xi^\mu$ around any point $P$ of a given spacetime. The associate Rindler horizon has bifurcate surface $\mathcal{B}$ passing through $P$. The system is perturbed by a small flux of matter crossing the past horizon and entering the left wedge. For the derivation to be valid, an infinite family of $\xi^\mu$ is actually considered, one per each direction.}}}
\label{Fig} 
\end{figure}

Given this geometric set-up, the first step of Jacobson's argument is to associate to the Rindler horizon its Unruh temperature: 
\be\label{Tunruh}
(i) \qquad T=\f\k{2\pi}, \qquad \k={\rm constant}.
\ee
 Next, three assumptions are made: first, that there is an energy flux through the horizon in the near past of $P$, see Fig.~\ref{Fig}, given by  a \emph{conserved} energy-momentum tensor $T_{\m\n}$: 
 \be
(ii) \qquad \D U := \int_{\cal H} T_{\m\n}\xi^\m l^\n d\l d^2S = - \k \int_{\cal H} T_{\m\n}l^\m l^\n \l d\l d^2S, \qquad \na_\m T^\m{}_\n = 0,
\ee
where  we used \Ref{xil} and the constancy of $\k$. This energy flux will be interpreted thermodynamically as a heat flux, $\D U=\D Q$. 
Second assumption, that there is a notion of entropy variation associated to the horizon, which is (universally, i.e. independently of the matter state) proportional to the area variation:
\be\label{iii}
(iii) \qquad \D S=\eta \,{\D A}  = \eta \int_{\cal H} \th d\l d^2S, 
\ee
where $\th$ is the expansion of horizon. This is controlled by the Raychadhuri equation for $l^\m$,
\be\label{Ray}
\f{d \theta}{d\l} = - \f{\theta^2}{2} - \sigma_{\mu \nu} \sigma^{\mu \nu} -R_{\mu \nu} l^\mu l^\nu.
\ee
The final, technical assumption made in \cite{PhysRevLett.75.1260} is that at $P$ one can take $\th=\s_{\m\n}=0$, and approximate the solution of the Raychadhuri equation simply by $\th=-\l R_{\mu \nu} l^\mu l^\nu+O(\l^2)$.
Using this approximation, 
\be\label{th95}
 \D S = -\eta \int_{\cal H}  \l\,R_{\m\n} l^\m l^\n d\l d^2S.
\ee

Finally, we observe that using $(i-iii)$ and the approximation \Ref{th95}, the Clausius first law of thermodynamics $\D Q=T\D S$ implies
\be\label{thermo}
\int_{\cal H} \left(\frac{2\pi}{\eta}\,T_{\m\n} -R_{\m\n}\right)l^\m l^\n \l d\l d^2S=0.
\ee
Since this is valid for an arbitrary direction of the Killing boost and at any point, we can remove the integral. The Einstein equations (with an undetermined cosmological constant) then follow by imposing the conservation law $\na^\m T_{\m\n}=0$. The Newton constant is determined by $G= 1/(4\eta)$.

\subsection{The torsional case}
In the Einstein-Cartan theory \Ref{SEC} the connection is a priori affine, and torsion can be present, affecting the geodesic and Raychaudhuri equations. One may then think that the argument above should be substantially revisited. As we now show, this is actually not the case. The first observation we make is that the starting point of Jacobson's argument, a Killing horizon, is
a purely metric notion:
\begin{align}
0 = \pounds_\xi g_{\m\n} &=\xi^\a\p_\a g_{\m\n}+g_{\m\a}\p_{\n}\xi^\a+g_{\n\a}\p_\m\xi^\a & \\\nn
& = 2\oge{\na}_{(\m}\xi_{\n)} = \na_{(\m} \xi_{\n)} + T_{(\m}{}^\r{}_{\n)} \xi_\r. 
\end{align}
Hence by definition, it does not depend on torsion, in spite of the apparent presence of the latter in the last expression above. The constancy of $\k$ on the approximate Rindler horizon also follows like in the metric case from the vanishing of the metric Riemann tensor.
Being Killing and null, $\xi^\m$ is automatically geodetic with respect to the Levi-Civita connection (which we recall is always well-defined and at disposal since we are only interested in the sector of Einstein-Cartan theory with invertible tetrads), so \Ref{geoLC} still holds. Hence, we can run most of the argument as in the metric case.
Step $(i)$ is unchanged. For step $(ii)$, we follow \cite{PhysRevLett.75.1260} and define the energy flux as the integral of the \emph{conserved} energy-momentum tensor. Proposition 1 identifies this object uniquely as $T^{\va eff}_{\m\n}$ defined in \Ref{Teff}, with its torsional dependence.
Step $(iii)$ is also unchanged: since 
the generators of the Killing horizon follow the Levi-Civita geodesics \Ref{geoLC},
the change of the expansion of the generators is governed by the Raychaudhuri equation \Ref{Ray} with the metric Ricci tensor $R_{\m\n}(e)$ appearing on the right-hand side. Imposing again the equilibrium Clausius relation $\D Q=T \D S$ with these $(i-iii)$, and using the same approximation \Ref{th95}, we arrive exactly at
\be
\int_{\cal H} \left(\frac{2\pi}{\eta} \,T^{\va eff}_{\m\n} -R_{\m\n}(e)\right)l^\m l^\n \l d\l d^2S.
\ee
We conclude that the torsion-full Einstein equations, in the form \Ref{FEEC}, can be derived \`a la Jacobson from the equilibrium Clausius relation. No need to consider a torsion-full Raychaudhuri equation, non-equilibrium terms and restrictions on torsion, as argued in \cite{Dey:2017fld} and reviewed in the next Section. It suffices to use the result of Proposition 1 to identify the correct energy-momentum tensor.

There is however an important caveat to our procedure: we are assuming the torsion equations to hold, since we used them to prove Proposition 1.
This may look unsatisfactory, since it is currently not known whether these equations can be derived from a thermodynamical description. 
Our logic is that if such a description of the torsion equations exists, then it is consistent to assume that they hold when deriving the tetrad Einstein equations. This said, it is also possible that Proposition 1 holds off-shell of the torsion equations, so that these are not needed to derive the Einstein equations.  Nonetheless, one would still need to be able to derive the torsion equations from thermodynamics for the whole framework to make sense. Assuming them to hold seems thus to us coherent if a complete thermodynamical framework exists.

In any case, the main problem if one does not want to use the conserved energy-momentum tensor is the ambiguity that one faces in defining it, see e.g. \cite{Hehl:1976vr}. 
The prescription used by the authors of \cite{Dey:2017fld}, for instance, is to take simply what would be the source \Ref{deftausigma} of the tetrad EC equations, namely the derivative of the matter Lagrangian with respect to the tetrad (or equivalently up to a symmetrization, the metric). Notice however that this can be tricky in the first-order theory, because one can either work with $(g,\G)$ or $(g,C)$ as fundamental variables. The field equations are completely equivalent since the actions are related by a (non-linear) field redefinition. However the sources differ as
\begin{align}\label{Tam}
& T^\G_{\m\n}:= \f 2{\sqrt{-g}}\f{\d L_m(g,\G)}{\d g^{\m\n}}, \\\label{Tam2}
& T^C_{\m\n} = \f 2{\sqrt{-g}}\f{\d L_m(g,C)}{\d g^{\m\n}}=T^\G_{\m\n} + \f 2{\sqrt{-g}}\f{\d L_m(g,\G)}{\d \G^\a_{\b\g}}\f{\d \G^\a_{\b\g}}{\d g^{\m\n}}.
\end{align}
Both coincide with the general relativity energy-momentum tensor when torsion vanishes.
This type of ambiguity reminds us that using a conserved energy-momentum tensor, when available, is always the best choice.
We now show how this ambiguity in turn affects the non-equilibrium approach to deriving the first-order tetrad equations.

\subsection{Non-equilibrium approach}\label{sec:arbitrio}

A more general setting including a non-vanishing shear has been considered in \cite{PhysRevLett.96.121301,Chirco:2009dc}. In this case the presence of additional terms on the right-most side of \Ref{iii} is incompatible with the equilibrium Clausius relation. Hence to run Jacobson's argument one must assume that there are non-equilibrium terms, 
\be\label{Cnonequi}
\D Q=T\D S+\D S_{\va non-equi}.
\ee 
The interpretation of the shear-squared terms as non-equilibrium is justified a priori from the horizon tidal heating effect \cite{Chirco:2009dc}. Notice, however, that the same shear-squared terms enter both the non-equilibrium contribution and the equilibrium $T\D S$ contribution, since one is still assuming that the entropy variation is proportional to the area variation. This feature seems to us unusual from a thermodynamical perspective.
In any cases, in the case of the tetrad EC equations, the application of the non-equilibrium approach is even more problematic, as we now explain.

We start as before from the observation that a Killing horizon is metric-geodetic, and use the same approximations leading to the integrated metric Raychaudhuri equation \Ref{th95}, but this time allowing a non-zero shear in \Ref{Ray}. 
Then from \Ref{Cnonequi} we obtain
\be\label{thermo?}
 \f{2\pi}{\eta} \int_{\cal H}T^{\rm ??}_{\m\n} l^\m l^\n \l d\l d^2S= \int_{\cal H} \Big(R_{\m\n}(e) l^\m l^\n+\sigma_{\m\n}\s^{\m\n}\Big) \l d\l d^2S+ \D S_{\va non-equi}.
\ee
The delicate point now is how to define the heat flux, namely what $T^{??}_{\m\n}$ needs to be used on the left-hand side of the above equation. Clearly, the identification of the non-equilibrium terms that will be needed to obtain the tetrad EC equations \Ref{FEEC} depends on how we define the energy-momentum tensor. If, as in the previous Section, the conserved one is used, the only non-equilibrium term comes from the shear, which can then be argued for as in the metric theory following \cite{PhysRevLett.96.121301,Chirco:2009dc}. 
This shows how the derivation of the EC tetrad equations from the conserved energy-momentum tensor and metric Raychaudhuri equation can be easily extended to the presence of shear.

If we chose instead to define the heat flux via a source tensor like in \Ref{Tam}, we would need additional non-equilibrium terms in order to fully reproduce the Einstein equations \Ref{FEEC}. The crucial point is whether they can be justified a priori as in the example of the tidal heating, else the construction is artificial. 
The authors of \cite{Dey:2017fld} argue that this is possible, if $(a)$ we choose $T^{??}_{\m\n}=T^C_{\m\n}$ for the heat flux, and $(b)$ we define the non-equilibrium terms as those arising from the torsion-full Raychaudhuri equation that include torsion-full derivatives of $l^\m$. 
There are three problems that we can see with this construction. First, since, as already stressed several times, a Killing vector is metric-geodesic, it is in general not geodesic with respect to the torsion-full connection. Using \Ref{defC}, indeed, we see that
\be\label{geoaff}
\xi^\n{\na}_\n\xi_\m = \k\, \xi_\m - C_{\n,\m\r}\,\xi^\n\xi^\r= \k\, \xi_\m - T_{\n,\m\r}\,\xi^\n\xi^\r.
\ee
For this reason, the authors of \cite{Dey:2017fld} restrict torsion to satisfy
\be\label{Cll}
C_{\n,\m\r}\xi^\n\xi^\r = 0.
\ee
This condition implies that metric and torsion-full geodesics coincides. Since the metric and torsion-full geodesic expansion also coincide,\footnote{In the presence of torsion, the displacement of a vector $q^\m$  Lie dragged along $\xi^\m$ is given by $$
\xi^\n\na_\n q^\m = B_{\m\n}q^\n, \qquad B_{\m\n}:=\na_\n \xi_\m + T_{\m,\l\n}\xi^\l = \oge\na_\n \xi_\m +C_{\r,\m\n}\xi^\r,
$$
hence introducing the usual projector $\perp^{\m\n}$ on a 2d space-like surface orthogonal to $\xi^\m$, we have $\th:=\perp^{\m\n}B_{\m\n}=\oge\th$.}
it follows that the torsion-full Raychaudhuri equation is \emph{identical} to the metric one. 
Therefore, it is unclear what one gains from this approach, except for a restriction on torsion that in the equilibrium approach presented in the previous Section is not necessary.\footnote{Since in order to recover the Einstein equations we will need to consider arbitrary boost Killing vectors, see discussion below \Ref{thermo}, the restriction on torsion should hold for any $\xi^\m$. This implies that the only non-vanishing part of the torsion field is its completely antisymmetric irreducible part. A priori it could be possible to consider a relaxation of \Ref{Cll}, allowing for a right-hand side proportional to $l_\m$ rather than vanishing, since this would only mismatch the inaffinity of metric and torsion-full geodesics. However we don't know whether the derivation of  \cite{Dey:2017fld} can be extended to this case.}

Second, the identification of the non-equilibrium contributions as torsion-full covariant derivatives of $l^\m$ does not appears to us to be well-grounded  a priori: We are not aware of any proof that in a spacetime with torsion it is the torsion-full shear that gives the tidal heating. Furthermore, the condition of vanishing initial expansion implies that at the point $P$ we have $\na_\m l^\m= \perp^{\m\n}\!T_{\m,\n\r}l^\r$, making some `non-equilibrium terms' indistinguishable from terms without derivatives, as the authors of  \cite{Dey:2017fld} acknowledge in a footnote. 

Third, the ambiguity in picking a non-conserved $T^{??}_{\m\n}$, as discussed before. Had we chosen the alternative source $T^\G$, which is also more natural from the perspective of a metric-connection action, the same identification of non-equilibrium contributions would not work, as it would miss the terms with covariant derivatives of the contorsion in \Ref{FEEC}.

Summarizing, although the non-equilibrium approach has the advantage of allowing to relax the assumption of an initial non-vanishing shear \cite{PhysRevLett.96.121301,Chirco:2009dc}, 
it is, in our opinion, inherently ambiguous when applied to first-order gravity.

\section{On the Laws of Black Hole Mechanics with Torsion}\label{SectionLaws}

As mentioned in the introduction, Jacobson's derivation is inspired by the laws of black hole thermodynamics. Having shown that the derivation works also in the presence of torsion, at least as far as recovering the tetrad Einstein equations, the next question we considered is what happens to the these laws. 

We have recalled earlier that the surface gravity of the Rindler horizon is constant simply because the Riemann tensor vanishes. For a general horizon, constancy of the surface gravity is the zeroth law, and its proof uses the Einstein equations and the dominant energy conditions.
In the presence of torsion, we can follow the proof with the equations \Ref{FEEC}, and the only modification is that the dominant energy condition will be a restriction on the effective energy-momentum tensor.

More interesting is the modification that occurs to the first law. 
To see this, let us consider the `physical process' version of the proof \cite{Wald:1995yp}, in which an initially stationary black hole is perturbed by some matter falling inside the horizon. For our generalization, we suppose that the in-falling matter has spin and sources torsion, and that the metric and connection satisfy the Einstein-Cartan field equations.

As in the metric case, we assume  that all matter falls into the black hole, and that the latter is not destroyed by the process, but settles down to a new stationary configuration \cite{Wald:1995yp,Gao:2001ut}. These assumptions are motivated by the no-hair theorem and the cosmic censorship conjecture, which keep their value also in a theory with non-propagating torsion.
For example, it is known that a compact ball of static or slowly spinning torsion-full Weyssenhoff fluid\footnote{This is a single component of torsion (the trace part) generated by the gradient of a scalar \cite{griffiths1982spin}.} admits a solution which satisfies the junction conditions with an external Schwarzschild or slowly rotating Kerr \cite{Prasanna:1975wx,arkuszewski1974linearized}.

Following \cite{Wald:1995yp}, we use the linearized Einstein equation to study the effect on the horizon geometry caused by the in-falling matter at first order in perturbation theory,
\be
g_{\m\n}=g_{\m\n}^0+h_{\m\n}, \qquad C_{\r,\m\n}=c_{\r,\m\n}.
\ee 
Being null and hypersurface orthogonal, the affine horizon generators are metric geodetic, and their expansion is governed by the Raychaudhuri equation \Ref{Ray}. The background generators $l^\m$ are proportional to the Killing generators $\xi^\m$ satisfying $l^\m=-(\l\k)^{-1}\xi^\m$, with constant $\k$ by the zeroth law. They have vanishing shear and expansion, giving therefore at first order  
\be\label{eq:rayFL}
\f{d}{d\l} \delta\theta = -\delta R_{\mu \nu}(h) l^\mu l^\nu. 
\ee
Integrating along the horizon $\cal H$  from the bifurcation surface $\cal B$ to a cut $S_\infty$ at future null infinity, we have for the total area variation
\be\label{DA}
\D A = \int_{\cal H} \d\th \, d\l d^2S 
= \int_{\cal H} \delta R_{\mu \nu}(h) l^\mu l^\nu \, \l d\l d^2S,
\ee
where we integrated by parts and used that $\l|_{\cal B}=0$ since $\xi^\mu|_{\cal B}=0$, and that $\th|_{S_\infty}=0$ by the late time settling down assumption.

In the standard particular case of torsion-less matter with conserved energy-momentum tensor $T_{\m\n}$, we have from the linearized Einstein equations 
\be
\int_{\cal H} \delta R_{\mu \nu}(h) l^\mu l^\nu \, \l d\l d^2S = 8\pi \int_{\cal H} \delta T_{\mu \nu}(h) l^\mu l^\nu \, \l d\l d^2S.
\ee
At this order, we can substitute $l^\m=-(\l\k)^{-1}\xi^\m$ in the right-hand side integrand 
\be
- \f{8\pi}\k \int_{\cal H} \delta T_{\mu \nu}(h) \xi^\mu l^\nu \, \l d\l d^2S = \f{8\pi}\k \int_{\cal H} \delta T_{\mu\n}(h) \xi^\mu \, d H^\n =
\f{8\pi}\k ( \D M -\Omega_H\D J),
\ee
where in the first equality we used that fact the future-pointing volume form on $\cal H$ is \linebreak $d H_\m=-l_\m d\l d^2S$, and in the second the explicit expression $\xi^\m=\p_t^\m+\Omega_H\p_\phi^\m$ as well as the definitions of $\D M$ and $\D J$ used in \cite{Wald:1995yp}.
 We conclude that the linearized Einstein equations imply the first law of perturbations around a stationary black hole,\footnote{To make contact between this `physical process' version of the first law, and the one in terms of ADM charges, recall that since we are assuming all matter to be falling in the black hole, the integral along the horizon equals the integral on a space-like hypersurface $\Sigma$ extending from $\cal B$ to a 2-sphere $S_\infty$ at spatial infinity $i^0$. Using again the Einstein equations and the explicit form of the conserved Noether current (see \cite{Iyer:1994ys}, here $\k$ is the Komar charge and $\Theta$ the Einstein-Hilbert symplectic potential) we find
\be\nn
\int_{\cal H} \delta T_{\mu \nu}(h) \xi^\mu d H^\n = \int_{\Sigma} \delta T_{\mu \nu}(h) \xi^\mu d\Sigma^\n =
\int_{\cal S_\infty} (k_\xi-\Theta\lrcorner\xi) - \int_{\cal B} k_\xi= \D M_{ADM} - \Omega_H\D J_{ADM},
\ee
 where 
 the final result follows from a standard calculation with $\xi^\m=\p_t^\m+\Omega_H\p_\phi^\m$. See \cite{DePaoli:2018erh} for a derivation of the first law with covariant Hamiltonian methods for Einstein-Cartan theory.}
 \be
 \D M = \f\k{8\pi}\D A + \Omega_H\D J.
 \ee
 
For torsion-generating matter, we can follow exactly the same procedure, the only difference being that we use the Einstein-Cartan equations \Ref{FEEC} with the conserved effective energy-momentum tensor on the right-hand side. The first law follows as before but with new mass and angular momentum variations
\be
 \D M -\Omega_H\D J = \int_{\cal H} \delta T^{\va eff}_{\mu\n}(h) \xi^\mu \, d H^\n 
\ee
determined by the torsion-dependent $T^{\va eff}_{\m\n}$.
This is consistent with the results of \cite{arkuszewski1974linearized} mentioned above, where the mass of the external Schwarzschild has a torsion contribution from an effective energy density profile of the static Weyssenhoff fluid compatible with the formula above.

Following the same approach of treating the effect of torsion as an effective energy-momentum tensor, we can conclude that also the second law of black hole mechanics is still valid, provided the required restrictions on the energy-momentum tensor of matter \cite{Bardeen1973} are applied to the effective tensor \Ref{Teff}.

The third law remains as elusive as in the metric case, and we do not discuss it here.

\section{Conclusions}
Motivated by the analysis of \cite{Dey:2017fld}, we looked at one aspect of conservation laws in Einstein-Cartan theory. In the sector of invertible tetrads, where one can choose to split the connection into the Levi-Civita one plus a contorsion tensor, it is possible to identify a conserved energy-momentum tensor for matter even without using Einstein equations. Thanks to this result, we were able to reproduce Jacobson's thermodynamical argument \cite{PhysRevLett.75.1260}, and derive the (tetrad) Einstein equations from the equilibrium Clausius relation. 
Our derivation is much simpler than the one proposed in \cite{Dey:2017fld}, and does not require any restriction on torsion. On the other hand, like in \cite{Dey:2017fld}, we are only able to derive the tetrad Einstein equations from a thermodynamical argument, and not the torsion equations. This remains the crucial open question in order to truly extend Jacobson's argument to theories with independent metric and connection.

The set-up we use is the same of \cite{PhysRevLett.75.1260}, in particular the initial shear and expansion vanish, which is argued to be in line with a notion of thermodynamical equilibrium. 
An initial non-vanishing shear can be nonetheless allowed and interpreted as a non-equilibrium term \cite{PhysRevLett.96.121301,Chirco:2009dc}. 
Non-equilibrium terms are also necessary to extend Jacobson's argument to derive the field equations of modified theories of gravity with higher order terms \cite{Guedens:2011dy}. 
In the Einstein-Cartan formalism, higher-order terms typically introduce propagating torsion, see e.g. \cite{Tseytlin:1981nu}. 
While in our paper we showed that the equilibrium Clausius relation is enough to derive the Einstein-Cartan equations (with vanishing initial shear), in which torsion is non propagating, an extension to higher-order tetrad-connection theories with propagating torsion will likely require non-equilibrium terms. It could be interesting if the dissipation present in this case would be associated with dissipation of energy to the torsional degrees of freedom. From this perspective, as well as the perspective of possibly recovering the torsion field equations from a thermodynamical argument, it could be intriguing to consider existing condensed matter models in which dissipating  lattice defects introduce torsion \cite{kroner1981description}.

%% file: Appendices/appPart1.tex
\begingroup
\renewcommand\thechapter{A}
\setcounter{section}{0}
\markboth{Addenda to Part I}{}
\addcontentsline{toc}{chapter}{Addenda to Part I}
\chapter*{Addenda to Part I}

\section[Extremal Surfaces have Vanishing Mean Extrinsic Curvature]{Extremal Surfaces have Vanishing Mean \\Extrinsic Curvature}\label{app:Knull}
In the notation of Section \ref{sec:maxSurAs2Dgeod}, the mean extrinsic curvature is defined by
\be
K =\nabla_\alpha n^\alpha= h^{ab} e^\alpha_a e^\beta_b \nabla_\alpha n_\beta 
\ee 
where $\nabla$ is the covariant derivative in $g_{\alpha \beta}$ and $n_\alpha$ is the normal to $\Sigma$. The Levi-Civita connections of $g_{A B}$ and $\tilde{g}_{A B}$ are related by:
\be
\Gamma^B{}_{AC} = \tilde{\Gamma}^B{}_{AC}- \frac{2}{r}\left(\delta_{Cr}\, \delta^B_A +\delta_{Ar}\, \delta^B_C - g^{Br} g_{AC} \right)
\ee
For the calculation that follows, keep in mind the following: $e^\alpha_\phi=\delta^\alpha_\phi$ , $e^\alpha_\theta=\delta^\alpha_\theta$, $h^{\phi \phi} = g^{\phi \phi}$ , $h^{\theta \theta} = g^{\theta \theta}$, $h^{\lambda \lambda} = (g_{A B}e^A_\lambda e^B_\lambda)^{-1} $\,, $n_\alpha e^\alpha_a =0$. Also, notice that $n_\alpha$ and $e^\alpha_\lambda$ can be replaced by $n_A$ and $e^A_\lambda$ when contracted since they have vanishing angular components. 

We then have
\begin{eqnarray}
-K &=& -h^{ab} e^\alpha_a e^\beta_b \nabla_\alpha n_\beta  \nonumber \\ 
&=&   n_B (h^{ab} e^A_a \nabla_\alpha e^B_b) \nonumber \\
&=& n_B (g^{\phi \phi} \Gamma^B{}_{\phi \phi} + g^{\theta \theta} \Gamma^B{}_{\theta \theta} + h^{\lambda \lambda} e^A_\lambda \nabla_A \, e^B_\lambda  ) \nonumber \\ 
&=&  n_B (g^{\phi \phi} \Gamma^B{}_{\phi \phi} + g^{\theta \theta} \Gamma^B{}_{\theta \theta}+ \frac{2}{r} h^{\lambda \lambda} g^{Br}g_{AC} e^A_\lambda e^C_\lambda  ) \nonumber \\
&=&  n_B g^{Br} (-g^{\phi \phi} \frac{g_{\phi \phi,r}}{2}   - g^{\theta \theta} \frac{g_{\theta \theta,r}}{2} + \frac{2}{r}) \nonumber \\
&=&  n_B g^{Br} (-\frac{1}{r}   - \frac{1}{r} + \frac{2}{r}) \nonumber \\
&=& 0 
\end{eqnarray}
where we used $\Gamma^B{}_{\phi \phi}=- \frac{1}{2} g^{Br}g_{\phi \phi,r}$, $\Gamma^B{}_{\theta \theta}=- \frac{1}{2} g^{Br}g_{\theta \theta,r}$ and that the surfaces are defined as $\Sigma \sim \gamma \times S^2$ with $\gamma$ a solution of eq.~\eqref{eq:geodEq}.

\section{Lifetimes}\label{app:lifetime}
In this Appendix we compare natural time scales that appear in the collapse models. As recalled in Section~\ref{sec:halcarlomodel}, there is the time scale,  introduced in \cite{Haggard:2014rza},  defined as the proper time $\tau$ that an observer seating just outside the horizon at $r_{\Delta}$ has to wait in order for allow quantum gravitational effects to pile up until $q=1$ in that region. This time scale, of order $m^2$, is referred to as the lifetime of the black hole in  \cite{Haggard:2014rza}. However,  when one talks about \emph{lifetime} in black hole physics,  one would rather refer to the retarded time elapsed at $\sI^{+}$ between an initial $u_0$ (roughly defined by detection of the first Hawking quantum), and the complete evaporation of the hole $u_s$ (in our case). 
More precisely $u_0$ can be identified with the retarded time at which the entanglement entropy at $\sI^+$ starts departing significantly from zero. The results of \cite{Bianchi:2014bma} show that this happens for the retarded time corresponding to the collapsing shell shrinking to $r=3m$. We can write  
\be\label{eq:life}
\tau_{\rm life} = u_s - u_0 =  \Delta u+(u_{\Delta}-u_0).
\ee 
The second term can be calculated from the diagrams (the result is the same in different models); the result is
\be
\tau_{\rm life}=\Delta u+\Delta v+\frac{4}{3} m+4m \log(3).
\ee
This means that, to leading scaling order, the lifetime defined in this way coincides with the one used in \cite{Haggard:2014rza}. It is driven by $\Delta v$ when it is chosen to scale with $m$ faster than linearly. In the present models we have
$\tau_{\rm life} \sim \tau \sim m^2$ if $\Delta v\sim m^2$.

\section{The Hartle-Hawking state}\label{app:HH}

In this Appendix we recall the basic formulae (used in the main text) that allow to compute the renormalized expectation value of the energy momentum tensor in $1+1$ dimensions. Moreover, we compute for completeness the analog of the Hartle-Hawking quantum state in the fireworks background. This state leads to a regular expectation value of the energy momentum tensor in the semiclassical part of the spacetime. It has the well-known thermal properties outside the collapsing shell. However, this state does not represent the physics of gravitational collapse as it does not satisfies the vacuum boundary conditions neither at  $\sI^{-}$ nor inside the collapsing shell as the following calculation shows.

To do this, let us first recall some basic relations \cite{fabbri2005modeling}. Any $1+1$ spacetime is conformally flat and can therefore be written as
\be
ds^2 = - e^{2\rho}dx_+dx_-
\ee
for some function $\rho$ and a double null coordinate system $x_\pm$. The mean value of the covariant stress-energy tensor on some state $\ket{\Psi}$ can be defined to be
\be\label{eq:mink}
\bra{\Psi}T_{\pm\pm}\ket{\Psi}= -\frac{\hbar}{12\pi}\big((\partial_\pm \rho)^2-\partial^2_\pm \rho\big)+\bra{\Psi}:T_{\pm\pm}:\ket{\Psi}
\ee
where $:T_{\pm\pm}:$ is the normal ordered stress-energy tensor. The off-diagonal term is given by
\be\label{eq:minky}
\bra{\Psi}T_{+-}\ket{\Psi}= -\frac{\hbar}{12\pi}\partial_+\partial_- \rho.
\ee

While $\bra{\Psi}T_{\mu\nu}\ket{\Psi}$ is covariant under a coordinate transformation $x_\pm \to \xi_\pm$, the normal ordered stress tensor transforms as
\be
\bra{\Psi}:T_{\xi_\pm \xi_\pm}:\ket{\Psi} =\bra{\Psi}:T_{x_\pm x_\pm}:\ket{\Psi} - \frac{\hbar}{24\pi}\big\{x_\pm,\xi_\pm\big\}
\ee
where
\be
\big\{x_\pm,\xi_\pm \big\} = \frac{d^3 x_\pm/ d\xi_\pm^3}{dx_\pm/d\xi_\pm}-\frac{3}{2}\left(\frac{d^2x_\pm/d\xi_\pm^2}{dx_\pm/d\xi_\pm}\right)^2
\ee
is the Schwarzian derivative.

The terms that are independent of the state $\ket{\Psi}$ are vacuum polarization contributions 
stemming from the conformal anomaly. For example, by identifying $x_+\to v$ and $x_-\to u$ in the Schwarzschild region 
with metric (\ref{schw}), they become:
\be\label{emi}
\begin{split}
 -\frac{\hbar}{12\pi}\big((\partial_\pm \rho)^2-\partial^2_\pm \rho\big)&=\frac{\hbar}{24 \pi}\left[ -\frac{m}{r^3} + \frac{3}{2}\frac{m^2}{r^4} \right] \\
-\frac{\hbar}{12\pi}\partial_+\partial_- \rho&=-\frac{\hbar}{24 \pi}\left( 1- \frac{2m}{r} \right) \frac{m}{r^3}.
\end{split}
\ee 

\vspace{2ex}\noindent{\scshape\bfseries The in-state}\hspace{3ex}

The $\ket{in}$ state is defined with respect to the mode expansion in terms of 
\be
\phi_{in}=e^{i \omega v}, \qquad \phi_{out}=e^{i \omega u_{in}} \;.
\ee
Inside the collapsing shell this state coincides with the Minkowski vacuum: the vacuum polarization vanishes and the normal ordered contribution vanishes.
Outside the collapsing shell we have  
\be
\begin{split}
\bra{in}T_{uu}\ket{in} &= \frac{\hbar}{24 \pi}\left[ -\frac{m}{r^3} + \frac{3}{2}\frac{m^2}{r^4} \right]- \frac{\hbar}{24\pi}\big\{u_{in},u\big\}\\
\bra{in}T_{vv}\ket{in} &= \frac{\hbar}{24 \pi}\left[ -\frac{m}{r^3} + \frac{3}{2}\frac{m^2}{r^4} \right]\\
\bra{in}T_{uv}\ket{in} &= -\frac{\hbar}{24 \pi}\left( 1- \frac{2m}{r} \right) \frac{m}{r^3} \;,
\end{split}
\ee
where we have explicitly written the vacuum polarization terms (\ref{emi}). Using equation (\ref{doce}) one can compute the Schwarzian derivative
term and obtain (\ref{jiji}). 

\vspace{2ex}\noindent{\scshape\bfseries The Hartle-Hawking-like state}\hspace{3ex}

Take the vacuum state $\ket{H}$ of the Fock space where positive frequencies are defined with respect to the mode expansion of solutions of \eqref{uno} of the form
\be
\phi_{in}=e^{i \omega V}, \qquad \phi_{out}=e^{i \omega U} \;,
\ee
where $U$ and $V$ are the Kruskal coordinates for the black hole geometry.
We compute are the  components of the covariant stress-energy tensor of this state 
in the Minkowski patch of the spacetime defining the inside of the collapsing shell (at least the one connected with the Schwarzschild one without touching the quantum region) which is described by the metric
\be
ds^2 = - dv du_{\rm in}\;.
\ee

\vspace{2ex}\noindent{\scshape\bfseries Outside the collapsing shell.}\hspace{3ex}
Outside the collapsing shell and all over its classical chronological future one has
\be
\begin{split}
\bra{H}T_{uu}\ket{H} 
&= \bra{H}T_{vv}\ket{H}=\frac{\hbar}{768\pi m^2}\left(1-\frac{2m}{r}\right)^2\left[1+\frac{4m}{r}+\frac{12m^2}{r^2}\right]\\
\bra{H}T_{uv}\ket{H}&=-\frac{\hbar}{24 \pi}\left(1-\frac{2m}{r}\right)\frac{m}{r^3}
\end{split}
\ee
Notice that these are well behaved in regular coordinates at the past horizon; see \eqref{eq:Tkrus}.
At large $r\to \infty$ we recover the energy momentum tensor of a thermal bath
 \be
\begin{split}
\bra{H}T_{uu}\ket{H} 
&= \bra{H}T_{vv}\ket{H}=\frac{\hbar}{768\pi m^2}\\
\bra{H}T_{uv}\ket{H}&=0.\end{split}
\label{lista10}
\ee

\vspace{2ex}\noindent{\scshape\bfseries Inside the collapsing shell.}\hspace{3ex}
In the Minkowski patch of the spacetime the first term on the right-hand side of eq.~\eqref{eq:mink} is zero and, moreover, by definition the state $\ket{H}$ is such that $\bra{H}:T_{UU}:\ket{H}=\bra{H}:T_{VV}:\ket{H}=\bra{H}:T_{UV}:\ket{H}=0$. Therefore we find
\be
\begin{split}
\bra{H}T_{u_{\rm in}u_{\rm in}}\ket{H} &= -\frac{\hbar}{24\pi}\big\{U,u_{\rm in}\big\}\\
&= \frac{\hbar}{768\pi m^2}\left(1-\frac{8m}{(u_{\rm in}-v_s)}+\frac{48m^2}{(u_{\rm in}-v_s)^2}\right)\\
\bra{H}T_{vv}\ket{H} &= -\frac{\hbar}{24\pi}\big\{V,v\big\}\\
&= \frac{\hbar}{768 \pi m^2}\\
\bra{H}T_{u_{\rm in}v}\ket{H}&=0 
\end{split}
\ee
where we used the matching conditions
\be
\begin{split}
u=-4m \log \left(-\frac{U}{4m}\right)&=u_{\rm in}-4m \log \left( \frac{v_s-u_{\rm in}-4m}{4m}\right)\\
V&=4m \exp\left(\frac{v}{4m}\right)\;.
\end{split}
\ee
For large $r\to \infty$ we recover the thermal fluid in \eqref{lista10}. The collapsing shell in this state is initially 
filled up with radiation at hawking temperature. Due to the contraction of the shell one gets a divergence
of the energy momentum tensor when the shell crosses the origin at $u_{in}=v_s$.
\endgroup

%% file: Appendices/appPart2.tex
\begingroup
\renewcommand\thechapter{A}
\markboth{Addenda to Part II}{}
\addcontentsline{toc}{chapter}{Addenda to Part II}
\chapter*{Addenda to Part II}

\section{Coordinate  Transformations}\label{app:coordinate}
The radial Conformal Killing Field in Minkowski spacetime $\xi$ naturally divides the space in six regions. For each of these regions, there exists a coordinate transformation $(t,r,\vartheta,\varphi) \to (\tau,\rho,\vartheta,\varphi)$ adapted to the MCKF in the sense that $\xi(\tau)=-1$. In this Appendix, we write down such transformations explicitly. 

\subsection*{The non-extremal case $\Delta\not=0$.}
\noindent
\textbf{Region I} {(\em the diamond)}{\bf.} The coordinate transformation is given by \cite{HalILQGS}
\be
\begin{split}
t &= \frac{\sqrt{\Delta}}{2a} \frac{\sinh(\tau \sqrt{\Delta})}{\cosh(\rho \sqrt{\Delta}) + \cosh(\tau \sqrt{\Delta})}\\
r &=\frac{\sqrt{\Delta}}{2a} \frac{\sinh(\rho \sqrt{\Delta})}{\cosh(\rho \sqrt{\Delta}) + \cosh(\tau \sqrt{\Delta})}
\end{split}
\ee
with $-\infty< \tau < + \infty$ and $0 < \rho < +\infty$.

\be
\begin{split}
v &= t+r = \frac{\sqrt{\Delta}}{2a} \tanh \frac{\bar{v}\sqrt{\Delta}}{2} \\ 
u &= t-r = \frac{\sqrt{\Delta}}{2a} \tanh \frac{\bar{u}\sqrt{\Delta}}{2} \, .
\end{split}
\ee
where we have defined the null coordinates $\bar{v}=\tau + \rho$ and $\bar{u}=\tau-\rho$.
The Minkowski metric \eqref{eq:min}  in the new coordinates reads
\be
ds^2 = \Omega_{\rm I}^2 \left(-d\tau^2 + d\rho^2 +\Delta^{-1} \sinh^2(\rho \sqrt{\Delta})dS^2\right)
\ee
with
\be
\Omega_{\rm I} = \frac{\Delta/2a}{\cosh(\rho \sqrt{\Delta})+\cosh(\tau \sqrt{\Delta})}.
\ee
\noindent
\textbf{Regions II} {\em (the causal complement of the diamond)}{\bf , III and IV.} Region II, III and IV can be described by the same coordinate transformation given by \cite{HalILQGS}
\be
\label{eq:cootranApp}
\begin{split}
t &=\frac{\sqrt{\Delta}}{2a} \frac{\sinh(\tau \sqrt{\Delta})}{\cosh(\rho \sqrt{\Delta}) - \cosh(\tau \sqrt{\Delta})}\\
r &=\frac{\sqrt{\Delta}}{2a} \frac{\sinh(\rho \sqrt{\Delta})}{\cosh(\rho \sqrt{\Delta}) - \cosh(\tau \sqrt{\Delta})}\,,
\end{split}
\ee
with $-\infty< \tau < + \infty$ and $0 \leq \rho < +\infty$. Region II is now given by the restriction $|\tau|<\rho$, Region III by $\tau > 0$ and $\tau > \rho$, while Region IV by $\tau < 0$ and $|\tau| > \rho$. In this case we have
\be\label{eq:uvapp}
\begin{split}
v &= t+r = -\frac{\sqrt{\Delta}}{2a} \coth\frac{\bar{u}\sqrt{\Delta}}{2}\\ 
u &= t-r = -\frac{\sqrt{\Delta}}{2a} \coth\frac{\bar{v}\sqrt{\Delta}}{2}\,.
\end{split}
\ee
The metric \eqref{eq:min} is now
\be
ds^2 = \Omega_{\rm II}^2 \left(-d\tau^2 + d\rho^2 +\Delta^{-1} \sinh^2(\rho \sqrt{\Delta})dS^2\right)
\ee
with
\be
\Omega_{\rm II} = \frac{\Delta/2a}{\cosh(\rho \sqrt{\Delta})-\cosh(\tau \sqrt{\Delta})}.
\ee
For Region II, given the above mentioned restrictions on the coordinate, we have $\bar{u} \in (-\infty,0)$ and $\bar{v} \in (0,+\infty)$. This is the transformation used in Section~\ref{HRCT}.

\noindent
\textbf{Region V.}
In the upper of the two regions where $\xi$ is spacelike, the coordinate transformation can be found to be
\be
\begin{split}
t &= \frac{\sqrt{\Delta}}{2a} \frac{\cosh(\tau \sqrt{\Delta})}{\sinh(\rho \sqrt{\Delta}) + \sinh(\tau \sqrt{\Delta})}\\
r &=\frac{\sqrt{\Delta}}{2a} \frac{\cosh(\rho \sqrt{\Delta})}{\sinh(\rho \sqrt{\Delta}) + \sinh(\tau \sqrt{\Delta})}
\end{split}
\ee
with $0< \tau < + \infty$ and $0 < \rho < +\infty$.
The double null coordinates are here given by
\be
\begin{split}
v &= t+r = \frac{\sqrt{\Delta}}{2a} \coth \frac{\bar{v}\sqrt{\Delta}}{2} \\ 
u &= t-r = \frac{\sqrt{\Delta}}{2a} \tanh \frac{\bar{u}\sqrt{\Delta}}{2} \, .
\end{split}
\ee

\noindent
\textbf{Region VI.}
Finally, for Region VI we have
\be
\begin{split}
t &= \frac{\sqrt{\Delta}}{2a} \frac{\cosh(\tau \sqrt{\Delta})}{\sinh(\rho \sqrt{\Delta}) - \sinh(\tau \sqrt{\Delta})}\\
r &=\frac{\sqrt{\Delta}}{2a} \frac{\cosh(\rho \sqrt{\Delta})}{\sinh(\rho \sqrt{\Delta}) - \sinh(\tau \sqrt{\Delta})}
\end{split}
\ee
with $-\infty< \tau < 0$ and $0 < \rho < +\infty$. This gives
\be
\begin{split}
v &= t+r = -\frac{\sqrt{\Delta}}{2a} \tanh \frac{\bar{u}\sqrt{\Delta}}{2} \\ 
u &= t-r = -\frac{\sqrt{\Delta}}{2a} \coth \frac{\bar{v}\sqrt{\Delta}}{2} \, .
\end{split}
\ee
In both last two cases, the metric \eqref{eq:min} becomes
\be
ds^2 = \Omega_{V/VI}^2 \left(-d\tau^2 + d\rho^2 +\Delta^{-1} \cosh^2(\rho \sqrt{\Delta})dS^2\right)
\ee
where, for Region V
\be
\Omega_{\rm V} = \frac{\Delta/2a}{\sinh(\rho \sqrt{\Delta})+\sinh(\tau \sqrt{\Delta})}\,,
\ee
and for Region VI
\be
\Omega_{\rm VI} = \frac{\Delta/2a}{\sinh(\rho \sqrt{\Delta})-\sinh(\tau \sqrt{\Delta})}\,.
\ee

\subsection*{The extremal case $\Delta = 0$}

In the $\Delta=0$ case, we have only Region II, III and IV and $\xi$ is everywhere timelike. The coordinate transformation in this extremal case can be obtained from the previous one by taking the limit $\Delta\to 0$ in all expressions. The result is
\be
\begin{split}
t=&\frac{\tau }{a \left(\tau ^2-\rho ^2\right)}\\
r=&\frac{\rho }{a \left(\rho ^2-\tau ^2\right)}
\end{split}
\ee
with $-\infty< \tau < + \infty$ and $0 \leq \rho < +\infty$. Region II is now given by the restriction $|\tau|<\rho$, Region III by $\tau > 0$ and $\tau > \rho$, while Region IV by $\tau < 0$ and $|\tau| > \rho$. In this case we have
\be
\begin{split}
v &= t+r = \frac{1}{a \bar v} \\ 
u &= t-r =  \frac{1}{a \bar u},
\end{split}
\ee
where, given the above mentioned restrictions on the coordinate, we have $\bar{u} \in (-\infty,0)$ and $\bar{v} \in (0,+\infty)$.
The Minkowski metric  in the new coordinates reads
\be
ds^2 = \Omega_{\rm ext}^2 \left(-d\tau^2 + d\rho^2 +\rho^2 dS^2\right)
\ee
with
\be
\Omega_{\rm ext} = \frac{\rho }{a \left(\rho ^2-\tau ^2\right)}.
\ee
This coincides with Eq.~(12) in \cite{doi:10.1063/1.532903}.

\subsection*{Near bifurcate sphere approximation}
In the non-extremal case, the bifurcate sphere is located at $\rho \to +\infty$ and $\tau=0$. Eq.~\eqref{eq:cootran} can therefore be expanded in the approximation $\rho >> 1/\sqrt{\Delta}$. This gives a Rindler-like coordinate transformation
\be
\begin{split}
t \sim -\sqrt{\frac{c}{a}} \;e^{-\rho\sqrt{\Delta}} \cosh(\tau \sqrt{\Delta})\\
r \sim \sqrt{\frac{c}{a}} \;e^{-\rho\sqrt{\Delta}} \sinh(\tau \sqrt{\Delta})
\end{split}
\ee
with the would-be proper distance given by $D = \sqrt{c/a} \;e^{-\rho\sqrt{\Delta}}$. The above approximation is inconsistent in the case $\Delta = 0$.

\section[Static FRW Spacetime and Region II]{Static FRW Spacetime and Region II}\label{app:FRW}
The coordinate transformations above show that the Regions I to IV in Minkowski spacetime are conformally related to pieces of a static FRW spacetime with negative spatial curvature $k=-|\Delta|$.
This fact was used in the computation of Bogoliubov coefficients in Section~\ref{HRCT}. In this Appendix we give some more details on the geometry of static FRW spacetime and its relation with Region II. The static FRW spacetime is a solution to the Einstein equation with zero cosmological constant and the energy stress tensor of a perfect fluid satisfying the state equation \cite{hawking1973large}
\be
\mu = - 3 p\,.
\ee
Here $\mu$ and $p$ are the energy density and pressure of the fluid respectively. The metric takes the form
\be\label{eq:FRWmetric}
ds^2 = -d\tau^2 + d\rho^2 +\Delta^{-1} \sinh^2(\rho \sqrt{\Delta})dS^2\,,
\ee
with $-\infty< \tau < + \infty$ and $0 \leq \rho < +\infty$. As shown for example in \cite{Candelas:1978gf,hawking1973large}, there exists a coordinate transformation that conformally maps this space into the Einstein static universe. This transformation allows to draw the Penrose diagram for the static FRW spacetime, which results in a diamond shaped diagram depicted in Figure~\ref{fig:FRW} \footnote{In the cited references \cite{Candelas:1978gf,hawking1973large}, however, they consider the non-static FRW spacetime with zero cosmological constant $\Lambda = 0$ and zero pressure $p=0$. The resulting Penrose diagram is therefore slightly different, being only the upper triangle of the whole diamond of Figure~\ref{fig:FRW}, with a ``big bang singularity'' for $\tau = 0$.}. 
\begin{figure}[t]
\center
\includegraphics[height=10cm]{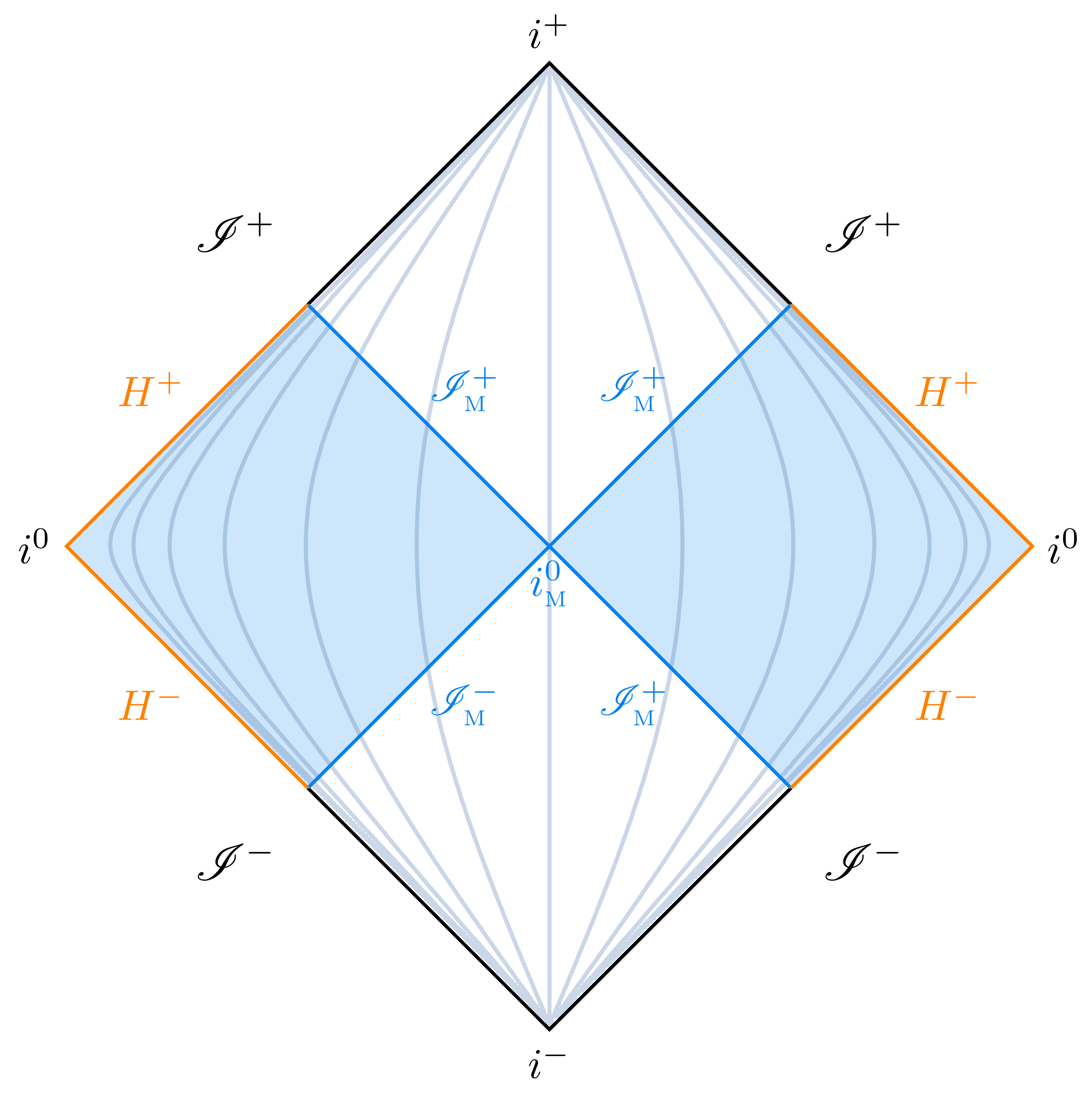}
\caption{The Penrose diagram for the static FRW spacetime of Eq.~\eqref{eq:FRWmetric}. The shaded region is the one conformally related to Region II in Minkowski spacetime. Its boundaries are in correspondence with pieces of Minkowskian future and past null infinities $\scri^\pm_M$, as well as with the bifurcate conformal Killing horizon $H^+ \cup H^-$. The light grey hyperbolas are radial flow lines of the field $\partial_\tau$, or in another words $\rho = const$ lines.}
\label{fig:FRW}
\end{figure}
The structure is very similar to the one of Minkowski, with past and future null infinities. There is however one main difference: here spacial infinity $i^0$ is a sphere and not a point as in the Minkowski case.

Region II is (conformally) given by the restriction $|\tau|<\rho$. This corresponds to the region outside the light cone shining from the origin; this is the shaded region in Figure~\ref{fig:FRW}. From Eq.~\eqref{eq:uvapp}, Minkowski future null infinity $v \to +\infty$ is mapped into the future light cone $\bar u = 0$. Analogously, Minkowski past null infinity $u \to - \infty$ is given by the past light cone $\bar v = 0$. In the same way, the future horizon $H^+$ located at $u = u_-$ is given by $\bar v \to +\infty$, namely the piece of FRW $\scri^+$ given by $\bar u < 0$. The past horizon $v = v_+$ is, in a similar way, mapped into the piece of FRW $\scri^-$ given by $\bar v > 0$. Finally, Minkowskian spacial infinite $i^0_M$ is mapped into the point given by the origin, while the bifurcate sphere $r=r_H$ at $t=0$ is given by the FRW spacial infinity $i^0$.
In Figure~\ref{fig:FRW}, the flow lines of the Killing field $\partial_\tau$, namely $\rho = const.$ surfaces are also plotted. From there, the behaviour in Region II of the conformal Killing field $\xi$ depicted in Figure~\ref{fig:families} becomes clear. The flow lines of $\xi$ start their life on $\scri^-$ to end on $\scri^+$. 

The above discussion shows also that the complete set of solutions \eqref{modes} to the Klein-Gordon equation \eqref{eq:KG} is a good complete set of solutions also in our region of interest. The set considered, indeed, is regular everywhere, at the origin too, where otherwise there could have been a problem in our setting.

\section[Another Conformal Mapping of Minkowski]{Another Conformal Mapping of Minkowski}\label{AP2}

The previous map to a suitable FRW spacetime is useful for the calculations in Section \ref{HRCT}. However, it is not the best suited for the geometrical interpretation due to the fact that the horizon $H$ is mapped to infinity in the FRW spacetime. Here we construct a new conformal mapping of flat spacetime where $\xi$ becomes a Killing field, and the horizon is mapped to a genuine Killing horizon embedded in the bulk of the host spacetime. In Minkowski spacetime we have
\be
\sL_{\xi} \eta_{ab}=\frac{\psi}{2} \eta_{ab}
\ee
where explicit calculation yields
\be \psi= \nabla_a\xi^a=8at=\frac{4(u+v)}{r_{\va O}^2-r_{\va H}^2}.\ee
Under a conformal transformation $g_{ab}=\Omega^2 \eta_{ab}$ one has
\be
\sL_{\xi}g_{ab}=\sL_{\xi}(\Omega^2 \eta_{ab})=\left[\frac{\psi}{2} + 2 \xi(\log(\Omega))\right]g_{ab}.
\ee
Therefore, in the new spacetime $g_{ab}$, $\xi$ will be a Killing field iff \be {\psi}+ 4 \xi(\log(\Omega))=0.\ee This equation does not completely fix $\Omega$: if $\Omega$ is a solution, then $\Omega^{\prime}=\omega \Omega$ is also a solution as long as $\xi(\omega)=0$. Writing explicitly the previous equation  using (\ref{eq:CKFb0}) we get:
\be\label{ew}
 u+v + (u^2-r_{\va H}^2) \partial_u(\log(\Omega))+ (v^2-r_{\va H}^2) \partial_v(\log(\Omega))=0.
\ee
It is easy to find solutions of the previous equation by separation of variables. Assuming that we want to preserve spherical symmetry then we can write
$\Omega(u,v)=V(v)U(u)$ and the previous system becomes
\ba
 \notag  u + (u^2-r_{\va H}^2) \partial_u(\log(U))&=&-\lambda,\\
  v + (v^2-r_{\va H}^2) \partial_v(\log(V))&=&\lambda,
\ea
where $\lambda$ is an arbitrary constant. If we choose $\lambda=0$ then the solution is
\be\label{OFRW}
\Omega=\frac{\Omega_0}{\sqrt{(u^2-r_{\va H}^2)(v^2-r_{\va H}^2)}}.
\ee
By fixing the integration constant $\Omega_0=r_{\va O}^2-r_{\va H}^2$, the previous solution corresponds to the  one that maps to the FRW spacetime studied in the previous section. This can be checked by recalling from (\ref{eq:FRWmetric}) that the conformal factor mapping to the FRW spacetime is $1/\sqrt{-\xi\cdot \xi}$ and  using (\ref{eq:norm}). The Killing vector $\xi$ in the FRW metric is normalized everywhere.

An alternative solution, which does not send the horizon to infinity, is obtained by choosing $\lambda=r_{\va H}$ which yields
\be\label{OBH}
\Omega_{\va BH}=\frac{4r^2_{\va H}}{(u-r_{\va H})(v+r_{\va H})},
\ee
where we have chosen the integration constant so that $\Omega_{\va BH}=1$ at the bifurcate surface $u=-r_{\va H}$ and $v=r_{\va H}$. 
In the new metric 
\be
g_{ab}=\frac{16 r^4_{\va H}}{(u-r_{\va H})^2(v+r_{\va H})^2}\, \eta_{ab}, 
\ee
the null surfaces $u=-r_{\va H}$ and $v=r_{\va H}$ are Killing horizons with constant cross-sectional area $A=4\pi r_{\va H}^2$.  These surfaces have the same geometric properties as black hole horizons which justifies the subindex BH in (\ref{OBH}).  At the inner horizons $u=r_{\va H}$  and $v=-r_{\va H}$ the conformal factor diverges. Therefore, in contrast with the FRW mapping, only these  horizons are pushed to infinity. If $r_O\le \sqrt{5} r_H$, then the Killing field $\xi$ is normalized on a timelike surface outside the horizon where stationary observers measure time and energy in agreement with those in the FRW mapping. 
More details on these geometries can be found in Chapter~\ref{chap:lightBH}. The previous conformal map plays a central role for the interpretation of the first law (\ref{firstLaw}) as discussed at the end of Section \ref{sec:firstlaw}.



\section{Coordinate Transformations [continued]}\label{CooTra}
In \cite{DeLorenzo:2017tgx}, a coordinate transformation $(t,r,\theta,\varphi) \to (\tau,\rho,\theta,\varphi)$ for each of the six regions Minkowski spacetime is divided into by the radial MCKF has been presented--see Appendix~\ref{app:coordinate}. The transformation was built in such a way the radial MCKF reduced to 
\be\label{eq:killtauapp}
\xi^\mu \frac{\partial}{\partial x^\mu} = \frac{\partial}{\partial \tau}\,.
\ee

Those six transformations can actually be grouped in one single transformation given by
\be
\begin{split}
\tau &= \frac{\rO^2-\rH^2}{4\rH}\log\frac{(u-\rH)(v-\rH)}{(u+\rH)(v+\rH)}\\
\rho &= \frac{\rO^2-\rH^2}{4\rH}\log\frac{(u+\rH)(v-\rH)}{(u-\rH)(v+\rH)}\,.
\end{split}
\ee
The coordinate $\tau$ is the same used in the main text, eq.~\eqref{eq:CooTra}.
Defining as in \cite{DeLorenzo:2017tgx}
\be
\begin{split}
\Delta &= \frac{4\rH^2}{(\rO^2-\rH^2)^2}\\
a &= \frac{1}{\rO^2-\rH^2}
\end{split}
\ee Minkowski metric becomes
\be
ds_\Mi^2=\left(\frac{\Delta/2a}{\cosh(\sqrt{\D}\, \tau)+\cosh(\sqrt{\D} \,\rho)}\right)^2\big(-d \tau^2+d\rho^2+\Delta^{-1}\sinh^2(\sqrt{\D}\,\rho)\,dS^2\big)\,.
\ee
The transformation is valid everywhere using the standard definition of the logarithm of a negative number, namely
\be
\log (-x) = i\,\pi + \log(x) \qquad x>0\,.
\ee
We can solve eq.~\eqref{eq:conftokill} in these coordinates finding
\be
\omega^2 = \left(\frac{\cosh(\sqrt{\D}\, \tau)+\cosh(\sqrt{\D} \,\rho)}{\D/2a}\frac{1}{G_{\rho}(\rho)}\right)^2\,,
\ee
which 
in terms of the Minkowskian double-null coordinates $(u,v)$
\be\label{eq:omega2}
\omega^2(u,v) = \frac{4r_{\va H}^4}{(u^2-\rH^2)(v^2-\rH^2)} \frac{1}{G^2(\rho,\theta,\phi)}\,.
\ee
A conformally flat metric $g_{ab}$ such that the radial MCKF becomes a Killing field can therefore also be written as
\be\label{metrho}
ds^2=\frac{1}{G^2_\rho(\rho,\theta,\phi)}(-d \tau^2+d\rho^2+\Delta^{-1}\sinh^2(\sqrt{\D}\rho)\,dS^2)\,,
\ee
Choosing the function $G_{\rho}$ to be a normalisation constant given by
\be
G_{\rho}= \frac{\rO^2-\rH^2}{2\rH^2}
\ee
one finds 
\be
\omega_{\va FRW} = \frac{\rO^2-\rH^2}{\sqrt{(u^2-\rH^2)(v^2-\rH^2)}} 
\ee
namely 
the conformal factor of Eq.~\eqref{OFRW}. The choice
\be\label{PBH}
G_{\rho}(\rho)=1/4\,e^{-2\sqrt{\D}\,\rho}\,,
\ee
instead, gives
\be
\omega_{\va BH}=\frac{4r^2_{\va H}}{(u-r_{\va H})(v+r_{\va H})}
\ee
defined in Eq.~\eqref{OBH}. The above two conformal factors were found in \cite{DeLorenzo:2017tgx} by separation of variables. 

Another interesting coordinates transformation is given by
\be
\begin{split}
\bar{\tau} &= \sqrt{\Delta}\,\tau\\
\rho &= \frac{1}{2\sqrt{\D}}\log (1-2z)\,.
\end{split}
\ee
which implies
\be
z=\frac12\left(1-\frac{(u+\rH)(v-\rH)}{(u-\rH)(v+\rH)}\right)\,.
\ee
The metric \eqref{metrho} takes 
the following Schwarzschild-like form
\be
ds^2=\frac{\Delta^{-1}}{G^2_z(z,\theta,\phi)} \left(-(1-2\,z)\,d\bar{\tau}^2+\frac{1}{1-2\,z} \,dz^2+z^2\,dS^2\right)\,,
\ee
where $G_z^2$ is a new function encoding the ambiguity in the conformal transformations, and the new coordinates are dimensionless. In these coordinates the horizon is located at $z=1/2$. $z$ is positive and greater then $1/2$ outside, and decreases to zero at the Minkowskian $i^0$ and origin. Inside the horizon, on the other hand, increases from $z=1/2$ to $z\to \infty$, the latter corresponding to the inner horizon. In Section~\ref{originalone}, these coordinates are used in the simplest case $G_z=1$.

\section{Conventions}\label{AppA}

We take $\ut{\eps}_{\m\n\r\s}$ as the completely antisymmetric spacetime density with $\ut{\eps}_{0123}=1$, and 
$\tl\eps^{\m\n\r\s}\ut{\eps}_{\m\n\r\s}=-4!$. It is related to the volume 4-form by
\be
\eps:=\f1{4!}\eps_{\m\n\r\s}dx^\m\w dx^\n\w dx^\r\w dx^\s, \qquad \eps_{\m\n\r\s}:=\sqrt{-g} \, \ut{\eps}_{\m\n\r\s}.
\ee
We define the Hodge dual in components as
\be
(\star \om^{(p)})_{\m_1..\m_{4-p}} := \f{1}{p!} \om^{(p)}{}^{\a_1..\a_p} \eps_{\a_1..\a_p\m_{1}..\m_{4-p}}.
\ee

For the internal Levi-Civita density $\eps_{IJKL}$ we refrain from adding the tilde. We use the same convention, ${\eps}_{0123}=1$, so the tetrad determinant is
\be\label{tetId2}
e = -\f1{4!}\eps_{IJKL}\tl\eps^{\m\n\r\s} e_\m^I e_\n^J e_\r^K e_\s^L,
\ee
and we take $e>0$ for a right-handed tetrad. 

Curvature and torsion are defined by
\be
F^{IJ}(\om) = d \om^{IJ} + \om^{IK}\w \om_K{}^J, \qquad \label{defT}
T^I(e,\om) = d_\om e^I,
\ee
where $d_\om$ is the covariant exterior derivative, whose components we denote by $D_\m$, to distinguish them from the spacetime covariant derivative $\na_\m$ with affine connection $\G^\r_{\m\n}$. The relation between the connections on the fiber and on the tangent space is given by
\be
D_\m e_\n^I = \G^\r_{\m\n} e_\r^I, \qquad \om^{IJ}_\m=e_\n^I{\na}_\m e^{\n J}
\ee
for $\om$ and $\G$ general affine connections, plus
the metricity condition $D_\m \eta^{IJ}=0$. The compatibility of the internal covariant derivative and the tetrad means that
$D_\m f^I=e^I_\n\na_\m f^\n$ and so on.

The commutators of the covariant derivatives satisfy: 
\begin{align}
& [D_\m,D_\n]f^I = F^I{}_{J\m\n}(\om)f^J, \\
& [D_\m,D_\n]f = - T^\r{}_{\m\n}(e,\om) \p_\r f, \\
& [\na_\m,\na_\n]f^\r = R^\r{}_{\s\m\n}(\G)f^\s - T^\s{}_{\m\n}\na_\s f^\r,
\end{align}
where
\begin{align}
R_{\r\s\m\n}(\G) = e_{I\r}e_{J\s} F^{IJ}_{\m\n}(\om) \qquad
T^\r{}_{\m\n}(\G) =e^\r_I \, T^I{}_{\m\n}(\om).
\end{align}
Finally, torsion and contorsion are related by 
\begin{align}
& T^\r{}_{\m\n}:=e^\r_I \, T^I{}_{\m\n}(e,C) = -2 C_{[\m,\n]}{}^\r=2\G^\r_{[\m\n]} 
\quad \Leftrightarrow \quad C_{\m,\n\r} = \f12 T_{\m,\n\r} - T_{[\n,\r]\m}. \label{CofT}
\end{align}
Both torsion and contorsion have spinorial decomposition
$
{\bf (\tfrac32,\tfrac12)\oplus(\tfrac12,\tfrac32)\oplus(\tfrac12,\tfrac12)\oplus(\tfrac12,\tfrac12)},
$
which corresponds to three irreducible components under Lorentz transformations (since the latter include parity). They can be defined as follows \cite{Hehl:1976kj},
\begin{align}
& C^{\m,\n\r} = \bar{C}^{\m,\n\r} + \f23 g^{\m[\rho} \check{C}^{\n]} + \eps^{\m\n\r\s} \hat C_{\s}, \\
& g_{\m\n} \bar{C}^{\m,\n\r}=0=\eps_{\m\n\r\s} \bar{C}^{\m,\n\r}, \qquad \check C^\m:=C_{\n,}{}^{\m\n}, \qquad \hat C_\s:=\f16\eps_{\s\m\n\r}C^{\m,\n\r}.
\end{align}

\section{Index Jugglers}
In this Appendix we prove Proposition 1, namely that the matter Noether identities \Ref{NM} on-shell of the matter field equations, plus the torsion field equation \Ref{FFET}, imply the conservation law for the effective energy-momentum tensor \Ref{dt0}, reported here for convenience
\be
\label{BI1}
d_{\om(e)}\left[\star\tau_I+\f1{16\pi\,}P_{IJKL} (e^J\w d_{\om(e)} C^{KL}+ e^J\w C^{KM} \w C_M{}^L)\right] = 0,
\ee
namely,
\be\label{BI2}
d_{\om(e)} \star\!\tau_I = \f1{8\pi\,}P_{IJKL} \Big(e^J\w C^K{}_M\w F^{ML}(e)+ e^J\w d_{\om(e)} C^{KM} \w C_M{}^L\Big).
\ee

To prove this identity, we start from \Ref{NM1}. On the left-hand side, we split the connection into Levi-Civita plus contorsion, see \Ref{defC}, obtaining
\be\label{LHS1}
d_\om\star\!\t_I = d_{\om(e)}\star\!\t_I-(C^{JK}\lrcorner e_I)e_K \w \star\t_J + T^J\lrcorner e_I \w \star\t_J
\ee
where we used
\be \label{tc}
T^I=C^{IJ}\w e_J\quad \rightarrow \quad C_I{}^{J}= -(C^{JK}\lrcorner e_I)e_K + T^J\lrcorner e_I.
\ee
In the second term of the right-hand side of \Ref{LHS1} we use the second Noether identity \Ref{NM2}, whereas  
the last term cancels the corresponding one on the right-hand side of \Ref{NM1}, which then reads
\begin{align}
d_{\om(e)} \star\!\t_I &=  \f12 F^{JK}(\om)\lrcorner e_I\w \star \s_{JK}-\f12 (C^{JK}\lrcorner e_I)d_\om \star\!\s_{JK} \nn\\
&=  \f1{16\pi\,} \left[F^{JK}(\om)\lrcorner e_I\w - (C^{JK}\lrcorner e_I)d_\om \right]P_{JKLM} e^L\w C^{M}{}_{N}\w e^N \nn \\
& = \f1{16\pi\,} \left[\big(F^{JK}(e)+ d_{\om(e)}C^{JK}\big)\lrcorner e_I\w - (C^{JK}\lrcorner e_I) d_{{\om}(e)} \right] P_{JKLM} e^L\w C^{M}{}_{N}\w e^N.
\label{BI3} 
\end{align}
In the second equality above we eliminated the torsion source using the corresponding field equation \Ref{FEC2}.
 In the third equality we expanded the curvature using the contorsion, see \Ref{FdC}, and observed that the piece quadratic in $C$ cancels the contorsion part of the exterior derivative in the last term.
 
Having performed these simplifications, our goal is to show the equivalence of the right-hand sides of \Ref{BI2} and \Ref{BI3}. This will  follow from the equivalence of the terms with the Riemann tensor $F^{IJ}(e)$, and the equivalence of the terms involving the Levi-Civita exterior derivatives. Both are consequences of trivial algebraic symmetries. Let us show them one by one. We notice in advance the following useful cycling identities:
\begin{align}
& P_{IJKL} e^K\w F^{LM}\w e_M = -P_{ABC[I}e^A\w F^{BC}\w e_{J]}, \label{cyc1} \\
& P_{IJKL}  F^{KM}\w C_{M}{}^L = -P_{ABC[I}F^{AB}\w C^C{}_{J]}, \label{cyc2}
\end{align}
which are easy to check.

To show the equivalence of the terms with the curvature, we start hooking a cotetrad vector field on a trivially vanishing 5-form,
\begin{align}\label{5form}
0 &= \Big(P_{JKLM }F^{JK}(e)\w e^L \w C^{M}{}_{N} \w e^N\Big)\lrcorner e_I \nn\\
& = P_{JKLM}F^{JK}(e)\lrcorner e_I\w e^L \w C^{M}{}_N \w e^N + P_{JKIM }F^{JK}(e) \w C^M{}_N \w e^N 
\nn\\&\quad - P_{JKLM } (C^{M}{}_N \lrcorner e_I) F^{JK}(e)\w e^L \w e^N + P_{JKLM }F^{JK}(e) \w e^L \w C^{M}{}_I.
\end{align}
Of these four terms, the third vanishes identically: its $1/\g$ part directly through the algebraic Bianchi identities for the Riemann tensor, the other part because of the antisymmetry in the $LP$ indices. The second and fourth terms recombine giving the left-hand side of \Ref{cyc2}, hence \Ref{5form} gives
 \be
2 P_{IJKL}\, F^{KM}(e)\w C_M{}^L\w e^J = P_{JKML} \,(F^{JK}(e)\lrcorner e_I)\w e^L \w C^M{}_N \w e^N,
\ee
which proves the equality of the curvature terms of \Ref{BI2} and \Ref{BI3}.

The equivalence of the $d_{\om(e)}C$ terms follows analogously. We hook the following 5-form,
\begin{align}\label{5form2}
0 &= \Big(P_{JKLM }C^{JK}\w e^L \w d_{\om(e)} C^{M}{}_{N} \w e^N\Big)\lrcorner e_I \nn\\
& = P_{JKLM} ( C^{JK}\lrcorner e_I) e^L \w d_{\om(e)} C^{M}{}_N \w e^N - P_{JKIM }C^{JK} \w d_{\om(e)} C^M{}_N \w e^N 
\nn\\&\quad + P_{JKLM } C^{JK}\w e^L\w d_{\om(e)} C^{M}{}_N \lrcorner e_I \w e^N + P_{JKLM }C^{JK} \w e^L \w d_{\om(e)}C^{M}{}_I.
\end{align}
Using an identity like \Ref{cyc2}, the second and fourth term give
\be
P_{JKM[I }C^{JK} \w d_{\om(e)}C^{M}{}_{N]} \w e^N = -2 P_{IJKL }\w e^J \w d_{\om(e)}C^{K}{}_M \w C^{ML}. 
\ee
For the third term we have 
\begin{align}
& P_{JKLM } C^{JK}\w e^L\w d_{\om(e)} C^{M}{}_N \lrcorner e_I \w e^N = -P_{JKLM} d_{\om(e)} C^{MN}\lrcorner e_I \w e^L \w C^{JK} \w e_N \\\nn
& \qquad = P_{JKLM} d_{\om(e)} C^{JK}\lrcorner e_I \w e^L \w C^{M}{}_N \w e^N,
\end{align}
which follows from a similar cycling identity as before.
Hence, \Ref{5form2} gives
\begin{align}
2 P_{IJKL }\w e^J \w d_{\om(e)}C^{K}{}_M \w C^{ML} &=  P_{JKLM} d_{\om(e)} C^{JK}\lrcorner e_I \w e^L \w C^{M}{}_N \w e^N \\\nn
&\quad + P_{JKLM} ( C^{JK}\lrcorner e_I) e^L \w d_{\om(e)} C^{M}{}_N \w e^N,
\end{align}
which proves precisely the equivalence between the $d_{\om(e)}C$ terms in \Ref{BI2} and \Ref{BI3}.